\documentclass[12pt]{article}
\usepackage[a4paper]{geometry}
\usepackage{amsfonts,amssymb,amsmath,enumerate,epsfig,lscape}

\renewcommand{\thesection}{\Roman{section}}
\renewcommand{\theequation}{\thesection.\arabic{equation}}

\newenvironment{mysubsection}[1]
{  
\addcontentsline{toc}{subsection}{\emph{\textbf{#1}} } { \emph{\textbf{#1}} }\nopagebreak
}

\newenvironment{mysubsubsection}[1]
{  
\addcontentsline{toc}{subsubsection} {\emph{#1}}  { \emph{#1} }\nopagebreak
}

\newenvironment{mytable}[1]
{  
\addcontentsline{toc}{subsubsection}{#1} { }\nopagebreak
}

\newenvironment{myfigure}[1]
{  
\addcontentsline{toc}{subsubsection}{\emph{#1}} { }\nopagebreak
}

\newenvironment{myappendix}[1]
{\renewcommand{\thesection}{\Alph{section}}
\addcontentsline{toc}{subsection}{\textbf{\emph{\appendixname}}\
  \emph{\textbf{\thesection :}}\   \textbf{\emph{{#1}}}}  {}
}

\numberwithin{equation}{section}
\begin{document}
\newcommand{\HRule}{\rule{\linewidth}{0.1mm}}
\newcommand{\as}{\alpha_{\textrm{s}}\,}
\newcommand{\ds}{\displaystyle}
\newcommand{\aB}{a_{\textrm{B}}\,}
\newcommand{\Bstar}{ \overset{*\!\!}{B} }
\newcommand{\Gstar}{ \overset{*\!\!}{G} }
\newcommand{\lP}{l_{\wp}}
\newcommand{\nP}{n_{\wp}}
\newcommand{\SP}{S_\wp}
\newcommand{\GG}{\textnormal I \! \Gamma}
\newcommand{\A}{\mathcal{A}}
\newcommand{\B}{\mathcal{B}}
\newcommand{\D}{\mathcal{D}}
\newcommand{\F}{\mathcal{F}}
\newcommand{\M}{\mathcal{M}}
\newcommand{\vr}{\vec{r}}
\newcommand{\sdot}{\,{\scriptscriptstyle{}^{\bullet}}\,}
\newcommand{\R}{\mathcal{R}}
\newcommand{\J}{\mathcal{J}}
\renewcommand{\S}{\mathcal{S}}
\newcommand{\Z}{\mathcal{Z}}
\newcommand{\ZP}{\mathcal{Z}_{\wp}}

\newcommand{\eklo}[2]{\, {}^{[\mathrm{#1}]}{\!#2}}

\newcommand{\rklo}[2]{\, {}^{(\mathrm{#1})}{ {#2} }}

%
\newcommand{\ekloi}[3]{\, {}^{[\mathrm{#1}]}{#2}_\mathrm{#3}}

%
\newcommand{\rkloi}[3]{\, {}^{(\mathrm{#1})}{#2}_\mathrm{#3}}

%
\newcommand{\ekru}[2]{#1_\mathrm{[#2]}}

%
\newcommand{\tekru}[2]{\tilde{#1}_\mathrm{[#2]}}

%
\newcommand{\orrk}[2]{ \mathbb{#1}^{(\textrm{#2})}  }

%
\newcommand{\vri}[2]{ \vec{#1}_{\textrm{#2}}   }

%
\newcommand{\nrft}[2]{ \tilde{\mathbb{#1}}_{[\mathrm{#2}]} }

%
\newcommand{\nrf}[2]{ {\mathbb{{#1}}}_{\mathrm{ {#2}}} }

%
\newcommand{\en}[1]{\varepsilon_{\textrm{#1}}}

%
\newcommand{\Ea}[3]{ #1^{(\mathrm{#2})}_{\mathrm{#3}}}

%
\newcommand{\Eb}[3]{ #1^{(\mathrm{#2})}_{\mathrm{[#3]}}}

%
\newcommand{\ru}[2]{#1_\mathrm{#2}}

%
\newcommand{\tru}[2]{\tilde{#1}_\mathrm{#2}}

%
\newcommand{\roek}[2]{ \mathbb{#1}^{[ \mathrm{#2} ] }  }

%
\newcommand{\rork}[2]{ #1^{( \mathrm{#2} ) }  }

%
\newcommand{\rorkt}[2]{ \tilde{#1}^{( \mathrm{#2} ) }  }

%
\newcommand{\gkloit}[3]{\, {}^{\{\mathrm{#1}\}}{ { \tilde{#2}  }  }_\mathrm{#3}}

\newcommand{\nsn}{\nu_*^{\{n\}}}

%
\newcommand{\rugk}[2]{#1_{\{\rm #2\}} }


\newcommand{\slo}[1]{\,'{\!#1}}

%
\newcommand{\gkloi}[3]{\, {}^{\{\mathrm{#1}\}}{\!{#2}}_\mathrm{#3}}

%
\newcommand{\gklo}[2]{\, {}^{\{\mathrm{#1}\}}{\!{#2}}}

%
\newcommand{\rrklo}[2]{\, {}^{(\mathrm{#1})}{\! {#2} }}
%
\newcommand{\srklo}[2]{\, {}^{({\sf #1})}{\! {#2} }}


\newcommand{\e}{\operatorname{e}}
\newcommand{\mustbe}{\stackrel{!}{=}}
\newcommand{\rt}{(r,\vartheta)}
\newcommand{\tablesize}{\scriptstyle}

\newcommand{\dn}{\delta_{0}} 
\newcommand{\dO}{\delta_{\Omega}}

\newcommand{\tanu}{\tilde{a}_\nu}
\newcommand{\tanuw}{\tilde{a}_{\nu w}}

\newcommand{\bjn}{{}^{(b)}\!j_0(\vec{r})} 

\newcommand{\bko}{{}^{(b)}k_0}
\newcommand{\bbko}{{}^{[b]}k_0}
\newcommand{\bkn}{{}^{(b)}\!k_0\left(\vec{r}\right)}
\newcommand{\bkv}{\vec{k}_b(\vec{r})} 
\newcommand{\bkphi}{{}^{(b)\!}k_\phi}
\newcommand{\bk}{\vec{k}_b} 
\newcommand{\bgkn}{{}^{\{b\}}\!k_0} 

\newcommand{\bpp}{{}^{(b)}\varphi_{+}(\vec{r})}
\newcommand{\bpm}{{}^{(b)}\varphi_{-}(\vec{r})}
\newcommand{\bppm}{{}^{(b)}\varphi_{\pm}(\vec{r})}
\newcommand{\bppk}{{}^{(b)}\varphi_{+}^\dagger(\vec{r})}
\newcommand{\bpmk}{{}^{(b)}\varphi_{-}^\dagger(\vec{r})}

\newcommand{\tNGe}{{\tilde{N}_\textrm{G}^\textrm{(e)}}}
\newcommand{\tNNO}{\tilde{\mathbb{N}}_{\Omega}}
\newcommand{\tNNGe}{{\tilde{\mathbb{N}}_\textrm{G}^\textrm{e}}} 
\newcommand{\tNNGee}{{\tilde{\mathbb{N}}_\textrm{G}^\textrm{[e]}}}
\newcommand{\tNNGer}{{\tilde{\mathbb{N}}_\textrm{G}^\textrm{(e)}}} 
\newcommand{\tNNGeg}{{\tilde{\mathbb{N}}_\textrm{G}^\textrm{\{e\}}}} 
\newcommand{\tNNegan}{{\tilde{\mathbb{N}}^\textrm{\{e\}}_{\,\text{an}}}} 

\newcommand{\Mee}{M^\textrm{(e)}}
\newcommand{\tMe}{\tilde{M}^\textrm{(e)}}
\newcommand{\tMMee}{\tilde{\mathbb{M}}^\textrm{[e]}}
\newcommand{\tMMeg}{\tilde{\mathbb{M}}^\textrm{\{e\}}}

\newcommand{\bMRpm}{{}^{(b)}\!\mathcal{R}_\pm}
\newcommand{\bMRp}{{}^{(b)}\!\mathcal{R}_+}
\newcommand{\bMRm}{{}^{(b)}\!\mathcal{R}_-}
\newcommand{\bMRpS}{{}^{(b)}\!{\overset{*}{\mathcal{R}}{} }_+}
\newcommand{\bMRmS}{{}^{(b)}\!{\overset{*}{\mathcal{R}}{} }_-}
\newcommand{\bMSpm}{{}^{(b)}\!\mathcal{S}_\pm}
\newcommand{\bMSp}{{}^{(b)}\!\mathcal{S}_+}
\newcommand{\bMSm}{{}^{(b)}\!\mathcal{S}_-}
\newcommand{\bMSpS}{{}^{(b)}\!{\overset{*}{\mathcal{S}}{} }_+}
\newcommand{\bMSmS}{{}^{(b)}\!{\overset{*}{\mathcal{S}}{} }_-}

\newcommand{\Kb}{K_{\{b\}}} 
\newcommand{\peAo}{{}^{[p]}\!A_0} 
\newcommand{\peAw}{{}^{[p]}\!A_w} 
\newcommand{\bAe}{{}^{[b]}\!A_0}
\newcommand{\bAn}{{}^{(b)}\!A_0}
\newcommand{\bgAe}{{}^{\{b\}}\!A_0} 
\newcommand{\bgAan}{{}^{\{b\}}\!A^{\text{an}}} 
\newcommand{\bgA}{{}^{\{b\}}\!A} 
\newcommand{\bgAiii}{\bgA^{\textsf{III}}} 
\newcommand{\bgAv}{\bgA^{\textsf{V}}} 
\newcommand{\bAmu}{{}^{(b)}\!A_\mu} 
\newcommand{\Acnuiii}{\mathcal{A}_\nu^\mathsf{III}} 
\newcommand{\Aciii}{\mathcal{A}^\mathsf{III}} 
\newcommand{\bgAcnuiii}{{}^{\{b\}}\mathcal{A}_\nu^\mathsf{III}} 
\newcommand{\bgAciii}{{}^{\{b\}}\mathcal{A}^\mathsf{III}} 
\newcommand{\MMeg}{\mathbb{M}^\textrm{\{e\}}} 
\newcommand{\tMMegan}{\tilde{\mathbb{M}}^\textrm{\{e\}}_{\ \text{an}}} 
\newcommand{\muegan}{\mu^\textrm{\{e\}}_{\ \text{an}}} 
\newcommand{\muegiii}{\mu^\textrm{\{e\}}_{\mathsf{III}}} 
\newcommand{\LLnu}{\mathbb{L}_\nu} 
\newcommand{\KKnu}{\mathbb{K}_\nu} 

\newcommand{\EE}{\mathbb{E}}
\newcommand{\eER}{E_\textrm{R}^\textrm{(e)}}
\newcommand{\ERe}{{E_\textrm{R}^\textrm{(e)}}}
\newcommand{\ERee}{{E_\textrm{R}^\textrm{[e]}}}
\newcommand{\eeER}{E_\textrm{R}^\textrm{[e]}} 
\newcommand{\egER}{E_\textrm{R}^\textrm{\{e\}}} 
\newcommand{\ET}{E_{\textrm{T}}}
\newcommand{\ETT}{\mathbb{E}_\textrm{T}} 
\newcommand{\tETT}{\tilde{E}_{\textrm{[T]}}}
\newcommand{\EET}{\mathbb{E}_\textrm{[T]}} 
\newcommand{\EETT}{\mathbb{E}_\textrm{[T]}} 
\newcommand{\EEivbn}{\mathbb{E}^\textrm{(IV)}(\beta,\nu)}
\newcommand{\EEgiv}{\mathbb{E}^{\{\textsf{IV}\}}} 
\newcommand{\tEEgiv}{\tilde{\mathbb{E}}^{\{\textsf{IV}\}}} 
\newcommand{\EEgivbn}{\mathbb{E}^{\{\textsf{IV}\}}(\beta,\nu)} 
\newcommand{\EEeiv}{\mathbb{E}^{[\textsf{IV}]}} 
\newcommand{\EEeivbn}{\mathbb{E}^{[\textsf{IV}]}(\beta,\nu)} 
\newcommand{\EEO}{\mathbb{E}_\Omega} 
\newcommand{\tEEO}{\tilde{\mathbb{E}}_{[\Omega]}}
\newcommand{\EEOe}{\mathbb{E}_{[\Omega]}} 
\newcommand{\GtEEO}{{}^{(G)}\!\tEEO} 
\newcommand{\GetEO}{{}^{[G]}\!\tEO} 
\newcommand{\DtEEO}{{}^{(D)}\!\tEEO} 
\newcommand{\DEEO}{{}^{(D)}\!\EEO} 
\newcommand{\DEEOe}{{}^{(D)}\!\EEOe} 
\newcommand{\etEEO}{{}^{(e)}\!\tEEO} 
\newcommand{\eetEEO}{{}^{[e]}\!\tEEO} 
\newcommand{\antEEO}{{}^{\{\text{an}\}}\!\tEEO} 
\newcommand{\anrtEEO}{{}^{(\text{an})}\!\tEEO} 
\newcommand{\eetEO}{{}^{[e]}\!\tEO} 
\newcommand{\Egiv}{E_\textrm{\{T\}}^{\textrm{(IV)}}} 

\newcommand{\Ekin}{{E_\textrm{kin}}}
\newcommand{\EKIN}{{E_\textrm{KIN}}} 
\newcommand{\ekin}{ {\varepsilon}_{\textrm{kin}} } 
\newcommand{\eKIN}{ {\varepsilon}_{\textrm{KIN}} } 
\newcommand{\rEkin}{{}^{(r)}\!\Ekin} 
\newcommand{\rEKIN}{{}^{(r)}\!\EKIN} 
\newcommand{\rekin}{{}^{(r)}\!\ekin} 
\newcommand{\reKIN}{{}^{(r)}\!\eKIN} 
\newcommand{\thEKIN}{{}^{(\vartheta)}\!\EKIN} 
\newcommand{\thekin}{{}^{(\vartheta)}\!\ekin} 
\newcommand{\theKIN}{{}^{(\vartheta)}\!\eKIN} 

\newcommand{\epot}{ {\varepsilon}_{\textrm{pot}} }
\newcommand{\ePOT}{ {\varepsilon}_{\textrm{POT}} }

\newcommand{\eegtot}{ {\varepsilon}^{\{e\}}_{\textrm{tot}} }

\newcommand{\tEEOo}{\tilde{\mathbb{E}}_\Omega} 
\newcommand{\EReg}{{E_\textrm{R}^\textrm{\{e\}}}} 
\newcommand{\Eegan}{{E^\textrm{\{e\}}_\text{an}}} 
\newcommand{\Ew}{{E_w}}
\newcommand{\Ewee}{E_w^{{}\textrm{[e]}}}
\newcommand{\cEk}{{\cal{E}_\textrm{kin}}}

\newcommand{\tPhi}{\tilde{\Phi}}
\newcommand{\tPhipm}{\tilde{\Phi}_\pm}
\newcommand{\tPhib}{\tPhi_b} 
\newcommand{\tPhinuw}{\tPhi_{\nu w}} 

\newcommand{\tO}{\tilde{\Omega}}
\newcommand{\tOpm}{\tilde{\Omega}_\pm}
\newcommand{\tOp}{\tilde{\Omega}_+}
\newcommand{\tOm}{\tilde{\Omega}_-}

\newcommand{\lO}{\ell_\mathcal{O}}
\newcommand{\dlO}{\dot{\ell}_\mathcal{O}}
\newcommand{\ddlO}{\ddot{\ell}_\mathcal{O}}

\newcommand{\lGe}{ {\lambda_\textrm{G}^{(\textrm{e})}}\!}

\title{\textbf{ Spherically Symmetric Approximation}  \Large \mbox{\textrm{\emph{(and beyond)}}}\\
  \LARGE \textbf{in}\\ \textbf{Relativistic Schr\"odinger Theory}}
\author{M.\ Mattes and M.\ Sorg} 
\date{ }
\maketitle
\begin{abstract}

  The energy eigenvalue problem of non-relativistic positronium is considered within the
  framework of Relativistic Schr\"odinger Theory (RST), and the results are compared to
  those of the conventional quantum theory. For the range of principal quantum
  numbers~$n=2,3,\ldots, 30$, the RST predictions for the non-relativistic positronium
  energies deviate now from the corresponding predictions of the conventional quantum
  theory at an average of (roughly) 3\%. These results suggest that the deviations will be
  further diminished in the higher orders of approximation.

  The emphasis aims at the role played by the assumption of \emph{spherical symmetry} of
  the gauge potential. A new approximation procedure is established in order to regard
  also the \emph{anisotropic} character of the interaction potential. This reduces the
  deviations of the RST predictions from the corresponding conventional results by 50
  percent, whereas the spherically symmetric corrections of higher order do amount only to
  (roughly) 0,2\%. Therefore further improvements of the RST predictions may be expected
  by considering more rigorously the \emph{anisotropy} of both the interaction potential
  and of the wave functions. Such an analysis (however complicated it may be) seems
  inevitable in order to decide whether perhaps the exact RST predictions do practically
  coincide with those of the conventional quantum theory.
 
  \vspace{2.5cm}
 \noindent


\end{abstract}


\tableofcontents
\section{Introduction and Survey of Results}
\indent

The present paper studies the numerical influence of the \emph{anisotropy} of the
electron-positron interaction potential on the energy spectrum of non-relativistic
positronium. In some preceding papers it has been found that (even up to very highly
excited states) the corresponding RST predictions come close to their conventional
counterparts up to some 10 percent, already by very rough approximation methods
being based on the complete neglection of the anisotropy of the interaction force. And
therefore one wishes to know now whether perhaps this unexpected agreement with the
conventional predictions becomes still better through including the anisotropy
effect. Naturally, the physical relevance of \emph{Relativistic Schr\"odinger Theory} is
to be tested via the experimental verification of its numerical predictions in the various
fields of possible applications (e.g. the presently considered positronium
system). Concerning this specific case of demonstration, one is tempted to suppose that
the \emph{exact} RST predictions could perhaps agree (nearly or exactly) with the
corresponding predictions of the conventional quantum theory. If such a result (i.e.\ the
numerical equivalence of RST and conventional predictions) could be shown to hold for the
whole field of atomic and molecular physics, this would then imply a partial dethronement
of the conventional quantum mechanics. But since the latter is rather of probabilistic
nature whereas RST is essentially a fluid-dynamic theory, it seems adequate and desirable
to first clarify the philosophical difference of both quantum approaches, where Bohr's
\emph{complementarity principle} plays a dominant role.  \vspace{4ex}
\begin{center}
\mysubsection{1.\ Particle-Wave Duality and Complementarity}
\end{center}

In the literature, one frequently encounters the viewpoint that the conventional quantum
theory represents a logically perfect system which cannot be further improved by "small"
modifications, neither concerning the formalism nor its predictive power. Such arrogant
judgment arose early during the development of quantum theory when Born and Heisenberg
maintained 80 years ago that "\emph{quantum mechanics is a complete theory; its basic
  physical and mathematical hypotheses are not further susceptible of
  modifications}"[1,2].

It seems that this viewpoint represents nowadays the general conviction of the community;
however it seems to us that some criticism is well-suited here. Firstly, even if presently
nobody is able to imagine how further improvements of the conventional theory could look
like, one cannot be sure that future workers in this field will not invent more efficient
logical systems which on the one hand do include their Born-Heisenberg precursor, but on
the other hand do also considerably exceed it. And secondly, the present-day form of the
conventional quantum mechanics as the offspring of the Born-Heisenberg construction may
admittedly represent a closed logical system but it nevertheless could perhaps miss
certain features of quantum matter which on principle cannot be grasped by \emph{any}
logical system of the \emph{probabilistic} type! If this latter supposition should be
true, these additional features (inaccessible to the purely probabilistic world view)
would then be well-hidden behind the statistical predictions of the conventional theory.

Concerning now a possible signal for the presence of such an additional structure beyond
the purely probabilistic approach, one could perhaps think of the wave-like behavior of
quantum matter; and this would suggest that for the description of certain (but surely not
all) quantum phenomena a fluid-dynamic approach would be better suited than the
conventional probabilistic viewpoint. This \emph{particle-wave duality} of microscopic
matter has ever been thought to be the origin of those strange quantum effects which
formerly did not only bother the fathers of quantum theory[3] but even nowadays do appear
as a "mystery" for most of the workers in this field[4,5]. Indeed, despite the widely
celebrated concept of \emph{quantum logic}, one nevertheless likes to discuss many quantum
interference effects in the classical terms of \emph{optical imagery}~\cite{ms}. Naturally, such
a dichotomic (or sometimes even contradictious) view on the quantum world must have
provoked attempts which try to erect some unifying framework embracing
\emph{simultaneously} both the fluid-dynamic \emph{and} the probabilistic aspects of quantum
mechanics. However, it seems that nobody was (or is) able to construct such a unified quantum
formalism; and therefore most physicists seem to accept now Bohr's original idea of
complementarity[2,6].  Or, summarizing this in the words of Omnes[7]: the wave logic and
the particle logic could not be united to a larger consistent logic which contains both of
them as sublogics and consequently, according to Omnes[7], we have to learn to live with two
mutually excluding logics and nevertheless can have a true theory!
 
It seems that Bohr[6] was the first who grasped this logical dilemma to its full extent by
establishing his "\emph{principle of complementarity}", whereas others weakened or
misinterpreted (or even misunderstood) this original idea of Bohr (for a historical
account see ref.~\cite{ku}). For instance, Pauli
"\emph{in his article for the} Handbuch der Physik~\cite{wp} \emph{called two classical concepts
  - and not two modes of desription - complementary, if the applicability of the one
  (e.g., position coordinate) stands in relation of exclusion to that of the other (e.g.,
  momentum) in the sense that any experimental setup for measuring the one interferes
  destructively with any experimental setup for measuring the other. Pauli, as we see, in
  contrast to Bohr, ascribed complementarity to two notions which belong to the same
  classical mode of description (e.g., the particle picture) and not to two mutually
  exclusive descriptions}"(ref.[2], p. 369). 

An other example of distortion of Bohr's complementarity proposal is given by C.F. von
Weizs\"acker who said "\emph{The complementarity between space-time description and the
  claim of causality is therefore precisely the complementarity between the description of
  nature in classical notions and in terms of the $\psi$ function}" (ref.[2],
p.369). Thus, von Weizs\"acker in contrast to Pauli, interpretes correctly Bohr's
complementarity principle, namely as referring to two mutually excluding "modes of
description" (or "logics", as Omnes terms it [7]); but in contrast to Bohr, von
Weizs\"acker obviously thinks that these two (mutually excluding) modes should refer to
the classical and the quantum-mechanical description of nature.

However, concerning that notorious concept of \emph{complementarity}, we do neither join
here to the misunderstanding of Pauli nor to that of v. Weizs\"acker; but rather we cling
to Bohr's original idea which says that the notion of complementarity must refer to two
mutually excluding logical systems (or "pictures") which both are to be used for the
description of \emph{quantum~(!)}  systems, i.e.\ the \emph{probabilistic} point-particle
picture in opposition to the \emph{fluid-dynamic} wave picture of the elementary
matter. If Bohr's complementarity principle is understood in this way (i.e. more
precisely: is understood in the sense of Omnes[7]), the logically coherent and causal
space-time descriptions of classical particles and waves (e.g. "Classical
Electrodynamics"~\cite{ja}) must be broken up for the transition to the quantum-mechanical
description into two mutually excluding modes of description (or "logics") for the massive
particles, namely into the probabilistic point-particle picture with its lack of a
physical mechanism for the local but ``indeterministic'' collapse of the probability
distribution and into the fluid-dynamic wave picture with its unability to predict the
instantaneous statistical correlations between widely separated regions of three-space
(see the EPR phenomena [4,5]). This particle-wave duality is in the first line thought to
refer to the quantum mechanics of the classical massive particles, whereas the classical
massless fields are left untouched in the first step of quantization. In this context, one
could perhaps mention in favour of RST that this theory provides a \emph{dynamical}
mechanism for the (otherwise only \emph{kinematically} describable) exchange phenomena!
These are brought forth by the exchange potential~$\B_\mu$ as that part of the bundle
connection~$\A_\mu$ which is responsible for the non-Abelian character of the theory.

Thus, each of the two mutually exclusive \emph{quantum} fragments (i.e.probabilistic
particle vs.\ fluid-dynamic wave picture), being left behind from the classical causal
space-time description, suffers from a certain deficiency if viewn from the side of the
original classical approach. And consequently one will hesitate to attribute to one or the
other of the two competing quantum approaches the status of \emph{completeness} (being
mostly claimed in favour of the probabilistic particle picture alone, i.e. the
"conventional quantum theory"~\cite{gv}).  
\pagebreak \vspace{4ex}
\begin{center}
\mysubsection{2.\ Relativistic Schr\"odinger Theory}
\end{center}

Accepting now the necessity of two mutually excluding logical systems (albeit only as the
preliminary state of the art), one might think that these two pictures have been
elaborated in a comparably symmetric manner during the past development of quantum
theory. However, it seems to us that just the contrary did occur: the generally accepted
and almost exclusively applied form of quantum theory is based upon the probabilistic
particle picture (i.e. the "conventional quantum theory"), whereas a fluid-dynamic
description of quantum matter has not been tried at all, apart from a few isolated (but in
the meantime forgotten) attempts. Such a situation may now be taken as sufficient
motivation in order to establish a fresh fluid-dynamic approach to quantum matter which is
able to counterbalance the conventional probabilistic particle logic, from both the
mathematical and physical point of view.  As such a competitor of the fluid-dynamic type,
there has recently been proposed the \emph{Relativistic Schr\"odinger Theory} (RST), see
the precedent paper~\cite{ms1} and the other papers cited therein. Indeed, this
fluid-dynamic theory differs from its probabilistic competitor (i.e. the conventional
probabilistic theory) in both the mathematical and physical respect: \textbf{(i)} the
many-particle systems are described in RST by the Whitney sum of one-particle fibre
bundles, not by the tensor product of one-particle Hilbert spaces (as in the conventional
theory); and \textbf{(ii)} the RST wave functions $\psi$ (as the sections of complex
vector or spinor bundles) are used in order to construct \emph{physical densities} of
charge, current, energy-momentum etc., not in order to construct probabilities as in the
conventional theory!

However, the essential point with such two mutually excluding competitors is now that they
both are expected to describe the same quantum world; and therefore there must exist
certain intersecting domains of application for which both proposals make definite
predictions.  Clearly, if both theoretical frameworks should pretend to the same physical
relevance, they are required to produce identical numerical predictions in that
intersecting domain of applications.  \vspace{4ex}
\begin{center}
\mysubsection{3.\ Test Case: Non-Relativistic Positronium}
\end{center}

One such common field of competition certainly refers to the energy (E) of bound
systems. This physical quantity emerges in the probabilistic particle picture as the
eigenvalue of the Hamiltonian~$\hat{H}$ of the considered system 
\begin{equation}
  \label{eq:i.1}
  \hat{H}\psi = E\psi\ .
\end{equation}
On the other hand, the fluid-dynamic character of RST lets emerge the total energy
$\ru{E}{T}$ of any RST field configuration as the spatial integral of
the total energy density $\rklo{T}{T}_{00}(r)$ 
\begin{equation}
  \label{eq:i.2}
  \ru{E}{T} = \int d^3\vec{r}\,\rklo{T}{T}_{00}(\vec{r}) 
\end{equation}
where the energy density $\rklo{T}{T}_{00}(r)$ is the time-component of the total energy-momentum density
$\rklo{T}{T}_{\mu\nu}$. Now it has already been demonstrated explicitly for the (relativistic and
non-relativistic) hydrogen atom that the particle-like energy~$E$ (\ref{eq:i.1}) is numerically
identical to the fluid-dynamic RST energy~$\ru{E}{T}$ (I.2); and this identity holds for all bound
\emph{one-particle} systems~\cite{ms2}.

Therefore the desired comparison of the RST and conventional predictions must refer to the
\emph{many-particle} systems which receive a quite different treatment by each of both
theoretical frameworks. Perhaps the simplest one of this non-trivial class of test cases
is the non-relativistic positronium system which consists of two oppositely charged Dirac
particles of the same rest mass M. The corresponding (non-relativistic) Hamiltonian~$\hat{H}$
 (\ref{eq:i.1}) of this two-body arrangement reads 
\begin{equation}
  \label{eq:i.3}
  \hat{H} = \frac{\ru{p}{e}^2}{2m} + \frac{\ru{p}{p}^2}{2M} -
  \frac{e^2}{|\vec{r}_e-\vec{r}_{\mathrm{p}}|}\ ,
\end{equation}
and the conventional Schr\"odinger equation (\ref{eq:i.1}) of this two-body problem can be
separated into the center-of-mass motion and the relative motion of the electron and
positron (see any standard textbook about quantum mechanics , e.g.~ref.~\cite{me}). The
conventional energy spectrum of the relative motion is essentially the same as for the
ordinary hydrogen problem (with infinitely heavy proton) where merely the mass M of the
electron must be replaced by the reduced mass M/2; i.e. the conventional non-relativistic
energy spectrum of positronium is given by
\begin{gather}
  \label{eq:i.4}
  \Ea{ E }{{n_c}}{conv} = -\frac{e^2}{4\aB}\cdot\frac{1}{n_c^2} \simeq
  - \frac{6,8029\ldots}{n_c^2}\,[eV]\\*
  \left(\aB=\frac{\hbar^2}{Me^2}\ldots \text{Bohr radius} \right)\notag
\end{gather}
where~$n_c=1,2,3,4,5\ldots$ is the principal quantum number of the
conventional theory. On the other hand, a first tentative RST calculation on the basis of
a two-parameter trial amplitude has reproduced this conventional spectrum (I.4) up to some
few percents of deviation~\cite{ms2,ms3}. Thus, a first crucial test for the physical relevance of the
RST predictions may be referred to the accuracy of the reproduction of the conventional spectrum (I.4)
when better approximation techniques are applied in RST; and this is just the main goal
of the present paper.

However, as will readily become evident, the RST treatment of positronium is much more
complicated than is its counterpart in the conventional theory, even in the
non-relativistic domain. But this complication is not an intrinsic feature of RST but is
rather a consequence of the (purely technical) fact that exact solutions of the RST
eigenvalue problem are very hard (or even impossible) to obtain. The RST eigenvalue
problem itself is a simple matter from the purely conceptual viewpoint since it consists
merely in a simple (albeit coupled) system of two equations, i.e. the Schr\"odinger-like
equation (III.34) and the Poisson equation (III.36).But despite its structural simplicity,
this system needs considerable work in order to construct approximate solutions (see
below). In contrast to this, the conventional eigenvalue problem (I.1) plus (I.3) admits
exact solutions which can easily be worked out because the internal (i.e. relative) motion
agrees with the standard hydrogen problem whose solutions can be read in any textbook of
elementary quantum mechanics. But in view of such a fortunate situation with the
conventional theory, on should not forget the fact that the conventional eigenvalue
problem (I.1) plus (I.3) cannot be deduced from a generally valid \emph{relativistic
  quantum mechanics} for many-particle systems (not to be confused with quantum field
theory!). It is true, the well-known Bethe-Salpeter equations are intended to represent
such a relativistic quantum mechanics for many-particle systems, but they seem to be
afflicted by many deficiencies and are therefore not generally
accepted~\cite{bs,la}. Presumably, these deficiencies in the relativistic domain must
necessarily emerge as a consequence of the probabilistic nature of those equations.

On the other hand, RST provides a consistent logical framework for relativistic N-particle
systems where the particles interact via the principle of minimal coupling which then
identifies RST as a relativistic gauge field theory (being Abelian for different particles
and Non-Abelian for identical particles). Within such a framework, one can easily deduce a
well-defined non-relativistic limit of the relativistic wave equations but the
corresponding non-relativistic interaction potential depends now on the quantum state of
the two-body system (in contrast to the state-independent Coulomb potential
(\ref{eq:i.3})); and additionally the interaction potential is now also anisotropic (in
contrast to the spherical symmetry of the conventional Coulomb potential (I.3)). This
entails that we have to solve \emph{simultaneously both} the wave equations for the
material particles \emph{and} the gauge field equations for the interaction potentials! As
a consequence, such simple exact solutions as for the conventional Coulomb problem (I.1)-(1.3)
are not available for the RST case; and therefore one has to put much effort in the
development of appropiate approximation techniques. This will require a large part of the
paper.  \vspace{4ex}
\begin{center}
\mysubsection{4.\ Spherically Symmetric Approximation}
\end{center}

But here we can resort to the fact that the coupled matter and gauge field equations may
be obtained by extremalizing the RST energy functional~$\tekru{E}{T}$, both in the
relativistic and non-relativistic situations. This provides us with the possibility to
invent trial configurations with a certain number of variational parameters whose values
become then fixed through extremalization of the corresponding value of the energy
functional~$\tekru{E}{T}$ on just those trial configurations. Moreover, this method allows
us to try spherically symmetric configurations in order to avoid the necessity to deal
with anisotropic interaction potentials. Of course, one might suppose here that such a
spherically symmetric approximation could eventually produce unacceptable large deviations
from the true RST result. But amazingly enough, the contrary seems to be true: by use of
hydrogen-like trial amplitudes of SO(3) symmetry one gets already in lowest approximation
order RST predictions for the non-relativistic positronium spectrum, which come close to
the conventional prediction (\ref{eq:i.4}) up to 10\% or less~\cite{ms1,ms3}, even for the
highly excited states.

From these tentative results we conclude that the RST predictions could eventually come
still closer to their conventional counterparts if one would admit anisotropic  trial
configurations and would also apply trial amplitudes with a larger number of variational
parameters. Thus, there naturally arises the question which of both effects is the dominant
one: Is it more effective to improve the spherically symmetric approximation by
resorting to trial amplitudes with more than two variational parameters; or is it more
promising to take into account the anisotropy of the interaction potential? Or are both
effects of the same order of magnitude?

The answer will turn out in favour of the anisotropic corrections which do diminish the
RST deviations up to 50\%; whereas the higher-order corrections of the spherically
symmetric type amount to merely 0,2\%.

These results are worked out now in the following arrangement:

\vspace{4ex}
\begin{center}
\mysubsubsection{A.\ RST Fundamentals}
\end{center}

In \textbf{Sect.II}, the RST fundamentals are briefly presented to that extent which is
necessary for the subsequent discussion of the positronium level system. Here the point of
departure is the coupled system of matter equations (\ref{eq:ii.1}) and gauge field
equations (\ref{eq:ii.20}). By splitting off the usual exponential time-factor from the
Dirac four-spinors~$\psi_a(\vec{r},t)$, cf.~(\ref{eq:ii.10}), one gets the \emph{mass
  eigenvalue equations} which are written here in Pauli form, i.e.\ in terms of
2-spinors~$\varphi_\pm(\vec{r})$, see equation (\ref{eq:ii.17}) below. The corresponding
time-independent form of the gauge field equations turns out as the ordinary Poisson
equation for the electrostatic interaction potential~$\rklo{p}{A}_0(\vec{r})$, see
equation (\ref{eq:ii.27}) below. Thus the combination of the mass eigenvalue equation
(\ref{eq:ii.17}) and the Poisson equation (\ref{eq:ii.27}) just represents the RST
eigenvalue problem for positronium.

At this stage, the eigenvalue system is still fully relativistic with the \emph{mass
  eigenvalue}~$M_*$ to be determined in one step together with the Pauli
eigenspinors~$\rklo{p}{\varphi}_\pm(\vec{r})$ and the gauge
potential~$\rklo{p}{A}_0(\vec{r})$. However, these positronium eigenvalue equations do
not represent the most general situation, for \textbf{(i)} the magnetic effects
are neglected ($\leadsto$~\emph{electrostatic approximation}) and \textbf{(ii)} the
\emph{exchange interactions} are also omitted since this kind of force can be active only
for \emph{identical} particles. Of course, this absence of the magnetic and exchange
interactions does considerably simplify the eigenvalue problem which, however, is still
left too complicated in order that exact solutions could be found.

\vspace{4ex}
\begin{center}
\mysubsubsection{B.\ Spherical Symmetry}
\end{center}

But fortunately, there does exist an energy functional~$\tekru{E}{T}$ (\ref{eq:ii.32})
whose extremal equations do just coincide with those coupled mass eigenvalue and gauge
field equations which define the RST eigenvalue problem. The corresponding variational
principle (\ref{eq:ii.40}) is the \emph{RST principle of minimal energy} which can now in
\textbf{Sect.III} be exploited for obtaining approximate solutions for the considered positronium
eigenvalue problem ($\leadsto$ \emph{spherically symmetric approximation}). This
approximation method works as follows:

In order to explicitly solve the mass eigenvalue equations one has in any case to parameterize the
Pauli spinors by means of both the spinor basis and the associated spinor components. The
RST hypothesis for this parameterization procedure is now that the fermionic or bosonic
character of the bound system as a whole is present already in any of its
constituents. Since positronium (as a whole) is a boson, both constituents (electron
and positron) do therefore also adopt bosonic character, i.e.\ the constituent Dirac four-spinors
must be decomposed with respect to a \emph{bosonic}  spinor basis which then turns out to
be double-valued, see equation (\ref{eq:iii.5}) below.

But once the desired decomposition of the Pauli spinors~$\rklo{p}{\varphi}_\pm(\vec{r})$
into its components~$\rklo{p}{\R}_\pm,\rklo{p}{\S}_\pm$ (i.e.\ the ``wave amplitudes'')
has been attained, see equations (\ref{eq:iii.9a})-(\ref{eq:iii.9b}) below, then one can
write down the eigenvalue equations in terms of these wave amplitudes,
(\ref{eq:iii.15a})-(\ref{eq:iii.16}). This set of equations is now the point of departure
for separating off the angular part of the wave amplitudes; however, this separation
process can be performed \emph{exactly} only if the electrostatic gauge
potential~$\rklo{p}{A}_0(\vec{r})$ is \emph{spherically symmetric}. But on the other hand,
this symmetry for the gauge potential can not be exactly realized, if the angular
separation process for the wave amplitudes is required to be exactly feasible, as
described by equations (\ref{eq:iii.17a})-(\ref{eq:iii.17b}) below. The reason for this is
that the gauge potential~$\rklo{p}{A}_0(\vec{r})$ must be anisotropic in any case even if
the wave amplitudes are adopted to exactly obey the spherical symmetry, see the Poisson
equation (\ref{eq:iii.16}). Therefore we have to put in by hand the \emph{spherically
  symmetric approximation} which consists just in the postulate that the interaction
potential~$\rklo{p}{A}_0(\vec{r})$ (but not the wave amplitudes themselves) be spherically
symmetric, see equation (\ref{eq:iii.1}) below. But once this approximation assumption is
accepted, it becomes a rather straightforward procedure to get the desired approximate
energy spectrum (for the present purpose we are satisfied with the non-relativistic
situation). Namely, after the selection of a nearby trial
amplitude~$\tilde{\Phi}(r)$ (see equation (\ref{eq:iii.50}) below) and computation
of the associated gauge potential~$\rklo{p}{A}_0(\vec{r})$ (\textbf{App.A}) one substitutes
this trial configuration~$\{\tilde{\Phi}(r),\rklo{p}{A}_0(\vr) \}$ into the
non-relativistic energy functional~$\nrft{E}{\Phi}$ (\ref{eq:iii.41}) and then one
minimalizes the resulting energy function~$\roek{E}{IV}(\beta,\nu)$ (\ref{eq:iii.54'}) with
respect to the two variational parameters~$\beta$ and~$\nu$ which are contained in the
chosen trial amplitude~$\tilde{\Phi}(r)$.

This minimalization process yields then the energy spectrum displayed in \textbf{table~1} on
p.~\pageref{table1}, from where it is obvious that these RST predictions for the
positronium energy levels deviate from the conventional predictions~$\ru{E}{conv}$
(\ref{eq:i.4}) mostly by less than 10\%. Indeed, this is a somewhat amazing result in view
of such a simple trial amplitude~$\tilde{\Phi}(r)$ (\ref{eq:iii.50}) with only two
variational parameters. Moreover, the deviation from the conventional
predictions~$\ru{E}{conv}$ decreases with increasing principal quantum number~$\ru{n}{c}$,
so that for high quantum numbers~$\ru{n}{c}\lesssim 100$ one arrives at deviations by less
than~4\%, see \textbf{table~1} on page~\pageref{table1}.

Clearly, such a result must evoke now some urgent questions:
\begin{itemize}
\item[\textbf{i)}] Is that RST deviation of some few percent an \emph{intrinsic} feature
  of RST (and therefore not improvable); or is it merely due to the applied approximation
  method? In the latter case, the \emph{exact} RST predictions could possibly coincide with their
  conventional counterparts (\ref{eq:i.4}).
\item[\textbf{ii)}] If the deviation is due to the approximation method, is this deviation
  then an inevitable consequence of the use of the spherically symmetric approximation or
  is it merely due to the selection of a very rough trial amplitude~$\tilde{\Phi}(r)$
  (\ref{eq:iii.50}) with too few variational parameters? In the latter case it would be
  worth while to try O(3) symmetric functions with more than two variational parameters.
\item[\textbf{iii)}] If the deviation is due to the use of spherically symmetric
    trial configurations, can then the RST predictions be improved by resorting to
    \emph{anisotropic} gauge potentials?
\end{itemize}
These questions will receive the following answers by the subsequent elaborations:
\begin{itemize}
\item[\textbf{(i)}] the former 10\% deviations~\cite{ms1} from the conventional results
  (\ref{eq:i.4}) can be pressed down to roughly~3\% by resorting to the higher orders of
  the chosen approximation method. This supports the expectation that even higher orders
  of approximation will shift the RST predictions further towards their conventional
  counterparts (\ref{eq:i.4}).
\item[\textbf{(ii)}] the RST predictions do receive an improvement of merely~0,2\% if the
  trial configurations are restricted to obey the spherical symmetry
  \item[\textbf{(iii)}] The reduction of the deviations from (roughly) 10\% to 3\% is due
    to the use of \emph{anisotropic} interaction potentials which therefore provide the
    possibility for even further improvements.
\end{itemize}
\section{RST Eigenvalue Problem}
\indent

From a more philosophical viewpoint, the subsequent relativistic mass eigenvalue problem is
to be conceived as the RST counterpart of the non-relativistic Schr\"odinger energy
eigenvalue problem, i.e.\ the eigenvalue problem of the Hamiltonian~$\hat{H}$. The latter
problem appears as the time-independent specialization of the general time-dependent
Schr\"odinger equation when a certain factorization ansatz for the wave function is
adopted so that the usual exponential time factor can be splitted off. The solutions of
the remaining time-independent equation describe then the groundstate and the excited
states of the considered bound system. The same logical arrangement does apply also to the
corresponding RST eigenvalue problem where, however, the gauge field equations for the
interaction potentials of the material particles must complete the matter wave equations! It
should be obvious that such a highly interactive system will necessitate to solve
simultaneously for \emph{both} the matter fields \emph{and} the interaction
potentials. Since exact solutions of such an intricate system are very difficult to obtain
we resort to an approximation method. This consists in adopting some physically plausible trial
function which contains a certain set of variational parameters serving for extremalizing
the RST energy functional. Indeed, the extremal equations of that energy functional are
identical to the system of coupled matter and gauge field equations forming the RST
eigenvalue problem.

Obviously this program requires to discuss the following items in due order:
\textbf{(i)} wave equations for the matter subsystem, \textbf{(ii)} equations for the
gauge field subsystem, \textbf{(iii)} stationary bound systems, \textbf{(iv)} energy
functional and principle of minimal energy, and \textbf{(v)} spherically symmetric vs.\
non-symmetric trial functions.  
\newpage
\vspace{4ex}
\begin{center}
\mysubsection{1.\ Matter Subsystem}
\end{center}

Quite generally, the dynamics of the spinning quantum matter is described in RST by the
N-particle Dirac equation
\begin{equation}
  \label{eq:ii.1}
  i\hbar c\GG^\mu\D_\mu\Psi = \M c^2\Psi\ .
\end{equation}

Here, for the considered two-particle system (i.e.\ positronium), the two-particle
\emph{velocity operator}~$\GG_\mu$ is the direct sum of the ordinary Dirac
matrices~$\gamma_\mu$
\begin{equation}
  \label{eq:ii.2}
  \GG_\mu = \left(-\gamma_\mu \right)\oplus\gamma_\mu
\end{equation}
where the minus sign refers (by convention) to the ``first'' particle (i.e.\ the
positron). Furthermore, the two-particle wave function~$\Psi$ is adopted in RST to be the
\emph{Whitney sum} of the two one-particle wave functions~$\psi_a\,(a=1,2)$
\begin{equation}
  \label{eq:ii.3}
  \Psi(\vec{r},t) = \psi_1(\vec{r},t)\oplus\psi_2(\vec{r},t)
\end{equation}
which says that each of the two particles does occupy a well-defined one-particle quantum
state.

The gauge-covariant derivative~$(\D)$ emerging in the basic equation (\ref{eq:ii.1}) is
defined as usual in the gauge theories, i.e.
\begin{equation}
  \label{eq:ii.4}
  \D_\mu\Psi = \partial_\mu\Psi + \A_\mu\Psi\ ,
\end{equation}
where the gauge potential~$\A_\mu$ takes here its value in the Lie algebra of the
\emph{structure group} U(2) and may therefore be decomposed as follows
\begin{equation}
  \label{eq:ii.5}
  \A_\mu = \sum_{a=1}^2 {A^a}_\mu\tau_a + B_\mu\chi - \Bstar_\mu\bar{\chi}\ .
\end{equation}
Strictly speaking, this general form of the gauge potential does apply for a system of
\emph{identical} particles which are subjected to the \emph{exchange forces} (being
described by the \emph{exchange potential}~$B_\mu$). But since the presently considered
positronium system consists of two \emph{different} particles (i.e.\ positron and
electron), the exchange potentials~$B_\mu,\Bstar_\mu$ as the components of~$\A_\mu$ with
respect to the \emph{exchange generators} $\chi,\bar{\chi}$ must vanish $(B_\mu\equiv
0)$. Thus we are left alone with the purely electromagnetic interactions which are
described by the electromagnetic potentials ${A^a}_\mu\,(a=1,2)$ as the components of
$\A_\mu$ with respect to the \emph{electromagnetic generators} $\tau_a$:
\begin{equation}
  \label{eq:ii.6}
  \A_\mu \Rightarrow \sum_{a=1}^2 {A^a}_\mu \tau_a = \rklo{1}{A}_\mu\cdot
  \tau_1 + \rklo{2}{A}_\mu\cdot\tau_2\ .
\end{equation}
And finally, the \emph{mass operator} $\M$ in equation (\ref{eq:ii.1}) can be taken to
be proportional to unity~\textbf{(1)} for the present situation where both particles have
the same rest mass ($M$, say)
\begin{equation}
  \label{eq:ii.7}
  \M = M\cdot \mathbf{1}\ .
\end{equation}

In order to further proceed towards the time-independent eigenvalue equations,
one first decomposes the original two-particle equation~(\ref{eq:ii.1}) into two
one-particle equations for the one-particle wave functions~$\psi_1(\vec{r},t)$
and~$\psi_2(\vec{r},t)$:
\begin{subequations}
  \begin{align}
    \label{eq:ii.8a}
    i\hbar c\gamma^\mu D_\mu\psi_1 &= -Mc^2\psi_1\\*
    \label{eq:ii.8b}
    i\hbar c\gamma^\mu D_\mu\psi_2 &= Mc^2\psi_2\ .
  \end{align}
\end{subequations}
Here, the gauge-covariant derivative~$(D)$ for each wave
function~$\psi_a\;(a=1,2)$ is defined as follows:
\begin{subequations}
  \begin{align}
    \label{eq:ii.9a}
    D_\mu \psi_1 &= \partial_\mu\psi_1 - i\rklo{2}{A}_\mu\cdot \psi_1\\*
    \label{eq:ii.9b}
    D_\mu \psi_2 &= \partial_\mu\psi_2 - i\rklo{1}{A}_\mu\cdot \psi_2
  \end{align}
\end{subequations}
which expresses the fact that the two different positronium constituents are not able to
feel the exchange forces~$(\leadsto B_\mu\equiv 0)$ but are subjected exclusively to the
electromagnetic interactions!

Next, one has to face the problem of time-dependence of both wave
functions~$\psi_a(\vec{r},t)$ where one naturally will expect that the relevant physical
objects must be time-independent for a stationary bound system. In this sense, one tries
the usual factorizing ansatz for the one-particle wave
functions~$\psi_a(\vec{r},t)\,(a=1,2)$
\begin{equation}
  \label{eq:ii.10}
  \psi_a(\vec{r},t) = \exp\left(-i\frac{M_a c^2}{\hbar}\,t \right)\cdot\psi_a(\vec{r})\ ,
\end{equation}
where the \emph{mass eigenvalues}~$M_a$ are to be determined just by solving the mass
eigenvalue problem below. Furthermore, one observes the fact that the Dirac
four-spinors~$\psi_a(\vec{r})$ may be conceived as the Whitney sums of Pauli
two-spinors~$\rklo{a}{\varphi}_\pm(\vr)$, i.e.\ we put
\begin{equation}
  \label{eq:ii.11}
  \psi_a(\vr) = \rklo{a}{\varphi}_+(\vr)\oplus \rklo{a}{\varphi}_-(\vr)\ ;
\end{equation}
and for these Pauli spinors one deduces from the original Dirac equations
(\ref{eq:ii.8a})-(\ref{eq:ii.8b}) the following \emph{mass eigenvalue equations}~\cite{bs2}
\begin{subequations}
  \begin{align}
    \label{eq:ii.12a}
    i\vec{\sigma}\sdot\vec{\nabla}\rklo{1}{\varphi}_\pm(\vr) + \rklo{2}{A}_0(\vr)\cdot\rklo{1}{\varphi}_\mp(\vr)
    &= \frac{\pm M + M_*}{\hbar}\, c\cdot\rklo{1}{\varphi}_\mp(\vr)\\[2ex]
    \label{eq:ii.12b}
    i\vec{\sigma}\sdot\vec{\nabla}\rklo{2}{\varphi}_\pm(\vr) + \rklo{1}{A}_0(\vr)\cdot\rklo{2}{\varphi}_\mp(\vr)
    &= -\frac{M_* \pm M }{\hbar}\, c\cdot\rklo{2}{\varphi}_\mp(\vr)\ .
  \end{align}
\end{subequations}
Here, the common rest mass is denoted by M, cf. (\ref{eq:ii.7}), and the common mass
eigenvalue by~$M_*\,(=-M_1=M_2)$. Furthermore, we have also neglected the magnetic
potentials (i.e.\ putting~$\vec{A}_a(\vr)\equiv 0\leadsto$ \emph{electrostatic
  approximation}). An essential point with this neglection of magnetism aims at the
well-known ortho/para dichotomy of positronium~\cite{ms1}. Namely, it should be self-evident
that the physical difference of ortho- and para-positronium becomes annihilated through
disregarding the magnetic interaction energy, see the discussion of this in
ref.~\cite{ms1}. But observe on the other hand that the present mass eigenvalue system
(\ref{eq:ii.12a})-(\ref{eq:ii.12b}) is still of truly relativistic nature, though it is
written in terms of Pauli 2-spinors, not in terms of Dirac 4-spinors!

A further crucial point of the eigenvalue system (\ref{eq:ii.12a})-(\ref{eq:ii.12b})
concerns the plausible assumption that, from symmetry reasons, both particles (i.e.\
electron and positron) will occupy the same physical quantum state (for a more detailed
discussion see ref.~\cite{bs2}). A nearby consequence of this assumption is that the Dirac
four-densities~$\rklo{a}{k}_\mu(\vr)$ must be the same for both particles
\begin{equation}
  \label{eq:ii.13}
  \rklo{1}{k}_\mu(\vr) = \rklo{2}{k}_\mu(\vr)\doteqdot   \rklo{p}{k}_\mu(\vr)\ ,
\end{equation}
provided the two particles do combine to the \emph{ para-state} (for the case of
\emph{ortho-positronium} see ref.s~\cite{ms1,bs2}). On the other hand, the Dirac
densities~$\rklo{a}{k}_0(\vr)$ read in terms of the Pauli
spinors~$\rklo{a}{\varphi}_\pm(\vr)$ (\ref{eq:ii.11}) 
\begin{equation}
  \label{eq:ii.14}
  \rklo{a}{k}_0(\vr) = \rklo{a}{\varphi}^\dagger_+(\vr) \rklo{a}{\varphi}_+(\vr) +
   \rklo{a}{\varphi}^\dagger_-(\vr) \rklo{a}{\varphi}_-(\vr)
\end{equation}
and similarly for the Dirac currents~$\vec{k}_a(\vr)$
\begin{equation}
  \label{eq:ii.15}
  \vec{k}_a(\vr) = \rklo{a}{\varphi}^\dagger_+(\vr)\vec{\sigma} \rklo{a}{\varphi}_-(\vr) +
   \rklo{a}{\varphi}^\dagger_-(\vr)\vec{\sigma} \rklo{a}{\varphi}_+(\vr)\ .
\end{equation}
Therefore, the physical equivalence of both one-particle states allows us to parametrize
both states by only one set of Pauli spinors~$\rklo{p}{\varphi}_\pm(\vr)$ which then
lets appear both original spinors~$\rklo{1}{\varphi}_\pm(\vr)$
and~$\rklo{2}{\varphi}_\pm(\vr)$ as follows:
\begin{subequations}
  \begin{align}
    \label{eq:ii.16a}
    \rklo{p}{\varphi}_+(\vr) &\doteqdot \rklo{1}{\varphi}_+(\vr) =
    i\left(\hat{\vec{k}}\sdot\vec{\sigma}\right) \rklo{2}{\varphi}_+(\vr)\\*
    \label{eq:ii.16b}
    \rklo{p}{\varphi}_-(\vr) &\doteqdot \rklo{1}{\varphi}_-(\vr) =
    i\left(\hat{\vec{k}}\sdot\vec{\sigma}\right) \rklo{2}{\varphi}_-(\vr)\\*
    &\Big(\hat{\vec{k}}=\frac{ \vri{k}{p} }{ ||\vri{k}{p}||  } \Big)\ .\notag
  \end{align}
\end{subequations}

By use of the usual spinor algebra one can easily verify that the algebraic requirements
(\ref{eq:ii.13}) are actually satisfied through the present arrangements
(\ref{eq:ii.16a})-(\ref{eq:ii.16b}); and additionally the \emph{two} equations
(\ref{eq:ii.12a})-(\ref{eq:ii.12b}) for the individual
spinors~$\rklo{1}{\varphi}_\pm(\vr)$ and $\rklo{2}{\varphi}_\pm(\vr)$ become reduced to
\emph{one} equation for the para-spinor~$\rklo{p}{\varphi}_\pm(\vr)$:
\begin{equation}
  \label{eq:ii.17}
  i\vec{\sigma}\sdot\vec{\nabla} \rklo{p}{\varphi}_\pm(\vr) - \rklo{p}{A}_0(\vr)\cdot
  \rklo{p}{\varphi}_\mp(\vr) = \frac{M_*\pm M}{\hbar}\,c\cdot \rklo{p}{\varphi}_\mp(\vr)\ .
\end{equation}
Here, both electrostatic potentials~$\rklo{1}{A}_0(\vr)$ and~$\rklo{2}{A}_0(\vr)$ have
also been identified up to sign
\begin{equation}
  \label{eq:ii.18}
  \rklo{1}{A}_0(\vr) \equiv -  \rklo{2}{A}_0(\vr)\doteqdot  \rklo{p}{A}_0(\vr)
\end{equation}
because they are generated by the same Dirac density~$\rklo{p}{k}_0(\vr)$
\begin{equation}
  \label{eq:ii.19}
    \rklo{p}{k}_0(\vr) \doteqdot \rklo{1}{k}_0(\vr) \equiv \rklo{2}{k}_0(\vr)\ .
\end{equation}
The precise form of the link between the potential~$\rklo{p}{A}_0(\vr)$ and the Dirac
density~$\rklo{p}{k}_0(\vr)$ must now be deduced from the original gauge field equations.

\vspace{4ex}
\begin{center}
\mysubsection{2.\ Gauge Field Subsystem}
\end{center}

The gauge field counterpart of the basic matter field equation (\ref{eq:ii.1}) is the
(generally non-Abelian) Maxwell equation
\begin{gather}
  \label{eq:ii.20}
  \D^\mu \F_{\mu\nu} = -4\pi i\as\J_\nu \\*
  \Big(\as\doteqdot\frac{e^2}{\hbar c}\Big)\ ,\notag
\end{gather}
with the curvature~$\F_{\mu\nu}$ of the bundle connection~$\A_\mu$ (\ref{eq:ii.5}) being
defined as usual in the gauge field theories
\begin{equation}
  \label{eq:ii.21}
  \begin{split}
  \F_{\mu\nu} &\doteqdot\nabla_\mu\A_\nu -\nabla_\nu\A_\mu + \left[\A_\mu,\A_\nu
  \right]\\*
  & = \sum^2_{a=1}{F^a}_{\mu\nu}\tau_a+G_{\mu\nu}\chi-\Gstar_{\mu\nu}\bar{\chi}\ .
  \end{split}
\end{equation}
However, for the present situation of \emph{different} particles the exchange potential~$B_\mu$
vanishes identically~$(B_\mu\equiv 0)$ and consequently the theory becomes Abelian with
the two field strengths~${F^a}_{\mu\nu}$ being defined in terms of the residual
electromagnetic potentials~${A^a}_\mu$ through~$(a=1,2)$
\begin{equation}
  \label{eq:ii.22}
  {F^a}_{\mu\nu} = \nabla_\mu{A^a}_\nu - \nabla_\nu{A^a}_\mu\ .
\end{equation}

Furthermore, since we are presently satisfied with the \emph{electrostatic approximation},
the relativistic relation (\ref{eq:ii.22}) reduces to a simple gradient link of the
electric field strengths~$\vec{E}_a(\vr)$ to the electrostatic
potentials~$\rklo{a}{A}_0(\vr)$
\begin{equation}
  \label{eq:ii.23}
  \vec{E}_a(\vr) = -\vec{\nabla}\rklo{a}{A}_0(\vr)\ .
\end{equation}
From the same reason, the original Maxwell equations (\ref{eq:ii.20}) do reappear as simple
source relations
\begin{equation}
  \label{eq:ii.24}
  \vec{\nabla}\cdot \vec{E}_a = 4\pi\as\cdot\rklo{a}{j}_0(\vr)
\end{equation}
where however the Maxwell densities~${j^a}_\mu=\{\rklo{a}j_0,-\vec{j}_a \}$ are related to
the Dirac densities~$k_{a\mu}$ through
\begin{subequations}
  \begin{align}
    \label{eq:ii.25a}
    {j^1}_\mu &= k_{1\mu} = \{\rklo{1}{k}_0,-\vec{k}_1 \}\\*
    \label{eq:ii.25b}
    {j^2}_\mu &= -k_{2\mu} = \{-\rklo{2}{k}_0,\vec{k}_2 \}\ .
  \end{align}
\end{subequations}
Thus the source equations (\ref{eq:ii.24}) ultimately appear as the well-known Poisson
equations
\begin{subequations}
  \begin{align}
    \label{eq:ii.26a}
    \Delta\rklo{1}{A}_0 &= -4\pi\as\rklo{1}{k}_0\\*
    \label{eq:ii.26b}
    \Delta\rklo{2}{A}_0 &= 4\pi\as\rklo{2}{k}_0\ ,
  \end{align}
\end{subequations}
which however are contracted to only one equation:
\begin{equation}
  \label{eq:ii.27}
  \begin{split}
    \Delta\rklo{p}{A}_0(\vr) &= -4\pi\as\cdot\rklo{p}{k}_0(\vr)\\*
    &= -4\pi\as\left[\rklo{p}{\varphi}^\dagger_+(\vr) \rklo{p}{\varphi}_+(\vr) + 
      \rklo{p}{\varphi}^\dagger_-(\vr) \rklo{p}{\varphi}_-(\vr) \right]\ ,
  \end{split}
\end{equation}
namely as a consequence of the former identifications (\ref{eq:ii.13}) and (\ref{eq:ii.18}).

Summarizing, the RST eigenvalue problem for positronium in the electrostatic approximation
consists of the mass eigenvalue equation (\ref{eq:ii.17}) in combination with the present
Poisson equation (\ref{eq:ii.27}). It is true, this is a \emph{closed} system for the Pauli
spinors~$\rklo{p}{\varphi}_\pm(\vr)$ and the interaction potential~$\rklo{p}{A}_0(\vr)$
but it demands the specification of certain boundary conditions. Clearly, for the
matter fields one demands their sufficiently rapid vanishing at infinity~($r\to\infty$),
i.e.
\begin{equation}
  \label{eq:ii.28}
  \lim_{r\to\infty}\rklo{p}{\varphi}_\pm(\vr)=0\Rightarrow  \lim_{r\to\infty}
  \rklo{p}{k}_0(\vr)=0\ ;
\end{equation}
and moreover one wishes also to have thereout the asymptotic Coulomb form for the interaction
potential~$\rklo{p}{A}_0(\vr)$
\begin{equation}
  \label{eq:ii.29}
  \lim_{r\to\infty}\rklo{p}{A}_0(\vr) = \frac{\as}{r}\ .
\end{equation}
Both conditions (\ref{eq:ii.28})-(\ref{eq:ii.29}) lead us to the standard solution of the
Poisson equation (\ref{eq:ii.27}), i.e.
\begin{equation}
  \label{eq:ii.30}
  \rklo{p}{A}_0(\vr)=\as\int d^3\vr\;'\frac{\rklo{p}{k}_0(\vr\;')}{||\vr-\vr\;' ||}\ ,
\end{equation}
where the normalization condition for the Pauli spinors
\begin{equation}
  \label{eq:ii.31}
  \int d^3\vr\,\rklo{p}{k}_0(\vr) = 1
\end{equation}
(cf.~(\ref{eq:ii.14})) actually ensures the asymptotic Coulomb form (\ref{eq:ii.29}) of the
potential~$\rklo{p}{A}_0(\vr)$ (\ref{eq:ii.30}). 
\vspace{4ex}
\begin{center}
\mysubsection{3.\ Energy Functional}
\end{center}

Even if one would be able to find exact solutions of the eigenvalue equations
(\ref{eq:ii.17}) plus (\ref{eq:ii.27}), with both constraints (\ref{eq:ii.29}) and
(\ref{eq:ii.31}) being obeyed, one nevertheless would be forced to face the problem of the
energy content carried by that elaborated solutions. In other words, one needs an energy
functional~($\tekru{E}{T}$, say) whose value upon the constructed solution yields its
physically relevant and observable energy. Surely, the wanted energy~$\ru{E}{T}$ cannot be
identified with the mass eigenvalue~$M_* c^2$ because this quantity refers separately to
each of both particles and therefore is a one-particle quantity whereas we would like to
know the total energy~$\ru{E}{T}$ of the \emph{interactive} two-particle system. It would
also not help to take~$2M_*c^2$ as the desired energy~$\ru{E}{T}$ because~$M_* c^2$ alone
does already contain the whole electrostatic interaction energy which then would be counted
twice.

The solution of this energy problem has been already worked out in some precedent
papers~\cite{ms1} and for the sake of brevity it may suffice here to simply quote the
result:
\begin{equation}
  \label{eq:ii.32}
  \tekru{E}{T} = \Eb{E}{IV}{T} + 2\ru{\lambda}{D}\cdot\ru{N}{D} +
  \Ea{\lambda}{e}{G}\cdot\Ea{N}{e}{G}\ .
\end{equation}
This general form of the energy functional says that there are essentially  two parts
which are equipped with a rather different meaning: the first part~$\Eb{E}{IV}{T}$ is a
collection of the truly physical energy contributions 
\begin{equation}
  \label{eq:ii.33}
  \Eb{E}{IV}{T} = 2Mc^2\cdot \Z^2_\wp + 4\rkloi{p}{T}{kin} + \Ea{E}{e}{R}
\end{equation}
while the second part does refer to the constraints.

Here the first constraint~$(\ru{N}{D})$ refers to the wave function normalization
(\ref{eq:ii.31}):
\begin{equation}
  \label{eq:ii.34}
  \ru{N}{D} \doteqdot \int d^3\vr\left( \rklo{p}{\varphi}^\dagger_+ \rklo{p}{\varphi}_+ +
    \rklo{p}{\varphi}^\dagger_- \rklo{p}{\varphi}_-\right) -1 \equiv 0\ ,
\end{equation}
and~$\ru{\lambda}{D}$ is the associated Lagrangean multiplier. Similarly, the second
constraint~$(\Ea{N}{e}{G})$ does refer to the \emph{Poisson identity}
\begin{equation}
  \label{eq:ii.35}
  \Ea{N}{e}{G} \doteqdot \Ea{E}{e}{R} - \rork{M}{e} c^2 \equiv 0
\end{equation}
with the gauge field energy~$\Ea{E}{e}{R}$ being defined through
\begin{equation}
  \label{eq:ii.36}
  \Ea{E}{e}{R} = -\frac{\hbar c}{4\pi\as}\int d^3\vr\,\left(\vri{E}{p}\sdot\vri{E}{p} \right)
\end{equation}
and its mass equivalent~$(\rork{M}{e} c^2)$ through
\begin{equation}
  \label{eq:ii.37}
  \rork{M}{e} c^2 = -\hbar c\int d^3\vr\,\rklo{p}{A}_0(\vr)\cdot\rklo{p}{k}_0(\vr)\ .
\end{equation}
The Poisson identity (\ref{eq:ii.35}) itself is an immediate consequence of the Poisson
equation (\ref{eq:ii.27}) and therefore is automatically satisfied not only for the exact
solutions of the RST eigenvalue problem but also for its approximate solutions (see below)
if only the approximate potential~$\rklo{p}{A}_0(\vr)$ is linked to the (approximate)
Dirac density~$\rklo{p}{k}_0(\vr)$ via the Poisson equation (\ref{eq:ii.27}). Thus, if one
takes care of satisfying simultaneously both constraints (\ref{eq:ii.34}) and
(\ref{eq:ii.35}) the energy functional~$\tekru{E}{T}$ (\ref{eq:ii.32}) becomes reduced to
its physical part~$\Eb{E}{IV}{T}$ (\ref{eq:ii.33}).

This latter part (\ref{eq:ii.33}) represents the proper physical content of the energy
functional~$\tekru{E}{T}$ and consists of the (renormalized) rest mass energy, the
kinetic energy of both particles and the gauge field energy~$\Ea{E}{e}{R}$
(\ref{eq:ii.36}). The mass renormalization factor~$\ZP$ is given by
\begin{equation}
  \label{eq:ii.38}
  \ZP^2 = \int d^3\vr\,\left( \rklo{p}{\varphi}^\dagger_+ \rklo{p}{\varphi}_+ -
  \rklo{p}{\varphi}^\dagger_- \rklo{p}{\varphi}_- \right)\ ,
\end{equation}
and the one-particle kinetic energy~$\rkloi{p}{T}{kin}$ reads
\begin{equation}
  \label{eq:ii.39}
  \rkloi{p}{T}{kin} = i\frac{\hbar c}{2}\int d^3\,\vr\left[
    \rklo{p}{\varphi}^\dagger_+(\vr)\left(\vec{\sigma}\sdot\vec{\nabla}\right)
    \rklo{p}{\varphi}_-(\vr) +
    \rklo{p}{\varphi}^\dagger_-(\vr)\left(\vec{\sigma}\sdot\vec{\nabla}\right)
    \rklo{p}{\varphi}_+(\vr)    
  \right]\ .
\end{equation}
After all, it is a nice consistency check to carry explicitly through that standard recipe of
variational calculus as it is required by the \emph{principle of minimal energy}
\begin{equation}
  \label{eq:ii.40}
  \delta\tekru{E}{T} = 0\ .
\end{equation}
Or in other words, the extremalization of the energy functional~$\tekru{E}{T}$
(\ref{eq:ii.32}) with respect to the Pauli
spinors~$\rklo{p}{\varphi}^\dagger_\pm,\rklo{p}{\varphi}_\pm$ yields just the mass
eigenvalue equations (\ref{eq:ii.17}) as the first part of the RST eigenvalue problem; and
analogously the extremalization of~$\tekru{E}{T}$ (\ref{eq:ii.32}) with respect to the
electrostatic potential~$\rklo{p}{A}_0(\vr)$ lets one recover the Poisson equation
(\ref{eq:ii.27}), provided both constraints (\ref{eq:ii.34}) and (\ref{eq:ii.35}) are duly
regarded.


\section{\ Spherically Symmetric Approximation}
\indent

From the preceding presentation of the RST eigenvalue problem it should be evident that it
is (almost) impossible to construct exact solutions of such a peculiar kind of eigenvalue
problem. On the other hand, it is perhaps not even necessary to know those exact solutions
since their main features could possibly be grasped already by studying certain
\emph{approximate} solutions. A nearby simplification as the basis for such an
approximative approach refers to those anisotropic effects which are invading the
calculations via the spin degree of freedom inherent in the Dirac equations. In order to
get rid of this type of complication, one may neglect the anisotropic influence of the
spin on the electrostatic interaction potential and may assume that the interaction
potential~$\rklo{p}{A}_0(\vr)$ is nearly spherically symmetric, i.e. we put
\begin{gather}
  \label{eq:iii.1}
  \rklo{p}{A}_0(\vr) \Rightarrow \eklo{p}{A}_0(r)\\*
  \Big(r = ||\vr|| \Big)\ ,\notag
\end{gather}
where this assumption of spherical symmetry needs not be applied to the wave amplitudes of
the particles themselves. For such a first estimate of the positronium spectrum (on the level of
accuracy of the conventional treatment (\ref{eq:i.4}) ) it is also not necessary to stick to the relativistic
formulation of the theory, but rather one may again be satisfied with the non-relativistic
limit. Thus we will first eliminate the explicit spin anisotropy effect from the eigenvalue
problem and afterwards we will resort to the non-relativistic approximation.
\vspace{4ex}
\begin{center}
\mysubsection{1.\ Double-Valued Spinor Fields}
\end{center}

The manifest spin degree of freedom may be eliminated by first parametrizing the Pauli 
spinors $\rklo{p}{\varphi}_\pm(\vr)$ through the {\em wave amplitudes}\/~$\rklo{p}{\R}_\pm(\vr)$
and~$\rklo{p}{\S}_\pm(\vr)$ as the components of~$\rklo{p}{\varphi}_\pm(\vr)$ with respect to
a selected spinor basis; next, one separates off their angular parts and finally one impresses a
\emph{rigid} link on the remaining radial parts of both
amplitudes~$\rklo{p}{\R}_\pm(\vr)$ and~$\rklo{p}{\S}_\pm(\vr)$. This procedure yields (in its non-relativistic version) a scalar eigenvalue problem of the Schr\"odinger type, see equation (\ref{eq:iii.34}) below.

The selection of an appropriate spinor basis starts with the standard eigenspinors~$\zeta^{j,m}_{l}$ of total
angular momentum~$\hat{\vec{J}}\,(=\hat{\vec{L}}+\hat{\vec{S}})$ in two-dimensional
unitary space
\begin{subequations}
  \begin{align}
    \label{eq:iii.2a}
    \hat{\vec{J}}^2\,\zeta^{j,m}_{l} &= j(j+1)\hbar^2\cdot \zeta^{j,m}_{l}\\*
    \label{eq:iii.2b}
    \hat{\vec{L}}^2\,\zeta^{j,m}_{l} &= l(l+1)\hbar^2\cdot \zeta^{j,m}_{l}\\*
    \label{eq:iii.2c}
    \hat{\vec{S}}^2\,\zeta^{j,m}_{l} &= s(s+1)\hbar^2\cdot \zeta^{j,m}_{l} =
    \frac{1}{2}\left(\frac{1}{2}+1 \right)\hbar^2\cdot \zeta^{j,m}_{l} \\*
    \label{eq:iii.2d}
    \hat{\vec{J}}_z\,\zeta^{j,m}_{l} &= m\hbar\cdot\zeta^{j,m}_{l}\ .
  \end{align}
\end{subequations}
For the sake of simplicity, one may prefer here~$l=0,1$ for the orbital part of angular
momentum so that~$\{\zeta_0^{\frac{1}{2},\pm\frac{1}{2}}\}$ could be a basis for the
``positive'' Pauli spinors~$\rklo{p}{\varphi}_+(\vr)$ and similarly
$\{\zeta_1^{\frac{1}{2},\pm\frac{1}{2}}\}$ a basis for the ``negative''
spinors~$\rklo{p}{\varphi}_-(\vr)$. Now one can show that (for para-positronium
(\ref{eq:ii.16a})-(\ref{eq:ii.16b})) the action of the z-component~$\hat{J}_z$ of total
angular momentum~$\hat{\vec{J}}$ must annihilate both
spinors~$\rklo{p}{\varphi}_\pm(\vr)$, i.e.
\begin{subequations}
  \begin{align}
    \label{eq:iii.3a}
    \rork{\hat{J}}{+}_z \rklo{p}{\varphi}_+(\vr) &= 0\\*
    \label{eq:iii.3b}
    \rork{\hat{J}}{-}_z \rklo{p}{\varphi}_-(\vr) &= 0\ ,
  \end{align}
\end{subequations}
see ref.s~\cite{ms1,bs2}. Or in other words, the action of the total operator 
$\hat{\J}_z\,(\doteqdot\rork{\hat{J}}{+}_z\oplus \rork{\hat{J}}{-}_z)$ annihilates the
one-particle Dirac spinor field~$\psi_\wp\,(\doteqdot\rklo{p}{\varphi}_+\oplus\rklo{p}{\varphi}_-)$, i.\,e.
\begin{equation}
  \label{eq:iii.4}
  \hat{\J}_z\psi_\wp = 0\ .
\end{equation}

But clearly, such a one-particle state must be of rather exotic nature because an ordinary
(``fermionic'') Dirac eigenspinor of~$\hat{\J}_z$ has always half-integer eigenvalue;
and consequently such a strange (``bosonic'') state as (\ref{eq:iii.4}) must own some unusual property. Indeed
this concerns its uniqueness in the sense that the spinor field~$\psi_\wp$
(\ref{eq:iii.3a})-(\ref{eq:iii.4}) is \emph{double-valued}~\cite{ms1,bs2}
\begin{equation}
  \label{eq:iii.5}
  \psi_\wp(r,\vartheta,\phi+2\pi) = -  \psi_\wp(r,\vartheta,\phi)
\end{equation}
where~$r,\vartheta,\phi$ are the usual spherical polar coordinates of flat three-space. On
the other hand, both basis systems~$\{\zeta^{\frac{1}{2},\pm\frac{1}{2}}_0 \}$
and~$\{\zeta^{\frac{1}{2},\pm\frac{1}{2}}_1 \}$  (\ref{eq:iii.2a})-(\ref{eq:iii.2d}) are
unique over three-space; and thus if one would decompose such a spinor-like~$\psi_\wp$
(\ref{eq:iii.5}) with respect to a standard basis, the corresponding components would
have to be non-unique scalar fields over three-space. However, our option is here just the
other way round, namely to choose certain basis systems~$\rork{\omega}{\pm}_0$
and~$\rork{\omega}{\pm}_1$ which themselves carry the double-valuedness (\ref{eq:iii.5})
\emph{alone} so that the components of~$\psi_\wp$ can remain unique scalar fields, i.e.
\begin{subequations}
  \begin{align}
    \label{eq:iii.6a}
    \rork{\omega}{\pm}_0(r,\vartheta,\phi+2\pi) &= -\rork{\omega}{\pm}_0(r,\vartheta,\phi)\\*
    \label{eq:iii.6b}
    \rork{\omega}{\pm}_1(r,\vartheta,\phi+2\pi) &= -\rork{\omega}{\pm}_1(r,\vartheta,\phi)\ .
  \end{align}
\end{subequations}
Now our choice of basis system looks as follows:
\begin{subequations}
  \begin{align}
    \label{eq:iii.7a}
    \rork{\omega}{+}_0&= {\rm e}^{-i\,\frac{\phi}{2}}\cdot\zeta_0^{\frac{1}{2},\frac{1}{2}}\\*
    \label{eq:iii.7b}
    \rork{\omega}{-}_0&= {\rm e}^{i\,\frac{\phi}{2}}\cdot\zeta_0^{\frac{1}{2},-\frac{1}{2}}\\*
    \label{eq:iii.7c}
    \rork{\omega}{+}_1&= {\rm e}^{-i\,\frac{\phi}{2}}\cdot\zeta_1^{\frac{1}{2},\frac{1}{2}}\\*
    \label{eq:iii.7d}
    \rork{\omega}{-}_1&= {\rm e}^{i\,\frac{\phi}{2}}\cdot\zeta_1^{\frac{1}{2},-\frac{1}{2}}\ ,
  \end{align}
\end{subequations}
so that this $\omega$-basis becomes annihilated by $\rork{\hat{J}}{\pm}_z$:
\begin{subequations}
  \begin{align}
    \label{eq:iii.8a}
    \rork{\hat{J}}{+}_z\,\rork{\omega}{\pm}_0&=0\\*
    \label{eq:iii.8b}
    \rork{\hat{J}}{-}_z\,\rork{\omega}{\pm}_1&=0\ .
  \end{align}
\end{subequations}

The double-valuedness (\ref{eq:iii.6a})-(\ref{eq:iii.6b}) of the chosen basis system transcribes now to the Pauli spinors $\rklo{p}{\varphi}_\pm(\vec{r})$, provided one adopts the corresponding components $\rklo{p}{\R}_\pm$, $\rklo{p}{\S}_\pm$ to be unique. Accordingly, the desired decomposition reads
\begin{subequations}
  \begin{align}
    \label{eq:iii.9a}
    \rklo{p}{\varphi}_+(\vr)&=\rklo{p}{\R}_+(\vr)\cdot\rork{\omega}{+}_0+\rklo{p}{\S}_+(\vr)\cdot\rork{\omega}{-}_0\\*
    \label{eq:iii.9b}
    \rklo{p}{\varphi}_-(\vr)&= -i\left\{\rklo{p}{\R}_-(\vr)\cdot\rork{\omega}{+}_1+\rklo{p}{\S}_-(\vr)\cdot\rork{\omega}{-}_1\right\}
  \end{align}
\end{subequations}
with \emph{unique} scalar components
\begin{subequations}
  \begin{align}
    \label{eq:iii.10a}
    \rklo{p}{\R}_\pm(r,\vartheta,\phi+2\pi)&=\rklo{p}{\R}_\pm(r,\vartheta,\phi)\\*
    \label{eq:iii.10b}
    \rklo{p}{\S}_\pm(r,\vartheta,\phi+2\pi)&=\rklo{p}{\S}_\pm(r,\vartheta,\phi)\ .
  \end{align}
\end{subequations}
\vspace{4ex}
\begin{center}
\mysubsection{2.\ Eigenvalue Equations for the Wave Amplitudes}
\end{center}

Further information about the components $\rklo{p}{\R}_\pm(\vr)$ and $\rklo{p}{\S}_\pm(\vr)$ can be gained by inspection of how the annihilation process (\ref{eq:iii.3a})-(\ref{eq:iii.3b}) can be realized in detail. Indeed, a straightforward calculation yields
\begin{equation}
  \label{eq:iii.11}
  \rork{\hat{J}}{+}_z\,\rklo{p}{\varphi}_+(\vr)= (\hat{L}_z\,\rklo{p}{\R}_+)\cdot\rork{\omega}{+}_0+(\hat{L}_z\,\rklo{p}{\S}_+)\cdot\rork{\omega}{-}_0 +\rklo{p}{\R}_+\cdot(\rork{\hat{J}}{+}_z\,\rork{\omega}{+}_0)+\rklo{p}{\S}_+\cdot(\rork{\hat{J}}{+}_z\,\rork{\omega}{-}_0)
\end{equation}
and analogously for $\rklo{p}{\varphi}_-(\vr)$. Observing here the annihilation relations for the $\omega$-basis (\ref{eq:iii.8a})-(\ref{eq:iii.8b}) it is clear that one has to demand
\begin{equation}
  \label{eq:iii.12}
  \hat{L}_z\,\rklo{p}{\R}_+=\hat{L}_z\,\rklo{p}{\S}_+=0 
\end{equation}
in order to have the eigenvalue equations (\ref{eq:iii.3a})-(\ref{eq:iii.3b}) for angular momentum satisfied. However, the latter demand (\ref{eq:iii.12}) can trivially be satisfied by letting the wave amplitudes $\rklo{p}{\R}_\pm(\vr)$, $\rklo{p}{\S}_\pm(\vr)$ depend exclusively upon $r$ and $\vartheta$, but not on $\phi$. Thus one puts
\begin{subequations}
  \begin{align}
  \label{eq:iii.13a}
  \rklo{p}{\R}_\pm(\vr)&\Rightarrow\rklo{p}{R}_\pm(r,\vartheta)\\*
  \label{eq:iii.13b}
  \rklo{p}{\S}_\pm(\vr)&\Rightarrow\rklo{p}{S}_\pm(r,\vartheta)\ .
  \end{align}
\end{subequations}

Furthermore, it turns out that the equations for the wave amplitudes become considerably simplified if one resorts to a further transformation:
\begin{subequations}
  \begin{align}
  \label{eq:iii.14a}
  \rklo{p}{R}_\pm(r,\vartheta)&=\frac{\rklo{p}{\tilde{R}}_\pm(r,\vartheta)}{\sqrt{r\sin\vartheta}}\\*
  \label{eq:iii.14b}
  \rklo{p}{S}_\pm(r,\vartheta)&=\frac{\rklo{p}{\tilde{S}}_\pm(r,\vartheta)}{\sqrt{r\sin\vartheta}}\ .
  \end{align}
\end{subequations}
Namely, the relativistic Pauli equations (\ref{eq:ii.12a})-(\ref{eq:ii.12b}) are then recast to the following relatively simple eigenvalue equations for the new amplitudes $\rklo{p}{\tilde{R}}_\pm$, $\rklo{p}{\tilde{S}}_\pm$ \cite{ms1}
\begin{subequations}
  \begin{align}
  \label{eq:iii.15a}
  \frac{\partial\rklo{p}{\tilde{R}}_+}{\partial r}+\frac{1}{r}\frac{\partial\rklo{p}{\tilde{S}}_+}{\partial\vartheta} -\rklo{p}{A}_0\cdot\rklo{p}{\tilde{R}}_-&=\frac{M+M_*}{\hbar}\,c\cdot\rklo{p}{\tilde{R}}_-
  \begin{array}{c}{}\\{}\end{array}\\*
  \label{eq:iii.15b}
  \frac{\partial\rklo{p}{\tilde{S}}_+}{\partial r}-\frac{1}{r}\frac{\partial\rklo{p}{\tilde{R}}_+}{\partial\vartheta} -\rklo{p}{A}_0\cdot\rklo{p}{\tilde{S}}_-&=\frac{M+M_*}{\hbar}\,c\cdot\rklo{p}{\tilde{S}}_-
  \begin{array}{c}{}\\{}\end{array}\\*
  \label{eq:iii.15c}
  \frac{\partial\rklo{p}{\tilde{R}}_-}{\partial r}+\frac{1}{r}\cdot\rklo{p}{\tilde{R}}_- -\frac{1}{r}\frac{\partial\rklo{p}{\tilde{S}}_-}{\partial\vartheta} +\rklo{p}{A}_0\cdot\rklo{p}{\tilde{R}}_+&=\frac{M-M_*}{\hbar}\,c\cdot\rklo{p}{\tilde{R}}_+
  \begin{array}{c}{}\\{}\end{array}\\*
  \label{eq:iii.15d}
  \frac{\partial\rklo{p}{\tilde{S}}_-}{\partial r}+\frac{1}{r}\cdot\rklo{p}{\tilde{S}}_- +\frac{1}{r}\frac{\partial\rklo{p}{\tilde{R}}_-}{\partial\vartheta} +\rklo{p}{A}_0\cdot\rklo{p}{\tilde{S}}_+&=\frac{M-M_*}{\hbar}\,c\cdot\rklo{p}{\tilde{S}}_+\ .
  \end{align}
\end{subequations}
In order to deal with a closed system of equations for the new amplitudes, one substitutes the Pauli spinors (\ref{eq:iii.9a})-(\ref{eq:iii.9b}) into the Poisson equation (\ref{eq:ii.27}) which lets reappear this equation now in the following form
\begin{equation}
  \label{eq:iii.16}
  \Delta\rklo{p}{A}_0= -\as\,\frac{\rklo{p}{\tilde{R}}_+^2+\rklo{p}{\tilde{S}}_+^2+\rklo{p}{\tilde{R}}_-^2+\rklo{p}{\tilde{S}}_-^2}{r\sin\vartheta}\,.
\end{equation}

The coupled system (\ref{eq:iii.15a})-(\ref{eq:iii.16}) represents now the RST eigenvalue problem for para-positronium in the electrostatic approximation; but despite its simplified form it is hard (or even impossible) to elaborate the exact solutions hereof.
\vspace{4ex}
\begin{center}
\mysubsection{3.\ Spherical Symmetry}
\end{center}

In this situation, the approximative assumption of spherical symmetry (\ref{eq:iii.1})
allows further progress. Namely, it is just this assumption which suggests to try a
product ansatz for the wave amplitudes of the following form \cite{ms1}
\begin{subequations}
  \begin{align}
  \label{eq:iii.17a}
  \rklo{p}{\tilde{R}}_\pm(r,\vartheta)&=\slo{R}_\pm(r)\cdot f_R(\vartheta)\\
  \label{eq:iii.17b}
  \rklo{p}{\tilde{S}}_\pm(r,\vartheta)&=\slo{S}_\pm(r)\cdot f_S(\vartheta)\ .
  \end{align}
\end{subequations}
Indeed, substituting this ansatz into the mass eigenvalue equations (\ref{eq:iii.15a}-\ref{eq:iii.15d}) allows one to separate the variables so that there emerges one subset of equations for the radial functions $\slo{R}_\pm(r),\,\slo{S}_\pm(r)$ and one subset for the angular functions $f_R(\vartheta)$ and $f_S(\vartheta)$. Here, the angular equations introduce the quantum number $\lP$ of orbital angular momentum in the following way:
\begin{subequations}
  \begin{align}
  \label{eq:iii.18a}
  \frac{df_R(\vartheta)}{d\vartheta}&=\lP\cdot f_S(\vartheta)\\
  \label{eq:iii.18b}
  \frac{df_S(\vartheta)}{d\vartheta}&=-\lP\cdot f_R(\vartheta)\ ,
  \end{align}
\end{subequations}
where a second possibility does exist which, however, has merely the sign of $\lP$ reversed (i.\,e. $\lP\Rightarrow-\lP$). The separated angular problem (\ref{eq:iii.18a})-(\ref{eq:iii.18b}) is so simple that its solutions are immediately evident, namely either
\begin{equation}
  \label{eq:iii.19}
  f_R(\vartheta)=\cos(\lP\cdot\vartheta)\,,\quad f_S(\vartheta)=-\sin(\lP\cdot\vartheta)
\end{equation}
or
\begin{equation}
  \label{eq:iii.20}
  f_R(\vartheta)=\sin(\lP\cdot\vartheta)\,,\quad f_S(\vartheta)=\cos(\lP\cdot\vartheta)\ ,
\end{equation}
with the quantum number $\lP$ adopting (half-)integer values
\begin{equation}
  \label{eq:iii.21}
  \lP=0,\,(\frac{1}{2}),\,1,\,(\frac{3}{2}),\,2,\,(\frac{5}{2}),\,3,\,...
\end{equation}
This postulate of (half-)integrity follows from the demand that the Dirac current $\vec{k}_p$ remains finite on the $z$-axis ($\vartheta=0,\pi$). For the exclusion of the half-integers see below.

After the angular part is split off in that way (\ref{eq:iii.17a})-(\ref{eq:iii.17b}),
there remains a purely radial problem for the determination of the mass eigenvalue
$M_*$. Indeed, substituting the separation ansatz (\ref{eq:iii.17a})-(\ref{eq:iii.17b})
back into the original eigenvalue equations (\ref{eq:iii.15a})-(\ref{eq:iii.15d}) and
using also the angular equations (\ref{eq:iii.18a})-(\ref{eq:iii.18b}) yields the
following set of coupled radial equations:
\begin{subequations}
  \begin{align}
  \label{eq:iii.22a}
  \frac{d\slo{R}_+(r)}{dr}-\frac{\lP}{r}\cdot\slo{S}_+(r)-\eklo{p}{A}_0(r)\cdot\slo{R}_-(r)&= \frac{M+M_*}{\hbar}\,c\cdot\slo{R}_-(r)\\
  \label{eq:iii.22b}
  \frac{d\slo{S}_+(r)}{dr}-\frac{\lP}{r}\cdot\slo{R}_+(r)-\eklo{p}{A}_0(r)\cdot\slo{S}_-(r)&= \frac{M+M_*}{\hbar}\,c\cdot\slo{S}_-(r)\\
  \label{eq:iii.22c}
  \frac{d\slo{R}_-(r)}{dr}+\frac{1}{r}\cdot\slo{R}_-(r)+\frac{\lP}{r}\cdot\slo{S}_-(r)+ \eklo{p}{A}_0(r)\cdot\slo{R}_+(r)&= \frac{M-M_*}{\hbar}\,c\cdot\slo{R}_+(r)\\
  \label{eq:iii.22d}
  \frac{d\slo{S}_-(r)}{dr}+\frac{1}{r}\cdot\slo{S}_-(r)+\frac{\lP}{r}\cdot\slo{R}_-(r)+ \eklo{p}{A}_0(r)\cdot\slo{S}_+(r)&= \frac{M-M_*}{\hbar}\,c\cdot\slo{S}_+(r)\ .
  \end{align}
\end{subequations}
Surely, it appears somewhat strange here that one and the same eigenvalue $M_*$ enters all
four equations (\ref{eq:iii.22a})-(\ref{eq:iii.22d}); but this may be understood as a hint
at the circumstance that there exists a rigid link between the four variables
$\slo{R}_\pm(r),\,\slo{S}_\pm(r)$. And indeed, the following identifications
\begin{subequations}
  \begin{align}
  \label{eq:iii.23a}
  \slo{R}_+(r)\equiv\slo{S}_+(r)\doteqdot\tilde{\Phi}_+(r)\\
  \label{eq:iii.23b}
  \slo{R}_-(r)\equiv\slo{S}_-(r)\doteqdot\tilde{\Phi}_-(r)
  \end{align}
\end{subequations}
recast this system of four equations to only two equations
\begin{subequations}
  \begin{align}
  \label{eq:iii.24a}
  \frac{d\tilde{\Phi}_+(r)}{dr}-\frac{\lP}{r}\cdot\tilde{\Phi}_+(r)-\eklo{p}{A}_0(r)\cdot\tilde{\Phi}_-(r)&= \frac{M+M_*}{\hbar}\,c\cdot\tilde{\Phi}_-(r)\\
  \label{eq:iii.24b}
  \frac{d\tilde{\Phi}_-(r)}{dr}+\frac{\lP+1}{r}\cdot\tilde{\Phi}_-(r)+ \eklo{p}{A}_0(r)\cdot\tilde{\Phi}_+(r)&= \frac{M-M_*}{\hbar}\,c\cdot\tilde{\Phi}_+(r)\ .
  \end{align}
\end{subequations}

This spherically symmetric eigenvalue problem may finally be completed by the corresponding Poisson equation
\begin{equation}
  \label{eq:iii.25}
  \left(\frac{d^2}{dr^2}+\frac{2}{r}\frac{d}{dr}\right)\eklo{p}{A}_0(r)= -\frac{\pi}{2}\,\as\,\frac{\tilde{\Phi}_+^2(r)+\tilde{\Phi}_-^2(r)}{r}
\end{equation}
which is to be conceived as the spherically symmetric simplification of (\ref{eq:iii.16}). Notice here also the important fact that the quantum number $\lP$ does not explicitly enter the source terms of the spherically symmetric Poisson equation (\ref{eq:iii.25})!
The standard solution of this type of equation reads formally
\begin{equation}
  \label{eq:iii.26}
  \eklo{p}{A}_0(r)=\frac{\as}{8}\int \frac{d^3\vr\,'}{r'}\; \frac{\tilde{\Phi}_+^2(r')+\tilde{\Phi}_-^2(r')}{\|\vec{r}-\vec{r}\,'\|}
\end{equation}
and approaches the Coulomb potential at infinity ($r\rightarrow\infty$), i.e.
\begin{equation}
  \label{eq:iii.27}
  \lim_{r\rightarrow\infty}\eklo{p}{A}_0(r)=\frac{\as}{r}\ ,
\end{equation}
namely just on account of the normalization condition (\ref{eq:ii.31}) which reads in terms of the wave amplitudes $\tilde{\Phi}_\pm(r)$
\begin{equation}
  \label{eq:iii.28}
  \frac{\pi}{2}\int dr'\,r'\,\left\{\tilde{\Phi}_+^2(r')+\tilde{\Phi}_-^2(r')\right\}=1\ .
\end{equation}
Clearly, such a requirement can only be satisfied if the amplitudes $\tilde{\Phi}_\pm(r)$ obey the usual boundary condition
\begin{equation}
  \label{eq:iii.29}
  \lim_{r\rightarrow\infty}\tilde{\Phi}_\pm(r)=0\ .
\end{equation}

However, a somewhat more critical point concerning the boundary conditions refers to the
behaviour of the angular functions $f_R(\vartheta)$ and $f_S(\vartheta)$
(\ref{eq:iii.19})-(\ref{eq:iii.20}) on the $z$-axis (i.\,e. $\vartheta=0,\pi$). From
generally accepted arguments one would think that all physically relevant objects of the
theory should be non-singular everywhere. In the present context, the crucial object is
here the Dirac current $\vec{k}_p$ which for the presently considered kinematics of the
Dirac spinor field $\psi\!_\wp=\rklo{p}{\varphi}_+\oplus\rklo{p}{\varphi}_-$ encircles the
$z$-axis
\begin{equation}
  \label{eq:iii.30}
  \vec{k}_p(\vec{r})=\rklo{p}{k}_\phi(r,\vartheta)\cdot\vec{e}_\phi
\end{equation}
with the azimuthal component being given by
\begin{equation}
  \label{eq:iii.31}
  \rklo{p}{k}_\phi(r,\vartheta)=\pm\frac{\tilde{\Phi}_+(r)\cdot\tilde{\Phi}_-(r)}{2\pi r}\cdot \frac{\sin\big[(2\lP+1)\cdot\vartheta\big]}{\sin\vartheta}\ .
\end{equation}
Therefore, if one wishes to keep this current component finite on the $z$-axis then one is forced to demand that $\lP$ can adopt only (half-)integer values, cf. (\ref{eq:iii.21}). Furthermore, if one demands that $\rklo{p}{k}_\phi$ should be distributed symmetrically along the $z$-axis, i.\,e.
\begin{equation}
  \label{eq:iii.32}
  \rklo{p}{k}_\phi(r,0)\stackrel{!}{=}\rklo{p}{k}_\phi(r,\pi)\quad\forall r\ ,
\end{equation}
then one can admit only integer values of $\lP$
\begin{equation}
  \label{eq:iii.33}
  \lP=0,\,1,\,2,\,3,\,...
\end{equation}
(For a sketch of $\rklo{p}{k}_\phi$ for integer and half-integer $\lP$, see fig.~1 of
ref.~\cite{ms1}). Henceforth we will prefer the present quantization prescription
(\ref{eq:iii.33}) over the former one~\cite{ms1}).

But with the eigenvalue and Poisson equations (\ref{eq:iii.24a})-(\ref{eq:iii.25}) being fixed now together with all the boundary conditions, one could in principle look for the solutions of this eigenvalue problem and determine the corresponding binding energies. However, it seems meaningful to first resort to a further (but ultimate) approximation.
\vspace{4ex}
\begin{center}
\mysubsection{4.\ Non-Relativistic Approximation}
\end{center}

Besides the electrostatic and spherically symmetric approximations one can now take into
account a final simplification, i.\,e. the non-relativistic limit. The point with this
approximation is namely that the relativistic corrections may well be of the same order of
magnitude as the magnetic effects; and therefore it seems somewhat inconsequent to neglect
the magnetic effects on the one hand but on the other hand to cling to a fully
relativistic formulation. Thus one should be willing now to pass over to the
non-relativistic approximation of the relativistic eigenvalue system
(\ref{eq:iii.24a})-(\ref{eq:iii.25}). Here, the two mass eigenvalue equations
(\ref{eq:iii.24a})-(\ref{eq:iii.24b}) contract to only one equation for the ``positive''
Pauli spinor $\tilde{\Phi}_+(r)$ which for the sake of brevity is simply termed as
$\tilde{\Phi}(r)$ and has then to obey the following Schr\"odinger-like equation:
\begin{equation}
  \label{eq:iii.34}
  -\frac{\hbar^2}{2M}\left\{\frac{d^2\tilde{\Phi}(r)}{dr^2}+\frac{1}{r}\frac{d\tilde{\Phi}(r)}{dr}\right\}+ \frac{\hbar^2}{2M}\frac{\lP^2}{r^2}\,\tilde{\Phi}(r)-\hbar c\eklo{p}{A}_0(r)\cdot\tilde{\Phi}(r)=E_*\cdot\tilde{\Phi}(r)\ .
\end{equation}
The Schr\"odinger eigenvalue $E_*$ emerging here is related to the former relativistic mass eigenvalue $M_*$ by
\begin{equation}
  \label{eq:iii.35}
  E_*=(M_*-M)c^2\ ,
\end{equation}
and the meaning of $\lP$ as a quantum number of orbital angular momentum becomes now evident. In order to have the non-relativistic eigenvalue system closed, the Schr\"odinger equation (\ref{eq:iii.34}) must be accompanied by the non-relativistic version of the Poisson equation (\ref{eq:iii.25}), i.\,e.
\begin{equation}
  \label{eq:iii.36}
  \left\{\frac{d^2}{dr^2}+\frac{2}{r}\frac{d}{dr}\right\}\eklo{p}{A}_0(r)= -\frac{\pi}{2}\,\as\,\frac{\tilde{\Phi}^2(r)}{r}\ .
\end{equation}
Correspondingly, the non-relativistic approximation of the former solution (\ref{eq:iii.26}) for $\eklo{p}{A}_0(r)$ appears now as
\begin{equation}
  \label{eq:iii.37}
  \eklo{p}{A}_0(r)=\frac{\as}{8}\int \frac{d^3\vec{r}\,'}{r'}\;\frac{\tilde{\Phi}^2(r')}{\|\vec{r}-\vec{r}\,'\|}
\end{equation}
and thus continues to obey the same Coulomb-like boundary condition (\ref{eq:iii.27}), namely on account of the non-relativistic form of the normalization condition (\ref{eq:iii.28})
\begin{equation}
  \label{eq:iii.38}
  \frac{\pi}{2}\int dr\,r\;\tilde{\Phi}^2(r)=1\ .
\end{equation}
In this way, one finally has arrived at a well-defined non-relativistic eigenvalue problem.

But even if one were able to solve exactly this radically simplified problem, one would
still be left with the problem of the energy ($E_{\rm T}$, say) being concentrated in this
field configuration. This is the question of the right energy functional
$\tilde{E}_{\rm[T]}$ whose value upon the solutions of the eigenvalue problem is required
to not only yield the desired energy $E_{\rm T}$ but whose extremal equations are also
required to just coincide with the eigenvalue equations:
\begin{equation}
  \label{eq:iii.39}
  \delta\tilde{E}_{\rm[T]}=0\ .
\end{equation}
This {\em principle of minimal energy}\/ can then also be used in order to construct {\em
  approximate}\/ variational solutions of the eigenvalue problem; namely by guessing some
trial configuration with a certain number of variational parameters
$\{\beta;\nu_1,\nu_2,\,...\,\nu_k\}$. The value of the non-relativistic version
$\tilde{\mathbb{E}}_{\rm [T]}$ of the relativistic energy functional $\tilde{E}_{\rm[T]}$
upon this trial configuration yields then an {\em energy function}\/
$\mathbb{E}^{\rm(IV)}(\beta;\nu_1,\nu_2,\,...\,\nu_k)$ whose (local) extremal values are
determined by the vanishing of all the first-order derivatives
\begin{equation}
  \label{eq:iii.40}
  \begin{array}{c@{\;}c@{\;}l}
  \displaystyle
  \frac{\partial\mathbb{E}^{\rm(IV)}(\beta;\nu_1,\nu_2,\,...\,\nu_k)}{\partial\beta}&=&0
  \vspace{2ex}\\
  \displaystyle
  \frac{\partial\mathbb{E}^{\rm(IV)}(\beta;\nu_1,\nu_2,\,...\,\nu_k)}{\partial\nu_1}&=&0
  \vspace{2ex}\\
  \displaystyle
  \vspace{2ex}\vdots&&\\
  \displaystyle
  \frac{\partial\mathbb{E}^{\rm(IV)}(\beta;\nu_1,\nu_2,\,...\,\nu_k)}{\partial\nu_k}&=&0\ .
  \end{array}
\end{equation}
The corresponding extremal values $\mathbb{E}^{\rm(IV)}_*$ do then specify the energy levels of the bound system, albeit only in the {\em non-relativistic, electrostatic and spherically symmetric}\/ approximation!

The desired energy functional $\tilde{\mathbb{E}}_{\rm [T]}$ is the non-relativistic
version of the former $\tilde{E}_{\rm [T]}$ (\ref{eq:ii.32})-(\ref{eq:ii.33}) and has
already been deduced in full detail in the preceding paper (see equation (VI.50) of
ref.~\cite{ms1})
\begin{equation}
  \label{eq:iii.41}
  \tilde{\mathbb{E}}_{\rm [T]}\Rightarrow\tilde{\mathbb{E}}_{[\Phi]}= 2E_{\rm kin}+E_{\rm R}^{\rm [e]}+2\lambda_{\rm S}\,\tilde{\mathbb{N}}_\Phi+\lambda_{\rm G}^{(e)}\,\tilde{\mathbb{N}}_{\rm G}^{\rm (e)}\ .
\end{equation}
Here, the first term represents the kinetic energy for the two particles where the one-particle energy $E_{\rm kin}$ is the sum of the radial part $\rklo{r}{E}_{\rm kin}$ and the longitudinal part $\rklo{\vartheta}{E}_{\rm kin}$, i.\,e.
\begin{equation}
  \label{eq:iii.42}
  E_{\rm kin}=\rklo{r}{E}_{\rm kin}+\rklo{\vartheta}{E}_{\rm kin}\ ,
\end{equation}
with the radial part being given by
\begin{equation}
  \label{eq:iii.43}
  \rklo{r}{E}_{\rm kin}=\frac{\hbar^2}{2M}\cdot\frac{\pi}{2}\int_0^\infty dr\,r\,\left(\frac{d\tilde{\Phi}(r)}{dr}\right)^2
\end{equation}
and similarly the longitudinal part by
\begin{equation}
  \label{eq:iii.44}
  \rklo{\vartheta}{E}_{\rm kin}=\frac{\hbar^2}{2M}\,\lP^2\cdot\frac{\pi}{2}\int_0^\infty \frac{dr}{r}\;\tilde{\Phi}^2(r)\ .
\end{equation}
Next, the energy content $E_{\rm R}^{\rm [e]}$ of the spherically symmetric potential $\eklo{p}{A}_0(r)$ reads
\begin{equation}
  \label{eq:iii.45}
  E_{\rm R}^{\rm [e]}=-\frac{\hbar c}{\as}\int_0^\infty dr\,\left(r\cdot\frac{d\eklo{p}{A}_0(r)}{dr}\right)^2
\end{equation}
and thus the physical part of the functional $\tilde{\mathbb{E}}_{[\Phi]}$ (\ref{eq:iii.41}) consists just of these three energy contributions (\ref{eq:iii.43})-(\ref{eq:iii.45}).

The residual two terms in (\ref{eq:iii.41}) are constraints with Lagrangean parameters
$\lambda_{\rm S}$ and $\lambda_{\rm G}^{\rm (e)}$. The first constraint refers to the
normalization (\ref{eq:iii.38}) of the non-relativistic wave amplitude $\tilde{\Phi}(r)$
\begin{equation}
  \label{eq:iii.46}
  \tilde{\mathbb{N}}_{\Phi}\doteqdot\frac{\pi}{2}\int_0^\infty dr\,r\,\tilde{\Phi}^2(r)-1\equiv0\ ,
\end{equation}
and the second constraint is the {\em Poisson identity}\/
\begin{equation}
  \label{eq:iii.47}
  \tilde{\mathbb{N}}_{\rm G}^{\rm [e]}\doteqdot E_{\rm R}^{\rm [e]}-\tilde{\mathbb{M}}^{\rm [e]}c^2=0
\end{equation}
with the mass equivalent $\tilde{\mathbb{M}}^{\rm [e]}c^2$ due to the gauge field energy $E_{\rm R}^{\rm [e]}$ being given by
\begin{equation}
  \label{eq:iii.48}
  \tilde{\mathbb{M}}^{\rm [e]}c^2\doteqdot-\frac{\pi}{2}\hbar c\int_0^\infty dr\,r\,\eklo{p}{A}_0(r)\cdot\tilde{\Phi}^2(r)\ .
\end{equation}

A nice check of all these assertions consists now in deducing the present eigenvalue system (\ref{eq:iii.34}) plus (\ref{eq:iii.36}) from the non-relativistic version ($\delta\tilde{\mathbb{E}}_{[\Phi]}=0$) of the {\em principle of minimal energy}\/ (\ref{eq:iii.39}) by means of the standard variational techniques. Thereby the Lagrangean parameters turn out as
\begin{subequations}
  \begin{align}
  \label{eq:iii.49a}
  \lambda_{\rm S}&=-E_*\\
  \label{eq:iii.49b}
  \lambda_{\rm G}^{\rm (e)}&=-2\ .
  \end{align}
\end{subequations}
This calculus of variations may now be exploited in order to compute approximately the non-relativistic energy spectrum of positronium.
\pagebreak
\begin{center}
\mysubsection{5.\ Spherically Symmetric Spectrum}
\end{center}

In order to present such an example for the energy spectrum due to the spherically symmetric approximation, one may select as trial amplitude $\tilde{\Phi}(r)$ the following function:
\begin{equation}
  \label{eq:iii.50}
  \tilde{\Phi}(r)=\Phi_*r^\nu{\rm e}^{-\beta r}\ ,
\end{equation}
where the normalization condition (\ref{eq:iii.46}) fixes the constant $\Phi_*$ to
\begin{equation}
  \label{eq:iii.51}
  \Phi_*^2=\frac{2}{\pi}\cdot\frac{(2\beta)^{2\nu+2}}{\Gamma(2\nu+2)}\ .
\end{equation}
This trial amplitude (\ref{eq:iii.50}) has only two variational parameters (i.\,e. $\beta$
and $\nu$), and therefore one cannot expect that the corresponding approximate energy
spectrum will turn out to be more accurate than up to some few percent ($\lesssim10\%$,
say). Nevertheless this order of accuracy should be sufficient in order to estimate the
magnitude of the anisotropic corrections which, plausibly, will be found to be somewhat smaller
than the error induced by just the spherically symmetric approximation.  In any case, once
the trial amplitude $\tilde{\Phi}(r)$ is fixed, one can proceed to determine the value of
the non-relativistic energy functional $\tilde{\mathbb{E}}_{\rm [T]}$ upon this trial
configuration which then yields the corresponding energy function $\mathbb{E}^{\rm
  (IV)}(\beta,\nu)$ through the following steps:
\begin{enumerate}[\bf (i)]
  \item Calculate the kinetic energy $E_{\rm kin}$ (\ref{eq:iii.42})-(\ref{eq:iii.44}) by means of the selected trial function $\tilde{\Phi}(r)$ (\ref{eq:iii.50}) and find this quantity to be of the form
    \begin{eqnarray}
      \label{eq:iii.52}
      &\displaystyle E_{\rm kin}\;=\;\frac{e^2}{2a_{\rm B}}\,(2a_{\rm B}\beta)^2\cdot\varepsilon_{\rm kin}(\nu)§\\
      \nonumber
      &\displaystyle (a_{\rm B}\;\doteqdot\;\frac{\hbar^2}{Me^2}\quad\ldots\quad\mbox{Bohr radius})&
    \end{eqnarray}
  with the kinetic function $\varepsilon_{\rm kin}(\nu)$ being given by
    \begin{equation}
      \label{eq:iii.53}
      \varepsilon_{\rm kin}(\nu)=\frac{1}{2\nu+1}\left(\frac{1}{4}+\frac{\lP^2}{2\nu}\right)\,.
    \end{equation}
  \item Solve the Poisson equation (\ref{eq:iii.36}) for the spherically symmetric
    potential $\eklo{p}{A}_0(r)$ (\textbf{App.A}) and compute by means of this result the
    eloctrostatic gauge field energy $E_{\rm R}^{\rm [e]}$ (\ref{eq:iii.45}) which then
    emerges as follows
    \begin{equation}
      \label{eq:iii.54}
      E_{\rm R}^{\rm [e]}=-\frac{e^2}{\aB}\left(2\beta\aB\right) \cdot\varepsilon_{\rm pot}(\nu)\ .
    \end{equation}
    Here the potential function $\varepsilon_{\rm pot}(\nu)$ may be expressed in two
    alternative ways on account of the Poisson identity (\ref{eq:iii.47}), see equation
    (\ref{eq:a28}) of \textbf{App.A}. (For a sketch of the spherically symmetric trial
    potentials $\eklo{p}{A}_0(r)$ (\ref{eq:iii.37}), see \textbf{Fig.A.I}).
  \item Substituting both energy contributions $E_{\rm kin}$ (\ref{eq:iii.52}) and $E_{\rm
      R}^{\rm [e]}$ (\ref{eq:iii.54}) back into the energy functional
    $\tilde{\mathbb{E}}_{[\Phi]}$ (\ref{eq:iii.41}) yields a certain function
    $\mathbb{E}^{\rm (IV)}(\beta,\nu)$ of the two variational parameters $\beta$ and
    $\nu$, see equation (\ref{eq:a22}) of \textbf{App.A}:
    \begin{equation}
      \label{eq:iii.54'}
      \roek{E}{IV}(\beta,\nu)=\frac{e^2}{a_{\rm B}}\left\{\big(2a_{\rm B}\beta\big)^2\cdot\varepsilon_{\rm kin}(\nu)-\big(2a_{\rm B}\beta\big)\cdot\varepsilon_{\rm pot}(\nu)\right\}\,.
    \end{equation}
    (Observe here that both the normalization condition (\ref{eq:iii.46}) for the wave
    amplitude $\tilde{\Phi}(r)$ and the Poisson identity (\ref{eq:iii.47}) are satisfied,
    so that the energy function $\mathbb{E}^{\rm [IV]}(\beta,\nu)$ due to
    $\tilde{\mathbb{E}}_{\rm [T]}$ (\ref{eq:iii.41}) consists solely of the kinetic energy
    and the gauge field energy).
  \item The extremalization process (\ref{eq:iii.40}) can now be carried through for the
    obtained function $\mathbb{E}^{\rm [IV]}(\beta,\nu)$ (\ref{eq:iii.54'}). Here it is
    most convenient to perform this in two steps: After the extremalization with respect
    to $\beta$ (cf. (\ref{eq:a23})), there remains to look for the minimal values
    ($\mathbb{E}_{\rm T}(\nu_*)$, say) of the reduced energy function $\mathbb{E}_{\rm
      T}(\nu)$
    \begin{subequations}
      \begin{align}
      \label{eq:iii.55a}
      \mathbb{E}_{\rm T}(\nu)&\doteqdot-\frac{e^2}{4a_{\rm B}}\cdot S_\wp(\nu)\\
      \label{eq:iii.55b}
      S_\wp(\nu)&\doteqdot\frac{\varepsilon_{\rm pot}^2(\nu)}{\varepsilon_{\rm kin}(\nu)}\ .
      \end{align}
    \end{subequations}
    Such a minimalization process may be achieved by means of an appropriate computer
    program. The conventional predictions (\ref{eq:i.4}) would be obtained if the {\em spectral
      function}\/ $S_\wp(\nu)$ took its maximal values at
    \begin{eqnarray}
    \label{eq:iii.56}
    S_\wp\Big|_{\rm max}&\Rightarrow&\frac{1}{n_\wp^2}\ .\\
    \nonumber
    (n_\wp&=&1,\,2,\,3,\,...)
    \end{eqnarray}
    Of course, this dream result cannot be attained by use of such a simple trial
    amplitude as the selected $\tilde{\Phi}(r)$ (\ref{eq:iii.50}); but nevertheless the
    present approximation procedure yields predictions $\mathbb{E}_{\rm T}(\nu_*)$ which
    come amazingly close to the conventional predictions $E_{\rm conv}^{(n)}$
    (\ref{eq:i.4}), see the subsequent \textbf{table~1}.
\end{enumerate}


\begin{flushleft}\label{table1}
  \begin{tabular}{|c||c|c|c|c|}
  \hline
$\nP\ (=\lP+1)$ & $\Ea{E}{n}{conv}$\ [eV], (\ref{eq:i.4}) & $\nrf{E}{T}(\nu_*^{[n]})$\ [eV],(\ref{eq:iii.55a}) &
  $\nu_*^{[n]}$ & $\frac{\Ea{E}{n}{conv}-\nrf{E}{T}(\nu_*^{[n]})}{\Ea{E}{n}{conv}} [\%]$ \\
  \hline\hline
1 & $  -6.80290 \ldots $ & $ -7.23055\ldots$ & $-0.2049$ & -6.3\\ \hline 
2 & $  -1.70072 \ldots $ & $ -1.55087\ldots$ & $1.7942$ & 8.8\\ \hline 
3 & $  -0.75588 \ldots $ & $ -0.66914\ldots$ & $3.7528$ & 11.5\\ \hline 
4 & $  -0.42518 \ldots $ & $ -0.37366\ldots$ & $5.8740$ & 12.1\\ \hline 
5 & $  -0.27212 \ldots $ & $ -0.23906\ldots$ & $8.1307$ & 12.1\\ \hline 
6 & $  -0.18897 \ldots $ & $ -0.16637\ldots$ & $10.5044$ & 12.0\\ \hline 
10 & $  -0.06803 \ldots $ & $ -0.06069\ldots$ & $20.9538$ & 10.8\\ \hline 
15 & $  -0.03024 \ldots $ & $ -0.02735\ldots$ & $35.7017$ & 9.5\\ \hline 
20 & $  -0.01701 \ldots $ & $ -0.01555\ldots$ & $51.9196$ & 8.6\\ \hline 
25 & $  -0.01088 \ldots $ & $ -0.01003\ldots$ & $69.3583$ & 7.9\\ \hline 
30 & $  -0.00756 \ldots $ & $ -0.00701\ldots$ & $87.8627$ & 7.3\\ \hline 
35 & $  -0.00555 \ldots $ & $ -0.00518\ldots$ & $107.3506$ & 6.8\\ \hline 
40 & $  -0.00425 \ldots $ & $ -0.00398\ldots$ & $127.8215$ & 6.4\\ \hline 
45 & $  -0.00336 \ldots $ & $ -0.00316\ldots$ & $149.3660$ & 6.1\\ \hline 
50 & $  -0.00272 \ldots $ & $ -0.00256\ldots$ & $172.1617$ & 5.8\\ \hline 
60 & $  -0.00189 \ldots $ & $ -0.00179\ldots$ & $222.5187$ & 5.3\\ \hline 
70 & $  -0.00139 \ldots $ & $ -0.00132\ldots$ & $280.9499$ & 4.8\\ \hline 
80 & $  -0.00106 \ldots $ & $ -0.00102\ldots$ & $348.2861$ & 4.5\\ \hline 
90 & $  -0.00084 \ldots $ & $ -0.00081\ldots$ & $422.8567$ & 4.1\\ \hline 
100 & $  -0.00068 \ldots $ & $ -0.00065\ldots$ & $501.4648$ & 3.8\\ \hline 
\end{tabular}
\end{flushleft}
{\textbf{Table 1:}\hspace{0.5cm} \emph{\large\textbf{Conventional Predictions and
 Spherically Symmetric  \hspace*{6cm}Approximation}}  }

\mytable{ \emph{Table 1: Conventional Predictions and Spherically \\\hspace*{3mm} Symmetric Approximation}}

The minimalization of the RST energy function~$\nrf{E}{T}(\nu)$
(\ref{eq:iii.55a})-(\ref{eq:iii.55b}) with respect to the second variational
parameter~$\nu$ occurs at the values~$\nu_*^{[n]}$ (fourth column) and yields the
corresponding RST prediction~$\nrf{E}{T}(\nu_*^{[n]})$ (third column). The relative
deviation from the conventional prediction~$\Ea{E}{n}{conv}$ (\ref{eq:i.4}) is displayed
in the last column; this deviation decreases from (roughly) 10\% at low quantum
numbers~$\nP\lesssim 10$ up to 4\% for high quantum numbers~$(\nP\sim 100)$. Such a
decreasement of the relative deviation is necessary in order that no cross-over with the
conventional predictions~$\Ea{E}{n}{conv}$ can occur (see~\textbf{Fig.III.A}). The
negative deviation of the groundstate~($\nP=1$, first line) is presumably caused by the
weak singularity of the trial amplitude~$\tilde{\Phi}(r)$~\cite{ms1}) (here
$\nu_*=-0,204\ldots$, see \textbf{Fig.4b} of ref.~\cite{ms1}). Such a singularity signals
a certain crowding of the electric charge in the vicinity of the origin~$(r=0)$ with a
non-zero Dirac current~$\vri{k}{p}$ on the z-axis, see both \textbf{Fig.1} and equation
(VI.30) in ref.~\cite{ms1}. Consequently, \emph{magnetism} may not be neglected for this
situation very close to the origin, and the presently used \emph{electrostatic}
approximation is assumed to break down for the groundstate~$(\nP=1)$ from this reason. On
the other hand, the excited states~$(\nP>1)$ have vanishing Dirac current on the z-axis so
that no magnetic excess force can occur and the electrostatic approximation allows for
sufficiently realistic predictions.

The average deviation~($\overline{\ekru{\Delta}{T}}$, say) due to the presently considered
lowest order of the spherically symmetric approximation may be defined through
\begin{equation}
  \label{eq:iii.58}
  \overline{\ekru{\Delta}{T}} \doteqdot \frac{1}{20} \sum_{\nP}^{100}
  |\ru{\Delta}{T}(\nu_*^{[n]})|\ ,
\end{equation}
where the individual deviations~$\ru{\Delta}{T}(\nu_*^{[n]})$ are specified by the last
column of table~1, i.e.:
\begin{equation}
  \label{eq:iii.59}
  \ru{\Delta}{T}(\nu_*^{[n]}) \doteqdot \frac{\Ea{E}{n}{conv}-\nrf{E}{T}(\nu_*^{[n]})
  }{\Ea{E}{n}{conv}}\ .
\end{equation}
Thus the average deviation of the present predictions is found as~$
\overline{\ekru{\Delta}{T}}=7,7\%$. This will become halved through regarding the
anisotropic corrections (\textbf{Sect.V}).
\begin{center}
\epsfig{file=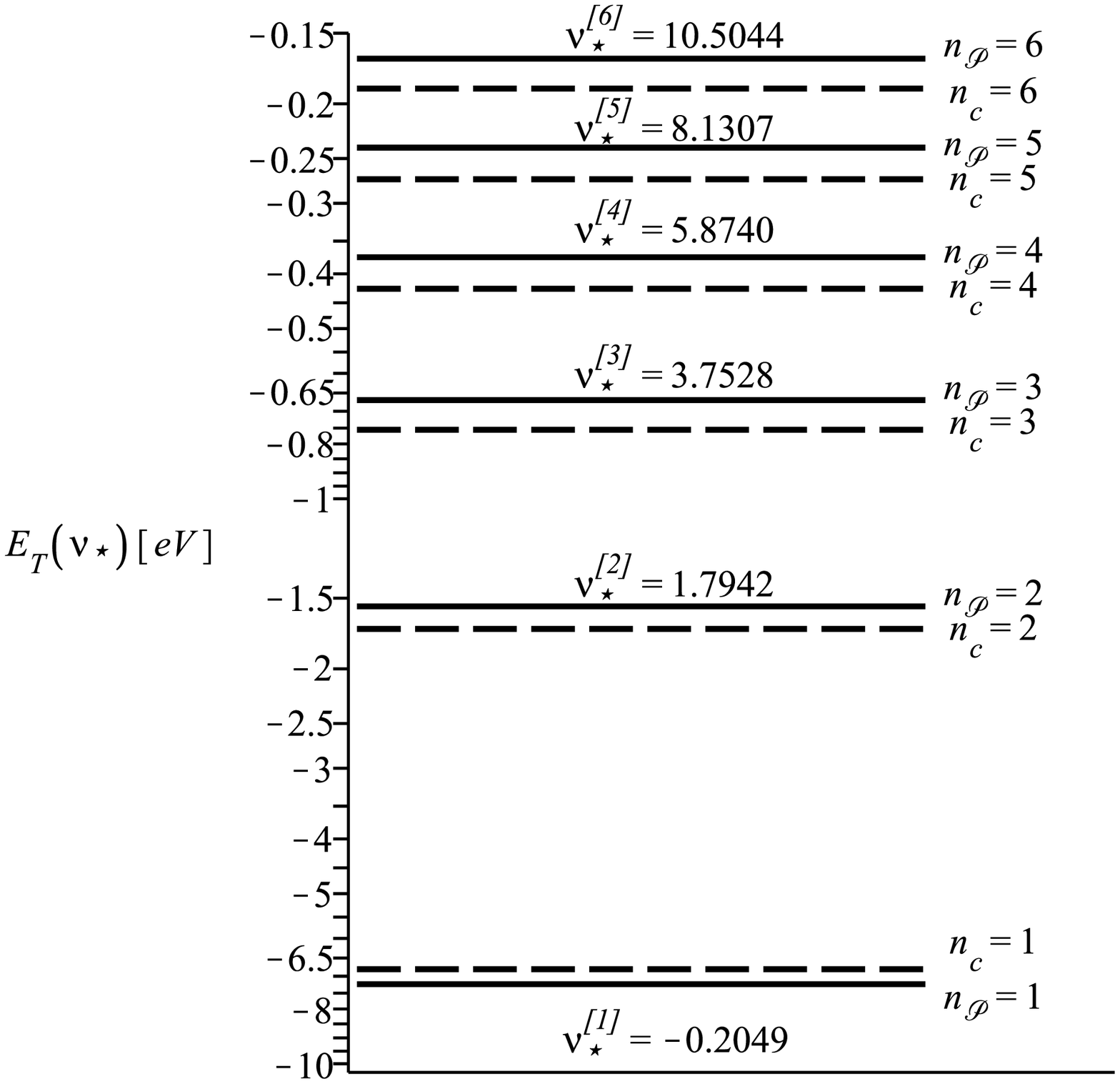,height=15cm}
\end{center}
{\textbf{Fig.~III.A}\hspace{5mm} \emph{\large\textbf{Energy Spectrum in the Spherically
    Symmetric \\ \centerline{Approximation}}}}
\indent
\myfigure{{Fig.~III.A: Energy Spectrum in the Spherically \\\hspace*{4mm}
    Symmetric Approximation}}

The minimal values~$\nrf{E}{T}(\nu_*^{[n]})$ ($\leadsto$ solid lines) of the energy
function~$\nrf{E}{T}(\nu)$ (\ref{eq:iii.55a}) are always unambiguously related to the
corresponding conventional predictions~$\Ea{E}{n}{conv}$ (\ref{eq:i.4}) (broken lines)
because the relative deviations do not amount to more than (roughly) 10\%, see also
\textbf{table~1} on p.~\pageref{table1}. Therefore there occurs no crossing-over of the
conventional and RST predictions which should provide a solid basis for estimating the
magnitude of the anisotropic corrections in \textbf{Sect.~IV}. The latter na\"ive type of
corrections turns out to be of the order of (roughly) 0.4\% for the groundstate, see
equations (\ref{eq:iv.45})-(\ref{eq:iv.46}) below, but would deteriorate the present
``isotropic'' RST predictions by (roughly) 8\% already for the first excited state, see
\textbf{App.C}. These unrealistic \emph{large} anisotropy corrections for the
\emph{excited} states necessitates then a reconsideration and improvement of the applied
approximation method (\textbf{Sect.~V}).

\section{Anisotropic Corrections of Gauge Potential}
\indent

The necessity to go beyond the spherically symmetric approximation has already been
stressed in the introduction ({\bf Sect.\,I}). But in view of the preceding difficulties
with obtaining exact solutions of the spherically symmetric type one will not expect that
{\em anisotropic exact}\/ solutions could ever be found. On the other hand, an estimate of
the magnitude of the anisotropic effects would be highly desirable and we will try to
attain this now at least partially, namely by expanding the formal solution
(\ref{eq:ii.30}) for the electrostatic potential $\rklo{p}{A}_0(\vr)$ with respect to the
longitudinal variable $\vartheta$ of the spherical polar coordinates
$(r,\vartheta,\phi)$. This means that we relax the spherically symmetric approximation
(\ref{eq:iii.1}) to the following weaker condition:
\begin{equation}
  \label{eq:iv.1}
  \rklo{p}{A}_0(\vr)\Rightarrow\gklo{p}{A}_0(r,\vartheta)\ .
\end{equation}
Here the $\vartheta$-dependent potential $\gklo{p}{A}_0(r,\vartheta)$ needs not be the {\em exact}\/ value of the integral in (\ref{eq:ii.30}) but we will be satisfied with some approximative value in lowest order due to a certain {\em expansion with respect to the anisotropy effect}\/.

In any case, the latter effect will not be fully taken into account since we still adopt that form of the wave amplitudes $\rklo{p}{\R}_\pm,\,\rklo{p}{\S}_\pm$ as it results from the combination of the transformations (\ref{eq:iii.13a})-(\ref{eq:iii.13b}), (\ref{eq:iii.14a})-(\ref{eq:iii.14b}), (\ref{eq:iii.17a})-(\ref{eq:iii.17b}), (\ref{eq:iii.19})-(\ref{eq:iii.20}) and (\ref{eq:iii.23a})-(\ref{eq:iii.23b}), i.\,e. in the last end:
\begin{subequations}
  \begin{align}
  \label{eq:iv.2a}
  \rklo{p}{\R}_\pm(\vr)&\Rightarrow\frac{\tilde{\Phi}_\pm(r)\cdot f_R(\vartheta)}{\sqrt{r\sin\vartheta}}\\
  \label{eq:iv.2b}
  \rklo{p}{\S}_\pm(\vr)&\Rightarrow\frac{\tilde{\Phi}_\pm(r)\cdot f_S(\vartheta)}{\sqrt{r\sin\vartheta}}\ .
  \end{align}
\end{subequations} 
Observing here the solutions (\ref{eq:iii.19})-(\ref{eq:iii.20}) for the angular functions $f_R(\vartheta)$ and $f_S(\vartheta)$, one finds for the Dirac density $\rklo{p}{k}_0(\vr)$ (\ref{eq:ii.19})
\begin{equation}
  \label{eq:iv.3}
  \rklo{p}{k}_0(\vr)\Rightarrow\gklo{p}{k}_0(r,\vartheta)=\frac{\tilde{\Phi}^2_+(r)+\tilde{\Phi}^2_-(r)}{4\pi r\sin\vartheta}\ .
\end{equation}
If this result is substituted into equation (\ref{eq:ii.30}), the electrostatic potential $\rklo{p}{A}_0(\vr)$ reappears as $\gklo{p}{A}_0(r,\vartheta)$ in the following form
\begin{equation}
  \label{eq:iv.4}
  \gklo{p}{A}_0(r,\vartheta)=\as\int\frac{d^3\vec{r}\,'}{4\pi r'\sin\vartheta'}\frac{\tilde{\Phi}^2_+(r')+\tilde{\Phi}^2_-(r')}{\|\vr-\vec{r}\,'\|}\ ,
\end{equation}
or in the non-relativistic approximation ($\tilde{\Phi}_+(r)\Rightarrow\tilde{\Phi}(r),\,\tilde{\Phi}_-(r)\Rightarrow0$)
\begin{equation}
  \label{eq:iv.5}
  \gklo{p}{A}_0(r,\vartheta)= \frac{\as}{4\pi}\int\frac{d^3\vec{r}\,'}{r'\sin\vartheta'}\frac{\tilde{\Phi}^2(r')}{\|\vr-\vec{r}\,'\|}\ .
\end{equation}

From here it is seen that the anisotropy of the {\em wave amplitudes}\/ themselves is taken into account only to that extent as it is assumed also in the spherically symmetric approximation for the potential $\eklo{p}{A}_0(r)$ (\ref{eq:iii.37})! Thus, the main anisotropic effect is thought to be due to the formal modification of the three-volume element in equation (\ref{eq:iv.5})
\begin{equation}
  \label{eq:iv.6}
  d^3\vr\Rightarrow\frac{d^3\vr}{r\sin\vartheta}=r\,dr\,d\vartheta\,d\phi\ ,
\end{equation}
where the denominator is a consequence of the former transformation(\ref{eq:iii.14a})-(\ref{eq:iii.14b}) of the wave amplitudes. Accordingly, our step beyond the spherically symmetric approximation consists now in computing the potential $\gklo{p}{A}_0(r,\vartheta)$
\begin{equation}
  \label{eq:iv.7}
  \gklo{p}{A}_0(r,\vartheta)=\frac{\as}{4\pi}\int dr'\,d\vartheta'\,d\phi'\;\frac{r'\cdot\tilde{\Phi}^2(r')}{\|\vr-\vec{r}\,'\|}
\end{equation}
in an approximate way (i.\,e. in lowest order beyond the spherically symmetric
approximation (\ref{eq:iii.1})) and then using the result for calculating the value of the
energy functional $\tilde{\mathbb{E}}_{[\Phi]}$ (\ref{eq:iii.41}) on the present trial
configuration $\{\tilde{\Phi}(r),\,\gklo{p}{A}_0(r,\vartheta)\}$. The corresponding
extremal values of $\tilde{\mathbb{E}}_{[\Phi]}$ within this class of trial configurations
do then determine again the approximative RST energy levels of the bound system which must
be opposed to both the conventional predictions and to the spherically symmetric RST
predictions of {\bf Sect.~III}. Anticipating the result of this comparison, we will find
that these anisotropy corrections are lowering the groundstate energy (first line of
table~1 on p.~\pageref{table1}) by 0.4\% so that the anisotropy correction appears to be
actually a negligeable pertubation of the spherical symmetry. However, this pleasant
result is readily shown to hold only for the groundstate (\textbf{App.B}) and even here is
spurious so that we are forced to develeop a more realistic pertubation procedure
\textbf{(Sect.V)}.  \vspace{4ex}
\begin{center}
\mysubsection{1.\ Anisotropic Potentials}
\end{center}

The wanted method of expanding the electrostatic potential
$\gklo{p}{A}_0(r,\vartheta)$~(\ref{eq:iv.7}) with respect to the magnitude of anisotropy
will (in the simplest case) consist in na\"ively expanding the denominator as follows:
\begin{eqnarray}
  \nonumber
  \frac{1}{\|\vr-\vec{r}\,'\|}&=& \frac{1}{(r^2+r'^2)^{\frac{1}{2}}}+\frac{rr'}{(r^2+r'^2)^{\frac{3}{2}}}\,(\hat{\vr}\sdot\hat{\vec{r}}\,')+\frac{3}{2}\frac{r^2r'^2}{(r^2+r'^2)^{\frac{5}{2}}}\,(\hat{\vr}\sdot\hat{\vec{r}}\,')^2\\
  \label{eq:iv.8}
  &&{}+\frac{5}{2}\frac{r^3r'^3}{(r^2+r'^2)^{\frac{7}{2}}}\,(\hat{\vr}\sdot\hat{\vec{r}}\,')^3+ \frac{35}{8}\frac{r^4r'^4}{(r^2+r'^2)^{\frac{9}{2}}}\,(\hat{\vr}\sdot\hat{\vec{r}}\,')^4+...\\
  \nonumber\\
  \nonumber
  \big(\hat{\vr}&\doteqdot&\frac{\vr}{\|\vr\,\|}=\frac{\vr}{r}\;\Rightarrow\;\|\hat{\vr}\,\|^2=\|\hat{\vr}\,'\|^2=1\big)\ .
\end{eqnarray}
Obviously, such a na\"ive expansion would induce a corresponding expansion of the potential
$\gklo{p}{A}_0(r,\vartheta)$ of the following product type:
\begin{eqnarray}
  \nonumber
  \gklo{p}{A}_0(r,\vartheta)&=&\gklo{p}{A}_{\sf I}(\vartheta)\cdot\gklo{p}{A}_{\sf I}(r)+\gklo{p}{A}_{\sf II}(\vartheta)\cdot\gklo{p}{A}_{\sf II}(r)+\gklo{p}{A}_{\sf III}(\vartheta)\cdot\gklo{p}{A}_{\sf III}(r)\\
  \label{eq:iv.9}
  &&{}+\gklo{p}{A}_{\sf IV}(\vartheta)\cdot\gklo{p}{A}_{\sf IV}(r)+\gklo{p}{A}_{\sf V}(\vartheta)\cdot\gklo{p}{A}_{\sf V}(r)+...
\end{eqnarray}
Here, the angular factors are found to look as follows
\begin{subequations}
  \begin{align}
  \label{eq:iv.10a}
  \gklo{p}{A}_{\sf I}(\vartheta)&\doteqdot\int\frac{d\vartheta'\,d\phi'}{4\pi}=\frac{\pi}{2}\\
  \label{eq:iv.10b}
  \gklo{p}{A}_{\sf II}(\vartheta)&\doteqdot\int\frac{d\vartheta'\,d\phi'}{4\pi}\,(\hat{\vr}\sdot\hat{\vec{r}}\,')=0\\
  \label{eq:iv.10c}
  \gklo{p}{A}_{\sf III}(\vartheta)&\doteqdot \frac{3}{2}\int\frac{d\vartheta'\,d\phi'}{4\pi}\,(\hat{\vr}\sdot\hat{\vec{r}}\,')^2=\frac{3}{16}\,\pi(1+\cos^2\vartheta)\\
  \label{eq:iv.10d}
  \gklo{p}{A}_{\sf IV}(\vartheta)&\doteqdot \frac{5}{2}\int\frac{d\vartheta'\,d\phi'}{4\pi}\,(\hat{\vr}\sdot\hat{\vec{r}}\,')^3=0\\
    \label{eq:iv.10e}
  \gklo{p}{A}_{\sf V}(\vartheta)&\doteqdot \frac{35}{8}\int\frac{d\vartheta'\,d\phi'}{4\pi}\,(\hat{\vr}\sdot\hat{\vec{r}}\,')^4=\frac{105}{128}\,\pi\{\cos^4\vartheta+\cos^2\vartheta\sin^2\vartheta+\frac{3}{8}\sin^4\vartheta\}\\
  \nonumber
  \vdots\quad\:&
  \end{align}
\end{subequations}
Evidently, these angular factors do not display the expected tendency of decreasing with increasing order of the expansion terms; therefore such a decreasing tendency must arise now with the radial factors, otherwise the series expansion (\ref{eq:iv.9}) of the potential $\gklo{p}{A}_0(r,\vartheta)$ would not be convergent.

The radial factors emerging in the expansion (\ref{eq:iv.9}) are found to look as follows:
\begin{subequations}
  \begin{align}
  \label{eq:iv.11a}
  \gklo{p}{A}_{\sf I}(r)&=\as\cdot\int_0^\infty dr'\,\frac{r'\cdot\tilde{\Phi}^2(r')}{(r^2+r'^2)^\frac{1}{2}}\\
  \label{eq:iv.11b}
  \gklo{p}{A}_{\sf II}(r)&=\as\cdot r\int_0^\infty dr'\,\frac{r'^2\cdot\tilde{\Phi}^2(r')}{(r^2+r'^2)^\frac{3}{2}}\\
  \label{eq:iv.11c}
  \gklo{p}{A}_{\sf III}(r)&=\as\cdot r^2\int_0^\infty dr'\,\frac{r'^3\cdot\tilde{\Phi}^2(r')}{(r^2+r'^2)^\frac{5}{2}}\\
  \label{eq:iv.11d}
  \gklo{p}{A}_{\sf IV}(r)&=\as\cdot r^3\int_0^\infty dr'\,\frac{r'^4\cdot\tilde{\Phi}^2(r')}{(r^2+r'^2)^\frac{7}{2}}\\
  \label{eq:iv.11e}
  \gklo{p}{A}_{\sf V}(r)&=\as\cdot r^4\int_0^\infty dr'\,\frac{r'^5\cdot\tilde{\Phi}^2(r')}{(r^2+r'^2)^\frac{9}{2}}\\
  \nonumber
  \pmb{\vdots}
  \end{align}
\end{subequations}
Observe here that both potentials $\gklo{p}{A}_{\sf II}(r)$ (\ref{eq:iv.11b}) and
$\gklo{p}{A}_{\sf IV}(r)$ (\ref{eq:iv.11d}) are irrelevant because they become eliminated
by their vanishing angular pre-factors $\gklo{p}{A}_{\sf II}(\vartheta)$ (\ref{eq:iv.10b})
and $\gklo{p}{A}_{\sf IV}(\vartheta)$ (\ref{eq:iv.10d}). Concerning the relative magnitude
of the remaining potentials $\gklo{p}{A}_{\sf I}(r)$ (\ref{eq:iv.11a}), $\gklo{p}{A}_{\sf
  III}(r)$ (\ref{eq:iv.11c}) and $\gklo{p}{A}_{\sf V}(r)$ (\ref{eq:iv.11e}) one could of
course explicitly compute these potentials by resorting to the former trial amplitude
$\tilde{\Phi}(r)$ (\ref{eq:iii.50}) which then would immediately yield the desired
relative magnitudes. However, one may circumvent those unwieldy integrals by establishing
a certain differential link between these radial auxiliary potentials
(\ref{eq:iv.11a})-(\ref{eq:iv.11e}) and then adopt a plausible guess for the lowest-order
potential $\gklo{p}{A}_{\sf I}(r)$. This guess can then be taken as the point of departure
for calculating the higher-order potentials $\gklo{p}{A}_{\sf III}(r),\,\gklo{p}{A}_{\sf
  V}(r),\,...$ just by means of that established differential link (see below).
\pagebreak

Concerning now the relative magnitude of the partial potentials $\gklo{p}{A}_{\sf
  I}(r,\vartheta)=\gklo{p}{A}_{\sf I}(\vartheta)\cdot\gklo{p}{A}_{\sf I}(r)$ etc.,
cf. (\ref{eq:iv.9}), it is obviously sufficient to consider their radial factors
$\gklo{p}{A}_{\sf I}(r)$ etc., cf. (\ref{eq:iv.11a})-(\ref{eq:iv.11e}). Here, the first
one (i.\,e. $\gklo{p}{A}_{\sf I}(r)$ (\ref{eq:iv.11a})) is singled out by its boundary
values for $r=0,\infty$. At the origin ($r=0$) one finds
\begin{equation}
  \label{eq:iv.12}
  \gklo{p}{A}_{\sf I}(r)\Big|_{r=0}=\as\int_0^\infty dr'\tilde{\Phi}^2(r')\ ,
\end{equation}
i.\,e. for the spherically symmetric trial function $\tilde{\Phi}(r)$ (\ref{eq:iii.50})
\begin{equation}
  \label{eq:iv.13}
  \gklo{p}{A}_{\sf I}(r)\Big|_{r=0}\Rightarrow\frac{2}{\pi}\,(2\beta\as)\cdot\frac{1}{2\nu+1}\ .
\end{equation}
Combinig this with its angular associate $\gklo{p}{A}_{\sf I}(\vartheta)$ (\ref{eq:iv.10a}) yields then at the origin
\begin{equation}
  \label{eq:iv.14}
  \gklo{p}{A}_{\sf I}(r,\vartheta)\Big|_{r=0}=\left[\gklo{p}{A}_{\sf I}(\vartheta)\cdot\gklo{p}{A}_{\sf I}(r)\right]_{r=0}\Rightarrow\frac{2\beta\as}{2\nu+1}\ ,
\end{equation}
which just agrees with the value of the spherically symmetric approximation $\eklo{p}{A}_0(r)$ at the origin ($r=0$), cf. the equation (\ref{eq:a10}) of {\bf App.\,A}. On the other hand, this value (\ref{eq:iv.12}) of the first auxiliary potential $\gklo{p}{A}_{\sf I}(r,\vartheta)$ (or its specialization (\ref{eq:iv.14}), resp.) at the origin ($r=0$) actually coincides with the corresponding value of the ``exact'' solution $\gklo{p}{A}_0(r,\vartheta)$ (\ref{eq:iv.7}):
\begin{equation}
  \label{eq:iv.15}
  \gklo{p}{A}_0(r,\vartheta)\Big|_{r=0}=\frac{\as}{4\pi}\int dr'\, d\vartheta'\,d\phi'\;\tilde{\Phi}^2(r')= \frac{\pi}{2}\,\as\int_0^\infty dr'\tilde{\Phi}^2(r')\equiv\frac{\pi}{2}\cdot\gklo{p}{A}_{\sf I}(r=0)\ ,
\end{equation}
see equation (\ref{eq:iv.12}). Thus the conclusion is now the following: since already the first auxiliary potential $\gklo{p}{A}_{\sf I}(r,\vartheta)=\gklo{p}{A}_{\sf I}(\vartheta)\cdot\gklo{p}{A}_{\sf I}(r)$ in the expansion (\ref{eq:iv.9}) for $\gklo{p}{A}_0(r,\vartheta)$ adopts the correct value of $\gklo{p}{A}_0(r,\vartheta)$ at the origin, all other auxiliary potentials must necessarily vanish at the origin ($r=0$); or expressed in terms of the radial potentials (\ref{eq:iv.11b})-(\ref{eq:iv.11e})
\begin{equation}
  \label{eq:iv.16}
  \gklo{p}{A}_{\sf III}(r)\Big|_{r=0}=\gklo{p}{A}_{\sf V}(r)\Big|_{r=0}=\pmb{\ldots}=0\ .
\end{equation}

A similar conclusion does also hold for the behaviour of the radial auxiliary potentials at infinity ($r\rightarrow\infty$). First, observe here that the ``exact'' anisotropic solution (\ref{eq:iv.7}) adopts the Coulomb form for $r\rightarrow\infty$
\begin{equation}
  \label{eq:iv.17}
  \lim_{r\rightarrow\infty}\gklo{p}{A}_0(r,\vartheta)=\frac{\as}{r}\cdot\frac{\pi}{2}\int_0^\infty dr'\;r'\tilde{\Phi}^2(r')=\frac{\as}{r}\ ,
\end{equation}
provided the non-relativistic wave amplitude $\tilde{\Phi}(r)$ obeys the normalization condition (\ref{eq:iii.38}). But on the other hand, the first-order radial potential $\gklo{p}{A}_{\sf I}(r)$ (\ref{eq:iv.11a}) has just the same limit form (apart from the angular factor (\ref{eq:iv.10a})):
\begin{equation}
  \label{eq:iv.18}
  \lim_{r\rightarrow\infty}\gklo{p}{A}_{\sf I}(r)=\frac{2}{\pi}\cdot\frac{\as}{r}\ ,
\end{equation}
and therefore the expansion (\ref{eq:iv.9}) tells us that the higher-order potentials must vanish faster than $\frac{1}{r}$ at infinity:
\begin{equation}
  \label{eq:iv.19}
  \lim_{r\rightarrow\infty}\gklo{p}{A}_{\sf III}(r)=\lim_{r\rightarrow\infty}\gklo{p}{A}_{\sf V}(r)=\,\ldots\,= {\scriptstyle\mathcal{O}}\!\:(\frac{1}{r^2})\ .
\end{equation}

Since both the second potential $\gklo{p}{A}_{\sf II}(r)$ (\ref{eq:iv.11b}) and the fourth potential $\gklo{p}{A}_{\sf IV}(r)$ (\ref{eq:iv.11d}) become annihilated by their angular counterparts $\gklo{p}{A}_{\sf II}(\vartheta)$ (\ref{eq:iv.10b}) and $\gklo{p}{A}_{\sf IV}(\vartheta)$ (\ref{eq:iv.10d}), it is sufficient to look for the announced link only between the non-zero potentials $\gklo{p}{A}_{\sf I}(r)$ (\ref{eq:iv.11a}), $\gklo{p}{A}_{\sf III}(r)$ (\ref{eq:iv.11c}) and $\gklo{p}{A}_{\sf V}(r)$ (\ref{eq:iv.11e}). For this purpose, one differentiates twice the first potential $\gklo{p}{A}_{\sf I}(r)$ (\ref{eq:iv.11a}) and thereby finds immediately for the third potential $\gklo{p}{A}_{\sf III}(r)$ (\ref{eq:iv.11c})
\begin{equation}
  \label{eq:iv.20}
  \gklo{p}{A}_{\sf III}(r)=-\frac{1}{3}\,r\,\frac{d^2}{dr^2}\left[r\cdot\gklo{p}{A}_{\sf I}(r)\right]\ .
\end{equation}
A similar procedure yields for the fifth potential $\gklo{p}{A}_{\sf V}(r)$ (\ref{eq:iv.11e})
\begin{equation}
  \label{eq:iv.21}
  \gklo{p}{A}_{\sf V}(r)=-\frac{1}{35}\,r^3\,\frac{d}{dr}\left[\frac{1}{r^4}\frac{d}{dr}\big(r^3\cdot\gklo{p}{A}_{\sf III}(r)\big)\right]\ ,
\end{equation}
etc. The importance of this procedure is now that one can guess some reasonable
approximation for the lowest-order potential $\gklo{p}{A}_{\sf I}(r)$ which is manageable
more conveniently than the integral on the right-hand side of equation (\ref{eq:iv.11a})
but which on the other hand comes sufficiently close to the value of that integral. If this
guess for $\gklo{p}{A}_{\sf I}(r)$ is substituted in equation (\ref{eq:iv.20}), one
obtains the corresponding approximative expressions for the higher-order potentials
$\gklo{p}{A}_{\sf III}(r)$ and $\gklo{p}{A}_{\sf V}(r)$ whose relative magnitude can then
be estimated.  \vspace{4ex}
\begin{center}
\mysubsection{2.\ Simplified Groundstate Potentials (\/{\boldmath$\nu=0$}\/)}
\end{center}

For the purpose of a simple demonstration of our approximation method, a {\em
  preliminary}\/ choice for the lowest-order potential $\gklo{p}{A}_{\sf I}(r)$ could
perhaps consist in the spherically symmetric potential $\eklo{p}{A}_0(r)$ ((\ref{eq:a2}) plus
(\ref{eq:a5}) for $\nu=0$) which has been treated extensively in the precedent {\bf
  Sect.\,III} (and also in {\bf App.\,A}, where the trial amplitude $\tilde{\Phi}(r)$
(\ref{eq:iii.50}) does underlie). Defering again the purely technical subtleties to {\bf
  App.\ C}, we can be satisfied here for the sake of demonstration with the special case
$\nu=0$ where the selected trial amplitude $\tilde{\Phi}(r)$ (\ref{eq:iii.50}) becomes
especially simple because there remains only one variational parameter ($\beta$):
\begin{subequations}
  \begin{align}
  \label{eq:iv.22a}
  \tilde{\Phi}(r)&\Rightarrow\Phi_*{\rm e}^{-\beta r}\\
  \label{eq:iv.22b}
  \Phi_*&=\sqrt{\frac{2}{\pi}}\ (2\beta)\ .
  \end{align}
\end{subequations}

The corresponding spherically symmetric solution $\eklo{p}{A}_0(r)$ of the Poisson equation (\ref{eq:iii.36}) looks as follows:
\begin{equation}
  \label{eq:iv.23}
  \eklo{p}{A}_0(r)=\frac{\as}{r}\,(1-{\rm e}^{-2\beta r})\ .
\end{equation}
Thus the asymptotic Coulomb form (\ref{eq:iv.18}) of the first auxiliary potential
$\gklo{p}{A}_{\sf I}(r,\vartheta)$ is safely adopted by the present spherically symmetric
approximation (\ref{eq:iv.23}). Furthermore, the value of the ``exact'' anisotropic
solution $\gklo{p}{A}_0(r,\vartheta)$ (\ref{eq:iv.7}) at the origin
\begin{equation}
  \label{eq:iv.24}
  \gklo{p}{A}_0(r,\vartheta)\Big|_{r=0}=\as\,\frac{\pi}{2}\int_0^\infty dr'\tilde{\Phi}^2(r')=2\beta\as
\end{equation}
also does agree with the corresponding value of that spherically symmetric approximation
$\eklo{p}{A}_0(r)$ (\ref{eq:iv.23}). Summarizing, the boundary values of the ``exact''
anisotropic potential $\gklo{p}{A}_0(r, \vartheta)$ (\ref{eq:iv.7}) are truly reproduced
by the present spherically symmetric candidate $\eklo{p}{A}_0(r)$ (\ref{eq:iv.23}); and
this justifies to {\em preliminarily}\/ adopt just that potential $\eklo{p}{A}_0(r)$ as
the starting point for generating the series of higher-order approximations via the
differential link (\ref{eq:iv.20})-(\ref{eq:iv.21}), i.\,e. we try
\begin{equation}
  \label{eq:iv.25}
  \gklo{p}{A}_{\sf I}(r)\Rightarrow\frac{2}{\pi}\cdot\eklo{p}{A}_0(r)=\frac{2}{\pi}\frac{\as}{r}\,(1-{\rm e}^{-2\beta r})\ .
\end{equation}

But once such a convention concerning the starting potential $\gklo{p}{A}_{\sf I}(r)$ has been attained, one can now substitute this into the differential links (\ref{eq:iv.20})-(\ref{eq:iv.21}) in order to calculate the higher-order potentials $\gklo{p}{A}_{\sf III}(r)$ and $\gklo{p}{A}_{\sf V}(r)$, etc. Indeed, by this simple process of differentiation one easily finds the following results
\begin{subequations}
  \begin{align}
  \label{eq:iv.26a}
  \gklo{p}{A}_{\sf III}(r)&\Rightarrow\frac{8}{3\pi}\,\as\beta^2r\,{\rm e}^{-2\beta r}\\
  \label{eq:iv.26b}
  \gklo{p}{A}_{\sf V}(r)&\Rightarrow -\frac{16}{105\pi}\,(2\beta\as)\,{\rm e}^{-2\beta r}\big\{\frac{(2\beta r)^3}{8}-\frac{1}{2}\,(2\beta r)^2-\frac{1}{2}\,(2\beta r)\big\}\ .
  \end{align}
\end{subequations}

\textbf{Fig.IV.A} below presents a sketch of all three radial potentials $\gklo{p}{A}_{\sf I}(r)$
(\ref{eq:iv.25}), $\gklo{p}{A}_{\sf III}(r)$ (\ref{eq:iv.26a}) and $\gklo{p}{A}_{\sf
  V}(r)$ (\ref{eq:iv.26b}) which clearly shows the hierarchy of their magnitudes. Thus it
is evident that the expansion (\ref{eq:iv.9}) of the anisotropic potential
$\gklo{p}{A}_0(r,\vartheta)$ yields a (more or less rapidly) converging series of radial
auxiliary potentials (\ref{eq:iv.11a}) etc. which constitutes a kind of {\em ``expansion
  with respect to the magnitude of anisotropy''}\/. For general variational parameter
$\nu$, this effect is studied in \textbf{App.B}. The result is that for all values of the
variational parameter $\nu$ ($> -\frac{1}{2}$) the magnitude of the radial potentials
$\gklo{p}{A}_{\sf I}(r)$, $\gklo{p}{A}_{\sf III}(r)$, $\gklo{p}{A}_{\sf V}(r)$, ... is
decreasing rapidly enough (see \textbf{Fig.B.I}) in order that one can tentatively restrict
oneself to the first anisotropic contribution $\gklo{p}{A}_{\sf III}(r,\vartheta)$ for an
estimate of the anisotropic effect upon the energy $\mathbb{E}_{\rm T}$ which is
concentrated in those anisotropic field configurations.

\newpage
\begin{center}
\epsfig{file=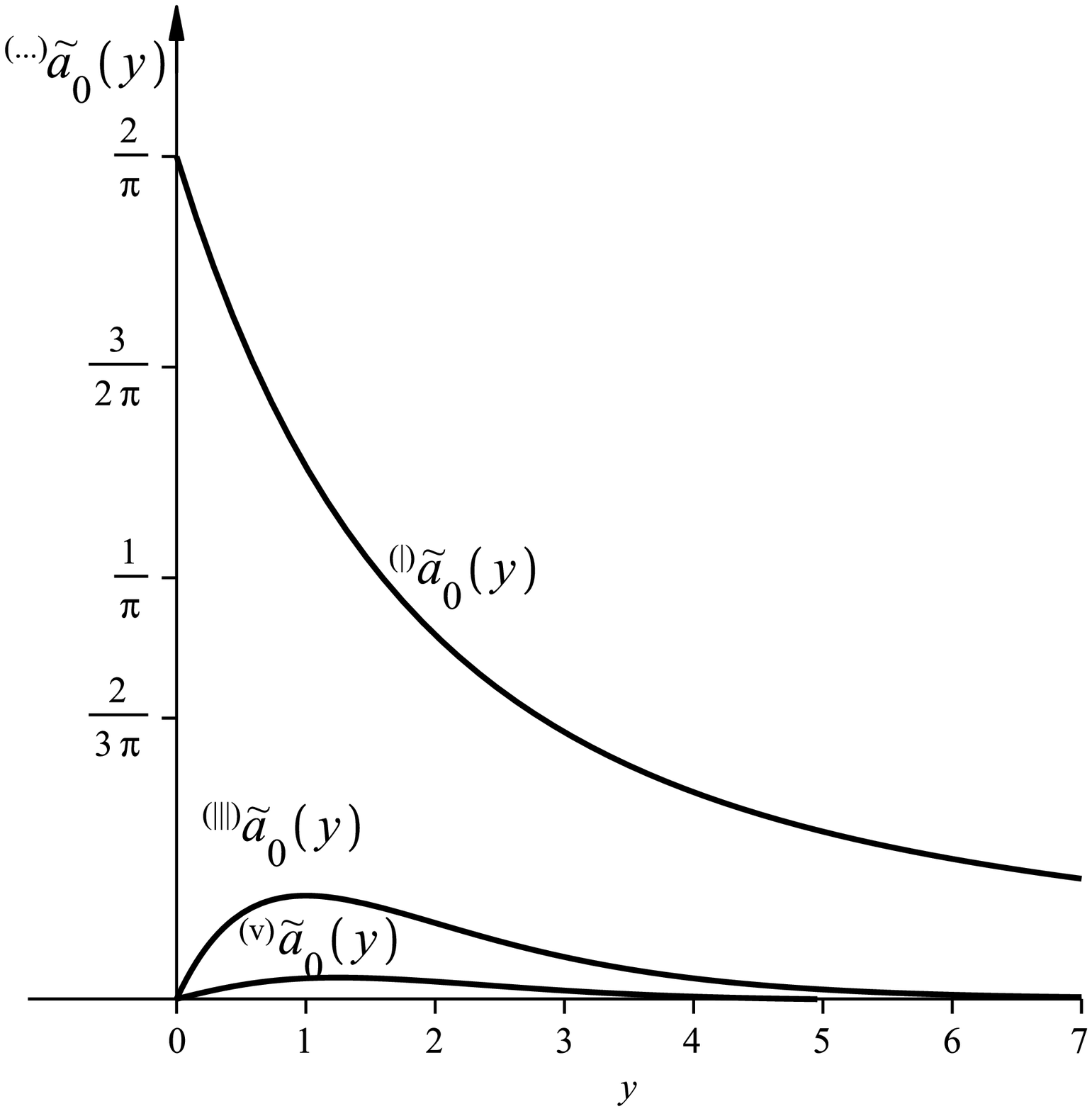,height=12cm}
\end{center}
{\textbf{Fig.~IV.A}\hspace{5mm} \emph{\large\textbf{Radial Auxiliary Potentials
 \boldmath$\gkloi{p}{A}{\sf I},\,\gkloi{p}{A}{\sf III},\,\gkloi{p}{A}{\sf V}$  }}}
\myfigure{Fig.~IV.A: Radial Auxiliary Potentials
 \boldmath$\gkloi{p}{A}{\sf I},\,\gkloi{p}{A}{\sf III},\,\gkloi{p}{A}{\sf V}$}
\indent

The potentials (\ref{eq:iv.25})-(\ref{eq:iv.26b}) for~$\nu=0$ are sketched in
dimensionsless units, cf.\ equations (\ref{eq:b5})-(\ref{eq:b9}) of \textbf{App.~B}. For
the present case~$\nu=0$ there is a distinct hierarchy of magnitudes $(\gkloit{V}{a}{0} <
\gkloit{|||}{a}{0} < \gkloit{|}{a}{0} )$ which lets the anisotropy corrections for the
simplified groundstate~$(\nu=0)$ appear as small as only 0.4\% (see the discussion below
equation (\ref{eq:iv.43})). For the excited states ($\nu>0$), this hierarchy of magnitudes
becomes flattened (see \textbf{Fig.B.I} on page~\pageref{figb1}) and consequently the
anisotropy corrections of the excited states do amount up to some unrealistic 10\% and
more, even for the first excited state (see the table on page~\pageref{tableC})
which necessitates to set up a more adequate series expansion than given by equations
(\ref{eq:iv.9})-(\ref{eq:iv.11e}), see \textbf{Sect.V}.

\newpage

\begin{center}
\mysubsection{3.\ Energy of the Simplified Anisotropic Groundstate $(\boldsymbol{\nu=0)}$}
\end{center}

After a choice of some trial amplitude $\tilde{\Phi}(r)$ (\ref{eq:iii.50}) has been made
together with the subsequent determination of the corresponding anisotropic potential
$\gklo{p}{A}_0(r,\vartheta)$ (\ref{eq:iv.7}) (to be used only up to a certain order of
approximation) one can now tackle more concretely the problem of the total energy
$\mathbb{E}_{\rm T}$ which is carried by such an anisotropic configuration. Here it may be
sufficient (for a first rough estimate of the anisotropy effect) to restrict oneself to
the inclusion of only the {\em first}\/ anisotropic correction term $\gklo{p}{A}_{\sf
  III}(r,\vartheta)=\gklo{p}{A}_{\sf III}(\vartheta)\cdot\gklo{p}{A}_{\sf III}(r)$. By
this restriction, one can represent the general approximation method in a rather
transparent way without the numerical calculations becoming too tedious. But observe here
that we have to admit a certain complication which consists in the fact that our adopted
approximative potential (cf. (\ref{eq:iv.9}))
\begin{equation}
  \label{eq:iv.27}
  \gklo{p}{A}_0(r,\vartheta)\Rightarrow\gklo{p}{A}_{\sf I}(r,\vartheta)+\gklo{p}{A}_{\sf III}(r,\vartheta)= \gklo{p}{A}_{\sf I}(\vartheta)\cdot\gklo{p}{A}_{\sf I}(r)+\gklo{p}{A}_{\sf III}(\vartheta)\cdot\gklo{p}{A}_{\sf III}(r)
\end{equation}
is {\em not}\/ an exact solution of the non-relativistic Poisson equation
\begin{equation}
  \label{eq:iv.28}
  \Delta\gklo{p}{A}_0(r,\vartheta)=-\as\,\frac{\tilde{\Phi}^2(r)}{r\sin\vartheta}\ ,
\end{equation}
i.\,e. the non-relativistic version of (\ref{eq:iii.16}). As a consequence, the
non-relativistic Poisson identity $\tilde{\mathbb{N}}_{\rm G}^{\rm (e)}\equiv0$
(\ref{eq:ii.35}) will not be satisfied, in contrast to the normalization condition
$\tilde{\mathbb{N}}_{\Phi}\equiv0$ (\ref{eq:iii.46}) which merely fixes the normalization
constant $\Phi_*$ (see equation (\ref{eq:iii.51})). Therefore the energy functional
$\tilde{\mathbb{E}}_{[\Phi]}$ (\ref{eq:iii.41}) does not reduce to its proper physical
part $\mathbb{E}^{\rm (IV)}_{[\Phi]}$ ($\doteqdot2E_{\rm kin}+E_{\rm R}^{\rm(e)}$) but
must be kept somewhat more general:
\begin{equation}
  \label{eq:iv.29}
  \tilde{\mathbb{E}}_{[\Phi]}\Rightarrow2E_{\rm kin}+E_{\rm R}^{\rm\{e\}}+\lambda_{\rm G}^{\rm (e)}\cdot\tilde{\mathbb{N}}_{\rm G}^{\rm \{e\}}\ .
\end{equation}

Thus the problem is now to look for the value of this functional on the class of trial
configurations for the groundstate which by assumption is parametrized by only one
variational parameter $\beta$ (because of putting $\nu=0$). This will yield a function
${\mathbb{\tilde{E}}}_\Phi(\beta)$ whose minimal value is just the desired total energy
($\mathbb{E}_{\rm T}^{\{1\}}$, say) of the anisotropic groundstate; and the crucial
questions with this result are now {\bf (i)} how close is this RST prediction
$\mathbb{E}_{\rm T}^{\{1\}}$ to the corresponding conventional prediction $E_{\rm
  conv}^{(1)}$ (\ref{eq:i.4}), and {\bf (ii)} how large is the {\em anisotropic}\/
contribution (beyond the spherically symmetric approximation) included in this RST
prediction $\mathbb{E}_{\rm T}^{\{1\}}$?

For working through this program, one evokes the trial amplitude $\tilde{\Phi}(r)$ for
the groundstate (\ref{eq:iv.22a})-(\ref{eq:iv.22b}) in combination with the associated
(but approximative) gauge potential $\gklo{p}{A}_0(r,\vartheta)$ (\ref{eq:iv.27}), with
the angular parts $\gklo{p}{A}_{\sf I}(\vartheta)$ and $\gklo{p}{A}_{\sf III}(\vartheta)$
given by equations (\ref{eq:iv.10a}) and (\ref{eq:iv.10c}) and the radial parts
$\gklo{p}{A}_{\sf I}(r)$ and $\gklo{p}{A}_{\sf III}(r)$ being given by (\ref{eq:iv.25})
and (\ref{eq:iv.26a}); i.\,e. the (one-parametric) trial potential due to the
(one-parametric) trial amplitude $\tilde{\Phi}(r)$ (\ref{eq:iv.22a}) reads ultimately
\begin{eqnarray}
  \label{eq:iv.30}
  \gklo{p}{A}_0(r,\vartheta)&\Rightarrow&\frac{\pi}{2}\cdot\frac{2}{\pi}\frac{\as}{r}\,(1-{\rm e}^{-2\beta r})+\frac{3}{16}\,\pi(1+\cos^2\vartheta)\cdot\frac{8}{3\pi}\,\as\beta^2r{\rm e}^{-2\beta r}\ .\qquad
\end{eqnarray}
Both these trial ans\"atze $\tilde{\Phi}(r)$ (\ref{eq:iv.22a}) and $\gklo{p}{A}_0(r,\vartheta)$ (\ref{eq:iv.30}) are now to be used in order to calculate any individual contribution to the energy function $\mathbb{\tilde{E}}_\Phi(\beta)$ as the value of the energy functional $\tilde{\mathbb{E}}_{[\Phi]}$ (\ref{eq:iv.29}) on this trial configuration.

The first contribution is the kinetic energy $E_{\rm kin}$ (\ref{eq:iii.42}) which itself
is the sum of the radial part $\rkloi{r}{E}{kin}$ (\ref{eq:iii.43}) and the longitudinal
part $\rkloi{\vartheta}{E}{kin}$ (\ref{eq:iii.44}). The latter, however, must vanish for
the groundstate because here we assume the quantum number~$\lP$ of angular momentum
(\ref{eq:iii.21}) to be zero ($\Rightarrow\lP=0$). Thus the (one-particle) kinetic energy
is of purely radial nature, cf. (\ref{eq:iii.52})-(\ref{eq:iii.53}) for $\lP=0$ and
$\nu=0$
\begin{gather}
  \label{eq:iv.31}
  E_{\rm kin}\Rightarrow\rkloi{r}{E}{kin}=\frac{\hbar^2}{2M}\cdot\frac{\pi}{2}\int_0^\infty dr\,r\left(\frac{d\tilde{\Phi}(r)}{dr}\right)^2=\frac{e^2}{2\aB}\,(\beta\aB)^2\\
  \nonumber
  (\aB=\frac{\hbar^2}{Me^2}\ ...\ \mbox{Bohr radius})\ .
\end{gather}

The next energy contribution is the electrostatic gauge field energy $E_{\rm R}^{\rm(e)}$ (\ref{eq:ii.36}) which reads in terms of our anisotropic potential $\gklo{p}{A}_0(r,\vartheta)$ (\ref{eq:iv.30})
\begin{equation}
  \label{eq:iv.32}
  E_{\rm R}^{\rm(e)}\Rightarrow E_{\rm R}^{\rm\{e\}}=-\frac{\hbar c}{4\pi\as}\int d^3\vr\,\|\vec{\nabla}\gklo{p}{A}_0(r,\vartheta)\|^2\doteqdot E_{\sf I}^{\rm\{e\}}+E_{\sf II}^{\rm\{e\}}+E_{\sf III}^{\rm\{e\}}\ ,
\end{equation}
with the three contributions being found by explicit integration as (cf. {\bf App.~C})
\begin{subequations}
  \begin{align}
  \label{eq:iv.33a}
  E_{\sf I}^{\rm\{e\}}&=-\hbar c\as\beta=-e^2\beta\\
  \label{eq:iv.33b}
  E_{\sf II}^{\rm\{e\}}&=-\frac{1}{6}\,\hbar c\as\beta=-\frac{1}{6}\,e^2\beta\\
  \label{eq:iv.33c}
  E_{\sf III}^{\rm\{e\}}&=-\frac{3}{160}\,\hbar c\as\beta=-\frac{3}{160}\,e^2\beta\ .
  \end{align}
\end{subequations}
Here, the first contribution $E_{\sf I}^{\rm\{e\}}$ (\ref{eq:iv.33a}) is due to the
spherically symmetric constituent $\gklo{p}{A}_{\sf I}(r,\vartheta)$ ($=\gklo{p}{A}_{\sf
  I}(\vartheta)\cdot\gklo{p}{A}_{\sf I}(r)$) of the potential (\ref{eq:iv.9}) with the
angular part $\gklo{p}{A}_{\sf I}(\vartheta)$ being given by (\ref{eq:iv.10a}) and the
radial part $\gklo{p}{A}_{\sf I}(r)$ by the postulate (\ref{eq:iv.25}). Therefore the
first expansion term $\gklo{p}{A}_{\sf I}(r,\vartheta)$ just coincides with the
spherically symmetric approximation $\eklo{p}{A}_0(r)$. Consequently, the present first
contribution $E_{\sf I}^{\rm\{e\}}$ (\ref{eq:iv.33a}) must be identical to the field
energy $E_{\rm R}^{\rm[e]}$ (\ref{eq:iii.54}) of the spherically symmetric approximation,
i.\,e. we must find for the sake of consistency:
\begin{equation}
  \label{eq:iv.34}
  E_{\sf I}^{\rm\{e\}}=-e^2\beta\stackrel{!}{=}-2\beta e^2\cdot\varepsilon_{\rm pot}(0)\equiv E_{\rm R}^{\rm[e]}\ .
\end{equation}
However, this consistency requirement
\begin{equation}
  \label{eq:iv.35}
  2\cdot\varepsilon_{\rm pot}(0)=1
\end{equation}
is easily validated, namely simply by specializing the general result (\ref{eq:a28}) of {\bf App.\,A} down to $\nu=0$ which yields
\begin{equation}
  \label{eq:iv.36}
  \varepsilon_{\rm pot}(0)=1-\frac{1}{4}\cdot\sum_{n=0}^\infty\frac{n}{2^n}=\frac{1}{2}\ .
\end{equation}

But whilst the first contribution $E_{\sf I}^{\rm\{e\}}$ (\ref{eq:iv.33a}) does refer to
the spherically symmetric approximation, the other two contributions $E_{\sf
  II}^{\rm\{e\}}$ and $E_{\sf III}^{\rm\{e\}}$ (\ref{eq:iv.33b})-(\ref{eq:iv.33c}) must
represent the {\em anisotropic correction}\/ ($E_{\rm a}^{\rm\{e\}}$, say)
\begin{equation}
  \label{eq:iv.37}
  E_{\rm a}^{\rm\{e\}}\doteqdot E_{\sf II}^{\rm\{e\}}+E_{\sf III}^{\rm\{e\}}=-(\frac{1}{6}+\frac{3}{160})\,e^2\beta= -\frac{89}{480}\,e^2\beta\ .
\end{equation}
However, the point with this anisotropic correction $E_{\rm a}^{\rm\{e\}}$
(\ref{eq:iv.37}) is now that it obviously cannot be considered as a {\em negligibly
  small}\/ correction of the spherically symmetric term $E_{\sf I}^{\rm\{e\}}$
(\ref{eq:iv.34})! Thus the problem arises now how it may come about that the anisotropy
corrections can be effectively smaller than the spherically symmetric contribution $E_{\sf
  I}^{\rm\{e\}}$ (\ref{eq:iv.34})! Otherwise one could in general not consider $E_{\sf
  I}^{\rm\{e\}}$ as the leading term of a meaningful perturbation expansion. Indeed, we
will readily show that this effect is based upon a certain peculiarity of the {\em
  principle of minimal energy}\/.

This question concerns now the last term of the energy functional
$\tilde{\mathbb{E}}_{[\Phi]}$ on the right-hand side of equation (\ref{eq:iv.29}). Namely,
this contribution is non-zero whenever the Poisson equation is not satisfied by the
selected trial configuration, i.\,e. more precisely: whenever the trial configuration
$\{\gklo{p}{A}_0(r,\vartheta),\,\tilde{\Phi}(r)\}$ does not obey the Poisson equation
(\ref{eq:iv.28}) for the gauge potential $\gklo{p}{A}_0(r,\vartheta)$. Since our selected
approximation (\ref{eq:iv.27}) for $\gklo{p}{A}_0(r,\vartheta)$ contains only the first
two non-vanishing terms of the general expansion (\ref{eq:iv.9}), it is not linked to the
chosen trial amplitude $\tilde{\Phi}(r)$ (\ref{eq:iv.22a})-(\ref{eq:iv.22b}) via the
Poisson equation (\ref{eq:iv.28}); and therefore the anisotropic Poisson constraint
(\ref{eq:ii.35}) is violated. This then necessitates to explicitly calculate the
constraint term $\tilde{\mathbb{N}}_{\rm G}^{\rm(e)}$
\begin{equation}
  \label{eq:iv.38}
  \mathbb{N}_{\rm G}^{\rm(e)}\Rightarrow\tilde{\mathbb{N}}_{\rm G}^{\rm\{e\}}=E_{\rm R}^{\rm\{e\}}-\tilde{\mathbb{M}}^{\rm\{e\}}c^2
\end{equation}
which enters the energy functional $\tilde{\mathbb{E}}_{[\Phi]}$ (\ref{eq:iv.29})
as the last term on the right-hand side.

Here, the gauge field energy $E_{\rm R}^{\rm\{e\}}$ has already been determined by
equations (\ref{eq:iv.32})-(\ref{eq:iv.33c}) so that we are left with the calculation of
the (non-relativistic) mass equivalent $\mathbb{M}^{\rm\{e\}}c^2$
(\ref{eq:ii.37}). If our approximative potential $\gklo{p}{A}_0(r,\vartheta)$
(\ref{eq:iv.27}), i.\,e. more precisely (\ref{eq:iv.30}) for the present groundstate
situation, is substituted therein, one finds that the mass equivalent appears as a sum of
two terms
\begin{equation}
  \label{eq:iv.39}
  \orrk{M}{e}c^2\Rightarrow \tilde{\mathbb{M}}^{\rm\{e\}}c^2=\tilde{\mathbb{M}}_{\sf I}^{\rm\{e\}}c^2+\tilde{\mathbb{M}}_{\sf III}^{\rm\{e\}}c^2\ ,
\end{equation}
where the first term contains the spherically symmetric approximation $\gklo{p}{A}_{\sf I}(r)$ (\ref{eq:iv.25}) of the potential $\gklo{p}{A}_0(r,\vartheta)$ and yields the following result
\begin{equation}
  \label{eq:iv.40}
  \tilde{\mathbb{M}}_{\sf I}^{\rm\{e\}}c^2=-\left(\frac{\pi}{2}\right)^2\hbar c\int_0^\infty dr\,r\,\gklo{p}{A}_{\sf I}(r)\cdot\tilde{\Phi}^2(r)=-\hbar c\as\beta\equiv E_{\sf I}^{\rm\{e\}}\ ,
\end{equation}
cf. (\ref{eq:iv.34}); and furthermore the second contribution is due to the first anisotropic correction (\ref{eq:iv.26a}) of $\gklo{p}{A}_0(r,\vartheta)$:
\begin{equation}
  \label{eq:iv.41}
  \tilde{\mathbb{M}}_{\sf III}^{\rm\{e\}}c^2=-\left(\frac{3\pi}{8}\right)^2\hbar c\int_0^\infty dr\,r\,\gklo{p}{A}_{\sf III}(r)\cdot\tilde{\Phi}^2(r)=-\frac{3}{32}\,\hbar c\as\beta\ .
\end{equation}
(For both results (\ref{eq:iv.40}) and (\ref{eq:iv.41}) the angular parts of the potentials are already integrated over). It should not come as a surprise that the lowest-order mass equivalent $\tilde{\mathbb{M}}_{\sf I}^{\rm\{e\}}c^2$ (\ref{eq:iv.40}) exactly agrees with the first contribution $E_{\sf I}^{\rm\{e\}}$ (\ref{eq:iv.33a}) to the gauge field energy $E_{\rm R}^{\rm\{e\}}$ (\ref{eq:iv.32})! The reason for this is of course that both quantities $E_{\sf I}^{\rm\{e\}}$ (\ref{eq:iv.33a}) and $\tilde{\mathbb{M}}_{\sf I}^{\rm\{e\}}c^2$ (\ref{eq:iv.40}) are based upon the spherically symmetric approximation $\eklo{p}{A}_0(r)$ for the potential $\gklo{p}{A}_0(r,\vartheta)$ and also upon the same trial amplitude $\tilde{\Phi}(r)$ (\ref{eq:iii.50}) with $\nu=0$, so that the spherically symmetric Poisson constraint is satisfied in first order (i.\,e. $E_{\sf I}^{\rm\{e\}}-\tilde{\mathbb{M}}_{\sf I}^{\rm\{e\}}c^2=0$). But the consequence of this peculiarity is now that the non-relativistic constraint term $\tilde{\mathbb{N}}_{\rm G}^{\rm\{e\}}$ (\ref{eq:iv.38}) is built up exclusively by the anisotropic corrections:
\begin{eqnarray}
  \label{eq:iv.42}
  \tilde{\mathbb{N}}_{\rm G}^{\rm\{e\}}&=&(E_{\sf I}^{\rm\{e\}}+E_{\sf II}^{\rm\{e\}}+E_{\sf III}^{\rm\{e\}})-(\tilde{\mathbb{M}}_{\sf I}^{\rm\{e\}}c^2+\tilde{\mathbb{M}}_{\sf III}^{\rm\{e\}}c^2)\\
  \nonumber
  &=&E_{\sf II}^{\rm\{e\}}+E_{\sf III}^{\rm\{e\}}-\tilde{\mathbb{M}}_{\sf III}^{\rm\{e\}}c^2\;=\; -\frac{11}{120}\,\hbar c\as\beta\ .
\end{eqnarray}

Collecting now all the partial results, i.\,e. the kinetic energy $E_{\rm kin}$ (\ref{eq:iv.31}) plus the gauge field energy (\ref{eq:iv.33a})-(\ref{eq:iv.33c}) plus the Poisson constraint term (\ref{eq:iv.42}), and then substituting this back into the energy functional $\tilde{\mathbb{E}}_{[\Phi]}$ (\ref{eq:iv.29}) yields the value of this functional on the chosen trial configuration $\{\tilde{\Phi}(r),\,\gklo{p}{A}_0(r,\vartheta)\}$ in form of an energy function ($\tilde{\mathbb{E}}_\Phi(\beta)$, say) which is found to be of the following shape:
\begin{equation}
  \label{eq:iv.43}
  \tilde{\mathbb{E}}_\Phi(\beta)=\frac{e^2}{\aB}\left\{(\beta\aB)^2-\frac{481}{480}\,(\beta\aB)\right\}\ .
\end{equation}
(Observe here also the value (\ref{eq:iii.49b}) for the Lagrangean parameter $\lambda_{\rm
  G}^{\rm(e)}$). It is now this result (\ref{eq:iv.43}) which is well-suited for a first
estimate of the influence of anisotropy. Namely, the {\em spherically symmetric
  approximation}\/ does also produce an energy function
(i.\,e. $\mathbb{E}^{\rm[IV]}(\beta,\nu)$, cf. (\ref{eq:a22})), which for the special case
$\nu=0$ looks as follows:
\begin{equation}
  \label{eq:iv.44}
  \mathbb{E}^{\rm[IV]}(\beta,0)=\frac{e^2}{\aB}\left\{(2\beta\aB)^2\cdot\varepsilon_{\rm kin}(0)-(2\beta\aB)\cdot\varepsilon_{\rm pot}(0)\right\}= \frac{e^2}{\aB}\left\{(\beta\aB)^2-(\beta\aB)\right\}
\end{equation}
where the special values of the functions $\varepsilon_{\rm kin}(\nu)$ and
$\varepsilon_{\rm pot}(\nu)$ may immediately be read off from the equations (\ref{eq:a27})
($\Rightarrow\varepsilon_{\rm kin}(0)=\frac{1}{4}$) and from (\ref{eq:iv.36})
($\Rightarrow\varepsilon_{\rm pot}(0)=\frac{1}{2}$). Comparing now the anisotropic result
(\ref{eq:iv.43}) to its present isotropic counterpart (\ref{eq:iv.44}) demonstrates
clearly that the anisotropic corrections for the groundstate will amount to roughly $4$
parts in $10^3$\,! More concretely, one determines the groundstate energy as the minimal
value of these energy functions (\ref{eq:iv.43}) and (\ref{eq:iv.44}) and thus finds for
the {\em ``anisotropic groundstate''}\/ (\ref{eq:iv.43})
\begin{equation}
  \label{eq:iv.45}
  \nrf{E}{T}^{\{1\}} \doteqdot\tilde{\mathbb{E}}_\Phi(\beta)\Big|_{\rm min}=-\frac{e^2}{4\aB}\cdot\left(\frac{481}{480}\right)^2\simeq -\frac{e^2}{4\aB}\cdot1.00417...
\end{equation}
and for the {\em ``isotropic groundstate''}\/ (\ref{eq:iv.44})
\begin{gather}
  \label{eq:iv.46}
  \mathbb{E}^{\rm[IV]}(\beta,0)\Big|_{\rm min}=-\frac{e^2}{4\aB}\equiv E_{\rm conv}^{(1)}\\
  \nonumber
  (\frac{e^2}{4\aB}=6.8029...\,{\rm eV})\ .
\end{gather}

From these results it is now obvious that the influence of the anisotropy on the
groundstate energy amounts to only $0.4\%$. It is true, our trial amplitude
$\tilde{\Phi}(r)$ (\ref{eq:iv.22a})-(\ref{eq:iv.22b}) with the fixation $\nu=0$ is not the
optimal choice for the groundstate; if one admits non-zero $\nu$ and carries through the
{\em isotropic\,(!)}\/ minimalization procedure, one will find the minimal value of the
corresponding energy function at \mbox{$-7.23...\,[{\rm eV}]$} for \mbox{$\nu=-0.2049...$}
(see \textbf{table 1} on p.~\pageref{table1} and fig.s 4a-4b of ref. \cite{ms1}). This
{\em anisotropic}\/ lowering (\ref{eq:iv.45}) of the groundstate energy below that simple
isotropic RST prediction (\ref{eq:iv.46}) (which despite its na\"ive deduction equals {\em
  exactly}\/ the conventional prediction $E_{\rm conv}^{(1)}$ (\ref{eq:i.4})) must now be
tested for the whole positronium spectrum. But since the excited states require here the
use of (at least) a two-parameter trial amplitude, we have to admit the more realistic
trial amplitudes $\tilde{\Phi}(r)$ (\ref{eq:iii.50}) with $\nu\neq0$ in connection with an
\emph{improved (!)} expansion with respect to the magnitude of anisotropy (see {\bf
  Sect.~V}). Namely, the point here is that the present anisotropic groundstate lowering
(\ref{eq:iv.45}) is actually false, as pleasant as its smallmess ($\sim 0,4\%$) may appear
in the light of a pertubation expansion. Indeed, we will readily demonstrate now that an
improved \emph{expansion with respect to the magnitude of anisotropy}, not so na\"ive as
the above proposal (\ref{eq:iv.9}), will for the excited states yield anisotropic
corrections in the range of some 6\% as opposed to the present 0,4\% due to our na\"ive
expansion (\ref{eq:iv.9}).


\section{Anisotropic Corrections for the Excited States}
\indent

Since, for the treatment of the excited states~(\textbf{App.C}), the expansion
(\ref{eq:iv.8})-(\ref{eq:iv.9}) of the anisotropic gauge potential
$\gklo{p}{A}_0(r,\vartheta)$ (\ref{eq:iv.5}) in combination with the lowest-order
postulate (\ref{eq:iv.25}) has been revealed as being too na\"ive, we have to develop now
a more systematic way of expanding the gauge potential around that spherically symmetric
approximation $\eklo{p}{A}_0(r)$ (\ref{eq:iv.23}). The crucial point here aims just at the
very concept of the spherically symmetric approximation, being termed
as~$\eklo{p}{A}_0(r)$. Naturally, one would associate to this
approximation~$\eklo{p}{A}_0(r)$ the property that it represents the average of the
original potential~$\gklo{p}{A}_0(r,\vartheta)$ on the 2-sphere, see equation
(\ref{eq:v.8}) below; or rephrased in other words: the average of the anisotropic
constituent~$(\gklo{p}{A}_{an}(r,\vartheta))$ of~$\gklo{p}{A}_0(r,\vartheta)$ on the
2-sphere must vanish, see equation (\ref{eq:v.7}) below. But these requirements are not
satisfied by our na\"ive expansion (\ref{eq:iv.8})-(\ref{eq:iv.11e}). Subsequently, we
will iron out this preliminary deficiency of our anisotropy expansion and thus will obtain
distinctly improved predictions for the positronium spectrum (cf.~\textbf{table~2} below,
p.~\pageref{table2}, to the former \textbf{table~1}, p.~\pageref{table1}).  \vspace{4ex}
\begin{center}
\mysubsection{1.\ Expansion around Spherically Symmetric Approximation}
\end{center}

To this end, we recall the integral representation (\ref{eq:iii.37}) of the spherically
symmetric approximation $\eklo{p}{A}_0(r)$ (for general $\nu$) and use this for the
desired expansion of $\gklo{p}{A}_0(r,\vartheta)$ around $\eklo{p}{A}_0(r)$, namely by
putting
\begin{eqnarray}
  \label{eq:v.1}
  \gklo{p}{A}_0(r,\vartheta)&=&\eklo{p}{A}_0(r)+\left(\gklo{p}{A}_0(r,\vartheta)-\eklo{p}{A}_0(r)\right)\\
  \nonumber
  &=&\frac{\as}{8}\int\frac{d^3\vec{r}\,'}{r'}\,\frac{\tilde{\Phi}^2(r')}{\|\vec{r}-\vec{r}\,'\|}+ \frac{\as}{4\pi}\int\frac{d^3\vec{r}\,'}{r'}\,\left(\frac{1}{\sin\vartheta'}-\frac{\pi}{2}\right)\frac{\tilde{\Phi}^2(r')}{\|\vec{r}-\vec{r}\,'\|}\ .
\end{eqnarray}
Here, the second part is obviously to be considered as the truly anisotropic constituent $\gklo{p}{A}^{\rm an}(r,\vartheta)$, i.\,e. we define
\begin{eqnarray}
  \label{eq:v.2}
  \gklo{p}{A}^{\rm an}(r,\vartheta)&\doteqdot&\gklo{p}{A}_0(r,\vartheta)-\eklo{p}{A}_0(r)\\
  \nonumber
  &=&\frac{\as}{4\pi}\int\frac{d^3\vec{r}\,'}{r'}\,\left(\frac{1}{\sin\vartheta'}-\frac{\pi}{2}\right) \frac{\tilde{\Phi}^2(r')}{\|\vec{r}-\vec{r}\,'\|}\ .
\end{eqnarray}
This anisotropic constituent of the gauge potential may now be expanded by use of equation (\ref{eq:iv.8}) which yields a series expansion quite similarly as shown by equation (\ref{eq:iv.9})
\begin{eqnarray}
  \nonumber
  \gklo{p}{A}^{\rm an}(r,\vartheta)&=&\gklo{p}{A}^{\sf I}(\vartheta)\cdot\gklo{p}{A}^{\sf I}(r)+\gklo{p}{A}^{\sf II}(\vartheta)\cdot\gklo{p}{A}^{\sf II}(r)+\gklo{p}{A}^{\sf III}(\vartheta)\cdot\gklo{p}{A}^{\sf III}(r)\\
  \label{eq:v.3}
  &&{}+\gklo{p}{A}^{\sf IV}(\vartheta)\cdot\gklo{p}{A}^{\sf IV}(r)+\gklo{p}{A}^{\sf V}(\vartheta)\cdot\gklo{p}{A}^{\sf V}(r)+...
\end{eqnarray}

Although this looks very similar to the former expansion (\ref{eq:iv.9}), the individual terms are somewhat different especially concerning the angular pre-factors:
\begin{subequations}
  \begin{align}
  \label{eq:v.4a}
  \gklo{p}{A}^{\sf I}(\vartheta)&\doteqdot \int\frac{d\Omega'}{4\pi}\left(\frac{1}{\sin\vartheta'}-\frac{\pi}{2}\right)=0\\
  \label{eq:v.4b}
  \gklo{p}{A}^{\sf II}(\vartheta)&\doteqdot \int\frac{d\Omega'}{4\pi}\left(\frac{1}{\sin\vartheta'}-\frac{\pi}{2}\right)(\hat{\vr}\sdot\hat{\vec{r}}\,')=0\\
  \label{eq:v.4c}
  \gklo{p}{A}^{\sf III}(\vartheta)&\doteqdot \frac{3}{2}\int\frac{d\Omega'}{4\pi}\left(\frac{1}{\sin\vartheta'}-\frac{\pi}{2}\right)(\hat{\vr}\sdot\hat{\vec{r}}\,')^2=\frac{3\pi}{16}\,(\cos^2\vartheta-\frac{1}{3})\\
  \label{eq:v.4d}
  \gklo{p}{A}^{\sf IV}(\vartheta)&\doteqdot \frac{5}{2}\int\frac{d\Omega'}{4\pi}\left(\frac{1}{\sin\vartheta'}-\frac{\pi}{2}\right)(\hat{\vr}\sdot\hat{\vec{r}}\,')^3=0\\
    \label{eq:v.4e}
  \gklo{p}{A}^{\sf V}(\vartheta)&\doteqdot \frac{35}{8}\!\int\!\frac{d\Omega'}{4\pi}\!\left(\!\frac{1}{\sin\vartheta'}-\frac{\pi}{2}\!\right)\!(\hat{\vr}\sdot\hat{\vec{r}}\,')^4=\frac{105}{128}\,\pi\{\cos^4\vartheta+\cos^2\vartheta\sin^2\vartheta+\frac{3}{8}\sin^4\vartheta-\frac{8}{15}\}\\
  \nonumber
  \vdots\quad\:&
  \end{align}
\end{subequations}
These angular factors $\gklo{p}{A}^\mathbb{N}(\vartheta)$ ($\mathbb{N}={\sf I},\,{\sf II},\,{\sf III},\,{\sf IV},\,{\sf V},\,...$) have zero average over the 2-sphere, i.\,e.
\begin{equation}
  \label{eq:v.5}
  \int\frac{d\Omega}{4\pi}\,\gklo{p}{A}^\mathbb{N}(\vartheta)=\frac{1}{4\pi}\int d\phi\,d\vartheta\,\sin\vartheta\,\gklo{p}{A}^\mathbb{N}(\vartheta)=0\ ,
\end{equation}
because of the commutativity of both angular integrations, i.\,e.
\begin{eqnarray}
  \nonumber
  \int\frac{d\Omega}{4\pi}\,\gklo{p}{A}^\mathbb{N}(\vartheta)&\simeq& \int\frac{d\Omega'}{4\pi}\,\left(\frac{1}{\sin\vartheta'}-\frac{\pi}{2}\right)\int\frac{d\Omega}{4\pi}\,(\hat{\vr}\sdot\hat{\vec{r}}\,')^{\mathbb{N}-1}\\
   \label{eq:v.6}
   &=&\int\frac{d\Omega'}{4\pi}\,\left(\frac{1}{\sin\vartheta'}-\frac{\pi}{2}\right)\cdot\frac{1}{\mathbb{N}}=0\ .
\end{eqnarray}
Therefore the average of the anisotropic potential $\gklo{p}{A}^{\rm an}(r,\vartheta)$ (\ref{eq:v.3}) is zero:
\begin{equation}
  \label{eq:v.7}
  \int\frac{d\Omega}{4\pi}\,\gklo{p}{A}^{\rm an}(r,\vartheta)\equiv0\ ;
\end{equation}
and thus the spherically symmetric approximation $\eklo{p}{A}_0(r)$ (\ref{eq:iii.37}), as the leading term of the series expansion (\ref{eq:v.1})-(\ref{eq:v.3}), is revealed as the angular average of the general anisotropic potential $\gklo{p}{A}_0(r,\vartheta)$ (\ref{eq:iv.7}), i.\,e.
\begin{equation}
  \label{eq:v.8}
  \eklo{p}{A}_0(r)\equiv\int\frac{d\Omega}{4\pi}\,\gklo{p}{A}_0(r,\vartheta)\ .
\end{equation}

Concerning the radial factors of the series expansion (\ref{eq:v.3}), one observes on account of the vanishing of the first two angular factors $\gklo{p}{A}^{\sf I}(\vartheta)$ (\ref{eq:v.4a}) and $\gklo{p}{A}^{\sf II}(\vartheta)$ (\ref{eq:v.4b}) that the first active non-trivial potential is that of third order
\begin{equation}
  \label{eq:v.9}
  \gklo{p}{A}^{\sf III}(r)=\as\int dr'\,r'\,\frac{(rr')^2\cdot\tilde{\Phi}^2(r')}{(r^2+r'^2)^\frac{5}{2}}\equiv \gklo{p}{A}_{\sf III}(r)\ ,
\end{equation}
cf. (\ref{eq:iv.11c}), and the next non-trivial one to be considered here is of fifth order (cf. (\ref{eq:v.4d})-(\ref{eq:v.4e}))
\begin{equation}
  \label{eq:v.10}
  \gklo{p}{A}^{\sf V}(r)=\as\int dr'\,r'\,\frac{(rr')^4\cdot\tilde{\Phi}^2(r')}{(r^2+r'^2)^\frac{9}{2}}\equiv \gklo{p}{A}_{\sf V}(r)\ ,
\end{equation}
cf. (\ref{eq:iv.11e}). Thus the present expansion (\ref{eq:v.1})-(\ref{eq:v.3}) of the
gauge potential is essentially the same as the na\"ive one (\ref{eq:iv.9}), and differs
from that former one only in the angular parts. Consequently, the hierarchical arrangement
of magnitudes, as expressed in \textbf{Fig.IV.A}, should {\em in principle}\/ apply also to the
present expansion (\ref{eq:v.1})-(\ref{eq:v.3}).
\vspace{4ex}
\begin{center}
\mysubsection{2.\ Energy Content of the Anisotropic Configurations}
\end{center}

The present expansion around the spherically symmetric approximation has also some useful consequences with respect to the anisotropic energy functional $\tilde{\mathbb{E}}_{[\Phi]}$ (\ref{eq:iv.29}). The crucial point here is the circumstance that with the specification of the angular pre-factors $\gklo{p}{A}^{\sf I}(\vartheta)$, $\gklo{p}{A}^{\sf II}(\vartheta)$, ... one can integrate over these angular functions so that the {\em principle of minimal energy}\/ becomes reduced to a purely radial problem for the remaining spherically symmetric fields $\eklo{p}{A}_0(r)$, $\gklo{p}{A}^{\sf III}(r)$, $\tilde{\Phi}(r)$. This reduction to a spherically symmetric problem is carried through as follows:

The first consequence of the representation of the anisotropic potential $\gklo{p}{A}_0(r,\vartheta)$ as a sum of two terms, cf. (\ref{eq:v.1})
\begin{equation}
  \label{eq:v.11}
  \gklo{p}{A}_0(r,\vartheta)=\eklo{p}{A}_0(r)+\gklo{p}{A}^{\rm an}(r,\vartheta)\ ,
\end{equation}
refers to its energy content $E_{\rm R}^{\rm(e)}$ (\ref{eq:ii.36})
\begin{equation}
  \label{eq:v.12}
  E_{\rm R}^{\rm(e)}\Rightarrow E_{\rm R}^{\rm\{e\}}=-\frac{\hbar c}{4\pi\as}\int d^3\vec{r}\ \|\vec{\nabla}\gklo{p}{A}_0(r,\vartheta)\|^2
\end{equation}
which appears now as a sum of three terms
\begin{equation}
  \label{eq:v.13}
  E_{\rm R}^{\rm\{e\}}=E_{\rm R}^{\rm[e]}+E_{\rm an}^{\rm\{e\}}+E_{\rm mx}^{\rm\{e\}}\ .
\end{equation}
Here, the first term is the energy content of the spherically symmetric approximation $\eklo{p}{A}_0(r)$
\begin{eqnarray}
  \nonumber
  E_{\rm R}^{\rm[e]}&=&-\frac{\hbar c}{4\pi\as}\int d^3\vec{r}\ \|\vec{\nabla}\eklo{p}{A}_0(r)\|^2\\
  \label{eq:v.14}
  &=&-\frac{\hbar c}{\as}\int_0^\infty dr\,\left(r\cdot\frac{d\eklo{p}{A}_0(r)}{dr}\right)^2
\end{eqnarray}
and is for our trial amplitude $\tilde{\Phi}(r)$ (\ref{eq:iii.50}) explicitly given by
\begin{equation}
  \label{eq:v.15}
  E_{\rm R}^{\rm[e]}=-\frac{e^2}{\aB}\,(2\aB\beta)\cdot\varepsilon_{\rm pot}(\nu)
\end{equation}
where the potential function $\varepsilon_{\rm pot}(\nu)$ is specified by equation (\ref{eq:a28}) of {\bf App.\,A}, see here also equation (\ref{eq:a22}) and {\bf App.\,D} of ref. \cite{ms2}. Analogously, the second energy contribution $E_{\rm an}^{\rm\{e\}}$ is the energy content of the anisotropic part $\gklo{p}{A}^{\rm an}(r,\vartheta)$ of the gauge potential and is given by
\begin{equation}
  \label{eq:v.16}
  E_{\rm an}^{\rm\{e\}}=-\frac{\hbar c}{4\pi\as}\int d^3\vec{r}\ \|\vec{\nabla}\gklo{p}{A}^{\rm an}(r,\vartheta)\|^2\ .
\end{equation}
This result says that the anisotropy energy is always negative and therefore must increase the binding energy!

And finally, there is a mixed contribution $E_{\rm mx}^{\rm\{e\}}$ which is built up cooperatively by both parts of the gauge potential
\begin{equation}
  \label{eq:v.17}
  E_{\rm mx}^{\rm\{e\}}=-\frac{\hbar c}{2\pi\as}\int d^3\vec{r}\,\left(\vec{\nabla}\eklo{p}{A}_0(r)\right)\!\!\sdot\!\!\left(\vec{\nabla}\gklo{p}{A}^{\rm an}(r,\vartheta)\right)\ .
\end{equation}
But the interesting point with this last contribution is now that it actually vanishes
\begin{equation}
  \label{eq:v.18}
  E_{\rm mx}^{\rm\{e\}}=0\ .
\end{equation}
This can easily be verified by simply observing the fact that the spherically symmetric
approximation $\eklo{p}{A}_0(r)$ does not depend upon the polar angles $\vartheta$ and
$\phi$ so that
\begin{equation}
  \label{eq:v.19}
  \vec{\nabla}\eklo{p}{A}_0(r)=\frac{d\eklo{p}{A}_0(r)}{dr}\,\vec{e}_r\ ,
\end{equation}
and consequently the mixed term depends exclusively on the derivatives in radial direction ($\sim\vec{e}_r$), i.\,e. we have by use of the anisotropy expansion (\ref{eq:v.3})
\begin{eqnarray}
  \label{eq:v.20}
  E_{\rm mx}^{\rm\{e\}}&=& -\frac{\hbar c}{2\pi\as}\int d^3\vec{r}\ \frac{d\eklo{p}{A}_0(r)}{dr}\cdot\frac{\partial\gklo{p}{A}^{\rm an}(r,\vartheta)}{\partial r}\\
  \nonumber
  &=&-\frac{\hbar c}{2\pi\as}\left\{\int_0^\infty dr\,r^2\,\frac{d\eklo{p}{A}_0(r)}{dr}\cdot\frac{d\gklo{p}{A}^{\sf I}(r)}{d r}\int d\Omega\gklo{p}{A}^{\sf I}(\vartheta)\right.\\
  \nonumber
  &&\hphantom{-\frac{\hbar c}{2\pi\as}}
  +\int_0^\infty dr\,r^2\,\frac{d\eklo{p}{A}_0(r)}{dr}\cdot\frac{d\gklo{p}{A}^{\sf II}(r)}{d r}\int d\Omega\gklo{p}{A}^{\sf II}(\vartheta)\\
  \nonumber
  &&\hphantom{-\frac{\hbar c}{2\pi\as}}
  +\int_0^\infty dr\,r^2\,\frac{d\eklo{p}{A}_0(r)}{dr}\cdot\frac{d\gklo{p}{A}^{\sf III}(r)}{d r}\int d\Omega\gklo{p}{A}^{\sf III}(\vartheta)+\ \ldots\\
  \nonumber
  &&\hphantom{-\frac{\hbar c}{2\pi\as}}
  \left.+\int_0^\infty dr\,r^2\,\frac{d\eklo{p}{A}_0(r)}{dr}\cdot\frac{d\gklo{p}{A}^\mathbb{N}(r)}{d r}\int d\Omega\gklo{p}{A}^\mathbb{N}(\vartheta)+\ \ldots\ \ \right\}\ .
\end{eqnarray}
Thus the vanishing mean value (\ref{eq:v.6}) of the angular factors $\gklo{p}{A}^\mathbb{N}(\vartheta)$ just validates the claim (\ref{eq:v.18}), and consequently the gauge field energy $E_{\rm R}^{\rm\{e\}}$ becomes composed exclusively by the two quadratic terms (\ref{eq:v.14}) and (\ref{eq:v.16})
\begin{equation}
  \label{eq:v.21}
  E_{\rm R}^{\rm\{e\}}=E_{\rm R}^{\rm[e]}+E_{\rm an}^{\rm\{e\}}\ .
\end{equation}

The intended reduction of the original angular-dependent variational problem to a purely
radial problem must here be carried through only for this gauge field contribution $E_{\rm
  R}^{\rm\{e\}}$ (\ref{eq:v.21}) because the isotropic part $E_{\rm R}^{\rm[e]}$ is
already defined as a purely radial integral, cf. (\ref{eq:iii.45}); and for the
anisotropic part $E_{\rm an}^{\rm\{e\}}$ (\ref{eq:v.16}) one integrates over the angular
pre-factor $\gklo{p}{A}^{\sf III}(\vartheta)$ (\ref{eq:v.4c}) so that a purely radial
integral over $\gklo{p}{A}^{\sf III}(r)$ is left over, see equation (\ref{eq:v.30}) below.

Clearly, if one could solve exactly the Poisson equation (\ref{eq:iv.28}) for the
anisotropic potential $\gklo{p}{A}_0(r,\vartheta)$, then the Poisson constraint
(\ref{eq:iv.38}) would not be a point of concern because it would disappear trivially from
the energy functional $\tilde{\mathbb{E}}_{[\Phi]}$ (\ref{eq:iv.29}). But since we will
not find such an exact solution for the anisotropic gauge potential, we are forced to
briefly inspect the Poisson constraint with reference to that splitting (\ref{eq:v.11}) of
the gauge potential into an isotropic and anisotropic part. Indeed, this splitting induces
a corresponding splitting of (the non-relativistic version of) the original mass
equivalent $M^{\rm(e)}c^2$ (\ref{eq:ii.37}) into a spherically symmetric part
$\tilde{\mathbb{M}}^{\rm[e]}c^2$ and an anisotropic part
$\tilde{\mathbb{M}}^{\rm\{e\}}_{\rm an}c^2$:
\begin{eqnarray}
  \label{eq:v.22}
  \tilde{\mathbb{M}}^{\rm\{e\}}c^2&\doteqdot&-\hbar c\int d^3\vec{r}\ \gklo{p}{A}_0(r,\vartheta)\,\frac{\tilde{\Phi}^2(r)}{4\pi r\sin\vartheta}\\
  \nonumber
  &=&\tilde{\mathbb{M}}^{\rm[e]}c^2+\tilde{\mathbb{M}}^{\rm\{e\}}_{\rm an}c^2\ ,
\end{eqnarray}
with the isotropic part being given by equation (\ref{eq:iii.48}) and analogously the anisotropic part by
\begin{equation}
  \label{eq:v.23}
  \tilde{\mathbb{M}}^{\rm\{e\}}_{\rm an}c^2=-\frac{\hbar c}{4\pi}\int\frac{d^3\vec{r}}{r\sin\vartheta}\, \gklo{p}{A}^{\rm an}(r,\vartheta)\cdot\tilde{\Phi}^2(r)\ .
\end{equation}
Concerning here the intended reduction to a purely radial problem, one integrates again over the angular part and is then left with a radial integral, see equation (\ref{eq:v.31}) below.

As a further consequence of the splitting (\ref{eq:v.11}), the Poisson constraint
$\tilde{\mathbb{N}}_{\rm G}^{\rm\{e\}}$ (\ref{eq:iv.38}) must then also be split up into
two terms
\begin{equation}
  \label{eq:v.24}
  \tilde{\mathbb{N}}_{\rm G}^{\rm\{e\}}=\tilde{\mathbb{N}}_{\rm G}^{\rm[e]}+\tilde{\mathbb{N}}_{\rm an}^{\rm\{e\}}\ ,
\end{equation}
with the isotropic part $\tilde{\mathbb{N}}_{\rm G}^{\rm[e]}$ being given by
\begin{equation}
  \label{eq:v.25}
  \tilde{\mathbb{N}}_{\rm G}^{\rm[e]}=E_{\rm R}^{\rm[e]}-\tilde{\mathbb{M}}^{\rm[e]}c^2
\end{equation}
and similarly the anisotropic part $\tilde{\mathbb{N}}_{\rm an}^{\rm\{e\}}$ by
\begin{equation}
  \label{eq:v.26}
  \tilde{\mathbb{N}}_{\rm an}^{\rm\{e\}}=E_{\rm an}^{\rm\{e\}}-\tilde{\mathbb{M}}_{\rm an}^{\rm\{e\}}c^2\ .
\end{equation}
Now the interesting point with this splitting (\ref{eq:v.24}) is the following: Since both isotropic parts $E_{\rm R}^{\rm[e]}$ (\ref{eq:v.14}) and $\tilde{\mathbb{M}}^{\rm[e]}c^2$ (\ref{eq:iii.48}) are built up by the spherically symmetric approximation $\eklo{p}{A}_0(r)$ and this potential is an {\em exact}\/ solution of the (albeit {\em approximate}\/) Poisson equation (\ref{eq:iii.36}), the spherically symmetric part $\tilde{\mathbb{N}}_{\rm G}^{\rm[e]}$ (\ref{eq:v.25}) of the general Poisson constraint $\tilde{\mathbb{N}}_{\rm G}^{\rm\{e\}}$ (\ref{eq:v.24}) is zero
\begin{equation}
  \label{eq:v.27}
  \tilde{\mathbb{N}}_{\rm G}^{\rm[e]}=0\ ;
\end{equation}
and consequently the general Poisson constraint $\tilde{\mathbb{N}}_{\rm G}^{\rm\{e\}}$ will consist exclusively of the properly anisotropic part $\tilde{\mathbb{N}}_{\rm an}^{\rm\{e\}}$:
\begin{equation}
  \label{eq:v.28}
  \tilde{\mathbb{N}}_{\rm G}^{\rm\{e\}}\Rightarrow\tilde{\mathbb{N}}_{\rm an}^{\rm\{e\}}\ ,
\end{equation}
whose vanishing must be considered separately (see below). Actually, this is the generalized reformulation of the former result (\ref{eq:iv.40}) for the simplified groundstate.

Thus, summarizing all the precedent partial results, the anisotropic energy functional $\tilde{\mathbb{E}}_{[\Phi]}$ (\ref{eq:iv.29}) can be recast to the following $SO(3)$ symmetric form (inclusive all constraints) which then neatly displays the separation into the spherically symmetric and the anisotropic gauge field degrees of freedom:
\begin{equation}
  \label{eq:v.29}
  \tilde{\mathbb{E}}_{[\Phi]}\Rightarrow\tilde{\mathbb{E}}_{\{\Phi\}}=2E_{\rm kin}+\big[E_{\rm R}^{\rm[e]}+\lambda_{\rm G}^{\rm(e)}\cdot\tilde{\mathbb{N}}_{\rm G}^{\rm[e]}\big]+\big\{E_{\rm an}^{\rm\{e\}}+\lambda_{\rm G}^{\rm(e)}\cdot\tilde{\mathbb{N}}_{\rm an}^{\rm\{e\}}\big\}+ 2\lambda_{\rm S}\cdot\tilde{\mathbb{N}}_\Phi\ .
\end{equation}
This result will readily become the point of departure for the approximate calculation of the ``anisotropic'' binding energy of the two-body system. But let us first remark here that the present functional $\tilde{\mathbb{E}}_{\{\Phi\}}$ (\ref{eq:v.29}) is to be understood in the sense that all angular factors are already integrated over so that we are left with a variational problem for the spherically symmetric fields $\tilde{\Phi}(r),\,\eklo{p}{A}_0(r),\,\gklo{p}{A}^{\sf III}(r),\,\gklo{p}{A}^{\sf V}(r),\,...$ The Poisson constraint term $\tilde{\mathbb{N}}_{\rm G}^{\rm[e]}$ is re-included here because this gives us the freedom to try (if we wish) also other configurations $\{\tilde{\Phi}(r),\eklo{p}{A}_0(r)\}$ which eventually do not obey the spherically symmetric Poisson constraint (\ref{eq:v.27}). The same does hold also with respect to the ``anisotropic'' Poisson constraint $\tilde{\mathbb{N}}_{\rm an}^{\rm\{e\}}$ (\ref{eq:v.26}).

Observe now that all contributions to the anisotropic energy $\tilde{\mathbb{E}}_{\{\Phi\}}$ (\ref{eq:v.29}) are well-defined by our precedent assumptions:
\begin{enumerate}[\bf (i)]
\item the kinetic energy $E_{\rm kin}$ is given in terms of the spherically symmetric wave amplitude $\tilde{\Phi}(r)$ by equations (\ref{eq:iii.42})-(\ref{eq:iii.44})
\item the ``spherically symmetric'' gauge field energy $E_{\rm R}^{\rm[e]}$ reads in terms of the $SO(3)$ symmetric part $\eklo{p}{A}_0(r)$ of the gauge potential as shown by equation (\ref{eq:v.14})
\item the ``spherically symmetric'' mass equivalent $\tilde{\mathbb{M}}^{\rm[e]}c^2$ is displayed by equation (\ref{eq:iii.48})
\item the ``anisotropic'' gauge field energy $E_{\rm an}^{\rm\{e\}}$ (\ref{eq:v.16}) becomes by the restriction to its lowest-order contribution $\gklo{p}{A}^{\sf III}(r)$, cf. (\ref{eq:v.3})
\begin{equation}
  \label{eq:v.30}
  E_{\rm an}^{\rm\{e\}}=-\frac{\pi^2}{320}\,\frac{\hbar c}{\as}\int_0^\infty dr\,r^2\,\left\{\left(\frac{d\gklo{p}{A}^{\sf III}(r)}{dr}\right)^2+6\left(\frac{\gklo{p}{A}^{\sf III}(r)}{r}\right)^2\right\}
\end{equation}
\item the ``anisotropic'' mass equivalent $\tilde{\mathbb{M}}^{\rm\{e\}}_{\rm an}c^2$ (\ref{eq:v.23}) is found to be of the following form
\begin{equation}
  \label{eq:v.31}
  \tilde{\mathbb{M}}^{\rm\{e\}}_{\rm an}c^2=-\frac{\pi^2}{64}\,\hbar c\int_0^\infty dr\,r\,\tilde{\Phi}^2(r)\cdot\gklo{p}{A}^{\sf III}(r)
\end{equation}
where we are satisfied again with the first anisotropic correction $\gklo{p}{A}^{\sf III}(r)$.
\end{enumerate}
\vspace{4ex}
\begin{center}
\mysubsection{3.\ Principle of Minimal Energy}
\end{center}

With the ({\em anisotropic!}\/) trial configuration being specified now through the three {\em spherically symmetric\,(!)}\/ fields $\tilde{\Phi}(r),\,\eklo{p}{A}_0(r),\,\gklo{p}{A}^{\sf III}(r)$ one can look for the extremal equations due to the functional $\tilde{\mathbb{E}}_{\{\Phi\}}$ (\ref{eq:v.29}) with respect to just these three fields. This may be done by prescribing the functional form of the three fields and equipping this form with certain variational parameters $\{\beta,\nu_k\}$ and then determining the extremal points of the corresponding energy function over the parameter space (see the former equations (\ref{eq:iii.40})). An other method would consist in leaving the three fields unspecified for the time being and deducing for them a coupled system of differential equations from the {\em principle of minimal energy}\/ (\ref{eq:iii.39}). After having solved this system, one may substitute back the corresponding solutions into the energy functional $\tilde{\mathbb{E}}_{\{\Phi\}}$ (\ref{eq:v.29}) in order to obtain its associated extremal value. An alternative ``mixed'' method would consist in prescribing the functional form (up to some variational parameters) of one or two of the three fields and determining the residual unprescribed field from its differential equation due to the {\em principle of minimal energy}\/.

In any case, it is very instructive to first deduce the system of extremal equations due to that functional $\tilde{\mathbb{E}}_{\{\Phi\}}$ (\ref{eq:v.29}). The extremalization with respect to the spherically symmetric approximation $\eklo{p}{A}_0(r)$ yields just again the former Poisson equation (\ref{eq:iii.36}). Therefore the former link (\ref{eq:iii.37}) of the spherically symmetric approximation $\eklo{p}{A}_0(r)$ to the wave amplitude $\tilde{\Phi}(r)$ still persists which entails the continued validity of the Poisson constraint (\ref{eq:v.27}). But the converse is not true, i.\,e. the Schr\"odinger-like wave equation (\ref{eq:iii.34}) receives an aditional coupling to the anisotropic part $\,\gklo{p}{A}^{\sf III}(r)$ of the gauge potential $\,\gklo{p}{A}_0(r,\vartheta)$ (\ref{eq:v.3}). Indeed, the extremalization of $\tilde{\mathbb{E}}_{\{\Phi\}}$ (\ref{eq:v.29}) with respect to the wave amplitude $\tilde{\Phi}(r)$ generates the following modification of the original equation (\ref{eq:iii.34}):
\begin{equation}
  \label{eq:v.32}
  -\frac{\hbar^2}{2M}\left(\frac{d^2\tilde{\Phi}(r)}{dr}+\frac{1}{r}\frac{d\tilde{\Phi}(r)}{dr}\right)+ \frac{\hbar^2}{2M}\frac{\lP^2}{r^2}\,\tilde{\Phi}(r)-\hbar c\left[\eklo{p}{A}_0(r)+\frac{\pi}{32}\gklo{p}{A}^{\sf III}(r)\right]\tilde{\Phi}(r)=E_*\cdot\tilde{\Phi}(r)\ .
\end{equation}
Thus it becomes evident that the gauge field influences the {\em spherically symmetric}\/
amplitude $\tilde{\Phi}(r)$ not only via its average potential $\eklo{p}{A}_0(r)$,
i.\,e. the spherically symmetric approximation (\ref{eq:v.8}), but also via its first
anisotropy correction $\gklo{p}{A}^{\sf III}(r)$. However, this influence occurs in a very
plausible way; namely the first anisotropy correction $\gklo{p}{A}^{\sf III}(r)$ is simply
added to the spherically symmetric approximation $\eklo{p}{A}_0(r)$ in order to build up a
slightly modified (but still spherically symmetric) potential, to be considered as some
kind of effective potential! But observe also that the coupling of the amplitude
$\tilde{\Phi}(r)$ to the first anisotropic correction $\gklo{p}{A}^{\sf III}(r)$ is
(roughly) ten times ($\sim\frac{\pi}{32}$) weaker than its coupling to the average
potential $\eklo{p}{A}_0(r)$! On the other hand, it will readily turn out that the
anisotropic correction $\gklo{p}{A}^{\sf III}(r)$ couples three times stronger to the
amplitude $\tilde{\Phi}(r)$ than does the average potential $\eklo{p}{A}_0(r)$, see
equation (\ref{eq:v.34}) below. Therefore it is reasonable to take into account the
anisotropy corrections for the gauge potential but not yet for the wave amplitude.

Finally, the extremal equation of $\tilde{\mathbb{E}}_{\{\Phi\}}$ (\ref{eq:v.29}) with respect to the anisotropic correction $\gklo{p}{A}^{\sf III}(r)$ must be determined. Since the latter field is contained only in two terms of $\tilde{\mathbb{E}}_{\{\Phi\}}$, namely the ``anisotropic'' energy $E_{\rm an}^{\rm\{e\}}$ (\ref{eq:v.30}) and the ``anisotropic'' mass equivalent $\tilde{\mathbb{M}}_{\rm an}^{\rm\{e\}}c^2$ (\ref{eq:v.31}), the variational proces ($\delta_{\sf III}$, say) with respect to $\gklo{p}{A}^{\sf III}(r)$ becomes very simple
\begin{equation}
  \label{eq:v.33}
  \delta_{\sf III}\tilde{\mathbb{E}}_{\{\Phi\}}\stackrel{!}{=}0\ \Rightarrow\ \delta_{\sf III}\left(E_{\rm an}^{\rm\{e\}}+\lambda_{\rm G}^{\rm(e)}\cdot\tilde{\mathbb{N}}_{\rm an}^{\rm\{e\}}\right)\stackrel{!}{=}0\ ,
\end{equation}
with the ``anisotropic'' energy $E_{\rm an}^{\rm\{e\}}$ given by equation (\ref{eq:v.30}) and the mass equivalent $\tilde{\mathbb{M}}_{\rm an}^{\rm\{e\}}c^2$ by (\ref{eq:v.31}). The result of this extremalization process is the following Poisson-like equation
\begin{equation}
  \label{eq:v.34}
  \left(\frac{d^2}{dr^2}+\frac{2}{r}\frac{d}{dr}-\frac{6}{r^2}\right)\gklo{p}{A}^{\sf III}(r)= -5\as\,\frac{\tilde{\Phi}^2(r)}{r}\ .
\end{equation}

Evidently, this equation owns some properties which are worthwhile to be mentioned
briefly. Firstly, the coupling of the anisotropy potential $\gklo{p}{A}^{\sf III}(r)$ to
the spherically symmetric wave amplitude $\tilde{\Phi}(r)$ is encanced by a factor of
$\frac{10}{\pi}\approx3$, cf. the corresponding Poisson equation (\ref{eq:iii.36}) for the
spherically symmetric approximation (i.\,e. the average potential)
$\eklo{p}{A}_0(r)$. This hints at the circumstance that the deviation from the spherical
symmetry is relatively large so that one (erroneously) tends to expect a correspondingly
large amount of anisotropy energy $E_{\rm an}^{\rm\{e\}}$ (\ref{eq:v.30}) in comparison to
the isotropic energy $E_{\rm R}^{\rm[e]}$ (\ref{eq:v.14}). Secondly, the last term
($\frac{\gklo{p}{A}^{\sf III}(r)}{r^2}$) on the left-hand side of equation (\ref{eq:v.34})
is necessary in order to ensure the former boundary conditions (\ref{eq:iv.16}) and
(\ref{eq:iv.19}) for the anisotropy {\em corrections}\/ of the gauge potential. Indeed, in
the asymptotic region of three-space ($r\rightarrow\infty,\,\tilde{\Phi}(r)\rightarrow0$)
the potential correction $\gklo{p}{A}^{\sf III}(r)$ has to obey the equation
\begin{equation}
  \label{eq:v.35}
  \left(\frac{d^2}{dr^2}+\frac{2}{r}\frac{d}{dr}-\frac{6}{r^2}\right)\gklo{p}{A}^{\sf III}(r)\cong0
\end{equation}
which imposes upon the solution $\gklo{p}{A}^{\sf III}(r)$ the condition to vanish faster ($\sim\frac{1}{r^3}$) than the Coulomb potential ($\sim\frac{1}{r}$), i.\,e. more precisely (see {\bf App.~D})
\begin{equation}
  \label{eq:v.36}
  \gklo{p}{A}^{\sf III}(r)\cong\frac{2}{\pi}\,(2\nu+3)(2\nu+2)\cdot\frac{2\beta\as}{(2\beta r)^3}\ ;
\end{equation}
of course, this hints at the \emph{quadrupole} character of this field (see \textbf{App.~D}).

Finally, there is a further interesting point in connection with that {\em ``quadrupole equation''}\/ (\ref{eq:v.34}). This concerns the fact that the ``anisotropic'' Poisson constraint $\tilde{\mathbb{N}}_{\rm an}^{\rm\{e\}}$ (\ref{eq:v.26}) vanishes whenever we can use an exact solution of the Poisson-like equation (\ref{eq:v.34}) as our trial potential $\gklo{p}{A}^{\sf III}(r)$, quite independently of the selection of a trial amplitude $\tilde{\Phi}(r)$ (see {\bf App.~D}). This is the same effect as does occur in connection with the ``isotropic'' Poisson constraint $\tilde{\mathbb{N}}_{\rm G}^{\rm[e]}$ (\ref{eq:v.25}). Thus, provided that we are able to use exact solutions of both gauge field equations (\ref{eq:iii.36}) and (\ref{eq:v.34}), then we can omit the corresponding constraint terms from the energy functional $\tilde{\mathbb{E}}_{[\Phi]}$ (\ref{eq:v.29}). In this case, this functional becomes cut down to the following simple form:
\begin{equation}
  \label{eq:v.37}
  \tilde{\mathbb{E}}_{\{\Phi\}}\Rightarrow2E_{\rm kin}+E_{\rm R}^{\rm[e]}+E_{\rm
    an}^{\rm\{e\}} \equiv 2\ru{E}{kin} + \ru{E}{R}^{\{ e\}}
  \ .
\end{equation}
Here it is also assumed that the selected trial amplitude is normalized to unity so that the normalization condition (\ref{eq:iii.46}) is satisfied.
\vspace{4ex}
\begin{center}
\mysubsection{4.\ Energy Spectrum of the Excited States}
\end{center}

After all these preparations, the calculation of the energy due to the excited states
($\nu>0$) becomes now a simple matter: since the kinetic energy $E_{\rm kin}$ is
proportional to the {\em square}\/ of the variational parameter $\beta$,
cf. (\ref{eq:iii.52})-(\ref{eq:iii.53}); and since furthermore the total gauge field
energy $E_{\rm R}^{\rm\{e\}}$ (\ref{eq:v.21}) is a {\em linear}\/ function of $\beta$ (see
equation (\ref{eq:e9a}) of {\bf App.\,E}), the extremalization process (\ref{eq:iii.40}),
or more concretely
\begin{subequations}
  \begin{align}
  \label{eq:v.38a}
  \frac{\partial\mathbb{E}^{   \rm\{IV\} }(\beta,\nu)}{\partial\beta}&=0\\
  \label{eq:v.38b}
  \frac{\partial\mathbb{E}^{\rm \{IV\} }(\beta,\nu)}{\partial\nu}&=0
  \end{align}
\end{subequations}
with
\begin{equation}
  \label{eq:v.39}
  \mathbb{E}^{\rm\{IV\}}(\beta,\nu)=\frac{e^2}{\aB}\left\{\big(2\aB\beta\big)^2\cdot\varepsilon_{\rm kin}(\nu)-\big(2\aB\beta\big)\cdot\varepsilon_{\rm tot}^{\rm\{e\}}(\nu)\right\}\ ,
\end{equation}
yields again the desired energy values~$\mathbb{E}^{\{n\}}_\wp$, namely as the minimal
values of the reduced energy function~$\mathbb{E}^\wp_{\{\nu\}}$ 
\begin{equation}
  \label{eq:v.40}
  \mathbb{E}^{\{n\}}_\wp \doteqdot \mathbb{E}^\wp_{\{\nu\}}\big|_{\mathrm{min}}\ .
\end{equation}
Here, the reduced energy function~$\mathbb{E}^\wp_{\{\nu\}}$ is again defined through
\begin{equation}
  \label{eq:v.41}
  \mathbb{E}^\wp_{\{\nu\}} \doteqdot -\frac{e^2}{4\aB}\cdot S_\wp^{\rm\{a\}}(\nu)
\end{equation}
with the ``anisotropic'' spectral function $S_\wp^{\rm\{a\}}(\nu)$ being itself defined through
\begin{equation}
  \label{eq:v.42}
  S_\wp^{\rm\{a\}}(\nu)\doteqdot\frac{\varepsilon_{\rm tot}^{\rm\{e\}}(\nu)^2}{\varepsilon_{\rm kin}(\nu)}\ .
\end{equation}
The (dimensionless) total potential function $\varepsilon_{\rm tot}^{\rm\{e\}}(\nu)$
is given by equation (\ref{eq:e9b}) of {\bf App.\,E} and is obviously nothing else than
the ``anisotropic'' generalization of the former $\varepsilon_{\rm pot}(\nu)$
(\ref{eq:a28}). Thus the present energy function $\mathbb{E}^{\rm\{IV\}}(\beta,\nu)$
(\ref{eq:v.39}) is the ``anisotropic'' generalization of the ``isotropic''
$\mathbb{E}^{\rm[IV]}(\beta,\nu)$ (\ref{eq:iii.54'}) so that it merely remains to determine
again the extremal values of the spectral function~$S^{\{a\}}_\wp(\nu)$ (\ref{eq:v.42}) for any principal
quantum number $n_\wp\ (=\lP+1)$; for the numerical details see {\bf App.\,F}.

The \textbf{table~2} (below) displays the RST predictions of the positronium
energy~$\mathbb{E}_\wp^{\{n\}}$~(\ref{eq:v.40}) in an arrangement quite similar to the
results of the spherically symmetric approximation in
\textbf{table~1}~(p.\pageref{table1}). Here it becomes obvious that the deviations from
the conventional results (last column in \textbf{tables~1} and~\textbf{2}) are
considerably reduced through the inclusion of the anisotropy corrections. Indeed, defining
the ``anisotropic `` deviations by
\begin{equation}
  \label{eq:v.43}
  \ru{\Delta}{T}(\nu_*^{\{n\}}) \doteqdot \frac{\ru{E}{conv}^{(n)}-\mathbb{E}_\wp^{\{n\}} }{\ru{E}{conv}^{(n)}}
\end{equation}
(last column of \textbf{table~2}) in a quite analogous manner to their ``isotropic'' counterparts
$\ru{\Delta}{T}(\nu_*^{[n]})$ (\ref{eq:f6}) (last column of \textbf{table~1}), one clearly realizes that
the ``\emph{anisotropic}'' deviations (\ref{eq:v.43}) in \textbf{table~2} amount to less than half of the
``\emph{isotropic}'' deviations (\textbf{table~1}). Observe here that the
average~$\overline{\rugk{\Delta}{T}}$ of the deviations (\ref{eq:v.43}), being defined by
\begin{equation}
  \label{eq:v.44}
  \overline{\rugk{\Delta}{T}} \doteqdot \frac{1}{14}\sum_{n_\wp}|\ru{\Delta}{T}(\nsn) |\ ,
\end{equation}
does amount now to~$ \overline{\rugk{\Delta}{T}}=3,1\%$ in comparison to the corresponding
former deviation $\overline{\ekru{\Delta}{T}}=7,7\%$ for the spherically symmetric
approximation (\textbf{table~1}). The groundstate~$(\nP=1)$ is disregarded here because the
presently applied \emph{electrostatic approximation} is supposed to fail for the lowest
energy eigenstate (see the discussion of this in the precedent paper~\cite{ms1}).

Clearly such a result suggests to consider the
higher-order approximations which eventually do shift the RST predictions even closer to
their conventional counterparts, see \textbf{Fig.V.A} below.


\begin{flushleft}\label{table2}
  \begin{tabular}{|c|c|c|c|c|c|c|}
  \hline
 $\nP $ & $\nu_n$ & $S_n^{\{a\}}(\nu=\nu_n)$ & $S_n|_{\textrm{max}}$ & $\mathbb{E}_\wp^{\{n\}}$\ [eV] & $\Ea{E}{n}{conv}$\ [eV] &
$\ru{\Delta}{T}(\nu_*^{\{n\}})$\\
 $ =(\lP+1)$ & &  (\ref{eq:v.41}) &\textbf{App.F} & (\ref{eq:v.40}) &  (\ref{eq:i.4}) &  (\ref{eq:v.43})  \\
  \hline\hline
2 & 2 & 2.42009e-01 & 2.42186e-01 & -1.647e+00 & -1.701e+00 & 3.1\\ \hline
3 & 4 & 1.05465e-01 & 1.05479e-01 & -7.175e-01 & -7.558e-01 & 5.1 \\ \hline
4 & 6 & 5.92349e-02 & 5.92388e-02 & -4.030e-01 & -4.251e-01 & 5.2\\ \hline 
5 & 17/2 & 3.80558e-02 & 3.80558e-02 & -2.589e-01 & -2.721e-01 & 4.9\\ \hline 
6 & 11 & 2.65700e-02 & 2.65701e-02 & -1.807e-01 & -1.890e-01 & 4.3\\ \hline 
10 & 45/2 & 9.77183e-03 & 9.77187e-03 & -6.647e-02 & -6.802e-02 & 2.3\\ \hline 
15 & 77/2 & 4.42916e-03 & 4.42916e-03 & -3.013e-02 & -3.023e-02 & 0.3\\ \hline 
20 & 113/2 & 2.52621e-03 & 2.52621e-03 & -1.718e-02 & -1.701e-02 & -1.0\\ \hline 
25 & 76 & 1.63360e-03 & 1.63360e-03 & -1.111e-02 & -1.088e-02 & -2.1\\ \hline 
30 & 193/2 & 1.14365e-03 & 1.14365e-03 & -7.779e-03 & -7.558e-03 & -2.9\\ \hline 
35 & 118 & 8.45729e-04 & 8.45729e-04 & -5.753e-03 & -5.553e-03 & -3.6\\ \hline 
40 & 141 & 6.51019e-04 & 6.51019e-04 & -4.428e-03 & -4.251e-03 & -4.2\\ \hline 
50 & 189 & 4.20206e-04 & 4.20206e-04 & -2.858e-03 & -2.721e-03 & -5.0\\ \hline 
  \end{tabular}
\end{flushleft}
{\textbf{Table 2:}\hspace{0.5cm} \emph{\large\textbf{RST
      predictions~$\mathbf{\mathbb{E}_\wp^{\{n\}}}$ (\ref{eq:v.40}) including the
      Anisotropic \\ \centerline{Corrections}
 }}  }
\mytable{\emph{Table 2: RST predictions}~$\mathbf{\mathbb{E}_\wp^{\{n\}}}$ (\ref{eq:v.40})
  \emph{including the\\\hspace*{2mm} Anisotropic Corrections}}

The ``\emph{anisotropic}'' RST predictions~$\mathbb{E}_\wp^{\{n\}}$ (\ref{eq:v.40})
deviate only half as much from the conventional predictions~$\Ea{E}{n}{conv}$ as is the
case for the spherically symmetric approximation of lowest order
(\textbf{table~1}). Indeed, the average deviation~$\overline{\rugk{\Delta}{T}}$
(\ref{eq:v.44}) is now $\overline{\rugk{\Delta}{T}}=3,1\%$ in place of
$\overline{\ekru{\Delta}{T}}=7,7\%$ for the spherically symmetric approximation. Observe
also that for quantum number~$\nP\approx 15$ the deviation is minimal as predicted by an
inspection of the ``isotropic'' and ``anisotropic'' deviations~$\Delta_{\{\nu\}}$
and~$\ru{\Delta}{T}(\nu_*^{[n]})$ in \textbf{Fig.F.II}, \textbf{App.F}


\enlargethispage{1cm}
\begin{center}
\epsfig{file=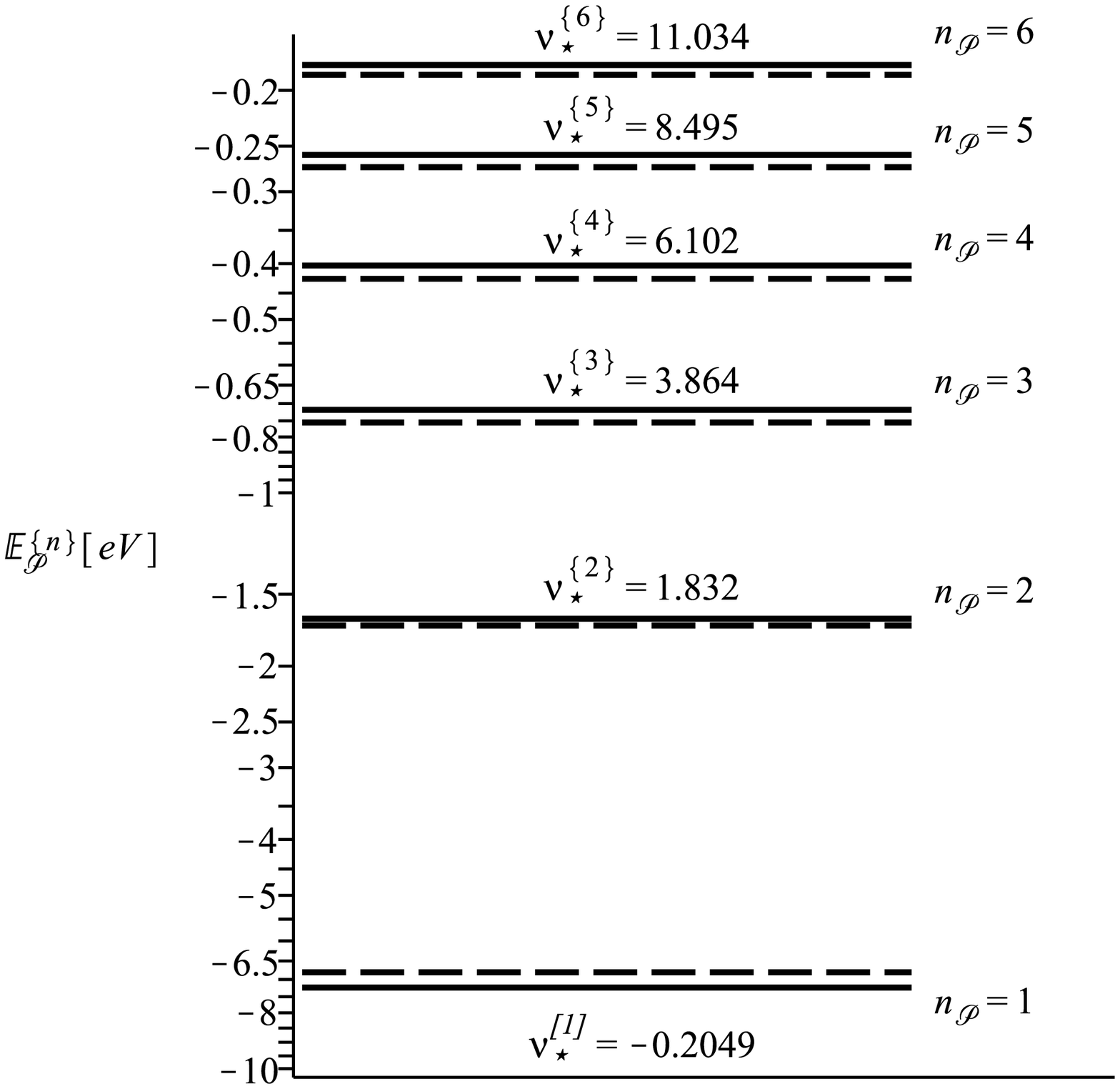,height=15cm}
\end{center}
{\textbf{Fig.~V.A}\hspace{5mm} \emph{\large\textbf{Energy
      Spectrum~$\mathbf{\mathbb{E}_\wp^{\{n\}}}$  (\ref{eq:v.40}) inclusive Anisotropy \\
      \centerline{Corrections} } }}
\myfigure{Fig.~V.A: Energy Spectrum~$\mathbf{\mathbb{E}_\wp^{\{n\}}}$  (\ref{eq:v.40}) inclusive\\\hspace*{2mm} Anisotropy Corrections}
\indent

The inclusion of the anisotropy corrections diminishes the deviations of the RST
predictions from their conventional counterparts by (roughly) 50\%,
cf.~\textbf{Fig.III.A}: This improvement suggests that the regard of the anisotropy
effects of higher approximation order will shift the RST predictions (solid lines) even
closer towards their conventional counterparts (broken lines).

\section[\ Improving the Spherically Symmetric Approximation]{Improving the Spherically Symmetric Approximation}
\indent

Concerning the essential point with the anisotropy corrections, it is instructive to first
recall those results being collected in \textbf{table~2} on p.~\pageref{table2}. These
results let appear the magnitude of the anisotropy corrections in the range from 3\% (for
the first excited state, $n_\wp=2$) up to 0.3\% (for the excited state with principal
quantum number $n_\wp=15$). Evidently, the consideration of the anisotropy corrections in
{\bf Sect.\,V} improves the ``isotropic'' RST predictions (\textbf{table~1}) by (roughly)
50\%. Surely, the higher-order corrections of the \emph{anisotropic} type can be expected
to shift the RST predictions further towards their conventional counterparts
(\ref{eq:i.4}); but for the moment we are rather interested in testing the extent to
which further improvements of the predictions may emerge \emph{within the framework of the
  spherically symmetric approximation} itself ($\leadsto$ {\em isotropic}\/ corrections).

In this sense, we study now the following three-parameter generalization
$\tilde{\Phi}_b(r)$ of the former two-parameter ansatz $\tilde{\Phi}(r)$
(\ref{eq:iii.50})-(\ref{eq:iii.51}):
\begin{equation}
  \label{eq:vi.1}
  \tilde{\Phi}_b(r)=\Phi_*r^\nu(b_0+b_1 r){\rm e}^{-\beta r}\ .
\end{equation} 
Here, the additional variational parameters are the (real) constants $b_0,b_1$ whose
presence entails a certain generalization of the normalization constant $\Phi_*$
(\ref{eq:iii.51}) so that the general normalization condition (\ref{eq:iii.38}) becomes
specified down to the following form
\begin{equation}
  \label{eq:vi.2}
p_0^2+4(\nu+1)\cdot p_0p_1+(2\nu+3)(2\nu+2)\cdot p_1^2 = 1\ .
\end{equation} 
Here, the parameters~$p_0,p_1$ are merely some convenient modification of the original
ansatz parameters~$b_0,b_1$ (\ref{eq:vi.1}), i.e.
\begin{subequations}
\begin{align}
  \label{eq:vi.3a}
 p_0 & \doteqdot \sqrt{\frac{\pi}{2}\cdot\Gamma(2\nu+2)}\,\frac{\Phi_*}{(2\beta)^{\nu+1}}\cdot b_0\\*
  \label{eq:vi.3b}
 p_1 &\doteqdot
 \sqrt{\frac{\pi}{2}\cdot\Gamma(2\nu+2)}\,\frac{\Phi_*}{(2\beta)^{\nu+2}}\cdot b_1\ .
\end{align}
\end{subequations}
Clearly, for $p_1\rightarrow 0,p_0\rightarrow 1$ one is led back to the former
two-parameter ansatz (\ref{eq:iii.50})-(\ref{eq:iii.51}). Observe also that the present
normalization condition (\ref{eq:vi.2}) is a slight generalization of the former condition
(IV.41) of ref.~\cite{ms3} which emerges from the present condition (\ref{eq:vi.2})
through putting~$\nu$ to zero~$(\nu\to 0)$.

\vspace{4ex}
\begin{center}
\mysubsection{1.\ Improved Energy Function}
\end{center}

This new trial amplitude $\tilde{\Phi}_b(r)$ (\ref{eq:vi.1}) must now be used in order to
calculate anew the kinetic energy $E_{\rm kin}$ (\ref{eq:iii.42})-(\ref{eq:iii.44}) and
the gauge field energy $E_{\rm R}^{\rm[e]}$ (\ref{eq:iii.45}) with its corresponding mass
equivalent $\tilde{\mathbb{M}}^{\rm[e]}c^2$ (\ref{eq:iii.48}). Both objects will then obey
the Poisson identity (\ref{eq:iii.47}) provided the electrostatic potential
$\eklo{p}{A}_0(r)$ is taken again as the exact solution of the spherically symmetric
Poisson equation (\ref{eq:iii.36}). Thus, both constraints (\ref{eq:iii.46}) and
(\ref{eq:iii.47}) will be automatically satisfied and therefore the non-relativistic
energy functional $\tilde{\mathbb{E}}_{[\Phi]}$ (\ref{eq:iii.41}) will consist exclusively
of its physical contributions (i.\,e. kinetic energy and gauge field energy):
\begin{equation}
  \label{eq:vi.4}
  \tilde{\mathbb{E}}_{[\Phi]}\Rightarrow\mathbb{E}^{\rm[IV]}_{[\Phi]}\doteqdot 2E_{\rm kin}+E_{\rm R}^{\rm[e]}\ .
\end{equation} 
Clearly, this is again the non-relativistic version of the former relativistic case $E_{\rm[T]}^{\rm(IV)}$ (\ref{eq:ii.33}).

Turning here first to the kinetic energy $E_{\rm kin}$, one recalls that this energy
contribution is the sum of the radial part $\rklo{r}{E}_{\rm kin}$ and the longitudinal
part $\rklo{\vartheta}{E}_{\rm kin}$, cf. equation (\ref{eq:iii.42}). Both parts must now
be calculated by use of the generalized trial amplitude $\tilde{\Phi}_b(r)$
(\ref{eq:vi.1}). This yields first for the radial part $\rklo{r}{E}_{\rm kin}$ the
usual form:
\begin{equation}
  \label{eq:vi.5}
  \rklo{r}{E}_{\rm kin}=\frac{e^2}{2\aB}\,(2\beta\aB)^2\cdot \rklo{r}{\varepsilon}_{\rm
    kin}(\nu,p_0,p_1)\ . 
\end{equation}
Here, the kinetic function of the radial type $\rklo{r}{\varepsilon}_{\rm
  kin}(\nu,p_0,p_1)$ is defined in terms of the dimensionless form $\tilde{\Phi}_{\nu
  p}(y)$ of the original trial amplitude $\tilde{\Phi}_b(r)$ (\ref{eq:vi.1}) through
\begin{gather}
  \label{eq:vi.6}
  \rklo{r}{\varepsilon}_{\rm kin}(\nu,p_0,p_1)=\frac{1}{\Gamma(2\nu+2)}\int
  dy\,y\left[\frac{d}{dy}\tilde{\Phi}_{\nu p}(y)\right]^2\\
  \nonumber
  (\,\tilde{\Phi}_{\nu p}(y)\doteqdot y^\nu(p_0+p_1 y){\rm e}^{-\frac{y}{2}},\ \mbox{see {\bf App.\,G}}\,)\ .
\end{gather}
The explicit calculation lets emerge this radial kinetic function in the following shape
\begin{equation}
  \label{eq:vi.7}
  \rklo{r}{\varepsilon}_{\rm kin}(\nu,p_0,p_1)=\frac{p_0^2}{4(2\nu+1)}+\frac{\nu}{2\nu+1}\cdot p_0p_1+\frac{\nu+1}{2}\cdot p_1^2\ ,
\end{equation}
i.e. a quadratic polynomial with respect to the variational parameters~$p_0,p_1$.
Clearly, when the parameter~$p_1$ tends to zero ($p_1\rightarrow0$) the present radial
function $\rklo{r}{\varepsilon}_{\rm kin}(\nu,p_0,p_1)$ (\ref{eq:vi.7}) collapses to the
corresponding {\em radial part}\/ of the result (\ref{eq:iii.53}) which is due to the
former two-parameter trial amplitude $\tilde{\Phi}(r)$ (\ref{eq:iii.50}).

A similar line of arguments may also be applied to the \emph{longitudinal} kinetic energy
$\rklo{\vartheta}{E}_{\rm kin}$ (\ref{eq:iii.44}) which then adopts an analogous form
\begin{equation}
  \label{eq:vi.8}
  \rklo{\vartheta}{E}_{\rm kin}=\frac{e^2}{2\aB} (2\beta\aB)^2 \lP^2\cdot\rklo{\vartheta}{\varepsilon}_{\rm kin}(\nu,p_0,p_1)
\end{equation}
with the longitudinal type of kinetic energy function $\rklo{\vartheta}{\varepsilon}_{\rm kin}(\nu,p_0,p_1)$ being found as
\begin{equation}
  \label{eq:vi.9}
  \rklo{\vartheta}{\varepsilon}_{\rm kin}(\nu,p_0,p_1)=\frac{p_0^2}{2\nu(2\nu+1)} +
  \frac{2p_0p_1}{2\nu+1} + p_1^2\ .
\end{equation}
Clearly, this quadratic form is again the longitudinal counterpart of the radial object
$\rklo{r}{\varepsilon}_{\rm kin}(\nu,p_0,p_1)$ (\ref{eq:vi.7}); and if the variational parameter
$p_1$ tends to zero ($\leadsto p_0\rightarrow 1$) one ends up here with the {\em longitudinal part}\/
of the former $\varepsilon_{\rm kin}(\nu)$ (\ref{eq:iii.53}). Thus summarizing the result
for the kinetic energy $E_{\rm kin}$ (\ref{eq:iii.42}) as the sum of the radial and
longitudinal parts, one ultimately arrives at
\begin{equation}
  \label{eq:vi.10}
  E_{\rm kin}=\frac{e^2}{2\aB}  (2\aB\beta)^2  \cdot\varepsilon_{\rm kin}(\nu,p_0,p_1)
\end{equation}
with the total kinetic function $\varepsilon_{\rm kin}(\nu,p_0,p_1)$ emerging here as the sum of
the radial and longitudinal parts:
\begin{equation}
  \begin{split}
  \label{eq:vi.11}
  \varepsilon_{\rm kin}(\nu,p_0,p_1)& \doteqdot\rklo{r}{\varepsilon}_{\rm
    kin}(\nu,p_0,p_1)+\rklo{\vartheta}{\varepsilon}_{\rm kin}(\nu,p_0,p_1)\cdot\lP^2\\
  &= \frac{\nu+2\lP^2}{4\nu(2\nu+1)}\cdot p_0^2 
  +\frac{\nu+2\lP^2}{2\nu+1}\cdot p_0p_1 
  +\left(\frac{\nu+1}{2}+\lP^2\right)\cdot p_1^2\ .    
  \end{split}
\end{equation}
As a consistency check one easily verifies that in the limit $p_1\rightarrow0$ this
three-parameter result (\ref{eq:vi.11}) collapses again to the former result
$\varepsilon_{\rm kin}(\nu)$ (\ref{eq:iii.53}) for the two-parameter variational
ansatz. For~$\nu=0$ and~$\lP=0$, the present kinetic energy~$\ru{E}{kin}$ (\ref{eq:vi.10})
agrees with the former kinetic energy~$\ru{E}{kin}$  (\ref{eq:iv.22a})-(\ref{eq:iv.22b}) of
ref.~\cite{ms3}.

The gauge field energy $E_{\rm R}^{\rm[e]}$ as the second constituent of the energy
functional $\mathbb{E}_{[\Phi]}^{\rm(IV)}$ (\ref{eq:vi.4}) is considerably more
complicated than the kinetic energy $E_{\rm kin}$. Here, the Poisson identity is of help
in order to find the simplest form of this energy contribution (see {\bf App.G} for the
details). Summarizing the results, one finds the desired gauge field energy $E_{\rm
  R}^{\rm[e]}$ (\ref{eq:iii.45}) to look as follows:
\begin{equation}
  \label{eq:vi.12}
  E_{\rm R}^{\rm[e]}=-\frac{e^2}{\aB}\,(2\aB\beta)\cdot\varepsilon_{\rm pot}(\nu,p_0,p_1)\ .
\end{equation}
Here, the {\em total potential function} $\varepsilon_{\rm pot}(\nu,p_0,p_1)$ is obviously the
generalization of the former potential function $\varepsilon_{\rm pot}(\nu)$
(\ref{eq:iii.54}) which for the simple trial amplitude $\tilde{\Phi}(r)$ (\ref{eq:iii.50})
is explicitly given by equation (\ref{eq:a28}) of {\bf App.\,A}. But clearly, the present
generalized trial amplitude $\tilde{\Phi}_b(r)$ (\ref{eq:vi.1}) must now entail a more
complicated potential function $\varepsilon_{\rm pot}(\nu,p_0,p_1)$; and indeed, the latter
function is found to be of the following form:
\begin{equation}
  \label{eq:vi.13}
 \begin{split}
\varepsilon_{\rm pot}(\nu,p_0,p_1) =\ p_0^4\cdot \varepsilon_{\rm
  pot}(\nu)+4p_0^3p_1\cdot\varepsilon_{\rm I}(\nu) 
 + 4p_0^2p_1^2\cdot\varepsilon_{\rm
    II}(\nu) &+ 4p_0p_1^3\cdot\varepsilon_{\rm III}(\nu)\\* & + p_1^4\cdot\varepsilon_{\rm IV}(\nu)\ ,
 \end{split}
\end{equation}
cf. equation (\ref{eq:g16}) of {\bf App.\,G}.

This is obviously a fourth-order polynomial with respect to the variational parameters
$p_0,p_1$  (\ref{eq:vi.3a})-(\ref{eq:vi.3b}) where the four additional coefficients $\varepsilon_{\rm
  I},...,\varepsilon_{\rm IV}$ are ordinary functions of the second variational parameter
$\nu$. Concerning their specific shape, one finds (see {\bf App.\,G}) that there are
certain interrelationships so that it is sufficient to specify in detail only three of
them (e.\,g. $\varepsilon_{\rm pot},\,\varepsilon_{\rm I},\,\varepsilon_{\rm II}$) and the
remaining two (i.\,e. $\varepsilon_{\rm III}$ and $\varepsilon_{\rm IV}$) are then related
to the first three in a simple way. More concretely, the first {\em auxiliary potential
  function}\/ $\varepsilon_{\rm I}(\nu)$ is given quite generally in terms of the
(dimensionless) potential $a_\nu(y)$ through
\begin{eqnarray}
  \label{eq:vi.14}
  \varepsilon_{\rm I}(\nu)&=&-\frac{1}{\Gamma(2\nu+2)^2}\int_0^\infty dy\,y^{2\nu+2}{\rm e}^{-y}\cdot a_\nu(y)
  \nonumber\\*
  &\equiv&-\frac{1}{\Gamma(2\nu+2)^2}\int_0^\infty dy\,y^{2\nu+1}{\rm e}^{-y}\cdot a_{\nu+\frac{1}{2}}(y)\ .
\end{eqnarray}
Here, the potential $a_\nu(y)$ is the solution of the Poisson-like equation
\begin{equation}
  \label{eq:vi.15}
  \frac{d^2a_\nu(y)}{dy^2}+\frac{2}{y}\frac{da_\nu(y)}{dy}=y^{2\nu-1}\cdot{\rm e}^{-y}
\end{equation}
which essentially is a dimensionless form of the original Poisson equation (\ref{eq:iii.36}). Therefore the solution can easily be specified as
\begin{equation}
  \label{eq:vi.16}
  a_\nu(y)=-\Gamma(2\nu+2)\cdot\left\{1-{\rm e}^{-y}\cdot\sum_{n=0}^\infty\frac{n}{\Gamma(2\nu+2+n)}\,y^{2\nu+n}\right\}\ ,
\end{equation}
cf. equation (\ref{eq:g13a}) of {\bf App.\,G}; and if this is used in order to explicitly calculate the auxiliary function $\varepsilon_{\rm I}(\nu)$ (\ref{eq:vi.14}) one ends up with
\begin{eqnarray}
  \nonumber
  \varepsilon_{\rm I}(\nu)&=&\frac{2\nu+2}{2\nu+1}\cdot\left\{1-\frac{1}{\Gamma(2\nu+3)\cdot2^{4\nu+3}}\sum_{n=0}^\infty\frac{n}{2^n}\cdot\frac{\Gamma(4\nu+3+n)}{\Gamma(2\nu+2+n)}\right\}\\
  \label{eq:vi.17}
  &=&1- \frac{1}{\Gamma(2\nu+2)\cdot2^{4\nu+3}}\sum_{n=0}^\infty\frac{n}{2^n}\cdot\frac{\Gamma(4\nu+3+n)}{\Gamma(2\nu+3+n)}\ .
\end{eqnarray}

Next, the second coefficient $\varepsilon_{\rm II}(\nu)$ of $\varepsilon_{\rm pot}(\nu,p_0,p_1)$
in (\ref{eq:vi.13}) turns out to be the sum of two contributions, i.\,e.
\begin{equation}
  \label{eq:vi.18}
  \varepsilon_{\rm II}(\nu)=\varepsilon'_{\rm II}(\nu)+\varepsilon''_{\rm II}(\nu)
\end{equation}
where the first part is related to the well-known $\varepsilon_{\rm pot}(\nu)$ through
\begin{equation}
  \label{eq:vi.19}
  \varepsilon'_{\rm II}(\nu)=(2\nu+2)^2\cdot\varepsilon_{\rm pot}\big|_{\nu+\frac{1}{2}}
\end{equation}
and the second part is given by
\begin{eqnarray}
  \nonumber
  \varepsilon''_{\rm II}(\nu)&=&\frac{(\nu+1)(2\nu+3)}{2\nu+1}\left\{1-\frac{1}{\Gamma(2\nu+4)\cdot2^{4\nu+4}}\sum_{n=0}^\infty\frac{n}{2^n}\cdot\frac{\Gamma(4\nu+4+n)}{\Gamma(2\nu+2+n)}\right\}\\
  \label{eq:vi.20}
  &=&2(\nu+1)\left\{1- \frac{1}{\Gamma(2\nu+2)\cdot2^{4\nu+4}}\sum_{n=0}^\infty\frac{n}{2^n}\cdot\frac{\Gamma(4\nu+4+n)}{\Gamma(2\nu+4+n)}\right\}\ .
\end{eqnarray}
Finally, as mentioned above, the last two coefficients $\varepsilon_{\rm III}(\nu)$ and $\varepsilon_{\rm IV}(\nu)$ in (\ref{eq:vi.13}) may be expressed in terms of the precedent potential functions which are to be shifted as follows:
\begin{subequations}
  \begin{align}
  \label{eq:vi.21a}
  \varepsilon_{\rm III}(\nu)&=(2\nu+2)^2\cdot\varepsilon_{\rm I}\big|_{\nu+\frac{1}{2}} \\
  \label{eq:vi.21b}
  \varepsilon_{\rm IV}(\nu)&=(2\nu+3)^2\cdot(2\nu+2)^2\cdot\varepsilon_{\rm pot}\big|_{\nu+1}\ .
  \end{align}
\end{subequations}

Thus the total potential function $\varepsilon_{\rm pot}(\nu,p_0,p_1)$ (\ref{eq:vi.13})
ultimately appears as a well-defined function of the variational parameters $\nu$ and
$p_0,p_1$; and if this is substituted back into the gauge field energy $E_{\rm R}^{\rm[e]}$
(\ref{eq:vi.12}) we obtain this quantity, as desired, as a function of all four
variational parameters $\{\beta;\nu,p_0,p_1\}$
\begin{equation}
  \label{eq:vi.22}
  \begin{split}
  E_{\rm R}^{\rm[e]} =-\frac{e^2}{\aB}(2\aB\beta)\cdot \left\{p_0^4\cdot\ru{\varepsilon}{pot}(\nu)
  + 4p_0^3p_1\cdot\ru{\varepsilon}{I}(\nu) +
  4p_0^2p_1^2\cdot\ru{\varepsilon}{II}(\nu)\right. &+ 4p_0p_1^3 \cdot\ru{\varepsilon}{III}(\nu)\\*
  &+ p_1^4\cdot \ru{\varepsilon}{IV}(\nu)\left.\right\} \ .
  \end{split}
\end{equation} 
Here, one is easily convinced that for $p_1\rightarrow0$ this four-parameter energy of the
gauge field actually collapses to the two-parameter result (\ref{eq:iii.54}) being due to
the simpler trial amplitude (\ref{eq:iii.50}).

But now that both the kinetic energy $E_{\rm kin}$ and the gauge field energy $E_{\rm
  R}^{\rm[e]}$ are at hand in their improved versions,
cf. (\ref{eq:vi.10})-(\ref{eq:vi.11}) and (\ref{eq:vi.22}), one can add them together in
order to find from the energy functional $\mathbb{E}_{[\Phi]}^{\rm(IV)}$ (\ref{eq:vi.4})
the corresponding energy function $\mathbb{E}^{\rm(IV)}(\beta;\nu,p_0,p_1)$ which then
appears in the form
\begin{equation}
  \label{eq:vi.23}
  \mathbb{E}^{\rm(IV)}(\beta;\nu,p_0,p_1)=\frac{e^2}{\aB}\left\{
    (2\aB\beta)^2\cdot\varepsilon_{\rm kin}(\nu,p_0,p_1) - (2\aB\beta)\cdot\varepsilon_{\rm pot}(\nu,p_0,p_1)\right\}\ .
\end{equation}
Of course, this is again the $p$-generalized form of the simple predecessor $\mathbb{E}^{\rm(IV)}(\beta,\nu)$ (\ref{eq:iii.54'}).
\vspace{4ex}
\begin{center}
\mysubsection{2.\ Principle of Minimal Energy}
\end{center}

With an improved energy function $\mathbb{E}^{\rm(IV)}(\beta;\nu,p_0,p_1)$ being at hand now,
one can next apply the principle of minimal energy (\ref{eq:iii.39})-(\ref{eq:iii.40}) in
order to look for the extremal points of that function. First, turn here to the
determination of the first variational parameter $\beta$ in terms of $\nu$ and $p$ by
exploiting the first of the equations (\ref{eq:iii.40}). More concretely, the requirement
\begin{equation}
  \label{eq:vi.24}
  \frac{\partial\mathbb{E}^{\rm(IV)}(\beta;\nu,p_0,p_1)}{\partial\beta}=0
\end{equation}
yields the equilibrium value ($\beta_*$) of $\beta$ as
\begin{equation}
  \label{eq:vi.25}
  2\aB\beta_*=\frac{1}{2}\cdot\frac{\varepsilon_{\rm pot}(\nu,p_0,p_1)}{\varepsilon_{\rm kin}(\nu,p_0,p_1)}\ ,
\end{equation}
and if this is substituted back into the original energy function
$\mathbb{E}^{\rm(IV)}(\beta;\nu,p_0,p_1)$ (\ref{eq:vi.23}) one obtains the reduced energy
function $\mathbb{E}_{\rm T}(\nu,p_0,p_1)$ in the following form:
\begin{equation}
  \label{eq:vi.26}
  \mathbb{E}_{\rm T}(\nu,p_0,p_1)\doteqdot\mathbb{E}^{\rm(IV)}(\beta;\nu,p_0,p_1)\Big|_{\beta=\beta_*} =-\frac{e^2}{4\aB}\cdot S_\wp(\nu,p_0,p_1)\ ,
\end{equation}
where the improved spectral function $S_\wp(\nu,p_0,p_1)$ is given now by
\begin{equation}
  \label{eq:vi.27}
  S_\wp(\nu,p_0,p_1)\doteqdot\frac{\varepsilon^2_{\rm pot}(\nu,p_0,p_1)}{\varepsilon_{\rm kin}(\nu,p_0,p_1)}\ .
\end{equation}
Clearly, this is again the $p$-generalization of the simple spectral function
(\ref{eq:iii.55b}) which is due to the simpler trial ansatz (\ref{eq:iii.50}). Indeed, for
$p_1\rightarrow0$ the present three-parameter result $S_\wp(\nu,p_0,p_1)$ (\ref{eq:vi.27})
becomes reduced to its one-parameter predecessor $S_\wp(\nu)$ (\ref{eq:iii.55b}).
\vspace{4ex}
\begin{center}
\mysubsection{3.\ Simplified Groundstate Demonstration}
\end{center}

The improving effect due to the presence of the additional variational parameters
$p_0,p_1$ may be roughly estimated by fixing the parameter $\nu$ in equation
(\ref{eq:vi.27}) and thus discussing the spectral function $S_\wp(\nu,p_0,p_1)$
exclusively as a function of $p_0,p_1$. If one knew the optimal value of $\nu$, one could find
in this way the desired extremal value of the energy function $\mathbb{E}_{\rm
  T}(\nu,p_0,p_1)$ (\ref{eq:vi.26}). But for the moment we will be satisfied with
only a moderate improvement of the results of {\bf Sect.\,III} which are due to the
simpler variational ansatz (\ref{eq:iii.50}), i.\,e. we will base our improvement upon
those values $\nu_*^{\rm [n]}$ which are displayed in \textbf{table~1}. Recall that those results
yielded a deviation of (roughly) 10\% from the conventional predictions within the
framework of the {\em two-parameter}\/ amplitude $\tilde{\Phi}(r)$ (\ref{eq:iii.50}). 

For the sake of a transparent but sufficiently detailed demonstration of our improvement,
we will restrict ourselves here to the groundstate situation ($n_\wp=1$) and, properly
speaking, this would imply that we have to choose $\nu_*=-0.2049...$ (see
\textbf{table~1}, first line). However, in order to keep our groundstate demonstration as
uncomplicated as possible, we are satisfied with choosing $\nu=0$. This choice reduces our
four-parameter amplitude $\tilde{\Phi}_b(r)$ (\ref{eq:vi.1}) to a three-parameter trial
ansatz which (for $p_1=0$) {\em incidentally}\/ predicts {\em exactly}\/ the same
groundstate energy as the conventional theory (\ref{eq:i.4}), see the discussion of this
in ref.s \cite{ms1,ms4}. It is true, a more realistic choice for the groundstate would be
$\nu_*=-0,2049$ (see the first line of \textbf{table~1}), but this would render the intended
demonstration unadequately complicated, inasmuch the groundstate
($n_\wp=1\Leftrightarrow\lP=0$) is in any case badly predicted by the present
\emph{electrostatic} approximation.

By these arrangements, the two-dimensional problem of looking for the extremal values of
the energy function $\mathbb{E}_{\rm T}(\nu,p_0,p_1)$ (\ref{eq:vi.26})-(\ref{eq:vi.27})
becomes reduced to a one-dimensional problem, namely to look for the extremal values of
the function $\mathbb{E}^{[0]}_{\rm T}(p_0,p_1)$
\begin{equation}
  \label{eq:vi.28}
  \mathbb{E}^{[0]}_{\rm T}(p_0,p_1)\doteqdot\mathbb{E}_{\rm T}(\nu,p_0,p_1)\Big|_{\nu=0}\ .
\end{equation}
Or equivalently, this means to look for the extremal values of the reduced spectral
function $S^{[0]}_\wp(p_0,p_1)\doteqdot S_\wp(0,p_0,p_1)$
\begin{equation}
  \label{eq:vi.29}
  S^{[0]}_\wp(p_0,p_1)\doteqdot \frac{\varepsilon^2_{\rm pot}(0,p_0,p_1)}{\varepsilon_{\rm kin}(0,p_0,p_1)}\ .
\end{equation}
Here, the general kinetic function $\varepsilon_{\rm kin}(\nu,p_0,p_1)$ (\ref{eq:vi.11}) becomes reduced to (observe $\lP=0$)
\begin{equation}
  \label{eq:vi.30}
  \varepsilon_{\rm kin}(0,p_0,p_1)=\frac{1}{4}p_0^2+\frac{1}{2}\,p_1^2\doteqdot\tilde{\varepsilon}_{\rm kin}(p_0,p_1)
\end{equation}
and finally the reduced potential function $\tilde{\varepsilon}_{\rm
  pot}(p_0,p_1)\doteqdot\varepsilon_{\rm pot}(0,p_0,p_1)$ is deduced from $\varepsilon_{\rm
  pot}(\nu,p_0,p_1)$ (\ref{eq:vi.13}) as follows
\begin{equation}
  \label{eq:vi.31}
  \tilde{\varepsilon}_{\rm pot}(p_0,p_1)= p_0^4\cdot\varepsilon_{\rm pot}(0) + 4p_0^3
  p_1\cdot\varepsilon_{\rm I}(0) + 4p_0^2p_1^2\cdot\varepsilon_{\rm II}(0) +
  4p_0p_1^3\cdot\varepsilon_{\rm III}(0) + p_1^4\cdot\varepsilon_{\rm IV}(0)\ ,
\end{equation}
i.\,e. by reference to the table in {\bf App.G}
\begin{equation}
  \label{eq:vi.32}
  \tilde{\varepsilon}_{\rm pot}(p_0,p_1)=\frac{1}{2}p_0^4 + 3p_0^3p_1 +
  \frac{17}{2}\,p_0^2p_1^2 + \frac{25}{2}\,p_0p_1^3 + \frac{33}{4}\,p_1^4\ .
\end{equation}
Thus the reduced spectral function $S_\wp^{[0]}(p)$ (\ref{eq:vi.29}) is ultimately found to be of the following form
\begin{equation}
  \label{eq:vi.33}
  S_\wp^{[0]}(p_0,p_1) = \frac{\tru{\varepsilon}{pot}(p_0,p_1)^2}{\tru{\varepsilon}{kin}(p_0,p_1)}
= \frac{ \left(p_0^4+6p_0^3p_1 + 17p_0^2 p_1^2 + 25 p_0p_1^3 + \frac{33}{2}p_1^4
  \right)^2}{p_0^2+2p_1^2}\ .
\end{equation}

For determining the local extrema of this spectral function (\ref{eq:vi.33}), one has to
observe the constraint (\ref{eq:vi.2}), For our present choice of~$\nu=0$, this constraint
implies the restriction to a one-dimensional compact subspace of the
two-dimensional~$\mathbb{R}^2$:
\begin{equation}
  \label{eq:vi.34}
  p_0^2+4p_0p_1 + 6p_1^2 = 1\ .
\end{equation}
Since such a closed subspace is topologically equivalent to a circle~$S^1$, one may
introduce a circular coordinate ($\alpha$,~say) in order to parameterize both real
numbers~$p_0,p_1$:
\begin{subequations}
  \begin{align}
    \label{eq:vi.35a}
    p_0 &= \cos\alpha - \sqrt{2}\sin\alpha\\*
    \label{eq:vi.35b}
    p_1 &= \frac{\sin\alpha}{\sqrt{2}}\ .
  \end{align}
\end{subequations}
Indeed, one is easily convinced that the constraint (\ref{eq:vi.34}) is automatically
satisfied by this parametrization. Furthermore, the kinetic energy
function~$\tru{\varepsilon}{kin}(p_0,p_1)$ (\ref{eq:vi.30}) becomes now also a function of
this angular parameter~$\alpha$, i.e.
\begin{subequations}
  \begin{align}
    \label{eq:vi.36a}
    \tru{\varepsilon}{kin}(p_0,p_1) &\Rightarrow \tru{\varepsilon}{kin}(\alpha)
    = \frac{1}{4}\left(1+\tilde{T}_1(\alpha) \right)\\*
    \label{eq:vi.36b}
    \tilde{T}_1(\alpha) &= 2\sin\alpha\left(\sin\alpha-\sqrt{2}\cos\alpha\right)\ .
  \end{align}
\end{subequations}
The same arguments do also convert the potential
function~$\tru{\varepsilon}{pot}(p_0,p_1)$ (\ref{eq:vi.31}) into a function of~$\alpha$
\begin{subequations}
  \begin{align}
    \label{eq:vi.37a}
    \tru{\varepsilon}{pot}(p_0,p_1) &\Rightarrow \tru{\varepsilon}{pot}(\alpha)
    = \frac{1}{2}\left(1+\tilde{V}_1(\alpha) \right)\\*
    \label{eq:vi.37b}
    \tilde{V}_1(\alpha) &= \sin\alpha\left(\frac{1}{2}\sin\alpha +
      \frac{\sqrt{2}}{4}\cos\alpha\cdot\sin^2\alpha -\sqrt{2}\cos\alpha - \frac{7}{8}\sin^3\alpha
    \right)\ .
  \end{align}
\end{subequations}
Consequently, the angular dependence of the spectral function~$S_\wp^{[0]}(\alpha)$
(\ref{eq:vi.33}) must look as follows
\begin{equation}
  \label{eq:vi.38}
S_\wp^{[0]}(p_0,p_1)\Rightarrow  S_\wp^{[0]}(\alpha) =
\frac{\left(1+\tilde{V}_1(\alpha)\right)^2}{1+\tilde{T}_1(\alpha)}\ .
\end{equation}
This function has been discussed in some detail in ref.~\cite{ms3} (see equation
(\ref{eq:iv.44}) of that reference); and it has been found that there occur two maxima and
two minima which then generate four energy values~$E_{1a}\ldots E_{1d}$ in the following
arrangement:
\parbox[c]{4cm}{\textbf{rel.\ maxima:}}
\parbox[c]{11,5cm}{
\begin{subequations}
  \begin{align}
    \label{eq:vi.39a}
    E_{1\mathrm{a}} &= - \frac{e^2}{4\aB}\cdot 1 \simeq -6,8029\ldots\ [eV]\\*
    \label{eq:vi.39b}
    E_{1\mathrm{b}} &= - \frac{e^2}{4\aB}\cdot 0,086717499\ldots\simeq -0,590\ [eV]
  \end{align}
\end{subequations}
}

\hrulefill
\\*
\parbox[c]{4cm}{\textbf{rel.\ minima:}}
\parbox[c]{11,5cm}{
\begin{subequations}
  \begin{align}
    \label{eq:vi.40a}
    E_{1\mathrm{c}} &= - \frac{e^2}{4\aB}\cdot 1,033319474 \ldots \simeq -7,030\ [eV]\\*
    \label{eq:vi.40b}
    E_{1\mathrm{d}} &= - \frac{e^2}{4\aB}\cdot 1,128194657\ldots\simeq -7,675\ [eV]\ .
  \end{align}
\end{subequations}
} 

Here, the first result~$E_{1a}$ (\ref{eq:vi.39a}) is clearly the most striking one,
because it agrees \emph{exactly} with both the zero-order approximation~$(\alpha=0)$ and the
conventional groundstate prediction (\ref{eq:i.4}), where the latter is commonly
considered to be exact within the framework of the generally accepted non-relativistic
quantum mechanics. However, the occurrence of two minima and two maxima in the first-order
spectrum demands an explanation: From the physical point of view there should occur
for~$\mathbb{N}=1$ just one maximum and one minimum, corresponding to the groundstate and
the first excited state. But evidently the present approximation procedure generates
spectral functions with additional \emph{spurious} extremal values which are not
reproduced in the next higher approximation steps. Therefore it may appear desirable to
get some confidence in the presently used approximation formalism. This is attained by
reproducing the above results through a different parametrization of the trial amplitude
(\textbf{App.H}).

Summarizing the simplified groundstate treatment, one gets a first estimate of the
groundstate energy at~$E_{1d}=-7,675\,[eV]$ (\ref{eq:vi.40b}) which deviates in an
unacceptable way from the conventional prediction
$\ru{E}{conv}^{(1)}=-\frac{e^2}{4\aB}\simeq-6,8029\,[eV]$, (cf.~equation
(\ref{eq:i.4})). However, this does not mean that RST is unable to reproduce the
conventional positronium spectrum (\ref{eq:i.4}). As we will readily see, the spectrum is
actually reproduced by the present approximation formalism with sufficient accuracy as far
as the \emph{excited} states are concerned. The \emph{groundstate} seems to represent an
exceptional situation which must be studied separately.
\vspace{4ex}
\begin{center}
\mysubsection{4.\ Excited States}
\end{center}

For the treatment of the excited states one has of course to relax the condition~$\nu=0$
which may be applicable exclusively for the (simplified) groundstate. Or otherwise, if one
wishes to stick to~$\nu=0$, one has to further generalize the present trial ansatz
(\ref{eq:vi.1}) in order to work with the more general hydrogen-like wave functions (the
latter approximation method has been applied in ref.~\cite{ms3}). However, for the present
elaborations, we rely upon the simpler variational ansatz  (\ref{eq:vi.1}) with
general~$\nu$ and integer quantum number~$\lP$ of angular momentum (\ref{eq:iii.21}). This
approach must generate the same energy spectrum as does the method of the hydrogen-like
wave functions in combination with vanishing angular momentum~$\lP=0$, provided the
angular momentum degeneracy emerges in RST quite analogously to the situation in the
conventional theory based upon the Hamiltonian~$\hat{H}$ (\ref{eq:i.3}).

Thus our procedure will be a slight generalization of the precedent treatment of the
simplified groundstate; i.e.\ for parametrizing the general spectral
function~$S_\wp(\nu,p_0,p_1)$ (\ref{eq:vi.27}) (with non-zero variational parameter~$\nu$)
by the~$S^1$-coordinate~$\alpha$ we first have to look for both
functions~$\ru{\varepsilon}{kin}(\nu,p_0,p_1)$ (\ref{eq:vi.11})
and~$\ru{\varepsilon}{pot}(\nu,p_0,p_1)$  (\ref{eq:vi.13}) as functions of that new
coordinate~$\alpha$, and afterwards one determines the maximal value of the corresponding
function~$S_\wp(\nu,\alpha)$.
\begin{equation}
  \label{eq:vi.41}
  S_\wp(\nu,\alpha)=\frac{\ru{\varepsilon}{pot}^2(\nu,\alpha)}{\ru{\varepsilon}{kin}(\nu,\alpha)}\ .  
\end{equation}
For such a reparametrization by the angle~$\alpha$ it is very convenient to first
eliminate from both functions~$\ru{\varepsilon}{kin}(\nu,p_0,p_1)$
and~$\ru{\varepsilon}{pot}(\nu,p_0,p_1)$ the parameter~$p_0$ as far as possible since its
reparametrization is somewhat more complicated than its associate~$p_1$:
\begin{subequations}
  \begin{align}
    \label{eq:vi.42a}
    p_0&\Rightarrow p_0(\alpha) = \cos\alpha-\sqrt{2(\nu+1)}\cdot\sin\alpha\\*
    \label{eq:vi.42b}
    p_1&\Rightarrow p_1(\alpha) = \frac{\sin\alpha}{\sqrt{2(\nu+1)}}\ .
  \end{align}
\end{subequations}
Clearly, this is the generalization of the predecessors
(\ref{eq:vi.35a})-(\ref{eq:vi.35b}) for the simplified groundstate~$(\nu=0)$; but indeed
it guarantees the validity of the more general normalization condition
(\ref{eq:vi.2}). Thus, the kinetic function~$\ru{\varepsilon}{kin}(\nu,p_0,p_1)$
  (\ref{eq:vi.11}) reappears now in terms of the~$S^1$-coordinate~$\alpha$ as
  \begin{equation}
    \label{eq:vi.43}
    \begin{split}
    \ru{\varepsilon}{kin}(\nu,p_0,p_1) \Rightarrow \ru{\varepsilon}{kin}(\nu,\alpha) &=
    \frac{1}{2\nu(2\nu+1)}
    \left[\left(\nu+2\lP^2\right)\cdot\left(\frac{1}{2} -
        \sqrt{\frac{2}{\nu+1}}\sin\alpha\cos\alpha\right)\right.\\*
      &+\left.\sin^2\alpha\cdot\left(\nu+\frac{\lP^2}{\nu+1} \right)\right]\ .
    \end{split}
  \end{equation}
Evidently, this result becomes reduced to the former case~$\ru{\varepsilon}{kin}(\nu)$
(\ref{eq:iii.53}) for~$\alpha\to 0$ (i.e.~$p_1\to 0,p_0\to 1$); and it becomes reduced to
the simplified groundstate situation~$\tru{\varepsilon}{kin}(\alpha)$
(\ref{eq:vi.36a})-(\ref{eq:vi.36b}) for~$\lP=0,\nu\to 0$.

The next step must consist in looking for the reparametrized
numerator~$\ru{\varepsilon}{pot}(\nu,\alpha)$, of the spectral
function~$S_\wp(\nu,\alpha)$ (\ref{eq:vi.41}); i.e.\ we have to elaborate the transition~$
\ru{\varepsilon}{pot}(\nu,p_0,p_1) \to \ru{\varepsilon}{pot}(\nu,\alpha)$ where the
original potential function~$\ru{\varepsilon}{pot}(\nu,p_0,p_1)$ is given by equation
(\ref{eq:vi.13}) and the angular parametrization of~$p_0,p_1$ by equations
(\ref{eq:vi.42a})-(\ref{eq:vi.42b}). The result then looks as follows:
\begin{equation}
  \label{eq:vi.44}
  \begin{split}
  \ru{\varepsilon}{pot}(\nu,p_0,p_1)&\Rightarrow  \ru{\varepsilon}{pot}(\nu,\alpha) =
  \ru{\varepsilon}{pot}(\nu)
  + \left(\varepsilon_2(\nu) + \frac{\varepsilon_4(\nu)}{2(\nu+1)}\cdot\sin^2\alpha \right)
  \cdot \frac{\sin^2\alpha}{2(\nu+1)}
  \\*
  &+ 2\left(\varepsilon_1(\nu) +
    \frac{\varepsilon_3(\nu)}{2(\nu+1)} \cdot \sin^2\alpha \right) \cdot
  \left(\frac{\sin 2\alpha}{\sqrt{2(\nu+1)}} + \cos 2\alpha-1 \right)\ .
  \end{split}
\end{equation}
Clearly, this is the generalization to arbitrary~$\nu$ of the former potential
function~$\tru{\varepsilon}{pot}(\alpha)$ (\ref{eq:vi.37a})-(\ref{eq:vi.37b}) which in turn
is recovered from the present more general result (\ref{eq:vi.44}) by putting~$\nu$ to
zero~$(\nu\to 0)$.
This claim can easily be validated by observing the definition of the auxiliary functions
(see also the table in \textbf{App.G})
\begin{subequations}
  \begin{align}
    \label{eq:vi.45a}
    \varepsilon_1(\nu) &\doteqdot\ru{\varepsilon}{I}(\nu) - 2(\nu+1)\cdot\ru{\varepsilon}{pot}(\nu)\\*
    \label{eq:vi.45b}
    \varepsilon_2(\nu) &\doteqdot 4\ru{\varepsilon}{II}(\nu) - 16(\nu+1)\cdot
    \ru{\varepsilon}{I}(\nu) + \left[16(\nu+1)^2 -4(\nu+1)(2\nu+3)\right] \cdot \ru{\varepsilon}{pot}(\nu)\\*
    \label{eq:vi.45c}
    \varepsilon_3(\nu) &\doteqdot  \ru{\varepsilon}{III}(\nu) - 4(\nu+1)\cdot
    \ru{\varepsilon}{II}(\nu) + 2(6\nu^2+11\nu+5)\cdot  \ru{\varepsilon}{I}(\nu)\\*
    &\hspace{6.5cm} - 4(2\nu^3+5\nu^2+4\nu+1)\cdot  \ru{\varepsilon}{pot}(\nu) \notag \\*
    \label{eq:45d}
    \varepsilon_4(\nu) &\doteqdot  \ru{\varepsilon}{IV}(\nu) - 8(\nu+1)(2\nu+3)\cdot
    \ru{\varepsilon}{II}(\nu) + 32(\nu+1)^2(2\nu+3)\cdot \ru{\varepsilon}{I}(\nu)\notag\\*
    &\hspace{6cm}- 4(\nu+1)^2 (2\nu+3)(6\nu+5)\cdot\ru{\varepsilon}{pot}(\nu)
  \end{align}
\end{subequations}
Thus, putting~$\nu$ to zero yields
\begin{subequations}
  \begin{align}
    \label{eq:vi.46a}
    \varepsilon_1(0) &= -\frac{1}{4}\\*
    \label{eq:vi.46b}
    \varepsilon_2(0) &= -\frac{3}{2}\\*
    \label{eq:vi.46c}
    \varepsilon_3(0) &= \frac{1}{8}\\*
    \label{eq:vi.46d}
    \varepsilon_4(0) &= -\frac{3}{4}
  \end{align}
\end{subequations}
and this simplifies the general~$\ru{\varepsilon}{pot}(\nu,\alpha)$ (\ref{eq:vi.44})
actually down to the former~$\tru{\varepsilon}{pot}(\alpha)$
(\ref{eq:vi.37a})-(\ref{eq:vi.37b}).

With both the numerator~$\ru{\varepsilon}{pot}(\nu,\alpha)$ and
denominator~$\ru{\varepsilon}{kin}(\nu,\alpha)$ of the spectral
function~$S_\wp(\nu,\alpha)$ (\ref{eq:vi.41}) being known now, one can let search the
maximum~$S_\wp^{[n]}(*)$ of~$S_\wp(\nu,\alpha)$ by means of an appropriate
numerical program and finally can then tabulate the corresponding positronium
spectrum~$\mathbb{E}^{[n]}_\wp$ 
\begin{subequations}
  \begin{align}
  \label{eq:vi.47a}
  \mathbb{E}^{[n]}_\wp &= -\frac{e^2}{4\aB}\cdot S_\wp^{[n]}(*)\\*
  \label{eq:vi.47b}
  S_\wp^{[n]}(*) &:= S_\wp(\nu_*^{[n]},p_*^{[n]})\ , 
  \end{align}
\end{subequations}
see \textbf{table~3} below.

As a check for the correct working of the program, one can alternatively determine the
energy spectrum also in the  reparametrized form (\ref{eq:H.26a})-(\ref{eq:H.26b}) which
must identically reproduce the spectrum due to the present form (\ref{eq:vi.47a})-(\ref{eq:vi.47b}).
{
\begin{flushleft}
\label{table3}
  \begin{tabular}{|c||c|c|c|c|c|}
  \hline
$n_P$ & $\Ea{E}{n}{conv}$ & $\mathbb{E}_p^{[n]}$ & $\nu_*^{[n]}$ &
$p_*^{[n]}$ & $ \Delta_\wp (\nu_*^{[n]}) $ \\
$(=lp+1)$ & (\ref{eq:i.4}) & (\ref{eq:vi.47a})-(\ref{eq:vi.47b}) & & &  (\ref{eq:vi.48}) \\ \hline\hline
1 & -6.80290 & -7.67499 & 0.00000 & 0.56525 & -12.82\\ \hline
2 & -1.70072 & -1.56251 & 1.25829 & 0.67798 & 8.13\\  \hline 
3 & -0.75588 & -0.67055 & 3.46267 & -0.04753 & 11.29\\  \hline 
4 & -0.42518 & -0.37411 & 5.42898 & -0.03560 & 12.01\\  \hline 
5 & -0.27212 & -0.23925 & 7.52949 & -0.02837 & 12.08\\  \hline 
6 & -0.18897 & -0.16647 & 9.74795 & -0.02351 & 11.91\\  \hline 
7 & -0.13883 & -0.12267 & 12.07237 & -0.02001 & 11.65\\  \hline 
8 & -0.10630 & -0.09423 & 14.49351 & -0.01738 & 11.35\\  \hline 
9 & -0.08399 & -0.07470 & 17.00395 & -0.01532 & 11.06\\  \hline 
10 & -0.06803 & -0.06070 & 19.59761 & -0.01367 & 10.77\\  \hline 
11 & -0.05622 & -0.05033 & 22.26932 & -0.01232 & 10.49\\  \hline 
12 & -0.04724 & -0.04241 & 25.01475 & -0.01120 & 10.22\\  \hline 
13 & -0.04025 & -0.03624 & 27.82998 & -0.01025 & 9.97\\  \hline 
14 & -0.03471 & -0.03133 & 30.71171 & -0.00944 & 9.74\\  \hline 
15 & -0.03024 & -0.02736 & 33.65704 & -0.00874 & 9.51\\  \hline 
 \end{tabular}
\end{flushleft}
}
{\textbf{Table 3:}\hspace{0.5cm} \emph{\large\textbf{Improved RST predictions (Spherical Symmetry)
 }}  }

\mytable{\emph{Table 3: Improved RST predictions (Spherical Symmetry)}}

The improved RST predictions~$\mathbb{E}_\wp^{[n]}$ (\ref{eq:vi.47a}) (spherically
symmetric approximation) do differ essentially from their conventional
counterparts~$\Ea{E}{n}{conv}$ (\ref{eq:i.4}) if compared to the corresponding
``anisotropic'' improvements of \textbf{table~2} (p.~\pageref{table2}). Defining here the
deviation~$\ekru{\bar{\Delta}}{\wp}$ quite analogously to the former
cases~$\ekru{\bar{\Delta}}{T}$ (\ref{eq:iii.58}) and~$\rugk{\bar{\Delta}}{T}$
(\ref{eq:v.44}) as
\begin{equation}
  \label{eq:vi.48}
  \ekru{\bar{\Delta}}{\wp} = \frac{1}{14}\sum_{n=2}^{14}
  \left(\frac{ \Ea{E}{n}{conv}-\mathbb{E}_\wp^{[n]} }{\Ea{E}{n}{conv}}\right)
  = \frac{1}{14}\sum_{n=2}^{14} \Delta_\wp(\nu_*^{[n]}) \ ,
\end{equation}
one finds for the present improvement of the spherically symmetric type
\begin{equation}
  \label{eq:vi.49}
  \ekru{\bar{\Delta}}{\wp} = 10,7\%
\end{equation}
This is an irrelevant improvement in comparison to the analogous average of 10,9\% due to
the simplest approximation (\textbf{table~1},~$2\leqq \nP \leqq 15$). Observe also that for the
groundstate~$(\nP=1)$ one finds~$\nu_*^{[1]}\simeq 0$ which is preliminarily assumed for
the sake of simplicity in order to exemplify our general approximation scheme, see the
precedent \textbf{Section VI.3}. Through this fortuitous agreement, the former
groundstate prediction~$\ru{E}{1d}$ (\ref{eq:vi.40b}) is practically identical with the
result~$\mathbb{E}_\wp^{[1]}$ of the present more rigorous calculation (first line,~$\nP=1$).


\section[\ \ Conclusion]{Conclusion}
\indent

The present results may give to us now a hint at the direction for searching further
improvements. It is true, both higher approximations (i.e.\ the ``anisotropic'' one of
\textbf{table~2} and the ``spherically symmetric'' one of \textbf{table~3}) yield certain
improvements in the sense that the corresponding RST predictions do better approach the
predictions~$\ru{E}{conv}^{(n)}$ (\ref{eq:i.4}) of the conventional theory. However, the
improvement of the spherically symmetric kind (\textbf{table~3}) are inconsiderable
because they amount to merely some 0,2 percent. Thus the conclusion is that the original
results of the spherically symmetric type (\textbf{table~1}) do already exhaust the
possibilities of this symmetry. But the ``anisotropic'' improvements (\textbf{table~2}) do
reduce the RST deviations (from the conventional predictions) by some 50\% in the
range~$2\leq\nP\leq 15$. Therefore one expects that the next higher approximation step of
the anisotropic type (i.e.\ the regard of the radial auxiliary potential~$\gklo{p}{A}^{\sf
  V}(r)$ in the expansion (\ref{eq:v.3})) will show a further clear step of the
RST predictions towards the conventional energy spectrum (\ref{eq:i.4}). Here, the regard
of the anisotropy can not be restricted to the interaction potential~$\rklo{p}{A}_0$ alone
but must refer also to the wave functions themselves. More concretely, this means that the
product ansatz (\ref{eq:iii.17a})-(\ref{eq:iii.17b}) for the wave amplitudes must receive
some generalization. Taking all these results together, it seems that the conventional
levels (\ref{eq:i.4}) play the part of a lower bound for the successively higher RST
approximations. In any case, further clarification of this question must be be left to a
more extensive investigation.


\addtocontents{toc}{\protect{\vspace*{5mm}}}
\renewcommand{\theequation}{\Alph{section}.\arabic{equation}}
  \setcounter{section}{1}
  \setcounter{equation}{0}
  \begin{center}
  {\textbf{\Large Appendix \Alph{section}:}}\\[2ex]
  \emph{ \textbf{\Large Spherically Symmetric Potential }}
  \end{center}
  \myappendix{Spherically Symmetric Potential}
  \vspace{2ex}

  The spherically symmetric approximation is defined by assuming the electric interaction
  potential~$\rklo{p}{A}_0(\vec{r})$ (\ref{eq:ii.30}) to be SO(3) symmetric, cf.\ equation
  (\ref{eq:iii.1}). The \emph{principle of minimal energy} (\ref{eq:ii.40}) then yields
  the Poisson equation (\ref{eq:iii.36}) for the symmetric potential~$\eklo{p}{A}_0(r)$ which
  may be recast to the following dimensionless form
\begin{gather}
  \label{eq:a1}
  \frac{d^2 \tilde{a}_\nu(y) }{dy^2} + \frac{2}{y}\frac{d\tilde{a}_\nu(y)}{dy} = -e^{-y}
  \frac{y^{2\nu-1}}{\Gamma(2\nu+2)}\\*
  \big(y \doteqdot 2\beta r \big)\ ,\notag
\end{gather}
provided the original potential~$\eklo{p}{A}_0(r)$ (\ref{eq:iii.1}) due to the trial
amplitude~$\tilde{\Phi}(r)$ is rescaled as follows
\begin{equation}
  \label{eq:a2}
  \eklo{p}{A}_0(r) \doteqdot 2\beta\as\cdot\tilde{a}_\nu(y)\ ,
\end{equation}
and similarly for the trial function~$\tilde{\Phi}(r)$ (\ref{eq:iii.50})
\begin{subequations}
  \begin{align}
  \label{eq:a3a}
  \tilde{\Phi}(r) &= \sqrt{\frac{2}{\pi}}\cdot\frac{2\beta}{\sqrt{\Gamma(2\nu+2)}}
  \tilde{\Phi}_\nu(y)\\*
  \label{eq:a3b}
  \tilde{\Phi}_\nu(y) &\doteqdot y^\nu\cdot e^{-\frac{y}{2}}\ .
  \end{align}
\end{subequations}

The solutions~$\tilde{a}_\nu(y)$ of the dimensionless Poisson equation (\ref{eq:a1}) can
be represented for integer values of ~$2\nu$ in form of a recurrence formula:
\begin{equation}
  \label{eq:a4}
  \tilde{a}_{\nu+\frac{1}{2}}(y) = \tilde{a}_\nu(y) - \frac{e^{-y}}{(2\nu+2){(2\nu+1)}} \cdot
  \sum_{n=0}^{2\nu}\frac{y^n}{n!}\ ,
\end{equation}
with the lowest-order potential~$\tilde{a}_0(y)$ for~$\nu=0$ being given by
\begin{equation}
  \label{eq:a5}
  \tilde{a}_0(y) = \frac{1}{y}\left(1-e^{-y}\right)\ .
\end{equation}
Thus, the lowest-order potentials (i.e.\ for $\nu=0,\frac{1}{2},1,\frac{3}{2}$) are found
to be of the following shape
\begin{subequations}
  \begin{align}
    \label{eq:a6a}
    \tilde{a}_{\frac{1}{2}}(y) &= \tilde{a}_0(y) - \frac{e^{-y}}{2}=\frac{1}{y}
    \left\{1-e^{-y}\left(1+\frac{y}{2}\right) \right\}\\*
    \label{eq:a6b}
    \tilde{a}_1(y) &= \tilde{a}_\frac{1}{2}(y) - \frac{e^{-y}}{6}\left(1+y\right) =\frac{1}{y}
    \left\{1-e^{-y}\left(1+\frac{2y}{3}+\frac{y^2}{6}\right) \right\}\ ,\\*
    \label{eq:a6c}
    \tilde{a}_{\frac{3}{2}}(y) &= \tilde{a}_1(y) - \frac{e^{-y}}{12}\left(1+y+\frac{y^2}{2}\right)
    =\frac{1}{y}\left\{1-e^{-y}\left(1+\frac{3}{4}y+\frac{1}{4}y^2+\frac{1}{24}y^3\right)
    \right\}\ ,
  \end{align}
\end{subequations}
see fig.~A.I below for a sketch of these lowest-order potentials. For general (but still integer)
values of~$2\nu$ the dimensionless potential~$\tilde{a}_\nu(y)$ (\ref{eq:a4}) looks as
follows:
\begin{equation}
  \label{eq:a7}
  \tilde{a}_\nu(y)=\frac{1}{y}\left(1-e^{-y}\cdot\sum_{n=0}^{2\nu}
    \frac{2\nu+1-n}{2\nu+1}\cdot\frac{y^n}{n!} \right)\ .
\end{equation}
Here it must be stressed that such a potential~$\tilde{a}_\nu(y)$ does actually exist for any
real-valued \mbox{$\nu\,(> -\frac{1}{2})$}, where the general expression for this situation is
then given by an \emph{infinite} sum 
\begin{equation}
  \label{eq:a8}
  \tilde{a}_\nu(y)=\frac{1}{2\nu+1}\left(1-e^{-y}\cdot\sum_{n=0}^\infty
    \frac{n}{\Gamma(2\nu+2+n)}\,y^{2\nu+n} \right)\ ,
\end{equation}
see appendix~D of ref.~\cite{ms1}.

The value~$\tilde{a}_\nu(0)$ of the potentials~$\tilde{a}_\nu(y)$
(\ref{eq:a5})-(\ref{eq:a6c}) at the origin~$(y=0)$ is easily found by expanding the
exponential function~$e^{-y}$ in the usual way which then yields
\begin{subequations}
  \begin{align}
    \label{eq:a9a}
    \tilde{a}_0(0) &= 1\\*
    \label{eq:a9b}
    \tilde{a}_{\frac{1}{2}}(0) &= \frac{1}{2}\\*
    \label{eq:a9c}
    \tilde{a}_1(0) &= \frac{1}{3}\\*
    \label{eq:a9d}
    \tilde{a}_{\frac{3}{2}}(0) &= \frac{1}{4}\ ,
  \end{align}
\end{subequations}
see also the sketch~A.I of potentials~$\tilde{a}_\nu(y)$ below. Obviously the general case is
given by
\begin{equation}
  \label{eq:a10}
  \tilde{a}_\nu(0) = \frac{1}{2\nu+1}\ ;
\end{equation}
and this result does hold for any real-valued power~$\nu>-\frac{1}{2}$. Indeed, this claim
can easily be verified by exploiting the general shape of the solution of the Poisson
equation (\ref{eq:a1})
\begin{equation}
  \label{eq:a11}
  \tilde{a}_\nu(y) = \frac{1}{4\pi\cdot\Gamma(2\nu+2)}\int d^3\vec{y}\,'\;
  \frac{e^{-y'}\cdot y\,'^{2\nu-1}}{||\vec{y}-\vec{y}\,' ||}
\end{equation}
which satisfies also the usual boundary condition at infinity~$(y\to\infty)$
\begin{equation}
  \label{eq:a12}
  \lim_{y\to\infty}\tilde{a}_\nu(y) = \frac{1}{y}\hspace{1cm}  \forall \nu\ .
\end{equation}
In terms of the original potential~$\eklo{p}{A}_0(r)$ (\ref{eq:a2}) this boundary
condition (\ref{eq:a12}) yields just the well-known Coulomb form
\begin{equation}
  \label{eq:a13}
  \lim_{r\to\infty}\eklo{p}{A}_0(r) = \frac{\as}{r}
\end{equation}
which thus is found to be the asymptotic limit form for all potentials~$\eklo{p}{A}_0(r)$ being
due to the class of trial functions~$\tilde{\Phi}(r)$ (\ref{eq:iii.50}), irrespective of
the special value of~$\nu$.
\begin{center}
\epsfig{file=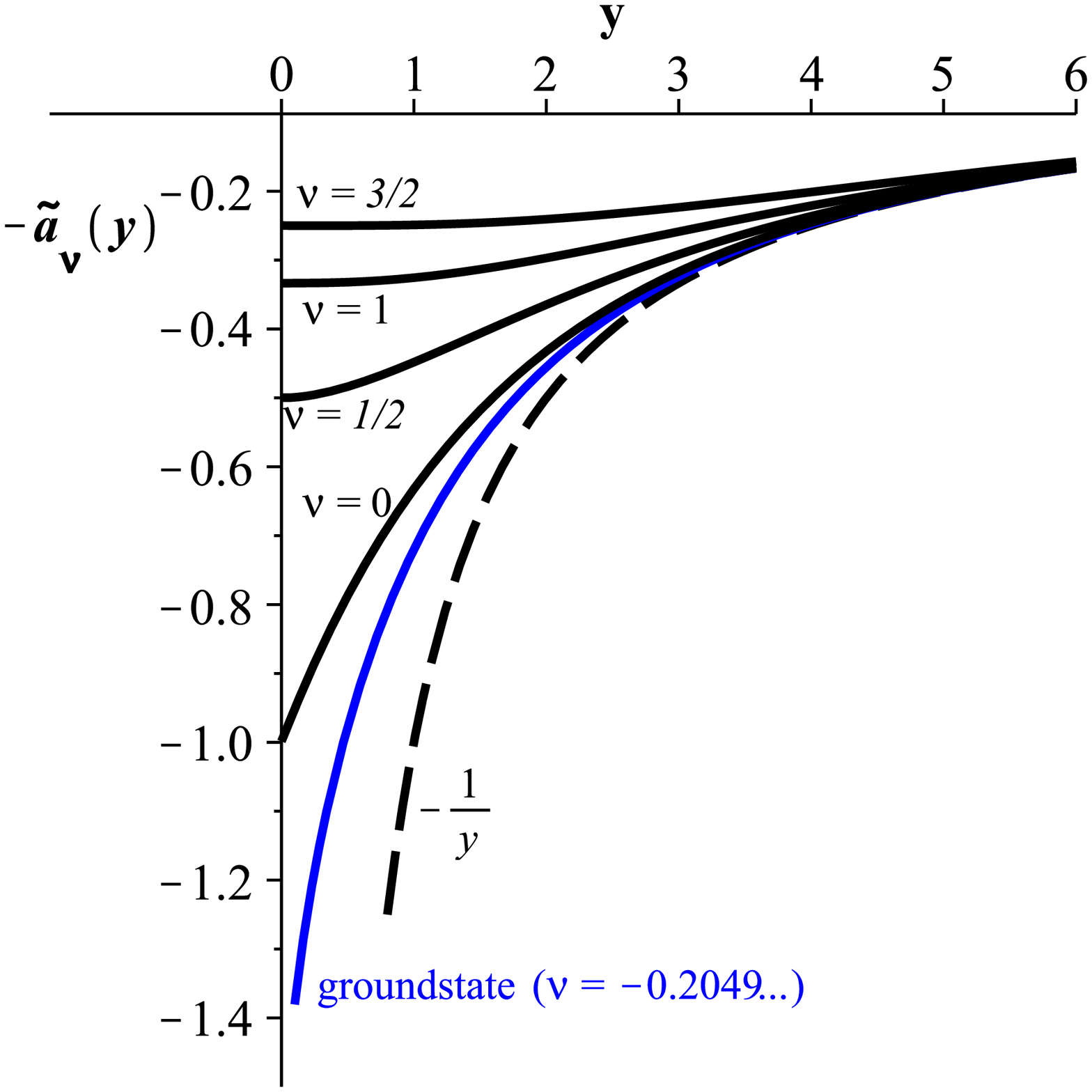,height=15cm}
\end{center}
{\textbf{Fig.\ A.I:}\hspace{0.5cm} \emph{\large\textbf{Spherically Symmetric Potentials}}
\emph{\large\textbf{(\ref{eq:a5})-(\ref{eq:a6c})}}  }
\myfigure{Fig.\ A.I: Spherically Symmetric Potentials (\ref{eq:a5})-(\ref{eq:a6c})}
\indent

The potentials~$\tilde{a}_\nu(y)$ (\ref{eq:a8}) are finite at the origin~$(y=0)$,
cf. (\ref{eq:a10}), and approach the Coulomb potential~$(\sim \frac{1}{y})$
for~$y\to\infty$, cf.~(\ref{eq:a12}). For the excited states the parameter~$\nu$ is
positive (see the \textbf{table~1} on p.~\pageref{table1} ) and therefore the
potentials~$\tilde{a}_\nu(y)$ have vanishing derivative at the origin~$(y=0)$, but for the
groundstate~$(\nu_*=-0,2049\ldots)$ the derivative becomes infinite, see equation
(\ref{eq:a20}) below.  \newpage The fact that the spherically symmetric
potentials~$\eklo{p}{A}_0(r)$ are finite at the origin $(r=0)$ does of course not mean
that the corresponding field strength~$\vri{E}{p}$ felt by the electron
\begin{equation}
  \label{eq:a14}
  \vri{E}{p}(\vec{r}) = \vec{\nabla}\eklo{p}{A}_0(r) = \frac{d\eklo{p}{A}_0(r)}{dr}\,\vec{e}_r
\end{equation}
is in any case a regular vector field for~$r\to 0$. Indeed, introducing the dimensionless
form~$\tilde{e}_\nu(y)$ of the radial component~$\eklo{p}{E}_r(r)$
\begin{equation}
  \label{eq:a15}
  \vri{E}{p}(\vec{r}) = \eklo{p}{E}_r(r)\vec{e}_r 
\end{equation}
through
\begin{equation}
  \label{eq:16}
  \eklo{p}{E}_r(r) =  \frac{d\eklo{p}{A}_0(r)}{dr} \doteqdot \left(2\beta \right)^2
  \as\tilde{e}_\nu(y)\ ,
\end{equation}
one arrives at the following link of (dimensionless) potentials~$\tilde{a}_\nu(y)$ and
field strengths~$\tilde{e}_\nu(y)$:
\begin{equation}
  \label{eq:a17}
  \tilde{e}_\nu(y)=\frac{d\tilde{a}_\nu(y)}{dy}\ .
\end{equation}

But since the spherically symmetric potentials~$\tilde{a}_\nu(y)$ are explicitly known
both for integer~$2\nu$ (\ref{eq:a7}) and general~$\nu$ (\ref{eq:a8}), the field
strengths~$\tilde{e}_\nu(y)$ are also known, e.g. for integer~$2\nu$:
\begin{equation}
  \label{eq:a18}
  \tilde{e}_\nu(y) = -\frac{1}{y^2}\left(1-e^{-y}\cdot\sum^{2\nu+1}_{n=0}\frac{y^n}{n!}
  \right)\ ,
\end{equation}
or similarly for general~$\nu$
\begin{equation}
  \label{eq:a19}
  \tilde{e}_\nu(y) = -e^{-y}\cdot\sum_{n=0}^\infty\frac{y^{2\nu+n}}{\Gamma(2\nu+3+n)}\ .
\end{equation}

From this latter result it is now easily seen that the field strength does behave at the
origin as follows:
\begin{equation}
  \label{eq:a20}
  \lim_{y\to 0}\tilde{e}_\nu(y) = -\lim_{y\to 0}\,\frac{y^{2\nu}}{\Gamma(2\nu+3)} =
  \begin{cases}
    -\infty\ , & \nu < 0\\*
    -\frac{1}{2}\ , & \nu = 0\\*
    \phantom{-}0 \ ,& \nu>0\ .
  \end{cases}
\end{equation}
This says that the field strength~$\vri{E}{p}(\vec{r})$ (\ref{eq:a14}) is a non-singular
unique vector field only for~$\nu > 0$, see the figure~A.II below. Thus it is only for the
groundstate~$\nu\simeq -0,2$ that the electric field strength turns out to be singular at
the origin~$(y=0)$. Observe also that (for all~$\nu$) the field
strength~$\tilde{e}_\nu(y)$ adopts the Coulomb form (broken line) at infinity, i.e.
\begin{equation}
  \label{eq:a21}
  \lim_{y\to\infty}\tilde{e}_\nu(y) = -\frac{1}{y^2}\ ,\ \ \forall \nu
\end{equation}
quite similarly as for the potential~$\tilde{a}_\nu(y)$ (\ref{eq:a12}).
\newpage
\begin{center}
\epsfig{file=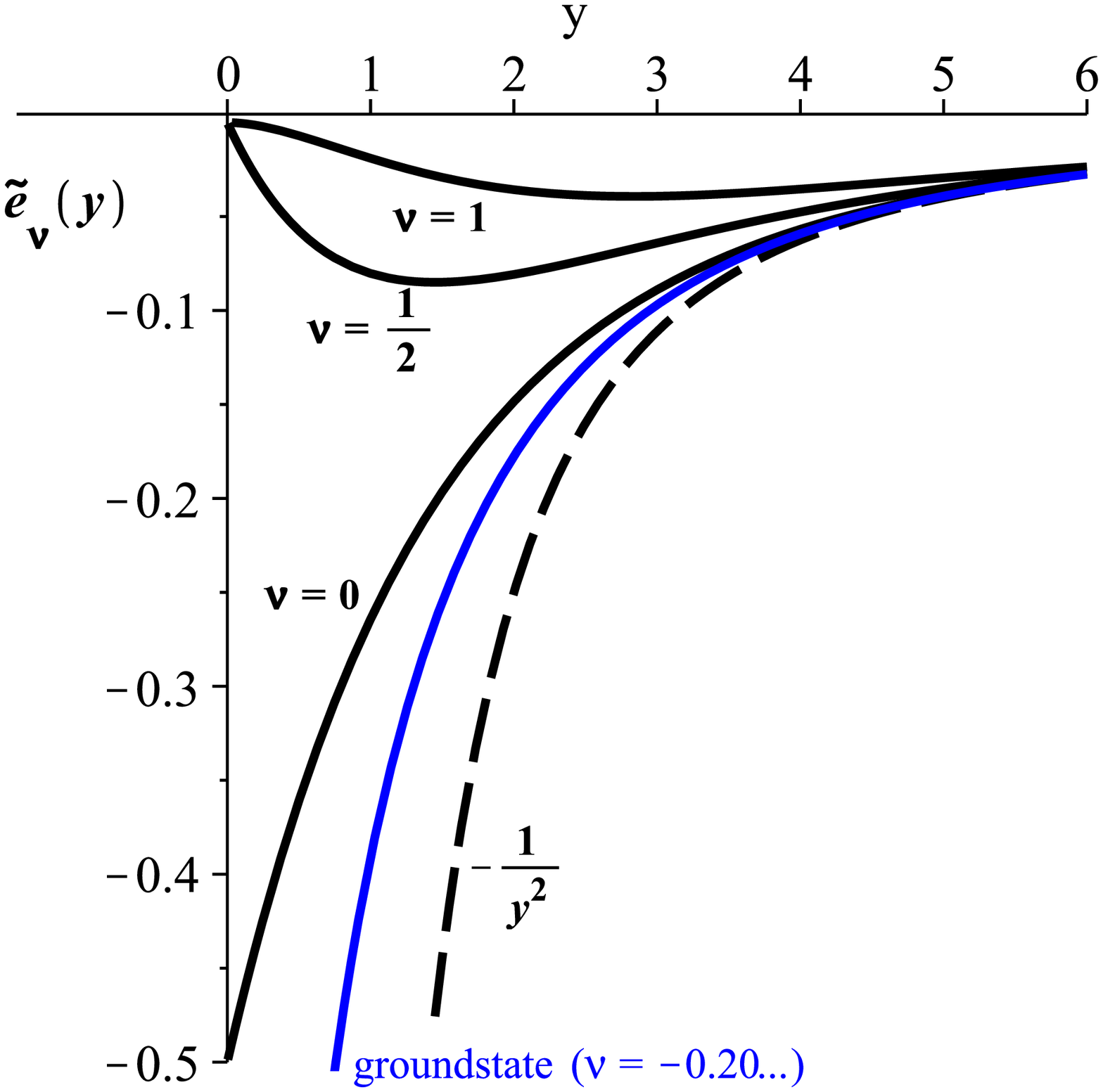,height=14cm}
\end{center}
{\textbf{Fig.\ A.II:}\hspace{0.5cm} \emph{\large\textbf{Electric Field Strength}}
\large{$\mathbf{\tilde{e}_{\nu}(y)}$} \emph{\large\textbf{(\ref{eq:a18})-(\ref{eq:a19})}}  }
\myfigure{Fig.\ A.II: Electric Field Strength $\mathbf{\tilde{e}_{\nu}(y)}$ (\ref{eq:a18})-(\ref{eq:a19})}

The electric field strength~$\tilde{e}_\nu(y)$ is non-singular only for the excited states
which have positive variational parameter~$\nu$. However, for the
groundstate~$(\nu\simeq-0.2)$ the field strength becomes singular, albeit less
singular than the Coulomb field~$(\sim -\frac{1}{y^2})$. This guarantees that the energy
content~$\Ea{E}{e}{R}$ (\ref{eq:ii.36}) of the gauge field remains finite.

\newpage
In the present context of positronium binding energy, the essential point refers now to
the value of the non-relativistic energy functional~$\nrft{E}{\Phi}$ (\ref{eq:iii.41}) upon the
class of trial functions~$\tilde{\Phi}(r)$ (\ref{eq:iii.50}). Indeed, the desired approximate
spectrum of binding energies emerges as the set of extremal values of~$\nrft{E}{\Phi}$
on that chosen class of trial functions~$\tilde{\Phi}(r)$ (\ref{eq:iii.50}). More concretely,
the value of~$\nrft{E}{\Phi}$ upon that class yields a
function~$\roek{E}{IV}(\beta,\nu)$ of the two variational parameters~$\beta$ and~$\nu$
which looks as follows~\cite{ms1}:
\begin{gather}
  \label{eq:a22}
  \roek{E}{IV}(\beta,\nu) = \frac{e^2}{\aB} \left[ \left(2\aB\beta\right)^2\cdot
    \en{kin}(\nu) - \left(2\aB\beta\right)\cdot\en{pot}(\nu) \right]\\*
    \Big(\aB \doteqdot \frac{\hbar^2}{M e^2}\Big)\ .\notag
\end{gather}
The extremalization procedure due to the principle of minimal energy (\ref{eq:ii.40})
requires now to look for the stationary points of this function~$\orrk{E}{IV}(\beta,\nu)$:
\begin{subequations}
  \begin{align}
    \label{eq:a23}
    \frac{\partial \roek{E}{IV}(\beta,\nu) }{\partial\beta}&=0\\*
    \label{eq:b23}
    \frac{\partial \roek{E}{IV}(\beta,\nu) }{\partial\nu}&=0\ .
  \end{align}
\end{subequations}

It is easy to see that these two requirements can be reduced to the one requirement
\begin{equation}
  \label{eq:a24}
  \frac{\partial \nrf{E}{T}(\nu)}{\partial\nu}=0
\end{equation}
with the reduced energy function~$\nrf{E}{T}(\nu)$ being defined through~\cite{ms1}
\begin{equation}
  \label{eq:a25}
  \nrf{E}{T}(\nu) \doteqdot -\frac{e^2}{4\aB}\cdot \SP(\nu)\ .
\end{equation}
The \emph{spectral function}~$\SP(\nu)$ emerging here is itself defined in terms of
the variational parameter~$\nu$ through
\begin{equation}
  \label{eq:a26}
  \SP(\nu) \doteqdot \frac{\en{pot}^2(\nu)}{\en{kin}(\nu)}\ ,
\end{equation}
with the \emph{kinetic function}~$\en{kin}(\nu)$ depending upon the variational
parameter~$\nu$ and upon the angular momentum quantum number~$\lP$ (\ref{eq:iii.33}) in the
following way~\cite{ms1}
\begin{equation}
  \label{eq:a27}
  \en{kin}(\nu) = \frac{1}{2\nu+1}\left(\frac{1}{4}+\frac{\lP^2}{2\nu}\right) \ .
\end{equation}
Furthermore, the \emph{potential function}~$\en{pot}(\nu)$ is found to be somewhat more
complicated but can be represented by virtue of the \emph{Poisson identity} in two
equivalent ways~\cite{ms1}
\begin{equation}
  \label{eq:a28}
  \begin{split}
    \en{pot}(\nu) &= \frac{1}{2^{4\nu+3}}\sum_{m,n=0}^\infty \frac{1}{2^{m+n}}\cdot
    \frac{\Gamma(4\nu+3+m+n)}{\Gamma(2\nu+3+m)\cdot\Gamma(2\nu+3+n)}\\*
    &=\frac{1}{2\nu+1}\left( 1-\frac{1}{2^{4\nu+2}}\cdot\sum_{n=0}^\infty
      \frac{n}{2^n}\cdot\frac{\Gamma(4\nu+2+n)}{\Gamma(2\nu+2)\cdot\Gamma(2\nu+2+n)}
      \right)\ .
  \end{split}  
\end{equation}

It follows from the present results in combination with the \emph{principle of minimal
  energy} (\ref{eq:ii.40}) that the non-relativistic positronium spectrum may be obtained
by looking for the stationary points of the spectral function~$\SP(\nu)$ (\ref{eq:a26})
for any quantum number~$\lP$. But since our trial function~$\tilde{\Phi}(r)$
(\ref{eq:iii.50}) is too simple in order to get the whole spectrum of excited states due
to the one chosen~$\lP$, one will obtain in this way merely the ``groundstate'' for
any~$\lP$, cf.~figure~2 of ref.~\cite{ms1}. The result of this extremalization procedure is
presented by \textbf{table~1} on page~\pageref{table1}.


  \stepcounter{section}
  \setcounter{equation}{0}
  \begin{center}
  {\textbf{\Large Appendix \Alph{section}:}}\\[2ex]
  \emph{ \textbf{\Large \hphantom{xxxxx}Auxiliary Potentials {\boldmath$\gkloi{p}{A}{\sf I},\,\gkloi{p}{A}{\sf III},\,\gkloi{p}{A}{\sf V}$}\hphantom{xxxxx} for general {\boldmath$\nu$}}}
  \end{center}
  \myappendix{ Auxiliary Potentials {\boldmath$\gkloi{p}{A}{\sf I},\,\gkloi{p}{A}{\sf
        III},\,\gkloi{p}{A}{\sf V}$} for general {\boldmath$\nu$}}
\vspace{2ex}
  
  For an estimate of the radial auxiliary potentials (\ref{eq:iv.11a})-(\ref{eq:iv.11e})
  due to {\em general}\/ $\nu$ it is most convenient to first recast these potentials to
  a dimensionless form, quite similarly as it was done for the spherically symmetric
  approximation $\eklo{p}{A}_0(r)$, cf. equation (\ref{eq:a2}); i.\,e. we put
\begin{subequations}
  \begin{align}
  \label{eq:b1a}
  \rrklo{\sf I}{\tilde{a}}(y)\;\doteqdot\;(2\beta\as)^{-1}\cdot\gkloi{p}{A}{\sf I}(r)&=\int_0^\infty dy'\,\frac{y'}{\left[y^2+{y'}^2\right]^\frac{1}{2}}\cdot\tilde{\Phi}^2(y')\\
  \label{eq:b1b}
  \rrklo{\sf III}{\tilde{a}}(y)\;\doteqdot\;(2\beta\as)^{-1}\cdot\gkloi{p}{A}{\sf III}(r)&=y^2\cdot\int_0^\infty dy'\,\frac{{y'}^3}{\left[y^2+{y'}^2\right]^\frac{5}{2}}\cdot\tilde{\Phi}^2(y')\\
  \label{eq:b1c}
  \rrklo{\sf V}{\tilde{a}}(y)\;\doteqdot\;(2\beta\as)^{-1}\cdot\gkloi{p}{A}{\sf V}(r)&=y^4\cdot\int_0^\infty dy'\,\frac{{y'}^5}{\left[y^2+{y'}^2\right]^\frac{9}{2}}\cdot\tilde{\Phi}^2(y')\\
  \nonumber
  &\;\:\vdots\\
  \nonumber
  \big(\,\tilde{\Phi}(y)\doteqdot\frac{\tilde{\Phi}(r)}{2\beta}&\;,\;y\doteqdot2\beta r\,\big)\ .
  \end{align}
\end{subequations}
The differential links (\ref{eq:iv.20})-(\ref{eq:iv.21}) read now in this dimensionless form
\begin{subequations}
  \begin{align}
  \label{eq:b2a}
  \rrklo{\sf III}{\tilde{a}}(y)&=-\frac{1}{3}\,y\frac{d^2}{dy^2}\Big(y\cdot\rrklo{\sf I}{\tilde{a}}(y)\Big)\\
  \label{eq:b2b}
  \rrklo{\sf V}{\tilde{a}}(y)&=
  -\frac{1}{35}\,y^3\frac{d}{dy}\left[\frac{1}{y^4}\frac{d}{dy}\Big(y^3\cdot\rrklo{\sf
      III}{\tilde{a}}(y)\Big)\right] \\
  \nonumber
  \vdots\quad\:& \hspace{6cm} . 
  \end{align}
\end{subequations}

For a treatment of the excited states it is necessary to admit the variational parameter
$\nu$ in the trial amplitude $\tilde{\Phi}(r)$ (\ref{eq:iii.50}) to adopt quite general
(real) values, not only the special value $\nu=0$ which is adopted in {\bf Sect.\,IV.2} for
the simplified treatment of the groundstate. For such a more general situation, one may
still take as the starting potential $\gkloi{p}{A}{\sf I}(r)$ again the spherically
symmetric potential $\eklo{p}{A}_0(r)$ which obeys the spherically symmetric Poisson
equation (\ref{eq:iii.36}) but now with non-zero $\nu$:
\begin{equation}
  \label{eq:b3}
  \frac{1}{(2\beta)^2}\left(\frac{d^2}{dr^2}+\frac{2}{r}\frac{d}{dr}\right)\eklo{p}{A}_0(r)= -\frac{2\beta\as}{\Gamma(2\nu+2)}\cdot(2\beta r)^{2\nu-1}\,{\rm e}^{-2\beta r}\ ,
\end{equation}
cf. also the dimensionless form (\ref{eq:a1}) of this Poisson equation and its solution (\ref{eq:a8}). Thus our starting potential $\rrklo{\sf I}{\tilde{a}}_\nu(y)$ is now just that spherically symmetric approximation (\ref{eq:a8}); i.\,e. we put
\begin{equation}
  \label{eq:b3'}
  \rrklo{\sf I}{\tilde{a}}_\nu(y)\equiv\frac{2}{\pi}\,(2\beta\as)^{-1}\,\eklo{p}{A}_0(r)= \frac{2}{\pi}\frac{1}{2\nu+1}\left(1-{\rm e}^{-y}\sum_{n=0}^\infty\frac{n}{\Gamma(2\nu+2+n)}\,y^{2\nu+n}\right)\,.
\end{equation}
For integer values of $2\nu$
(i.\,e. $\nu=0,\frac{1}{2},1,\frac{3}{2},2,\frac{5}{2},\,...$) this starting potential
appears as a finite sum, namely \cite{ms1}
\begin{equation}
  \label{eq:b4}
  \rrklo{\sf I}{\tilde{a}}_\nu(y)= \frac{2}{\pi}\frac{1}{y}\left(1-{\rm e}^{-y}\sum_{n=0}^{2\nu}\frac{2\nu+1-n}{2\nu+1}\cdot\frac{y^n}{n!}\right)\,.
\end{equation}
Clearly, for $\nu=0$ one recovers here the former starting potential $\gkloi{p}{A}{\sf I}(r)$ (\ref{eq:iv.25}).

But now that the more general starting potential $\rrklo{\sf I}{\tilde{a}}_\nu(y)$ has been fixed, one can use it in order to calculate the associated third auxiliary potential $\rrklo{\sf III}{\tilde{a}}_\nu(y)$ (\ref{eq:b2a}):
\begin{equation}
  \label{eq:b5}
  \rrklo{\sf III}{\tilde{a}}_\nu(y)=-\frac{1}{3}\,y\frac{d^2}{dy^2}\Big(y\cdot\rrklo{\sf I}{\tilde{a}}_\nu(y)\Big)= \frac{2}{3\pi}\frac{y^{2\nu+1}{\rm e}^{-y}}{\Gamma(2\nu+2)}\ .
\end{equation}
As a consistency check, one puts here $\nu=0$ which yields
\begin{equation}
  \label{eq:b6}
  \rrklo{\sf III}{\tilde{a}}_0(y)=\frac{2}{3\pi}\,y{\rm e}^{-y}=\frac{2}{3\pi}\,(2\beta r)\,{\rm e}^{-2\beta r}\ ,
\end{equation}
and thus the third radial potential $\gkloi{p}{A}{\sf III}(r)$ for $\nu=0$ becomes
\begin{equation}
  \label{eq:b7}
  \gkloi{p}{A}{\sf III}(r)=2\beta\as\cdot\rrklo{\sf III}{\tilde{a}}_0(y)=\frac{8}{3\pi}\,\as\beta^2r{\rm e}^{-2\beta r}
\end{equation}
which is nothing else than the former result (\ref{eq:iv.26a}) being used for the simplified treatment of the groundstate. Once the third auxiliary potential $\rrklo{\sf III}{\tilde{a}}_\nu(y)$ is known, cf. (\ref{eq:b5}), one can proceed to calculate by means of it the corresponding fifth auxiliary potential $\rrklo{\sf V}{\tilde{a}}_\nu(y)$ (\ref{eq:b2b}):
\begin{eqnarray}
  \label{eq:b8}
  \rrklo{\sf V}{\tilde{a}}_\nu(y)\!\!&=&\!\! -\frac{1}{35}\,y^3\frac{d}{dy}\left[\frac{1}{y^4}\frac{d}{dy}\Big(y^3\cdot\rrklo{\sf III}{\tilde{a}}_\nu(y)\Big)\right]\\
  \nonumber
  &=&\!\!-\frac{2}{105\pi}\frac{1}{\Gamma(2\nu+2)}\,{\rm e}^{-y}\Big\{y^{2\nu+3}-4(\nu+1)y^{2\nu+2}+2(\nu+2)(2\nu-1)y^{2\nu+1}\Big\}\ .
\end{eqnarray}
The consistency check for $\nu=0$ yields here
\begin{equation}
  \label{eq:b9}
  \rrklo{\sf V}{\tilde{a}}_0(y)=-\frac{16}{105\pi}\,{\rm e}^{-y}\left\{\frac{y^3}{8}-\frac{1}{2}\,y^2-\frac{1}{2}\,y\right\}\,,
\end{equation}
i.\,e. for the potential $\gkloi{p}{A}{\sf V}(r)$ due to $\nu=0$ one obtains
\begin{equation}
  \label{eq:b10}
  \gkloi{p}{A}{\sf V}(r)=2\beta\as\rrklo{\sf V}{\tilde{a}}_0(y)= -\frac{16}{105\pi}\,(2\beta\as){\rm e}^{-2\beta r}\left\{\frac{(2\beta r)^3}{8}-\frac{1}{2}\,(2\beta r)^2-\frac{1}{2}\,(2\beta r)\right\}
\end{equation}
and this again is just the former result (\ref{eq:iv.26b}) for the simplified groundstate.

Furthermore it is also rather obvious that it is the auxiliary potential of lowest order,
i.\,e. $\rrklo{\sf I}{\tilde{a}}_\nu(y)$~(\ref{eq:b3'}), which obeys the Coulomb-like
boundary condition (\ref{eq:iv.18}) at infinity ($\!\rrklo{\sf
  I}{\tilde{a}}_\nu(y)\rightarrow\frac{2}{\pi}\cdot\frac{1}{y}$) whereas all auxiliary
potentials of higher order (e.\,g.  $\rrklo{\sf III}{\tilde{a}}_\nu(y)$ (\ref{eq:b5}) and
$\rrklo{\sf V}{\tilde{a}}_\nu(y)$ (\ref{eq:b8})) tend to zero more rapidly than the
Coulomb potential ($\sim\frac{1}{y}$). Similarly, the boundary conditions at the origin
($y=0\,\Leftrightarrow\,r=0$) are also satisfied: the higher-order potentials such as
$\rrklo{\sf III}{\tilde{a}}_\nu(y)$ (\ref{eq:b5}) and $\rrklo{\sf V}{\tilde{a}}_\nu(y)$
(\ref{eq:b8}) do actually vanish at the origin ($y=0$) and thus validate the boundary
conditions (\ref{eq:iv.16}). On the other hand, it is only the auxiliary potential of
lowest order (i.\,e. $\rrklo{\sf I}{\tilde{a}}_\nu(y)$ (\ref{eq:b3'})) which adopts the
required non-zero values (\ref{eq:iv.13}) at the origin ($y=0$).

Summarizing, the expansion (\ref{eq:iv.9}) generates a hierarchical set of radial
auxiliary potentials (\ref{eq:iv.11a})-(\ref{eq:iv.11e}) etc. of ever decreasing
magnitude; for the three lowest-order potentials $\rrklo{\sf I}{\tilde{a}}_\nu(y)$
(\ref{eq:b4}), $\rrklo{\sf III}{\tilde{a}}_\nu(y)$ (\ref{eq:b5}) and $\rrklo{\sf
  V}{\tilde{a}}_\nu(y)$ (\ref{eq:b8}), see the figure B.I below. With increasing order of
this ``anisotropy expansion'' the deviation of the corresponding electrostatic potential
$\gkloi{p}{A}{0}(r,\vartheta)$ (\ref{eq:iv.9}) from the exact potential (\ref{eq:iv.5})
becomes reduced step by step. In the present context, the importance of such an expansion
refers to the gauge field energy $E_{\rm R}^{\rm(e)}$ (\ref{eq:ii.36}) which undergoes a
corresponding expansion with respect to the magnitude of the anisotropy and thus provides
us with an estimate of the relative weight of the anisotropic terms (compare, e.\,g., the
``anisotropic'' groundstate energy (\ref{eq:iv.45}) to its ``isotropic'' counterpart
(\ref{eq:iv.46}), where a deviation of only $0.4\%$ arises). Though the
treatment of the groundstate~$(\nu=0)$ on the basis of the present anisotropic correction 
$\gkloi{p}{A}{\sf III}(r)$ (\ref{eq:b7}) appears to be meaningful, the generalization to the excited
states~$\nu>0$ turns out to be unacceptable (see \textbf{App.~C}).


\begin{center}
\label{figb1}
\epsfig{file=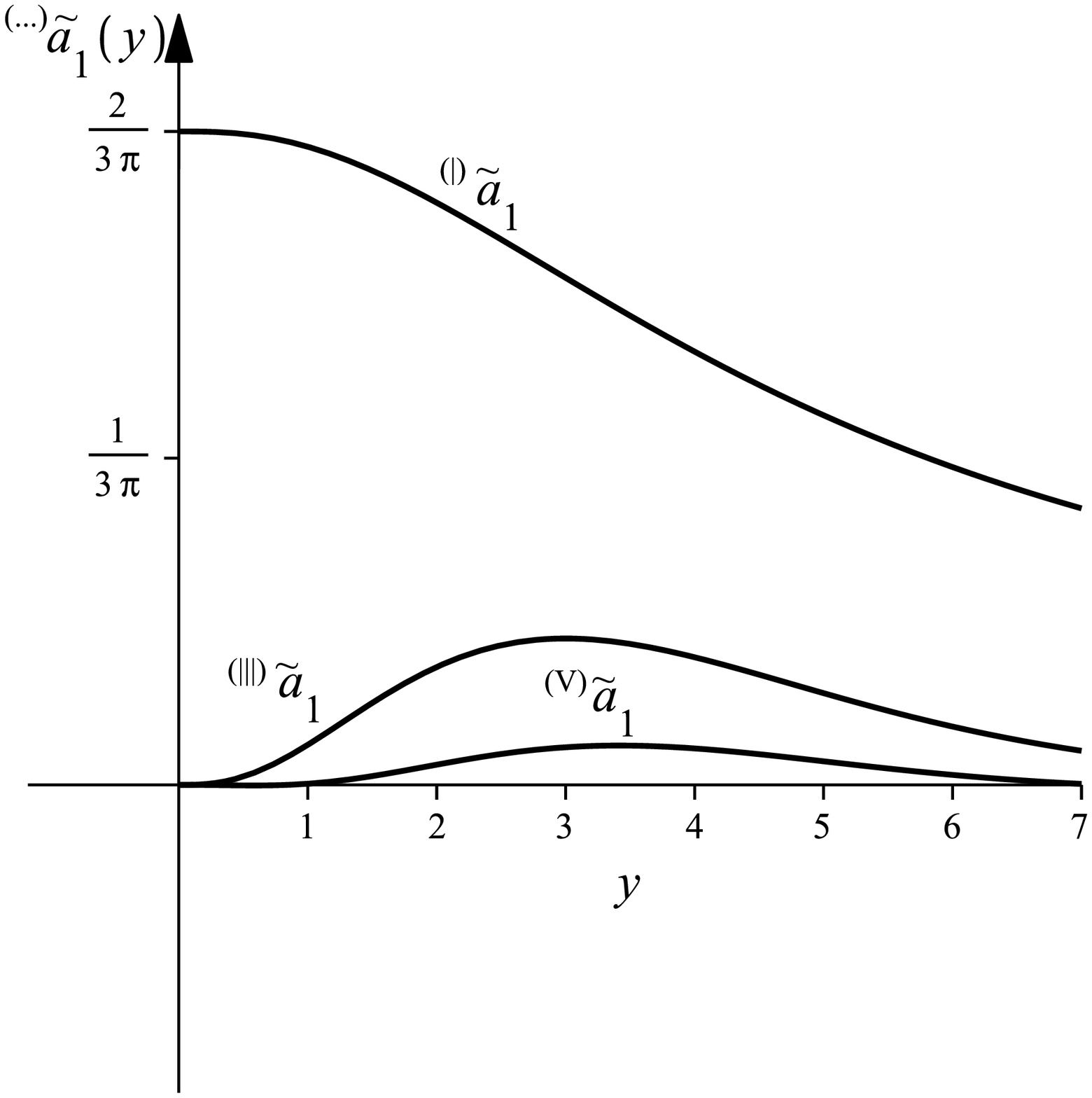,height=12cm}
\end{center}
{\textbf{Fig.~B.I}\hspace{5mm} \emph{\large\textbf{Radial Auxiliary Potentials
 \boldmath$\gkloit{\sf I}{a}{1},\,\gkloit{\sf III}{a}{1},\,\gkloit{\sf V}{a}{1}$\\
 \centerline{(\ref{eq:b4}), (\ref{eq:b5}) and (\ref{eq:b8}) } 
  }}}
\myfigure{Fig.~B.I: Radial Auxiliary Potentials
 \boldmath$\gkloit{\sf I}{a}{1},\,\gkloit{\sf III}{a}{1},\,\gkloit{\sf V}{a}{1}$
 (\ref{eq:b4}), (\ref{eq:b5})\\\hspace*{2mm} and (\ref{eq:b8})}
\indent

The hierarchy of magnitudes of the potentials $\gkloit{\sf I}{a}{\nu}(y), \gkloit{\sf
  III}{a}{\nu}(y), \gkloit{\sf V}{a}{\nu}(y),\ldots$ persists also for~$\nu>0$
(here~$\nu=1$) similarly as shown for~$\nu=0$ by \textbf{Fig.~IV.A}. Nevertheless the
associated series expansion (\ref{eq:iv.9}) of the anisotropic
potential~$\gklo{p}{A}_0(r,\vartheta)$ fails to describe the excited states correctly, see
\textbf{App.\,C}.

  \stepcounter{section}
  \setcounter{equation}{0}
  \begin{center}
  {\textbf{\Large Appendix \Alph{section}:}}\\[2ex]
  \emph{ \textbf{\Large 
  Failure of the Expansion (\ref{eq:iv.9}) for the Excited States 
  ({\boldmath$\nu>0$})}}
  \end{center}
 \myappendix{Failure of the Expansion (\ref{eq:iv.9}) for the Excited \\\hspace*{18mm} States ({\boldmath$\nu>0$})}
  \vspace{2ex}
  
  As plausible as the result of the simplified groundstate ($\nu=0$) may appear (see the
  discussion below (\ref{eq:iv.46})), this way of dealing with the anisotropic corrections
  cannot be transferred to the excited states! Indeed, we will readily see that the energy
  predictions of the excited states would come out in an unacceptable magnitude. Since the
  simplified groundstate discussion is based upon the ``anisotropy expansion''
  (\ref{eq:iv.9}), this expansion must be afflicted with a serious deficiency, at least as
  far as the excited states are concerned. Therefore, an improved expansion is presented
  in {\bf Sect.\,V}, but for the sake of completeness we will subsequently work out the
  reduced energy function $\mathbb{E}_*(\nu)$ due to the (rejectable) series expansion
  (\ref{eq:iv.9}), in order to see more clearly what is going wrong with this expansion.

To this end, we compute the corresponding energy function $\tilde{\mathbb{E}}_\Phi(\beta,\nu)$ which is to be understood as the generalization of the groundstate function $\tilde{\mathbb{E}}_\Phi(\beta)$ (\ref{eq:iv.43}), such that
\begin{equation}
  \label{eq:c1}
  \tilde{\mathbb{E}}_\Phi(\beta)=\tilde{\mathbb{E}}_\Phi(\beta,\nu)\Big|_{\nu=0}\ .
\end{equation}
Here it is convenient to scale off the atomic energy unit (a.\,u.) by introducing the spectral function $S^{\{*\}}_\wp$ through
\begin{gather}
  \label{eq:c2}
  \mathbb{E}_*(\nu)\doteqdot-\frac{e^2}{4\aB}\,S^{\{*\}}_\wp(\nu)\\
  \nonumber
  (\,\frac{e^2}{4\aB}=6.8029...\,[{\rm eV}]\,)\ ,
\end{gather}
so that we are left with the problem to elaborate this spectral function
$S^{\{*\}}_\wp(\nu)$ for general $\nu$. Clearly, its extremal values will then yield the
wanted energy spectrum of positronium (though to be rejected).

Fortunately, it is possible to explicitly calculate that spectral function
$S_\wp^{(*)}(\nu)$ (\ref{eq:c2}) for {\em integer}\/ $2\nu$. More concretely, the general
gauge field energy $E_{\rm R}^{\rm(e)}$ (\ref{eq:ii.36}) appears now in terms of the
dimensionless auxiliary potentials $\rrklo{\sf I}{\tilde{a}}_\nu(y)$ (\ref{eq:b3'}) and
$\rrklo{\sf III}{\tilde{a}}_\nu(y)$ (\ref{eq:b5}) as a sum of three terms, cf. equation
(\ref{eq:iv.32}), with each of the three terms being found to appear in the following
form:
\begin{subequations}
  \begin{align}
  \label{eq:c3a}
  E_{\sf I}^{\rm\{e\}}&\doteqdot -\hbar c\,(2\beta\as)\left(\frac{\pi}{2}\right)^2\int_0^\infty dy\,y^2\left(\frac{d\rrklo{\sf I}{\tilde{a}}_\nu(y)}{dy}\right)^{\!\!2}\doteqdot-2\beta e^2\cdot\rrklo{\sf I}{\varepsilon}_{\rm pot}(\nu)\\
  \label{eq:c3b}
  E_{\sf II}^{\rm\{e\}}&\doteqdot -\hbar c\,(2\beta\as)\left(\frac{\pi}{2}\right)^2\int_0^\infty dy\,y^2\,\frac{d\rrklo{\sf I}{\tilde{a}}_\nu(y)}{dy}\cdot\frac{d\rrklo{\sf III}{\tilde{a}}_\nu(y)}{dy}\doteqdot-2\beta e^2\cdot\rrklo{\sf II}{\varepsilon}_{\rm pot}(\nu)\\
  \label{eq:c3c}
  E_{\sf III}^{\rm\{e\}}&\doteqdot -\hbar c\,(2\beta\as)\,\frac{3\pi^2}{160}\int_0^\infty dy\,y^2\left\{\frac{7}{2}\left(\frac{d\rrklo{\sf III}{\tilde{a}}_\nu(y)}{dy}\right)^{\!\!2}\!+\left[\frac{\rrklo{\sf III}{\tilde{a}}_\nu(y)}{y}\right]^2\right\}\doteqdot-2\beta e^2\cdot\!\rrklo{\sf III}{\varepsilon}_{\rm pot}(\nu)\ .
  \end{align}
\end{subequations}
But since the radial auxiliary potentials $\rrklo{\sf I}{\tilde{a}}_\nu(y)$ and
$\rrklo{\sf III}{\tilde{a}}_\nu(y)$ are explicitly known,
cf. (\ref{eq:b3'})-(\ref{eq:b5}), one can calculate here also explicitly the three
auxiliary potential functions $\rrklo{\sf I}{\varepsilon}_{\rm pot}(\nu)$, $\rrklo{\sf
  II}{\varepsilon}_{\rm pot}(\nu)$ and $\rrklo{\sf III}{\varepsilon}_{\rm pot}(\nu)$ for
general $\nu$. The result is
\begin{subequations}
  \begin{align}
  \label{eq:c7a}
  \rkloi{\sf I}{\varepsilon}{pot}(\nu)&\equiv\varepsilon_{\rm pot}(\nu)\,,\quad{\rm cf.\ (\ref{eq:a28})}\\
  \label{eq:c7b}
  \rkloi{\sf II}{\varepsilon}{pot}(\nu)&=\frac{1}{4!\,2^{4\nu}}\cdot\frac{\Gamma(4\nu+3)}{\Gamma(2\nu+2)^2}\\
  \label{eq:c7c}
  \rkloi{\sf III}{\varepsilon}{pot}(\nu)&=\frac{9+7\nu}{1920}\cdot\frac{\Gamma(4\nu+3)}{2^{4\nu}\cdot\Gamma(2\nu+2)^2}\ .
  \end{align}
\end{subequations} 
These three terms add then up to the total ``anisotropic'' gauge field energy $E_{\rm R}^{\rm\{e\}}$
\begin{equation}
  \label{eq:c5}
  E_{\rm R}^{\rm\{e\}}=E_{\sf I}^{\rm\{e\}}+E_{\sf II}^{\rm\{e\}}+E_{\sf III}^{\rm\{e\}}\doteqdot -\hbar c\,(2\beta\as)\cdot\varepsilon_{\rm R}^{\rm\{e\}}(\nu)
\end{equation}
where the ``anisotropic`'' potential function $\varepsilon_{\rm R}^{\rm\{e\}}(\nu)$ is evidently given by
\begin{equation}
  \label{eq:c6}
  \varepsilon_{\rm R}^{\rm\{e\}}(\nu)\doteqdot \rkloi{\sf I}{\varepsilon}{pot}(\nu)+\rkloi{\sf II}{\varepsilon}{pot}(\nu)+\rkloi{\sf III}{\varepsilon}{pot}(\nu)\ .
\end{equation}

Obviously, the present gauge field energy $E_{\rm R}^{\rm\{e\}}$ (\ref{eq:c5}) with (\ref{eq:c7a})-(\ref{eq:c7c}) is nothing else than the generalization of the anisotropic groundstate result $E_{\rm R}^{\rm\{e\}}$ (\ref{eq:iv.32})-(\ref{eq:iv.33c}) to the excited states. Observe also that the Poisson identity $\tilde{\mathbb{N}}_{\rm G}^{\rm(e)}\equiv0$ (cf. (\ref{eq:iv.38})) is obeyed strictly for the spherically symmetric approximation, but this identity is of course {\em not}\/ obeyed by the present anisotropic approximation procedure, neither for the groundstate nor for the excited states!

Therefore we have to calculate now the non-relativistic version $\tilde{\mathbb{M}}^{\rm\{e\}}c^2$ of the (relativistic) mass equivalent $\tilde{M}^{\rm(e)}c^2$ (\ref{eq:ii.37}) in order to build up again the constraint term $\tilde{\mathbb{N}}_{\rm G}^{\rm(e)}$ (\ref{eq:iv.38})-(\ref{eq:iv.39}), but now for general values of the variational parameter $\nu$. This means that for the trial function $\tilde{\Phi}(r)$ in both parts $\tilde{\mathbb{M}}_{\sf I}^{\rm\{e\}}$ and $\tilde{\mathbb{M}}_{\sf III}^{\rm\{e\}}$ (\ref{eq:iv.40})-(\ref{eq:iv.41}) of the mass equivalent one has now to make use of the {\em general}\/ form (\ref{eq:iii.50})-(\ref{eq:iii.51}), i.\,e.
\begin{subequations}
  \begin{align}
  \label{eq:c8a}
  \tilde{\mathbb{M}}_{\sf I}^{\rm(e)}c^2\Rightarrow\tilde{\mathbb{M}}_{\sf I}^{\rm\{e\}}c^2&= -\left(\frac{\pi}{2}\right)^2\hbar c\int dr\,r\,\gklo{p}{A}_{\sf I}(r)\cdot\Phi_*^2r^{2\nu}{\rm e}^{-2\beta r}\\
  \label{eq:c8b}
  \tilde{\mathbb{M}}_{\sf III}^{\rm(e)}c^2\Rightarrow\tilde{\mathbb{M}}_{\sf III}^{\rm\{e\}}c^2&= -\left(\frac{3\pi}{8}\right)^2\hbar c\int dr\,r\,\gklo{p}{A}_{\sf III}(r)\cdot\Phi_*^2r^{2\nu}{\rm e}^{-2\beta r}\ ,
  \end{align}
\end{subequations}
where the radial auxiliary potentials $\gklo{p}{A}_{\sf I}(r)$ and $\gklo{p}{A}_{\sf III}(r)$  are specified for general $\nu$ (in dimensionless form ) by equations (\ref{eq:b4})-(\ref{eq:b5}) of {\bf App.\,B}. Thus one finds by explicitly carrying out the required integrations in (\ref{eq:c8a})-(\ref{eq:c8b})
\begin{subequations}
  \begin{align}
  \label{eq:c9a}
  \tilde{\mathbb{M}}_{\sf I}^{\rm\{e\}}c^2&= -\hbar c\,(2\as\beta)\cdot\varepsilon_{\rm pot}(\nu)\equiv E_{\sf I}^{\rm\{e\}}\\
  \label{eq:c9b}
  \tilde{\mathbb{M}}_{\sf III}^{\rm\{e\}}c^2&= -\hbar c\,(2\as\beta)\,\mu_{\rm R}^{\rm\{e\}}(\nu)\\
  \label{eq:c9c}	
  \mu_{\rm R}^{\rm\{e\}}(\nu)&\doteqdot \frac{3}{128}\cdot\frac{1}{2^{4\nu}}\cdot\frac{\Gamma(4\nu+3)}{\Gamma(2\nu+2)^2}\ .
  \end{align}
\end{subequations}

Clearly for $\nu=0$, this result leads us back to the previous groundstate results (\ref{eq:iv.40})-(\ref{eq:iv.41}). But even for $\nu\neq0$, the present results display a structure very similar to the former groundstate situation (\ref{eq:iv.42}). Namely, the striking feature concerns here the Poisson constraint term $\tilde{\mathbb{N}}_{\rm G}^{\rm\{e\}}$ (\ref{eq:iv.38}) which for the present situation of the excited states adopts the following form:
\begin{equation}
  \label{eq:c10}
  \tilde{\mathbb{N}}_{\rm G}^{\rm\{e\}}=\mathbb{E}_{\sf II}^{\rm\{e\}}+\mathbb{E}_{\sf III}^{\rm\{e\}}-\tilde{\mathbb{M}}_{\sf III}^{\rm\{e\}}c^2=-\hbar c\,(2\as\beta)\cdot\varepsilon_{\rm G}^{\rm\{e\}}
\end{equation}
with
\begin{equation}
  \label{eq:c11}
  \varepsilon_{\rm G}^{\rm\{e\}}(\nu)\doteqdot \srklo{II}{\varepsilon}_{\rm pot}(\nu)+\srklo{III}{\varepsilon}_{\rm pot}(\nu)-\mu_{\rm R}^{\rm\{e\}}(\nu)\ .
\end{equation}
This similarity comes about through the present identity relation (\ref{eq:c7a}) for the potential functions $\varepsilon_{\rm pot}(\nu)$ and $\rrklo{\sf I}{\varepsilon}_{\rm pot}(\nu)$ which entails for $\nu\neq0$ the same equality of the energy $E_{\sf I}^{\rm\{e\}}$ and mass equivalent $\tilde{\mathbb{M}}_{\sf I}^{\rm\{e\}}c^2$, cf. the remark above (\ref{eq:iv.42}), as is the case for the previous groundstate situation (\ref{eq:iv.40}) with $\nu=0$. Consequently, both electrostatic contributions (\ref{eq:c5}) and (\ref{eq:c10}) to the energy functional (\ref{eq:iv.29}) add up in order to yield now
\begin{subequations}
  \begin{align}
  \label{eq:c12a}
  E_{\rm R}^{\rm\{e\}}+\lambda_{\rm G}^{\rm(e)}\cdot\tilde{\mathbb{N}}_{\rm G}^{\rm\{e\}}&= -\hbar c\,(2\as\beta)\left[\varepsilon_{\rm R}^{\rm\{e\}}(\nu)-2\varepsilon_{\rm G}(\nu)\right]= -\hbar c\,(2\as\beta)\cdot\varepsilon^{\rm\{e\}}(\nu)\\
  \label{eq:c12b}
  \varepsilon^{\rm\{e\}}(\nu)&\doteqdot\varepsilon_{\rm R}^{\rm\{e\}}-2\varepsilon_{\rm G}(\nu)= \rrklo{\sf I}{\varepsilon}_{\rm pot}(\nu)-\rrklo{\sf II}{\varepsilon}_{\rm pot}(\nu)-\rrklo{\sf III}{\varepsilon}_{\rm pot}(\nu)+2\mu_{\rm R}^{\rm\{e\}}(\nu)\ .
  \end{align}
\end{subequations}

The crucial point with this total electrostatic contribution (\ref{eq:c12a}) to the energy function $\tilde{\mathbb{E}}_\Phi(\beta,\nu)$, to be conceived as the value of the energy functional $\tilde{\mathbb{E}}_{[\Phi]}$ (\ref{eq:iv.29}) on the chosen trial configuration, is now that it is of the same structure as in the case for the {\em spherically symmetric}\/ approximation (\ref{eq:a22}) and for the {\em anisotropic}\/ groundstate (\ref{eq:iv.43}); namely, both the gauge field energy $E_{\rm R}^{\rm\{e\}}$ and its mass equivalent $\tilde{\mathbb{M}}^{\rm\{e\}}c^2$ do contribute a linear function of the variational parameter $\beta$. On the other hand, the kinetic energy $E_{\rm kin}$ has been found to contribute a quadratic function of $\beta$, cf. (\ref{eq:iii.52})-(\ref{eq:iii.53}); and therefore the present energy function $\tilde{\mathbb{E}}_\Phi(\beta,\nu)$ for the anisotropic excited states displays again the formerly encountered shape for the isotropic situation (cf. (\ref{eq:a22})):
\begin{equation}
  \label{eq:c13}
  \tilde{\mathbb{E}}_\Phi(\beta,\nu)=\frac{e^2}{\aB}\left[(2\aB\beta)^2\cdot\varepsilon_{\rm kin}(\nu)-(2\aB\beta)\cdot\varepsilon^{\rm\{e\}}(\nu)\right]\ ,
\end{equation}
where merely the simple potential function $\varepsilon_{\rm pot}(\nu)$ (\ref{eq:a28}) is to be replaced by the present more complicated {\em electrostatic function}\/ $\varepsilon^{\rm\{e\}}(\nu)$ (\ref{eq:c12b}). Clearly, as a consistency check one puts here $\nu=0$ and is led back to the former groundstate result $\tilde{\mathbb{E}}_\Phi(\beta)$ (\ref{eq:iv.43}).

This similarity to the former configurations of spherical symmetry ({\bf Sect. III}) and of the anisotropic groundstate ({\bf Subsect. IV.3}) enables us to write down immediately the result for the present anisotropic situation. The extremalization process (\ref{eq:iii.40}), being exemplified for the spherically symmetric approximation through equations (\ref{eq:a23})-(\ref{eq:b23}) in {\bf App.\,A}, reduces now the two-parameter energy function $\tilde{\mathbb{E}}_\Phi(\beta,\nu)$ (\ref{eq:c13}) again to the wanted one-parameter energy function $\mathbb{E}_*(\nu)$ (\ref{eq:c2}). 
Here, the spectral function $S_\wp^{\{*\}}(\nu)$ is formally the same as previously for the spherically symmetric approximation, cf. (\ref{eq:iii.55a})-(\ref{eq:iii.55b}), with merely the former potential function $\varepsilon_{\rm pot}(\nu)$ being replaced now by its present ``anisotropic'' counterpart $\varepsilon^{\rm\{e\}}(\nu)$ (\ref{eq:c12b}), i.\,e.
\begin{equation}
  \label{eq:c14}
  S_\wp^{\{*\}}(\nu)=\frac{\big(\varepsilon^{\rm\{e\}}(\nu)\big)^2}{\varepsilon_{\rm kin}(\nu)}\ .
\end{equation}
Consequently, the last step consists again in looking for the maximal value of $S_\wp^{(*)}(\nu)$ ($\leadsto$ minimal value of $\mathbb{E}_*(\nu)$ (\ref{eq:c2})) by means of an appropriate computer program.

For our present purposes, it may be sufficient to display the result for some representative values of the variational parameter $\nu$, see the table below. For the sake of comparison with the spherically symmetric approximation, one can introduce here the relative shift ($\Delta$) due to the present anisotropic corrections
\begin{equation}
  \label{eq:c15}
  \Delta(\nu)\doteqdot\frac{\mathbb{E}_*(\nu)-\mathbb{E}_{\rm T}(\nu)}{\mathbb{E}_{\rm T}(\nu)}\equiv\frac{S_\wp(\nu)-S_\wp^{\{*\}}(\nu)}{S_\wp(\nu)}\ ,
\end{equation}
where the energy function $\mathbb{E}_{\rm T}(\nu)$ due to the spherically symmetric approximation is defined through equations (\ref{eq:iii.55a})-(\ref{eq:iii.55b}). From the results of this table it becomes obvious that the anisotropy effect lifts up the energy levels of the spherically symmetric approximation, where the shift increases with increasing variational parameter $\nu$. For instance, for $\nu=0$, $\lP=0$ one finds a shift of $-0.41\%$ (first line of the table); whereas for $\nu_*\cong1.8$ (i.\,e. the first excited state) the energy shift amounts to roughly $8\%$ which is the same deviation as is found for the difference between the spherically symmetric approximation and the conventional result (see the second line of table 1 on p.\,\pageref{table1}). However, the unacceptable feature of this result is the fact that for the excited states ($\nu>0$) the energy levels of the spherically symmetric approximation become {\em lifted}\/ instead of {\em lowered}\/, as required by the principle of minimal energy! Therefore the present anisotropy expansion (\ref{eq:iv.9}) must be replaced by a more systematic procedure ($\leadsto$ {\bf Sect.\,V}).

\begin{flushleft}
  \begin{tabular}{|c|c||c|c|c|} 
  \hline
$\nu $ & $\lP$ & $S_\wp(\nu)$\ (\ref{eq:iii.55b}) &
  $S_\wp^{\{*\}}(\nu)$\ (\ref{eq:c14}) & $\Delta(\nu)$\ (\ref{eq:c15})\ [\%] \\
  \hline\hline
                                        & 0 & 1.0000\hphantom{\ldots} & 1.0041\ldots & -0.41\hphantom{-} \\ \cline{2-5}
\raisebox{2.3ex}[-2.3ex]{0}             & 1 & -                       & -                       & -     \\ \hline
                                        & 0 & 0.7812\ldots            & 0.7715\ldots            & 1.24  \\ \cline{2-5}
\raisebox{2.3ex}[-2.3ex]{$\frac{1}{2}$} & 1 & 0.1562\ldots            & 0.1543\ldots            & 1.22  \\ \hline
                                        & 0 & 0.6302\ldots            & 0.6089\ldots            & 3.38  \\ \cline{2-5}
\raisebox{2.3ex}[-2.3ex]{1}             & 1 & 0.2100\ldots            & 0.2029\ldots            & 3.38  \\ \hline
                                        & 0 & 0.5278\ldots            & 0.4969\ldots            & 5.85  \\ \cline{2-5}
\raisebox{2.3ex}[-2.3ex]{$\frac{3}{2}$} & 1 & 0.2262\ldots            & 0.2129\ldots            & 5.88  \\ \hline
                                        & 0 & 0.4546\ldots            & 0.4153\ldots            & 8.64  \\ \cline{2-5}
\raisebox{2.3ex}[-2.3ex]{2}             & 1 & 0.2273\ldots            & 0.2076\ldots            & 8.67  \\ \hline
  \end{tabular}
\end{flushleft}
\label{tableC}


\renewcommand{\theequation}{\Alph{section}.\arabic{equation}}
  \stepcounter{section}
  \setcounter{equation}{0}
  \begin{center}
  {\textbf{\Large Appendix \Alph{section}:}}\\[2ex]
  \emph{ \textbf{\Large 
  Exact Solutions of the Quadrupole Equation (\ref{eq:v.34}) 
  }}
\end{center}
\myappendix{Exact Solutions of the Quadrupole Equation (\ref{eq:v.34})}
  \vspace{2ex}
  
For the construction of solutions $\gklo{p}{A}^{\sf III}(r)$ to the equation (\ref{eq:v.34}) it is very convenient to pass over again to dimensionless objects, i.\,e. we first define the dimensionless version of $\gklo{p}{A}^{\sf III}(r)$ through
\begin{equation}
  \label{eq:d1}
  \mathcal{A}^{\sf III}_\nu(y)\doteqdot(2\beta\as)^{-1}\,\gklo{p}{A}^{\sf III}(r)\ ,
\end{equation}
so that the quadrupole equation (\ref{eq:v.34}) reappears in the following form:
\begin{gather}
  \label{eq:d2}
  \left(\Delta_y-\frac{6}{y^2}\right)\mathcal{A}^{\sf III}_\nu(y)=-\frac{10}{\pi}\frac{y^{2\nu-1}\cdot{\rm e}^{-y}}{\Gamma(2\nu+2)}\\
  \nonumber
  (y\doteqdot2\beta r)\ .
\end{gather}

A first integral of this equation can easily be found here by observing that the (ordinary!) differential equation (\ref{eq:d2}) may be rewritten in the following form
\begin{equation}
  \label{eq:d3}
  \frac{1}{y^4}\frac{d}{dy}\left[y^6\cdot\frac{d}{dy}\left(\frac{\mathcal{A}^{\sf III}_\nu(y)}{y^2}\right)\right]=-\frac{10}{\pi}\frac{y^{2\nu-1}\cdot{\rm e}^{-y}}{(2\nu+1)!}\ , 
\end{equation}
where, for the present purposes, we restrict ourselves to integer values of $2\nu$ (i.\,e. $2\nu=1,\,2,\,3,\,4,\,...$). By a first integration step in (\ref{eq:d3}) one easily arrives at the intermediate result (for $\nu\geq1$)
\begin{equation}
  \label{eq:d4}
  \frac{d}{dy}\left(\frac{\mathcal{A}^{\sf III}_\nu(y)}{y^2}\right)= -\frac{10}{\pi}\,(2\nu+3)(2\nu+2)\,\frac{1-{\rm e}^{-y}\sum_{n=0}^{2\nu+3}\frac{y^n}{n!}}{y^6}\ ,
\end{equation}
which can be directly checked by means of a simple differentiation process. For the next
integration step it is very advantageous to first consider the special value $\nu=1$ for
the variational parameter. In this case, one obtains from the intermediate result
(\ref{eq:d4})
\begin{equation}
  \label{eq:d5}
  \mathcal{A}^{\sf III}_1(y)= \frac{40}{\pi}\,\frac{1-{\rm e}^{-y}\sum_{n=0}^{4}\frac{y^n}{n!}}{y^3}\ .
\end{equation}

This result for $\nu=1$ may now be used in order to further integrate in equation (\ref{eq:d4}) which yields for integer $2\nu$
\begin{gather}
  \nonumber
  \mathcal{A}^{\sf III}_\nu(y)=\frac{(2\nu+3)(2\nu+2)}{20}\,\mathcal{A}^{\sf III}_1(y)-\frac{10}{\pi}\,(2\nu+3)(2\nu+2)\,y^2\cdot\sum_{n=6}^{2\nu+3}\frac{1}{n!}\int_y^\infty dy'\,y'^{n-6}{\rm e}^{-y'}\\
  \label{eq:d6}
  (\nu\geq1)
\end{gather}
(for $\nu=1$, the second term is to be omitted). As a check, one lets act the differential operator ($\Delta_y-\frac{6}{y^2}$) upon the result in order to find that the differential equation (\ref{eq:d2}) is actually satisfied. In the last step, one explicitly calculates the residual integral on the right-hand side of equation (\ref{eq:d6}) and thus obtains the final result (for $\nu\geq\frac{3}{2}$)
\begin{equation}
  \label{eq:d7}
  \mathcal{A}^{\sf III}_\nu(y)=\frac{(2\nu+3)(2\nu+2)}{20}\,\mathcal{A}^{\sf III}_1(y)-\frac{10}{\pi}\,(2\nu+3)(2\nu+2)\,y^2{\rm e}^{-y}\sum_{n=6}^{2\nu+3}\frac{1}{n!}\sum_{m=0}^{n-6}\frac{d^m}{dy^m}\,(y^{n-6})
\end{equation}
(for $\nu=1$, the second term is to be omitted).

From this result it is clearly seen that the behaviour of the first anisotropic correction
$\mathcal{A}^{\sf III}_\nu(y)$ is determined by $\mathcal{A}^{\sf III}_1(y)$ (\ref{eq:d5})
in both the asymptotic region ($y\rightarrow\infty$) and around the origin ($y\to
0$). Since $\mathcal{A}^{\sf III}_1(y)$ (\ref{eq:d5}) behaves as follows
\begin{equation}
  \label{eq:d8}
  \lim_{y\rightarrow\infty}\mathcal{A}^{\sf III}_1(y)\simeq\frac{40}{\pi}\cdot\frac{1}{y^3}\ ,
\end{equation}
one obtains for the general $\mathcal{A}^{\sf III}_\nu(y)$ (\ref{eq:d7})
\begin{equation}
  \label{eq:d9}
  \lim_{y\rightarrow\infty}\mathcal{A}^{\sf III}_\nu(y)\simeq\frac{2}{\pi}\,(2\nu+3)(2\nu+2)\cdot\frac{1}{y^3}
\end{equation}
which is nothing else than the claim (\ref{eq:v.36}) at the end of {\bf Sect.V.3}, rewritten in dimensionless form.

In order to gain some insight into the properties of the anisotropy corrections $\mathcal{A}^{\sf III}_\nu(y)$ in the vicinity of the origin ($y=0$), one writes down explicitly some of the lowest-order cases:
\begin{subequations}
  \begin{align}
  \label{eq:d10a}
  \mathcal{A}^{\sf III}_1(y)&=\frac{40}{\pi}\,{\rm e}^{-y}\sum_{n=5}^\infty\frac{y^{n-3}}{n!}\cong \frac{1}{3\pi}\left(y^2-\frac{5}{6}\,y^3+\frac{5}{14}\,y^4-\frac{5}{48}y^5+\,...\right)\\
  \label{eq:d10b}
  \mathcal{A}^{\sf III}_{\frac{3}{2}}(y)&=\frac{3}{2}\,\mathcal{A}^{\sf III}_1(y)-\frac{5}{12\pi}\,y^2{\rm e}^{-y}\cong \frac{1}{12\pi}\left(y^2-\frac{5}{14}\,y^4+\,...\right)\\
  \label{eq:d10c}
  \mathcal{A}^{\sf III}_2(y)&=\frac{21}{10}\,\mathcal{A}^{\sf III}_1(y)-\frac{1}{12\pi}\,{\rm e}^{-y}(8y^2+y^3)\cong \frac{1}{30\pi}\left(y^2-\frac{5}{48}\,y^5+\,...\right).
  \end{align}
\end{subequations}
Observe here that with increasing $\nu$ ($\rightarrow1,\,\frac{3}{2},\,2,\,...$) the
anisotropic correction $\mathcal{A}^{\sf III}_\nu(y)$ becomes smaller and smaller in the
vicinity of the origin ($y\rightarrow0$), whereas at infinity ($y\rightarrow\infty$) the
correction $\mathcal{A}^{\sf III}_\nu(y)$ becomes larger and larger,
cf. (\ref{eq:d9}). For a sketch of the three lowest-order corrections
(\ref{eq:d10a})-(\ref{eq:d10c}) see \textbf{Fig.~D.I} below.


\begin{center}
\epsfig{file=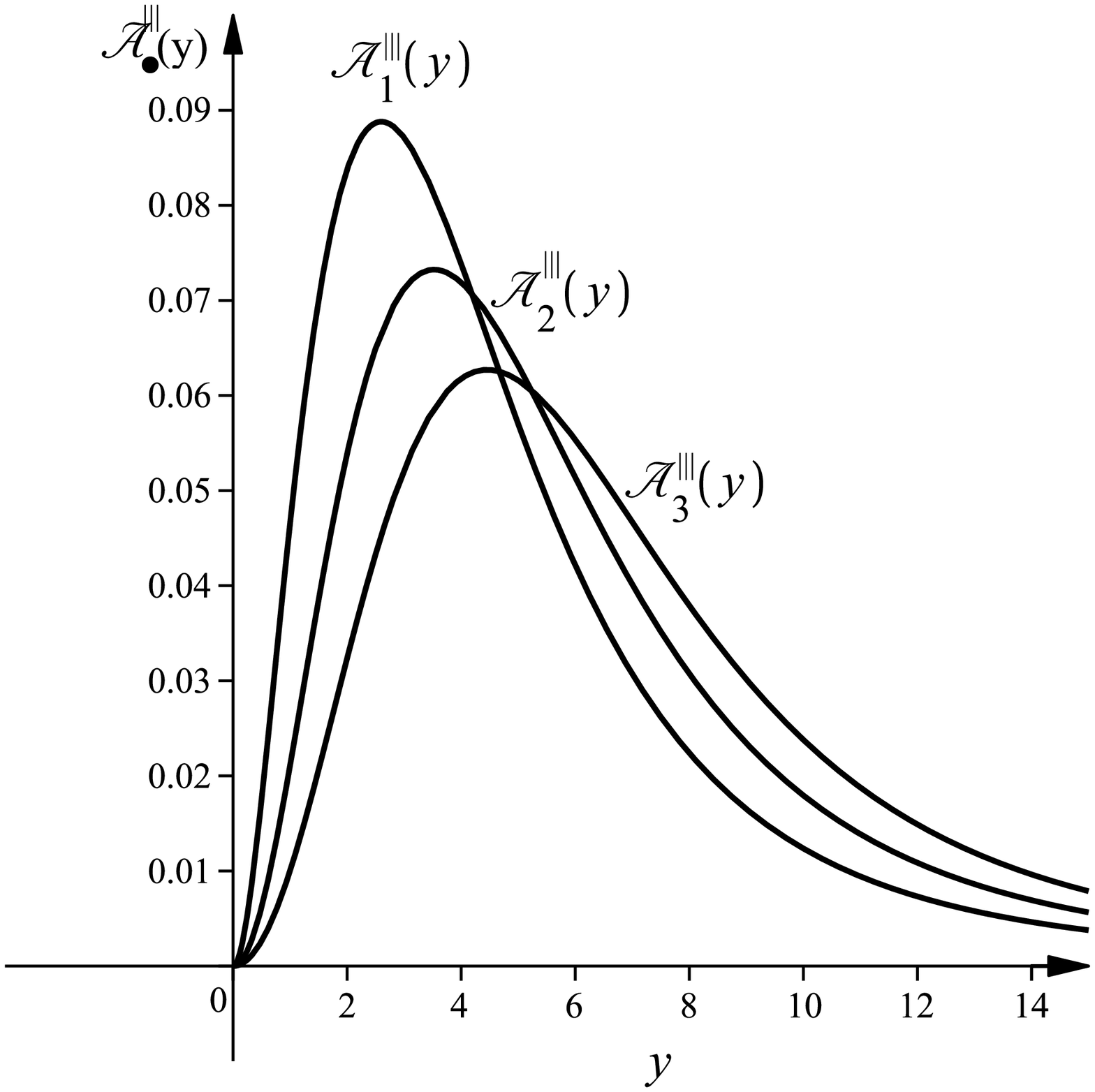,height=12cm}
\end{center}
{\textbf{Fig.~D.I}\hspace{5mm} \emph{\large\textbf{ Anisotropy Corrections
      \boldmath$\A^{\sf |||}_\nu(y)$ (\ref{eq:d6}) 
}}}
\myfigure{Fig.~D.I: Anisotropy Corrections \boldmath$\A^{\sf |||}_\nu(y)$ (\ref{eq:d6})}
The exact solution~$\A^{\sf |||}_\nu(y)$ (\ref{eq:d6}) of the quadrupole equation
(\ref{eq:d2}) behaves like~$y^{-3}$ in the asymptotic region~$(y\to\infty)$,
cf. (\ref{eq:v.36}); whereas in the vicinity of the origin~$(y\to 0)$, the behaviour is
for all~$\nu$ like~$y^2$, cf.~(\ref{eq:d10a})-(\ref{eq:d10b}). The magnitude of this
quadrupole correction~$\A^{\sf |||}_\nu(y)$ ($\le 1/10$) is relatively small in comparison
to the magnitude of the spherically symmetric approximation~$\tilde{a}_\nu(y)$,
cf.~(\ref{eq:a10}), so that the anisotropy approximation can indeed be treated as a
pertubation of the spherical symmetry, see the table of relative
deviations~$\Delta_{\{\nu\}}$ in \textbf{App.~F}.


  \stepcounter{section}
  \setcounter{equation}{0}
  \begin{center}
  {\textbf{\Large Appendix \Alph{section}:}}\\[2ex]
  \emph{ \textbf{\Large 
  Electrostatic Energy {\boldmath$E_{\rm R}^{\rm\{e\}}$} of the Anisotropic Configurations 
  }}
  \end{center}
\myappendix{Electrostatic Energy {\boldmath$E_{\rm R}^{\rm\{e\}}$} of the Anisotropic\\\hspace*{17mm} Configurations}
  \vspace{2ex}
  
  Since exact solutions of the ``quadrupole'' equation (\ref{eq:v.34}) are at hand (see
  {\bf App.\,D}), one can use them in order to calculate both the anisotropy energy
  $E_{\rm an}^{\rm\{e\}}$ (\ref{eq:v.30}) and its mass equivalent $\tilde{\mathbb{M}}_{\rm
    an}^{\rm\{e\}}c^2$ (\ref{eq:v.31}) which of course must be numerically identical. This
  identity is itself a helpful mutual check for the numerical correctness of the
  electrostatic field energy and its mass equivalent. Subsequently, we will present the
  results for some (half-)integer values of the variational parameter $\nu$ which in {\bf
    Sect.\,V} are then used in order to build up the ``anisotropic'' energy function
  $\mathbb{E}^{\rm\{IV\}}(\beta,\nu)$ for the excited states, i.\,e. the anisotropic
  generalization of the ``isotropic'' $\mathbb{E}^{\rm[IV]}(\beta,\nu)$
  (\ref{eq:iii.54'}).

First, it is very convenient to introduce here again the dimensionless objects, i.\,e. the anisotropy energy $E_{\rm an}^{\rm\{e\}}$ (\ref{eq:v.30}) reads in terms of the dimensionless potential correction $\mathcal{A}_\nu^{\sf III}(y)$ (\ref{eq:d1})
\begin{eqnarray}
  \label{eq:e1}
  E_{\rm an}^{\rm\{e\}}&=&-\frac{\pi^2}{160}\,(e^2\beta)\cdot\left\{\int_0^\infty dy\,y^2\left(\frac{d\mathcal{A}_\nu^{\sf III}}{dy}\right)^2+6\int_0^\infty dy\,\left(\mathcal{A}_\nu^{\sf III}\right)^2\right\}\doteqdot-e^2\beta\cdot\varepsilon_{\rm an}^{\rm\{e\}}\ .\quad
\end{eqnarray}
Similarly, the corresponding mass equivalent $\tilde{\mathbb{M}}_{\rm an}^{\rm\{e\}}c^2$ (\ref{eq:v.31}) also reappears in terms of that dimensionless $\mathcal{A}_\nu^{\sf III}(y)$ as follows:
\begin{equation}
  \label{eq:e2}
  \tilde{\mathbb{M}}_{\rm an}^{\rm\{e\}}c^2=-\frac{\pi}{16}\frac{e^2\beta}{(2\nu+1)!}\int_0^\infty dy\,y^{2\nu+1}\,{\rm e}^{-y}\cdot\mathcal{A}_\nu^{\sf III}(y)\ .
\end{equation}
The vanishing of the ``anisotropic'' Poisson identity $\tilde{\mathbb{N}}_{\rm
  an}^{\rm\{e\}}$ (\ref{eq:v.26}) says now that both right-hand sides of equations
(\ref{eq:e1}) and (\ref{eq:e2}) must be numerically identical! The subsequent table at the
end of this appendix displays the results in units of $(e^2\beta)$ for some values of the
variational parameter $\nu$.

Now in order to deduce a practicable recipe for calculating the anisotropy energy $E_{\rm an}^{\rm\{e\}}$ (\ref{eq:e1}) and its mass equivalent $\tilde{\mathbb{M}}_{\rm an}^{\rm\{e\}}c^2$ (\ref{eq:e2}) it is very convenient to refer the {\em general}\/ case (for integer values of $2\nu$) to the {\em special}\/ case for $\nu=1$, quite similarly as it has been done for the calculation of the anisotropy corrections $\mathcal{A}_\nu^{\sf III}(y)$, see equation (\ref{eq:d7}) of {\bf App.\,D}. First, one recasts that equation (\ref{eq:d7}) to a somewhat more concise form, i.\,e. we put
\begin{equation}
  \label{eq:e3}
  \mathcal{A}_\nu^{\sf III}(y)=(2\nu+3)(2\nu+2)\cdot\left\{\frac{\mathcal{A}_1^{\sf III}(y)}{20}-\frac{10}{\pi}\,\mathcal{B}_\nu^{\sf III}(y)\right\}\ ,
\end{equation}
with the short-range constituent $\mathcal{B}_\nu^{\sf III}(y)$ being defined through
\begin{equation}
  \label{eq:e4}
  \mathcal{B}_\nu^{\sf III}(y)\doteqdot y^2{\rm e}^{-y}\cdot\sum_{n=6}^{2\nu+3}\frac{1}{n!}\sum_{m=0}^{n-6}\frac{d^m}{dy^m}\,(y^{n-6})\ ,
\end{equation}
and obeying the following differential equation
\begin{gather}
  \label{eq:e5}
  \left(\Delta_y-\frac{6}{y^2}\right)\mathcal{B}_\nu^{\sf III}(y)=\frac{y^{2\nu-1}{\rm e}^{-y}}{(2\nu+3)!}-\frac{y{\rm e}^{-y}}{5!}\\
  \nonumber
  (\,\leadsto\mathcal{B}_\nu^{\sf III}(y)=0\;\mbox{for}\;\nu=1\,)\ .
\end{gather}

Next, one could substitute this new form $\mathcal{A}_\nu^{\sf III}(y)$ (\ref{eq:e3}) into the equation (\ref{eq:e1}) and would then find the desired anisotropy correction $E_{\rm an}^{\rm\{e\}}$ being expressed in terms of $\mathcal{A}_1^{\sf III}(y)$ and $\mathcal{B}_\nu^{\sf III}(y)$. We do not present here that equation, but rather prefer to calculate the corresponding mass equivalent $\tilde{\mathbb{M}}_{\rm an}^{\rm\{e\}}c^2$ (\ref{eq:e2}), namely on behalf of its more convenient handling. Indeed, if we introduve two auxiliary objects $\mathbb{K}_\nu$ and $\mathbb{L}_\nu$ through
\begin{subequations}
  \begin{align}
  \label{eq:e6a}
  \mathbb{K}_\nu&\doteqdot\int_0^\infty dy\,y^{2\nu+1}\,{\rm e}^{-y}\cdot\mathcal{B}_\nu^{\sf III}(y)\\
  \label{eq:e6b}
  \mathbb{L}_\nu&\doteqdot\frac{\pi}{40}\frac{1}{(2\nu-2)!}\int_0^\infty dy\,y^{2\nu+1}\,{\rm e}^{-y}\cdot\mathcal{A}_1^{\sf III}(y)\ ,
  \end{align}
\end{subequations}
then the mass equivalent $\tilde{\mathbb{M}}_{\rm an}^{\rm\{e\}}c^2$ (\ref{eq:e2}) reads in terms of these newly introduced objects
\begin{eqnarray}
  \label{eq:e7}
  \tilde{\mathbb{M}}_{\rm an}^{\rm\{e\}}c^2&=& -e^2\beta\cdot\frac{(2\nu+3)(2\nu+2)}{(2\nu+1)!}\left\{\frac{(2\nu-2)!}{8}\cdot\mathbb{L}_\nu-\frac{5}{8}\cdot\mathbb{K}_\nu\right\}\\
  \nonumber
  &\doteqdot&-e^2\beta\cdot\mu_{\rm an}^{\rm\{e\}}(\nu)\ .
\end{eqnarray}
The straightforward integration yields
\begin{subequations}
  \begin{align}
  \label{eq:e8a}
  \mathbb{K}_\nu&= \sum_{n=0}^{2\nu-3}\frac{n!}{(n+6)!}\sum_{m=0}^n\frac{1}{2^{2\nu+4+n-m}}\cdot\frac{(2\nu+3+n-m)!}{(n-m)!}\\
  \label{eq:e8b}
  \mathbb{L}_\nu&=1-\sum_{n=0}^4\frac{1}{n!\,2^{2\nu-1+n}}\cdot\frac{(2\nu-2+n)!}{(2\nu-2)!}\ .
  \end{align}
\end{subequations}

The subsequent table presents the value of the ``anisotropic'' mass equivalent
$\tilde{\mathbb{M}}_{\rm an}^{\rm\{e\}}c^2$ (\ref{eq:e7}) for some representative values
of the variational parameter $\nu$. Since the ``anisotropic'' Poisson identity
$\tilde{\mathbb{N}}_{\rm an}^{\rm\{e\}}$ vanishes for our exact solution
$\mathcal{A}_\nu^{\sf III}(y)$ (\ref{eq:e3})-(\ref{eq:e4}) of the (dimensionless)
quadrupole equation (\ref{eq:d3}), this table simultaneously displays the ``anisotropic''
gauge field energy $E_{\rm an}^{\rm\{e\}}$ (\ref{eq:e1}) for those specified values of
$\nu$. This then admits to determine also the total gauge field energy $E_{\rm
  R}^{\rm\{e\}}$ (\ref{eq:v.21}) as the sum of its ``isotropic'' part $E_{\rm R}^{\rm[e]}$
(\ref{eq:v.15}) and its ``anisotropic'' counterpart $E_{\rm an}^{\rm\{e\}}$ (\ref{eq:e1}):
\begin{subequations}
  \begin{align}
  \nonumber
  E_{\rm R}^{\rm\{e\}}=E_{\rm R}^{\rm[e]}+E_{\rm an}^{\rm\{e\}}&=-(e^2\beta)\left\{2\varepsilon_{\rm pot}(\nu)+\mu_{\rm an}^{\rm\{e\}}(\nu)\right\}\\
  \label{eq:e9a}
  &=-e^2\beta\cdot2\varepsilon_{\rm tot}^{\{e\}}(\nu)\\
  \label{eq:e9b}
  \varepsilon_{\rm tot}^{\{e\}}(\nu)&\doteqdot\varepsilon_{\rm pot}(\nu)+\frac{1}{2}\,\mu_{\rm an}^{\rm\{e\}}(\nu)\ .
  \end{align}
\end{subequations}
In order to better estimate the magnitude of the anisotropy corrections, it is useful to introduce the relative magnitude $\Delta_{\rm an}^{\rm\{e\}}(\nu)$ through
\begin{equation}
  \label{eq:e10}
  \Delta_{\rm an}^{\rm\{e\}}(\nu)\doteqdot\frac{E_{\rm an}^{\rm\{e\}}}{E_{\rm
      R}^{\rm\{e\}}}=\frac{\mu_{\rm an}^{\rm\{e\}}(\nu)}
  {2\varepsilon_{\rm tot}^{\{e\}}(\nu)}=\frac{\mu_{\rm an}^{\rm\{e\}}(\nu)}{2\varepsilon_{\rm pot}(\nu)+\mu_{\rm an}^{\rm\{e\}}(\nu)}\ .
\end{equation}

\begin{flushleft}
\footnotesize
  \begin{tabular}{|c||*{6}{@{\!\:\,}c@{\!\:\,}|}}
  \hline
 & $\mathbb{K}_\nu$ & $\mathbb{L}_\nu$ & $\mu_{\rm an}^{\rm\{e\}}(\nu)$ & $2\cdot\varepsilon_{\rm pot}(\nu)$ & $2\cdot\varepsilon_{\rm tot}^{\rm\{e\}}(\nu)$ & $\Delta_{\rm an}^{\rm\{e\}}(\nu)\,[\%]$ \\
\raisebox{2.3ex}[-2.3ex]{$\nu$} & (\ref{eq:e8a}) & (\ref{eq:e8b}) & (\ref{eq:e7}) & (\ref{eq:a28}) & (\ref{eq:e9b}) & (\ref{eq:e10}) \\
  \hline\hline

 &  & $\frac{1}{32}$ & $\frac{5}{384}$ & $\frac{11}{24}$ & $\frac{181}{384}$ & \\
\raisebox{2.3ex}[-2.3ex]{1}
 &  & $=3.125\cdot10^{-2}$ & $\simeq1.3020\cdot10^{-2}$ & $\simeq4.5833\cdot10^{-1}$ & $\simeq4.7135\cdot10^{-1}$ & \raisebox{2.3ex}[-2.3ex]{2.76} \\ \hline

 & $\frac{1}{128}$ & $\frac{7}{64}$ & $\frac{45}{4096}$ & $\frac{93}{256}$ & $\frac{1533}{4096}$ & \\
\raisebox{2.3ex}[-2.3ex]{$\frac{3}{2}$} 
 & $\simeq7.8125\cdot10^{-4}$ & $=1.09375\cdot10^{-1}$ & $\simeq1.09863\cdot10^{-2}$ & $\simeq3.6328\cdot10^{-1}$ & $\simeq3.7426\cdot10^{-1}$ & \raisebox{2.3ex}[-2.3ex]{2.93} \\ \hline

 & $\frac{3}{64}$ & $\frac{29}{128}$ & $\frac{49}{5120}$ & $\frac{193}{640}$ & $\frac{1593}{5120}$ & \\
\raisebox{2.3ex}[-2.3ex]{2}
 & $\simeq4.6875\cdot10^{-2}$ & $\simeq2.2656\cdot10^{-1}$ & $\simeq9.5703\cdot10^{-3}$ & $\simeq3.0156\cdot10^{-1}$ & $\simeq3.1113\cdot10^{-1}$ & \raisebox{2.3ex}[-2.3ex]{3.07} \\ \hline

 & $\frac{267}{1024}$ & $\frac{93}{256}$ & $\frac{2093}{245760}$ & $\frac{793}{3072}$ & $\frac{65533}{245760}$ & \\
\raisebox{2.3ex}[-2.3ex]{$\frac{5}{2}$}
 & $\simeq2.6074\cdot10^{-1}$ & $\simeq3.6325\cdot10^{-1}$ & $\simeq8.5164\cdot10^{-3}$ & $\simeq2.58138\cdot10^{-1}$ & $\simeq2.66654\cdot10^{-1}$ & \raisebox{2.3ex}[-2.3ex]{3.19} \\ \hline

 & $\frac{1575}{1024}$ & $\frac{1}{2}$ & $\frac{4413}{573440}$ & $\frac{1619}{7168}$ & $\frac{133933}{573440}$ & \\
\raisebox{2.3ex}[-2.3ex]{3}
 & $\simeq1.53808$ & $=0.5$ & $\simeq7.6956\cdot10^{-3}$ & $\simeq2.25864\cdot10^{-1}$ & $\simeq2.3356\cdot10^{-1}$ & \raisebox{2.3ex}[-2.3ex]{3.29} \\ \hline

4 & 7.0229e+01 &   7.2559e-01 & 6.4899e-03 & 1.8101e-01 & 1.8750e-01 & 3.46\\  \hline 
5 & 4.6795e+03 &   8.6658e-01 & 5.6388e-03 & 1.5124e-01 & 1.5688e-01 & 3.59\\  \hline 
6 & 4.4549e+05 &   9.4077e-01 & 5.0013e-03 & 1.3000e-01 & 1.3500e-01 & 3.70\\  \hline 
7 & 5.8810e+07 &   9.7548e-01 & 4.5034e-03 & 1.1407e-01 & 1.1857e-01 & 3.80\\  \hline 
8 & 1.0442e+10 &   9.9039e-01 & 4.1024e-03 & 1.0167e-01 & 1.0577e-01 & 3.88\\  \hline 
9 & 2.4216e+12 &   9.9640e-01 & 3.7717e-03 & 9.1728e-02 & 9.5500e-02 & 3.95\\  \hline 
10 & 7.1438e+14 &   9.9870e-01 & 3.4938e-03 & 8.3582e-02 & 8.7076e-02 & 4.01\\  \hline 
20 & 4.3817e+43 &   1.0000e+00 & 2.0511e-03 & 4.4495e-02 & 4.6547e-02 & 4.41\\  \hline 
30 & 1.6409e+77 &   1.0000e+00 & 1.4718e-03 & 3.0423e-02 & 3.1895e-02 & 4.61\\  \hline 
40 & 6.9176e+113 &   1.0000e+00 & 1.1543e-03 & 2.3146e-02 & 2.4300e-02 & 4.75\\  \hline 
50 & 5.1832e+152 &   1.0000e+00 & 9.5231e-04 & 1.8692e-02 & 1.9644e-02 & 4.85\\  \hline 
60 & 2.3612e+193 &   1.0000e+00 & 8.1198e-04 & 1.5682e-02 & 1.6494e-02 & 4.92\\  \hline 
70 & 3.2381e+235 &   1.0000e+00 & 7.0856e-04 & 1.3511e-02 & 1.4220e-02 & 4.98\\  \hline 
80 & 8.1332e+278 &   1.0000e+00 & 6.2905e-04 & 1.1870e-02 & 1.2499e-02 & 5.03\\  \hline 
90 & 2.5844e+323 &   1.0000e+00 & 5.6593e-04 & 1.0587e-02 & 1.1153e-02 & 5.07\\  \hline 
100 & 7.8031e+368 &   1.0000e+00 & 5.1458e-04 & 9.5545e-03 & 1.0069e-02 & 5.11\\  \hline 
  \end{tabular}
\end{flushleft}

\newpage

\indent

It is instructive to see from this table that the anisotropy correction in the vicinity of
the first excited state~$\nP=2$ (which has $\nu_*\simeq1.8$, see \textbf{table~1} on
p.~\pageref{table1}) lowers the electrostatic interaction energy $E_{\rm R}^{\rm[e]}$ of
the spherically symmetric approximation by some 3\%. The smallness of this anisotropy
effect is the reason why the spherically symmetric approximation produces preliminary
predictions with acceptable exactness, i.\,e. roughly 8\% deviation for the first excited
state~$\nP=2$ (second line of \textbf{table~1} on p.~\pageref{table1}). Observe, however,
that the present anisotropy corrections ($\simeq3\%$) must improve the RST predictions for
the binding energy by a factor of two (i.\,e. roughly 6\%) because the total binding
energy $\mathbb{E}_{\rm T}$ amounts to only half the value of the interaction energy
$E_{\rm R}^{\rm[e]}$. For this reason, the deviation of the RST prediction for the first
excited state ($n_\wp=2$) becomes reduced from 8.8\% (\textbf{table~1}) to 3\% (see {\bf
  App.\,F}).


  \stepcounter{section}
  \setcounter{equation}{0}
  \begin{center}
  {\textbf{\Large Appendix \Alph{section}:}}\\[2ex]
  \emph{ \textbf{\Large 
  Extremalization Process (\ref{eq:v.38a})-(\ref{eq:v.38b}) for the Anisotropic Configurations 
  }}
  \end{center}
\myappendix{Extremalization Process (\ref{eq:v.38a})-(\ref{eq:v.38b}) for the \\\hspace*{16mm} Anisotropic Configurations}
  \vspace{2ex}
  
  If all constraints are duly regarded, the original energy functional
  $\tilde{\mathbb{E}}_{\rm[T]}$ ultimately consists of two terms only,
  cf. (\ref{eq:v.37}), i.\,e. the kinetic energy $E_{\rm kin}$
  (\ref{eq:iii.52})-(\ref{eq:iii.53}) and the gauge field energy $E_{\rm R}^{\rm\{e\}}$
  (\ref{eq:v.21}) which is the sum of the ``isotropic'' part $E_{\rm R}^{\rm[e]}$
  (\ref{eq:v.14})and the ``anisotropic'' part $E_{\rm an}^{\rm\{e\}}$ (\ref{eq:v.30}). From
  the principal point of view, both kinds of energy $E_{\rm kin}$ an $E_{\rm
    R}^{\rm\{e\}}$ are of equivalent conceptual importance; but from a more technical
  viewpoint the gauge field part $E_{\rm R}^{\rm\{e\}}$ (and especially its anisotropic
  constituent $E_{\rm an}^{\rm\{e\}}$ (\ref{eq:v.30})) requires greater effort for
  its exact calculation (for general $\nu$). Therefore we are satisfied for the moment
  with specifying $E_{\rm an}^{\rm\{e\}}$ for integer $2\nu$ only, see {\bf App.\,E}. But
  this then implies that also the total energy function $\mathbb{E}^{\rm\{IV\}}(\beta,\nu)$
  (\ref{eq:v.39}) and its reduced form $\mathbb{E}_{\{\nu\}}$ (\ref{eq:v.40}) can be
  specified only for integer values of $2\nu$. On the other hand, for the extremalization
  procedure (\ref{eq:iii.40}), as being specified by equations
  (\ref{eq:v.38a})-(\ref{eq:v.38b}), we need a continuously differentiable energy function
  $\mathbb{E}_{\{\nu\}}$ whose stationary points ($\frac{d\mathbb{E}_{\{\nu\}}}{d\nu}=0$)
  determine the wanted energy spectrum. Therefore, as an auxiliary construction, we are
  satisfied for the present purposes with a supporting polynomial (of appropriate rank)
  which has in common with the spectral function $S_\wp^{\rm\{a\}}(\nu)$ (\ref{eq:v.42})
  its values on some suitable integers $2\nu$. The extremalization process
  ($\frac{dS^{\rm\{a\}}_\wp(\nu)}{d\nu}=0$) is then carried through with the help of this
  supporting polynomial.

  For given quantum number $\lP$ ($=0,\,1,\,2,\,3,\,...$) of angular momentum it will not
  be necessary to take into account the values of $S_\wp^{\rm\{a\}}(\nu)$ on all integers
  $2\nu$ but one merely considers a limited number of supporting points (6, say) around
  the expected stationary point of $S_\wp^{\rm\{a\}}(\nu)$. For an estimate of the latter
  stationary point one may resort to \textbf{table~1} (p.~\pageref{table1}) because the
  anisotropy correction will be sufficiently small so that the stationary points due to
  the spherically symmetric approximation will not be shifted too far away. Finally, it
  may also be interesting to see the extent to which the ``isotropic'' energy curve
  $\mathbb{E}_{\rm T}(\nu)$ (\ref{eq:iii.55a})-(\ref{eq:iii.55b}) becomes lowered to the
  ``anisotropic'' $\mathbb{E}^\wp_{\{\nu\}}$ (\ref{eq:v.41}) as consequence of the
  anisotropy effect. To this end, one defines the corresponding relative energy shift
  $\Delta_{\{\nu\}}$ through
\begin{eqnarray}
  \nonumber
  \Delta_{\{\nu\}}&\doteqdot&\frac{\mathbb{E}^\wp_{\{\nu\}}-\mathbb{E}_{\rm
      T}(\nu)}{|\mathbb{E}_{\rm T}(\nu)|}=
  \frac{S_\wp(\nu)-S_\wp^{\rm\{a\}}(\nu)  }{S_\wp(\nu)}\\
  \label{eq:f1}
  &=&\frac{\varepsilon_{\rm pot}^2(\nu)-\varepsilon_{\rm tot}^{\rm\{e\}}(\nu)^2}{\varepsilon_{\rm pot}^2(\nu)}=1-\left(\frac{\varepsilon_{\rm tot}^{\rm\{e\}}(\nu)}{\varepsilon_{\rm pot}(\nu)}\right)^2\ ,
\end{eqnarray}
see the table below.

Perhaps it is instructive to exemplify the proposed extremalization process by considering
in some detail the first excited state $n_\wp=2$ ($\leadsto\lP=1$). In order to write down
the values of the spectral function $S_\wp^{\rm\{a\}}(\nu)$ (\ref{eq:v.42}) on the
integers $2\nu$, one takes over the potential function $\varepsilon_{\rm
  tot}^{\{e\}}(\nu)$ from equations (\ref{eq:e9a})-(\ref{eq:e9b}) of {\bf App.\,E} in
combination with (\ref{eq:e7}), whereas the kinetic function $\varepsilon_{\rm kin}(\nu)$
has already been specified by equation (\ref{eq:iii.53}). In this way, one sets up the
subsequent table for the values of the spectral function $S_\wp^{\rm\{a\}}(\nu)$ (on the
integers of $2\nu$) in the vicinity of its extremal value $\nu_*^{\{2\}}$
($\nu^{[2]}_*\simeq1.7942$ for the first excited state~$(\nP=2)$ in the spherically
symmetric approximation, see \textbf{table~1} on p.~\pageref{table1}). For an evaluation of the
present RST results recall also the conventional prediction (\ref{eq:i.4}), which for the
value of the spectral function due to the first excited state ($n_c=2$) would yield
$\left(\frac{1}{2}\right)^2=0.25$.  (Compare this to the subsequent values of
$S_\wp^{\rm\{a\}}(\nu)$ for~$\nu=\frac{3}{2}$ and~$\nu=2$)

\begin{flushleft}
  \begin{tabular}{|c||c|c|c|c|c|}
  \hline
$\nu$ & 1 & $\frac{3}{2}$ & 2 & $\frac{5}{2}$ & 3 \\ \hline\hline
$S_\wp^{\rm\{a\}}(\nu)$ & $\left(\frac{181}{384}\right)^2$ & $\frac{12}{7}\cdot\left(\frac{1533}{4096}\right)^2$ & $10\cdot\left(\frac{1593}{10240}\right)^2$ & $\frac{10}{3}\cdot\left(\frac{65533}{245760}\right)^2$ & $\frac{84}{5}\cdot\left(\frac{133933}{1146880}\right)^2$ \\
(\ref{eq:v.41}) & $\simeq0.22217$ & $\simeq0.24013$ & $\simeq0.24200$ & $\simeq0.23701$ & $\simeq0.22911$ \\ \hline
$\Delta_{\{\nu\}}$ & & & & & \\
(\ref{eq:f1}) & \raisebox{2.3ex}[-2.3ex]{-5.76\%} & \raisebox{2.3ex}[-2.3ex]{-6.13\%} & \raisebox{2.3ex}[-2.3ex]{-6.44\%} & \raisebox{2.3ex}[-2.3ex]{-6.70\%} & \raisebox{2.3ex}[-2.3ex]{-6.93\%} \\ \hline
  \end{tabular}
\end{flushleft}

It is clearly seen from this table that the ``isotropic'' energy curve $\mathbb{E}_{\rm
  T}(\nu)$ (\ref{eq:iii.55a})-(\ref{eq:iii.55b}) becomes \emph{lowered} in the vicinity of the
first excited state (which has $\nu^{[2]}_*\simeq1.79$) by roughly 6\%; and this evidently
yields an important improvement of the 8.8\% deviation of the spherically symmetric
approximation from the conventional results ($\leadsto$ 2nd line of the \textbf{table~1} on
p.~\pageref{table1}). Recall here also the failure of the former series expansion
(\ref{eq:iv.9}) of the gauge potential $\gklo{p}{A}_0(r,\vartheta)$ as it is expressed by
the table in {\bf App.\,C}. Indeed, in place of the unrealistic {\em raising}\/ of the
energy curve by 8\% we observe now a \emph{lowering} by some 6\%! 

For exemplifying our approximative method of determining the binding energy for the selected
quantum number~$\nP=2$, we construct the supporting polynomial~$S_2(\nu)$ as
\begin{equation}
\begin{split}
  \label{eq:f2}
  S_2(\nu) = &
 - 0.000000111342528740459\cdot {\nu}^{10}+ 0.00000441158546612208\cdot{\nu}^{9}  \\*
& -0.0000783955928235642\cdot{\nu}^{8}+ 0.000825779502976810\cdot{\nu}^{7} \\*
& -0.00574182989858913\cdot{\nu}^{6}+ 0.0277779729669865\cdot{\nu}^{5}-
0.0959826304956328\cdot
{\nu}^{4}\\* 
&+0.238726544780364\cdot{\nu}^{3}- 0.419428837290107\cdot{\nu}^{2}+ 0.460369057103840
\cdot\nu\\*
&+ 0.015702789114075
\end{split}
\end{equation}
which can easily be checked to agree with $S_2^{\rm\{a\}}(\nu)$ (\ref{eq:v.42}) on the
values $\nu=1,\frac{3}{2},2,\frac{5}{2},3$ (see the above table for $S_\wp^{\rm\{a\}}(\nu)$
with the deviation~$\Delta\{\nu\}$). Next, the maximal value of $S_2(\nu)$
(\ref{eq:f2}) is found as
\begin{equation}
  \label{eq:f3}
  S_2\big|_{\rm max} = 0.2424\ldots
\end{equation}
at the value~$\nu^{\{2\}}_*$ of the variational parameter
\begin{equation}
  \label{eq:f4}
  \nu^{\{2\}}_* = 1.83299\ldots
\end{equation}
(see the figure below). Thus the RST prediction for the energy of the first excited
state~($\mathbb{E}_\wp^{\{2\}}$, say) is obtained as
\begin{equation}
  \label{eq:f5}
  \mathbb{E}_\wp^{\{2\}} = -\frac{e^2}{4\aB}\cdot  S_2\big|_{\rm max}  = -6,8029\cdot
  0.2424\ldots\, [\rm eV] = -1,649\ldots\, [\rm eV]\ .
\end{equation}
\enlargethispage{1cm}
\begin{center}
\epsfig{file=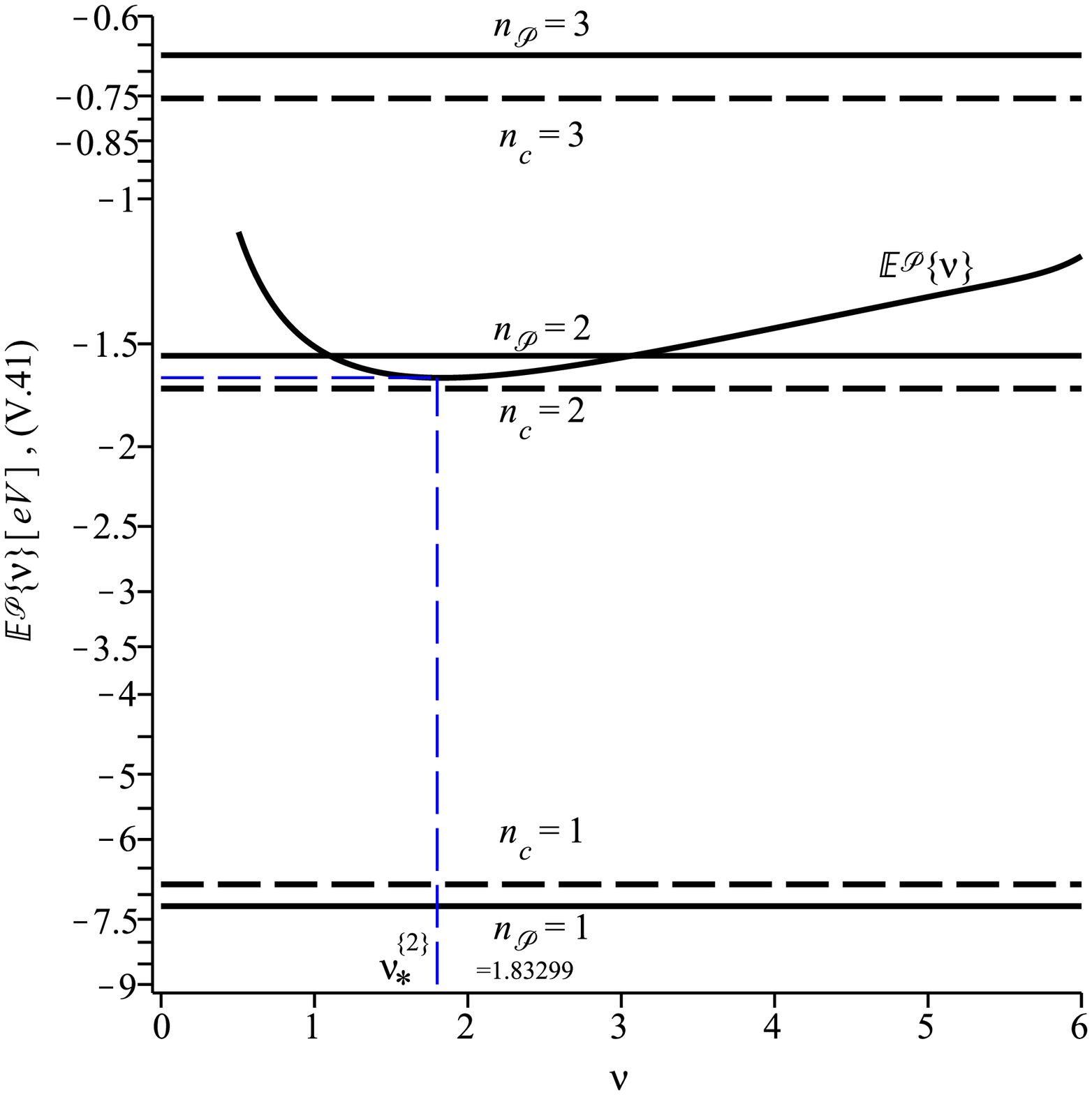,height=12cm}
\end{center}
{\textbf{Fig.~F.I}\hspace{5mm} \emph{\large\textbf{Positronium Energy and Maximal Value of\\
 \centerline{the Spectral Function~\boldmath$S_\wp^{\{a\}}(\nu)$ (\ref{eq:v.42})} 
 }}}
\myfigure{Fig.~F.I: Positronium Energy and Maximal Value of
 the Spectral\\\hspace*{2mm} Function~\boldmath$S_\wp^{\{a\}}(\nu)$ (\ref{eq:v.42})}
\indent

From the smallest (but integer) values of~$2\nu$ (see the table on p.\pageref{tableC}) one
constructs the interpolating polynomial~$S_2(\nu)$ (\ref{eq:f2}) as an approximation to
the spectral function~$S_2^{\{a\}}(\nu)$ (\ref{eq:v.42}). Its maximal value (\ref{eq:f3})
determines the energy~$\mathbb{E}^{\{2\}}_\wp$ (\ref{eq:f5}) of the first excited
state~$(n_\wp=2\leftrightarrow\lP=1)$. This ``anisotropic'' result deviates now from the
corresponding conventional prediction (\ref{eq:i.4}) by only 3\%. Thus the present
anisotropic correction \emph{improves} the ``isotropic'' RST prediction (solid line) from
8,8\% deviation (cf.~\textbf{table~1} on p.~\pageref{table1}) to 3\% deviation (with
respect to the conventional prediction, broken line) which is a reduction of the deviation
of roughly 66\%, see also \mbox{\textbf{table~2} on p.~\pageref{table2} for the case of
  higher excited states.}

\newpage

The 66\% reduction of the deviation (from the conventional results) naturally leads us to the
question of how large this decrease may be for the higher excited states~$(\lP>1)$? In
order to get some feeling for this, one may sketch the ``correction
function''~$\Delta\{\nu\}$~(F.1) over the variational parameter~$\nu$, see \textbf{Fig.F.II}
below. Here it is satisfying to observe a decrease of deviation which ranges from 5,76\%
for~$\nu=1$ up to~11\% for~$\nu=100$. The latter case~$(\nu\simeq 100)$ is due to a
quantum number of~$\nP\simeq 35$ (see table~1). It becomes obvious from this diagram that
the ``anisotropic'' correction~$\Delta\{\nu\}$ is countercurrent to the
``isotropic'' deviation~$\ru{\Delta}{T}[\nu]$ due to the spherically symmetric approximation 
\textbf{Sect.III}, being defined through
\begin{equation}
  \label{eq:f6}
  \ru{\Delta}{T}(\nu_*^{[n]}) \doteqdot
  \frac{\Ea{E}{n}{conv}-\ru{E}{T}(\nu_*^{[n]})}{\Ea{E}{n}{conv}}\ .
\end{equation}
Indeed, this object quantifies the extent of deviation of our ``isotropic'' RST
predictions from their conventional counterpart (\ref{eq:i.4}) and is displayed in the
last column of \textbf{table~1} on p.~\pageref{table1}; and when this deviation
function~$\ru{\Delta}{T}(\nu_*^{[n]})$ (\ref{eq:f6}) is also represented in
\textbf{Fig.F.II}, it becomes evident that for increasing~$\nu$ it runs countercurrently
to the present correction function~$\Delta\{\nu\}$ (\ref{eq:f1}). But this then entails
the consequence that both curves must intersect at a certain value of~$\nu$ ($\nu_x$,
say); and such an intersection means that the conventional and RST predictions will agree
in the vicinity of that~$\nu_x$ (see \textbf{Fig.F.II} below which yields~$\nu_x\simeq 30$
corresponding to quantum number~$\nP\simeq 15$). See also the discussion of this effect
below \textbf{table~2} on p.~\pageref{table2}.

\enlargethispage{1cm}
\setlength{\footskip}{50pt}
\begin{center}
\epsfig{file=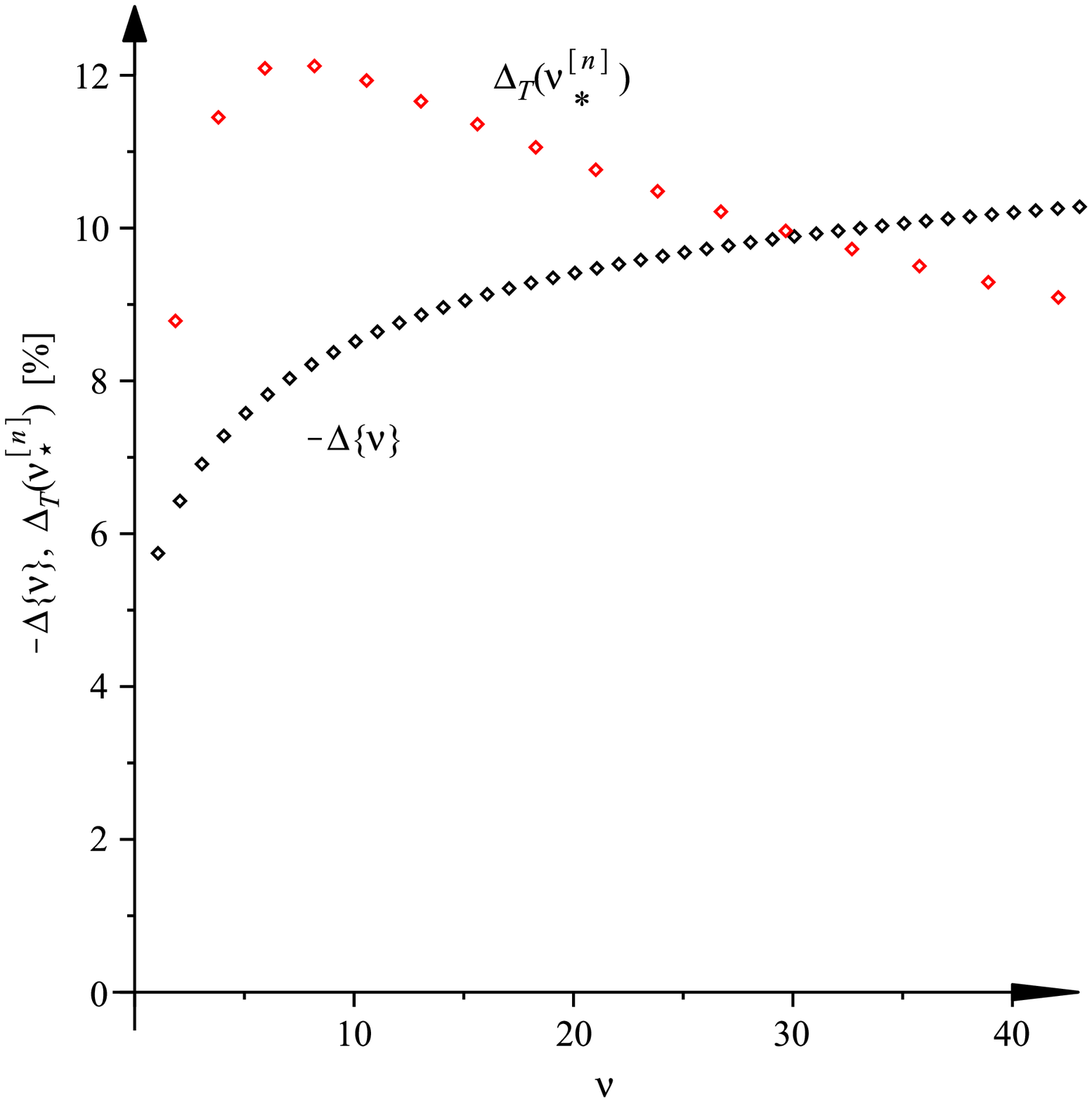,height=12cm}
\end{center}
{\textbf{Fig.~F.II}\hspace{5mm} \emph{\large\textbf{Relative Anisotropy Correction
\boldmath$\Delta\{\nu\}$  (\ref{eq:f1})\\* \phantom{Fig.~F.II} and ``Isotropic'' Deviations
\boldmath$\ru{\Delta}{T}(\nu_*^{[n]})$  (\ref{eq:f6})
}}}
\myfigure{Fig.~F.II: Relative Anisotropy Correction
\boldmath$\Delta\{\nu\}$  (\ref{eq:f1}) and \\\hspace*{2mm} ``Isotropic'' Deviations \boldmath$\ru{\Delta}{T}(\nu_*^{[n]})$  (\ref{eq:f6})}
\indent

The consideration of the anisotropy effect in the first approximation order~$(\sim
{}^{\{p\}}A^{|||}(r) )$ yields a relative lowering of the ``isotropic'' RST
predictions~$\nrf{E}{T}(\nu)$ (\ref{eq:iii.55a}) by~$\Delta\{\nu\}$ percent,
cf.~(\ref{eq:f1}). Since these ``\emph{isotropic}'' predictions yield always higher energy
than their conventional counterparts (\ref{eq:i.4}), their lowering by the anisotropy
effect yields considerable improvement of the RST predictions for the low excited states
(\textbf{table~2} on p.~\pageref{table2}). But since the ``anisotropic''
deviation~$\Delta\{\nu\}$ increases up to 11\% (for~$\nu\to\infty$), the decreasing
``isotropic'' deviation~$\ru{\Delta}{T}(\nu_*^{[n]})$ (last column of \textbf{table~1} on
p.~\pageref{table1}) must become compensated by~$\Delta\{\nu\}$ for some value~$\nu_x$
of~$\nu$ (here~$\nu_x\simeq 30$, i.e.~$\nP\simeq 15$); and in the vicinity of this~$\nu_x$
the RST predictions will then agree with the conventional results, see the discussion
below equ.~(\ref{eq:f6}) and \textbf{table~2} on p.~\pageref{table2}.

\normalsize
  \stepcounter{section}
  \setcounter{equation}{0}
  \begin{center}
  {\textbf{\Large Appendix \Alph{section}:}}\\[2ex]
  \emph{ \textbf{\Large 
  Poisson Identity and Generalized Trial Amplitude 
  }}
  \end{center}
\myappendix{Poisson Identity and Generalized Trial Amplitude }
  \vspace{2ex}
  
  For the numerical exploitation of the ``isotropic'' Poisson identity (\ref{eq:iii.47}) it is very
  convenient to rescale again both the gauge potential $\eklo{p}{A}_0(r)$ and the
  generalized trial amplitude $\tilde{\Phi}(r)$ (\ref{eq:vi.1})-(\ref{eq:vi.3b}). For the
  latter object, one puts
\begin{gather}
  \label{eq:g1}
  \tilde{\Phi}_b(r)=\sqrt{\frac{2}{\pi}}\,\frac{2\beta}{\sqrt{\Gamma(2\nu+2)}} \cdot \tilde{\Phi}_{\nu p}(y) \\
  \nonumber
  (y:=2\beta r)
\end{gather}
so that the dimensionless trial amplitude $\tilde{\Phi}_{\nu p}(y)$ appears in the following form
\begin{equation}
  \label{eq:g2}
  \tilde{\Phi}_{\nu p}(y)=y^\nu  (p_0+p_1y) {\rm e}^{-\frac{y}{2}}\ .
\end{equation}

Next, it is necessary to look for the (exact) solution of the Poisson equation
(\ref{eq:iii.36}) when the trial amplitude $\tilde{\Phi}(r)$ in that equation is given by
the three-parameter ansatz (\ref{eq:vi.1})-(\ref{eq:vi.3b}), i.\,e. in dimensionless form
(\ref{eq:g1})-(\ref{eq:g2}). The corresponding Poisson equation reads explicitly
\begin{gather}
  \label{eq:g3}
  \frac{d^2\tilde{a}_{\nu p}(y)}{dy^2}+\frac{2}{y}\frac{d\tilde{a}_{\nu p}(y)}{dy}=
  -\frac{(p_0^2\cdot y^{2\nu-1}+2p_0p_1\cdot y^{2\nu} + p_1^2\cdot y^{2\nu+1}){\rm e^{-y}}} {\Gamma(2\nu+2)}\\
  \nonumber
  (\eklo{p}{A}_0(r)\doteqdot2\beta\as\,\tilde{a}_{\nu p}(y)\,)\ .
\end{gather}
Obviously, the generalized potential $\tilde{a}_{\nu p}(y)$ collapses here to the former
potential $\tilde{a}_\nu(y)$ (\ref{eq:a8}) if the parameter~$p_1$ tends to zero, because
the present equation (\ref{eq:g3}) becomes in this limit ($p_1\rightarrow0$) identical to
the former Poisson equation (\ref{eq:a1}) for the original ansatz
(\ref{eq:iii.50})-(\ref{eq:iii.51}). Indeed, the present generalized Poisson equation
(\ref{eq:g3}) is closely related to its precursor (\ref{eq:a1}). This becomes readily more
obvious by exploiting the fact that (\ref{eq:g3}) is a linear inhomogeneous equation where
the inhomogeneity is essentially a quadratic polynomial with respect to the variational
parameters $p_0,p_1$. For this reason, the solution $a_{\nu p}(y)$ may be written as a
superposition of three solutions $a_\nu(y)$ of the simple precursor (\ref{eq:a1}):
\begin{equation}
  \label{eq:g4}
  \tilde{a}_{\nu p}(y)= - \frac{p_0^2\cdot a_\nu(y) + 2p_0p_1\cdot a_{\nu+\frac{1}{2}}(y)
    + p_1^2\cdot a_{\nu+1}(y) } {\Gamma(2\nu+2) }\ ,
\end{equation}
where the functions $a_\nu(y)$ are (exact) solutions of the (slightly modified) precursor equation (\ref{eq:a1})
\begin{equation}
  \label{eq:g5}
  \frac{d^2a_\nu(y)}{dy^2}+\frac{2}{y}\frac{da_\nu(y)}{dy}=y^{2\nu-1}\cdot{\rm e}^{-y}\ .
\end{equation}

Quite analogously to the former situation (\ref{eq:a17}) one introduces here the dimensionless field strength $\tilde{e}_{\nu p}(y)$ through
\begin{equation}
  \label{eq:g6}
  \tilde{e}_{\nu p}(y)\doteqdot\frac{d\tilde{a}_{\nu p}(y)}{dy}
\end{equation}
which then is subjected to the source equation
\begin{equation}
  \label{eq:g7}
  \frac{d\tilde{e}_{\nu p}(y)}{dy}+\frac{2}{y}\,\tilde{e}_{\nu p}(y) =
  -\frac{(p_0^2\cdot y^{2\nu-1}+2p_0p_1\cdot y^{2\nu}+p_1^2\cdot y^{2\nu+1}){\rm e}^{-y} } {\Gamma(2\nu+2) }\ .
\end{equation}
This is the same situation as with the Poisson equation (\ref{eq:g3}) so that the field strength $\tilde{e}_{\nu p}(y)$ can also be written as a sum of three terms, i.\,e. we may put
\begin{equation}
  \label{eq:g8}
  \tilde{e}_{\nu p}(y)=-\frac{p_0^2\cdot e_\nu(y) + 2p_0p_1\cdot e_{\nu+\frac{1}{2}}(y) +
    p_1^2\cdot e_{\nu+1}(y) } {\Gamma(2\nu+2)}\ ,
\end{equation}
so that the simple field strength $e_\nu(y)$ obeys the source equation
\begin{equation}
  \label{eq:g9}
  \frac{de_\nu(y)}{dy}+\frac{2}{y}\,e_\nu(y)=y^{2\nu-1}\cdot{\rm e}^{-y}\ .
\end{equation}

Comparing now this source equation for the field strength $e_\nu(y)$ to the Poisson equation (\ref{eq:g5}) for the potentials $a_\nu(y)$ one recovers the field strengths $e_\nu(y)$ as the simple derivatives of the corresponding potentials $a_\nu(y)$, i.\,e.
\begin{equation}
  \label{eq:g10}
  e_\nu(y)=\frac{da_\nu(y)}{dy}\ .
\end{equation}

Consequently, since the solutions of the Poisson equation (\ref{eq:g5}) can easily be obtained for integer $2\nu$, cf. (\ref{eq:a7})
\begin{equation}
  \label{eq:g11}
  a_\nu(y)=-\frac{(2\nu+1)!}{y}\left\{1-{\rm e}^{-y}\cdot\sum_{n=0}^{2\nu}\frac{2\nu+1-n}{2\nu+1}\cdot\frac{y^n}{n!}\right\}\,,
\end{equation}
and the solutions of the source equation (\ref{eq:g9}) for the field strengths $e_\nu(y)$
are obtained from this result by the simple differentiation process (\ref{eq:g10}) as
\begin{equation}
  \label{eq:g12}
  e_\nu(y)=\frac{(2\nu+1)!}{y^2}\left\{1-{\rm e}^{-y}\cdot\sum_{n=0}^{2\nu+1}\frac{y^n}{n!}\right\}\,,
\end{equation}
cf. (\ref{eq:a18}). Clearly, the latter two results for integer $2\nu$ may be generalized to arbitrary real values of the variational parameter $\nu$ and do then appear in the following form:
\begin{subequations}
  \begin{align}
  \label{eq:g13a}
  a_\nu(y)&=-\Gamma(2\nu+1)\cdot\left\{1-{\rm e}^{-y}\cdot\sum_{n=0}^{\infty}\frac{n}{\Gamma(2\nu+2+n)}\,y^{2\nu+n}\right\}\\
  \label{eq:g13b}
  e_\nu(y)&=\Gamma(2\nu+2)\,{\rm e}^{-y}\cdot\sum_{n=0}^{\infty}\frac{y^{2\nu+n}}{\Gamma(2\nu+3+n)}\ ,
  \end{align}
\end{subequations}
cf. (\ref{eq:a8}) and (\ref{eq:a19}).

But now that both the generalized potentials $\tilde{a}_{\nu p}(y)$ (\ref{eq:g4}) and $\tilde{e}_{\nu p}(y)$ (\ref{eq:g8}) are explicitly known, one can realize the Poisson identity (\ref{eq:iii.47}) within the present framework of the generalized trial ansatz (\ref{eq:vi.1})-(\ref{eq:vi.2}). First, consider the gauge field energy $E_{\rm R}^{\rm[e]}$ (\ref{eq:iii.45}) which may be rewritten in terms of the dimensionless objects as follows
\begin{eqnarray}
  \label{eq:g14}
  E_{\rm R}^{\rm[e]}&=&-\hbar c\,(2\as\beta)\int_0^\infty dy\,y^2\left(\frac{d\tilde{a}_{\nu p}(y)}{dy}\right)^2= -\hbar c(2\as\beta)\int_0^\infty dy\,y^2\tilde{e}_{\nu p}^2(y)\ .
\end{eqnarray}
Inserting here the dimensionless field strengths $\tilde{e}_{\nu p}(y)$ (\ref{eq:g8}) recasts the gauge field energy $E_{\rm R}^{\rm[e]}$ into the following shape
\begin{equation}
  \label{eq:g15}
  E_{\rm R}^{\rm[e]}=-\frac{e^2}{\aB}\,(2\beta\aB)\cdot\varepsilon_{\rm pot}(\nu,p_0,p_1)\ .
\end{equation}
This is indeed a result which looks very similar to the former case (\ref{eq:iii.54}) for
the simpler trial amplitude (\ref{eq:iii.50})-(\ref{eq:iii.51}); but the present potential
function $\varepsilon_{\rm pot}(\nu,p_0,p_1)$ emerges now as a polynomial of fourth order with
respect to the additional parameters $p_0$ and~$p_1$, i.\,e.
\begin{equation}
  \label{eq:g16}
  \varepsilon_{\rm pot}(\nu,p_0,p_1)=\varepsilon_{\rm pot}(\nu)\cdot p_0^4 + 
4p_0^3p_1\cdot\varepsilon_{\rm I}(\nu) + 4p_0^2p_1^2\cdot\varepsilon_{\rm II}(\nu) +
4p_0p_1^3\cdot\varepsilon_{\rm III}(\nu) + p_1^4\cdot\varepsilon_{\rm IV}(\nu)\ ,
\end{equation}
with the potential functions being given by
\begin{subequations}
  \begin{align}
  \label{eq:g17a}\vphantom{\Bigg|}
  \varepsilon_{\rm pot}(\nu)&=\int_0^\infty dy\,y^2\left(\frac{e_\nu(y)}{\Gamma(2\nu+2)}\right)^2\\
  \label{eq:g17b}\vphantom{\Bigg|}
  \varepsilon_{\rm I}(\nu)&=\int_0^\infty dy\,y^2\,\frac{e_\nu(y)\cdot e_{\nu+\frac{1}{2}}(y)}{\Gamma(2\nu+2)^2}\\
  \label{eq:g17c}\vphantom{\Bigg|}
  \varepsilon_{\rm II}(\nu)&=\int_0^\infty dy\,y^2\,\frac{e_{\nu+\frac{1}{2}}(y)\cdot e_{\nu+\frac{1}{2}}(y)+\frac{1}{2}\,e_\nu(y)\cdot e_{\nu+1}(y)}{\Gamma(2\nu+2)^2}\\
  \label{eq:g17d}\vphantom{\Bigg|}
  \varepsilon_{\rm III}(\nu)&=\int_0^\infty dy\,y^2\,\frac{e_{\nu+\frac{1}{2}}(y)\cdot e_{\nu+1}(y)}{\Gamma(2\nu+2)^2}\\
  \label{eq:g17e}\vphantom{\Bigg|}
  \varepsilon_{\rm IV}(\nu)&=\int_0^\infty dy\,y^2\left(\frac{e_{\nu+1}(y)}{\Gamma(2\nu+2)}\right)^2\,.
  \end{align}
\end{subequations}

Since the field strengths $e_\nu(y)$ are explicitly known, all the potential functions (\ref{eq:g17a})-(\ref{eq:g17e}) can now be explicitly calculated. For the first one, i.\,e. (\ref{eq:g17a}), one finds
\begin{eqnarray}
  \label{eq:g18}
  \varepsilon_{\rm pot}(\nu)&=& \frac{1}{2^{4\nu+3}}\sum_{m,n=0}^\infty\frac{1}{2^{m+n}}\cdot\frac{\Gamma(4\nu+3+m+n)}{\Gamma(2\nu+3+m)\cdot\Gamma(2\nu+3+n)}\\
  \nonumber
  &\stackrel{2\nu\,\rm int.\vphantom{\big|}}{\Longrightarrow}& \sum_{m,n=0\atop(m+n\geq2)}^{2\nu+1}\frac{1}{2^{m+n-1}}\cdot\frac{(m+n-2)!}{m!\,n!}-2\sum_{n=2}^{2\nu+1}\frac{(n-2)!}{n!}\ .
\end{eqnarray}
Here, the result in the first line refers to general (real) values of the variational
parameter $\nu$ and the second line presents the specialization to integer values of
$2\nu$. Clearly, this first potential function $\varepsilon_{\rm pot}(\nu)$ (\ref{eq:g18})
is nothing else than the corresponding result (\ref{eq:a28}) for the simple trial function
(\ref{eq:iii.50}) which has $p_1=0$; and therefore the present generalized potential
function $\varepsilon_{\rm pot}(\nu,p_0,p_1)$ (\ref{eq:g16}) must of course collapse to
$\varepsilon_{\rm pot}(\nu)$ (\ref{eq:a28}) for $p_1\rightarrow0$.

But for non-vanishing variational parameter $p_1$, the other potential functions
$\varepsilon_{\rm I},\,...,\,\varepsilon_{\rm IV}$ (\ref{eq:g17b})-(\ref{eq:g17e}) yield
important contributions and must be explicitly known for the extremalization process
(\ref{eq:iii.40}) due to the {\em principle of minimal energy}\/ (the role played by the
third variational parameter $\nu_2$ is played here by the parameter $p_1$,
cf. (\ref{eq:vi.1})-(\ref{eq:vi.3b})). Thus one finds by explicit integration for the first
\mbox{additional potential function $\varepsilon_{\rm I}(\nu)$ (\ref{eq:g17b}):}
\begin{eqnarray}
  \label{eq:g19}
  \varepsilon_{\rm I}(\nu)&=& \frac{2\nu+2}{2^{4\nu+4}}\sum_{m,n=0}^\infty\frac{1}{2^{m+n}}\cdot\frac{\Gamma(4\nu+4+m+n)}{\Gamma(2\nu+3+m)\cdot\Gamma(2\nu+4+n)}\\
  \nonumber
  &\stackrel{2\nu\,\rm int.\vphantom{\big|}}\Longrightarrow& (2\nu+2)\cdot\varepsilon_{\rm pot}(\nu) 
  - \frac{1}{2\nu+1}\left(
    1-\frac{1}{(2\nu)!\,2^{2\nu+1}}\cdot\sum_{n=0}^{2\nu+1}\frac{1}{2^n}
    \cdot\frac{(2\nu+n)!}{n!}\right)\ .
\end{eqnarray}

Next the second potential function $\varepsilon_{\rm II}(\nu)$ (\ref{eq:g17c}) obviously splits up into two contributions where the first one is closely related to the original function $\varepsilon_{\rm pot}(\nu)$ (\ref{eq:g17a}) with a shift $\nu\rightarrow\nu+\frac{1}{2}$ so that this second function appears as
\begin{subequations}
  \begin{align}
  \label{eq:g20a}
  \varepsilon_{\rm II}(\nu)&=\varepsilon'_{\rm II}(\nu)+\varepsilon''_{\rm II}(\nu)\\
  \label{eq:g20b}
  \varepsilon'_{\rm II}(\nu)&\doteqdot \int dy\,y^2\,\frac{e_{\nu+\frac{1}{2}}(y)\cdot e_{\nu+\frac{1}{2}}(y)}{\Gamma(2\nu+2)^2}=(2\nu+2)^2\cdot\varepsilon_{\rm pot}(\nu+\frac{1}{2})\vphantom{\Bigg|}\\
  \nonumber
  \varepsilon''_{\rm II}(\nu)&\doteqdot \frac{1}{2}\int dy\,y^2\,\frac{e_\nu(y)\cdot e_{\nu+1}(y)}{\Gamma(2\nu+2)^2}\vphantom{\Bigg|}\\
  \label{eq:g20c}
  &= \frac{(\nu+1)(2\nu+3)}{2^{4\nu+5}}\sum_{m,n=0}^\infty\frac{1}{2^{m+n}}\cdot\frac{\Gamma(4\nu+5+m+n)}{\Gamma(2\nu+5+m)\,\Gamma(2\nu+3+n)}\\
  \nonumber
  \stackrel{2\nu\,\rm int.\vphantom{\big|}}{\Longrightarrow}& \,(\nu+1)(2\nu+3)\!\left\{\sum_{m,n=0\atop(m+n\geq2)}^{n=2\nu+3\atop m=2\nu+1}\frac{1}{2^{m+n-1}}\cdot\frac{(m+n-2)!}{m!\,n!}- \sum_{n=2}^{2\nu+1}\frac{1}{n(n-1)}-\sum_{n=2}^{2\nu+3}\frac{1}{n(n-1)}\right\}.
  \end{align}
\end{subequations}

A similar situation does occur also for the third potential $\varepsilon_{\rm III}(\nu)$ (\ref{eq:g17d}). Indeed, comparing this to the first potential $\varepsilon_{\rm I}(\nu)$ (\ref{eq:g17b}) one easily realizes that both potentials differ essentially by the shift $\nu\rightarrow\nu+\frac{1}{2}$, i.\,e. one finds
\begin{equation}
  \label{eq:g21}
  \varepsilon_{\rm III}(\nu)=(2\nu+2)^2\cdot\varepsilon_{\rm I}(\nu+\frac{1}{2})\ .
\end{equation}
And finally, such a shift $\nu\rightarrow\nu+1$ of the variational parameter $p$ is also observed in connection with the original and the fourth potentials $\varepsilon_{\rm pot}(\nu)$ (\ref{eq:g17a}) and $\varepsilon_{\rm IV}(\nu)$ (\ref{eq:g17e}) so that one arrives at the following result:
\begin{equation}
  \label{eq:g22}
  \varepsilon_{\rm IV}(\nu)=(2\nu+3)^2(2\nu+2)^2\cdot\varepsilon_{\rm pot}(\nu+1)\ .
\end{equation}

The generalized potential function $\varepsilon_{\rm pot}(\nu,p)$ (\ref{eq:g16}) is an
important object for the calculation of the positronium spectrum since it is the essential
constituent of the gauge field energy $E_{\rm R}^{\rm[e]}$ (\ref{eq:g15}). The point here
is that $E_{\rm R}^{\rm[e]}$ must obey the Poisson identity (\ref{eq:iii.47}); otherwise
the energy functional $\tilde{\mathbb{E}}_{[\Phi]}$ (\ref{eq:iii.41}) could not be reduced
to its physical part $\mathbb{E}^{\rm[IV]}$ (\ref{eq:iii.54'}) alone which consists solely
of the kinetic and potential energy (recall here that for the approximative treatment of
the anisotropic configurations in {\bf Sect.\,IV} the Poisson identity is not satisfied so
that the energy functional must embrace also the Poisson constraint term in addition to
the kinetic and potential energy, cf. (\ref{eq:iv.29})). Therefore it is a satisfying
reinsurance to see the Poisson identity being satisfied here even for the
\emph{approximative}~(!)  configurations $\tilde{\Phi}_{\nu p}(y)$ (\ref{eq:g2}) and its
associated potential $\tilde{a}_{\nu p}(y)$ (\ref{eq:g4}).

In its rescaled form, the mass equivalent $\tilde{\mathbb{M}}^{\rm[e]}c^2$ (\ref{eq:iii.48}) reads
\begin{equation}
  \label{eq:g23}
  \tilde{\mathbb{M}}^{\rm[e]}c^2=-\frac{\hbar c\,(2\as\beta)}{\Gamma(2\nu+2)}\int_0^\infty dy\,y\,\tilde{a}_{\nu p}(y)\cdot\tilde{\Phi}^2_{\nu p}(y)
\end{equation}
with the trial amplitude $\tilde{\Phi}_{\nu p}(y)$ being given by equation (\ref{eq:g2})
and the corresponding gauge potential $\tilde{a}_{\nu p}(y)$ by equation
(\ref{eq:g4}). Thus, inserting the latter two objects into the present mass equivalent
$\tilde{\mathbb{M}}^{\rm[e]}c^2$ (\ref{eq:g23}) yields obviously a polynomial of fourth
order with respect to the variational parameters $p_0,p_1$, just as is the case with the gauge
field energy $E_{\rm R}^{\rm[e]}$ (\ref{eq:g15})-(\ref{eq:g16}). The Poisson identity
(\ref{eq:iii.47}) demands now that both fourth-order polynomials must be identical which
implies the identity of the coefficients in front of the various products of the parameters
$p_0,p_1$.

First, the lowest-order identification ($\sim p_0^4$) lets emerge the zero-order function
$\varepsilon_{\rm pot}(\nu)$ (\ref{eq:g17a}) in terms of the potential $a_\nu(y)$ in the
following alternative way
\begin{equation}
  \label{eq:g24}
  \varepsilon_{\rm pot}(\nu)=-\frac{1}{\Gamma(2\nu+2)^2}\int dy\,y^{2\nu+1}\,{\rm e}^{-y}\cdot a_\nu(y)\ ,
\end{equation}
i.\,e. for integer $2\nu$
\begin{equation}
  \label{eq:g25}
  \varepsilon_{\rm pot}(\nu)= \frac{1}{2\nu+1}\left\{1-\frac{1}{(2\nu+1)!\,2^{2\nu+1}}\sum_{n=0}^{2\nu}\frac{2\nu+1-n}{2^n}\cdot\frac{(2\nu+n)!}{n!}\right\}
\end{equation}
whose generalization to arbitrary $\nu$ is then given by equation (\ref{eq:a28}).

Next, the identification of the linear terms ($\sim p_0^3p_1$) lets emerge the first-order function
$\varepsilon_{\rm I}(\nu)$ (\ref{eq:g17b}) alternatively in terms of the potential
$a_\nu(y)$ as follows
\begin{eqnarray}
  \label{eq:g26}
  \varepsilon_{\rm I}(\nu)&=&-\frac{1}{\Gamma(2\nu+2)^2}\int dy\,y^{2\nu+2}\,{\rm e}^{-y}\cdot a_\nu(y)\\
  \nonumber
  &=&-\frac{1}{\Gamma(2\nu+2)^2}\int dy\,y^{2\nu+1}\,{\rm e}^{-y}\cdot a_{\nu+\frac{1}{2}}(y)\ ,
\end{eqnarray}
i.\,e. for integer $2\nu$:
\begin{eqnarray}
  \label{eq:g27}
  \varepsilon_{\rm I}(\nu)&=& 1-\frac{1}{(2\nu+1)\cdot(2\nu+1)!}\frac{1}{2^{2\nu+2}}\sum_{n=0}^{2\nu}\frac{2\nu+1-n}{2^n}\cdot\frac{(2\nu+1+n)!}{n!}\\
  \nonumber
  &=& \frac{2\nu+2}{2\nu+1}\left\{1-\frac{2\nu+1}{(2\nu+2)!\,2^{2\nu+1}}\sum_{n=0}^{2\nu+1}\frac{2\nu+2-n}{2^n}\cdot\frac{(2\nu+n)!}{n!}\right\}\,,
\end{eqnarray}
or more generally for arbitrary $\nu$:
\begin{eqnarray}
  \label{eq:g28}
  \varepsilon_{\rm I}(\nu)&=& \frac{2\nu+2}{2\nu+1}\left\{1-\frac{1}{\Gamma(2\nu+3)\cdot2^{4\nu+3}}\sum_{n=0}^\infty\frac{n}{2^n}\cdot\frac{\Gamma(4\nu+3+n)}{\Gamma(2\nu+2+n)}\right\}\\
  \nonumber
  &=& 1-\frac{1}{\Gamma(2\nu+2)\cdot2^{4\nu+3}}\sum_{n=0}^\infty\frac{n}{2^n}\cdot\frac{\Gamma(4\nu+3+n)}{\Gamma(2\nu+3+n)}\ .
\end{eqnarray}

Furthermore, the identification of the quadratic terms ($\sim p_0^2\cdot p_1^2$) confirms the former
splitting (\ref{eq:g20a}) into two parts $\varepsilon'_{\rm II}(\nu)$ and
$\varepsilon''_{\rm II}(\nu)$ with the first part $\varepsilon'_{\rm II}(\nu)$ being given
by equation (\ref{eq:g20b}). The second part $\varepsilon''_{\rm II}(\nu)$ is found in
terms of the potential $a_\nu(y)$ as follows
\begin{eqnarray}
  \label{eq:g29}
  \varepsilon''_{\rm II}(\nu)&=& -\frac{1}{2\cdot\Gamma(2\nu+2)^2}\int dy\,y^{2\nu+3}{\rm e}^{-y}\cdot a_\nu(y)\\
  \nonumber
  &=& -\frac{1}{2\cdot\Gamma(2\nu+2)^2}\int dy\,y^{2\nu+1}{\rm e}^{-y}\cdot a_{\nu+1}(y)\ ,
\end{eqnarray}
i.\,e. explicitly for integer $2\nu$
\begin{eqnarray}
  \label{eq:g30}
  \varepsilon''_{\rm II}(\nu)&=& (\nu+1)\left\{1-\frac{1}{(2\nu+2)!\cdot2^{2\nu+3}}\cdot\sum_{n=0}^{2\nu}\frac{1}{2^n}\frac{2\nu+1-n}{2\nu+1}\cdot\frac{(2\nu+2+n)!}{n!}\right\}\qquad\\
  \nonumber
  &=& \frac{(\nu+1)(2\nu+3)}{2\nu+1}\left\{1-\frac{1}{(2\nu)!\,2^{2\nu+1}}\cdot\sum_{n=0}^{2\nu+2}\frac{1}{2^n}\frac{2\nu+3-n}{2\nu+3}\cdot\frac{(2\nu+n)!}{n!}\right\}\,.
\end{eqnarray}
For arbitrary $\nu$, this result is generalized to
\begin{eqnarray}
  \label{eq:g31}
  \varepsilon''_{\rm II}(\nu)&=& \frac{(\nu+1)(2\nu+3)}{2\nu+1}\left\{1-\frac{1}{\Gamma(2\nu+4)\cdot2^{4\nu+4}}\cdot\sum_{n=0}^\infty\frac{n}{2^n}\cdot\frac{\Gamma(4\nu+4+n)}{\Gamma(2\nu+2+n)}\right\}\qquad\\
  \nonumber
  &=& (\nu+1)\left\{1-\frac{1}{\Gamma(2\nu+2)\cdot2^{4\nu+4}}\cdot\sum_{n=0}^\infty\frac{n}{2^n}\cdot\frac{\Gamma(4\nu+4+n)}{\Gamma(2\nu+4+n)}\right\}\,.
\end{eqnarray}

Finally, the last two auxiliary functions $\varepsilon_{\rm III}(\nu)$ (\ref{eq:g17d}) and
$\varepsilon_{\rm IV}(\nu)$ (\ref{eq:g17e}) are recovered by the identification process to
undergo the former relationships (\ref{eq:g21})-(\ref{eq:g22}) and therefore need not be
reproduced explicitly once more. Thus the overall result is that the total potential
function $\varepsilon_{\rm pot}(\nu,p)$ is built up by only {\em three}\/ ``independent''
auxiliary potential functions $\varepsilon_{\rm pot}(\nu)$ (\ref{eq:a28}),
$\varepsilon_{\rm I}(\nu)$ (\ref{eq:g28}) and $\varepsilon''_{\rm II}(\nu)$
(\ref{eq:g31}). The subsequent table displays the values of these auxiliary functions on the
relevant values of the variational parameter $\nu$.

\pagebreak
\begin{landscape}
{\scriptsize
\begin{flushright}
  \begin{tabular}{|c||c|c|c|c|c|}
  \hline
$\nu$ & $\boldsymbol\varepsilon_{\rm \mathbf{pot}}(\nu)$ & $\boldsymbol\varepsilon_{\rm \mathbf{I}}(\nu)$ & $\boldsymbol\varepsilon_{\rm \mathbf{II}}(\nu)$ &
$\boldsymbol\varepsilon_{\rm \mathbf{III}}(\nu)$ & $\boldsymbol\varepsilon_{\rm \mathbf{IV}}(\nu)$\\
 & (\ref{eq:a28}),(\ref{eq:g18}),(\ref{eq:g25}) & (\ref{eq:g19}), (\ref{eq:g21}), (\ref{eq:g28}) & (\ref{eq:g20a})-(\ref{eq:g20c}), (\ref{eq:g29})-(\ref{eq:g31}) & (\ref{eq:g21}) &  (\ref{eq:g22}) \\
\hline\hline
0 & 1/2 & 3/4 & 17/8 & 25/8 & 33/4\\  \hline 
1/2 & 5/16 & 25/32 & 27/8 & 231/32 & 837/32\\  \hline 
1 & 11/48 & 77/96 & 895/192 & 837/64 & 965/16\\  \hline 
3/2 & 93/512 & 837/1024 & 6115/1024 & 10615/512 & 59475/512\\  \hline 
1.7942 & 11410011/70398355 & 18307008/22197935 & 30171115/4469191 & 67286322/2579651 & 127537747/790294\\  \hline 
2 & 193/1280 & 2123/2560 & 37371/5120 & 30927/1024 & 101997/512\\  \hline 
5/2 & 793/6144 & 10309/12288 & 53081/6144 & 169995/4096 & 1290317/4096\\  \hline 
3 & 1619/14336 & 24285/28672 & 572891/57344 & 447661/8192 & 480429/1024\\  \hline 
7/2 & 26333/262144 & 447661/524288 & 5950665/524288 & 9128151/131072 & 87461775/131072\\  \hline 
3.758 & 0.095054241197409 & 0.857009038635841 & 12.054442498170000 & 78.107662353796800 & 788.549642958726000\\  \hline 
5.8740 & 0.066169154715619 & 0.876608961672518 & 17.889125526774400 & 166.345557346382000 & 2381.422474398040000\\  \hline 
8.1307 & 0.050122450469590 & 0.890244891770568 & 24.190554653983300 & 297.684754381852000 & 5602.879738648750000\\  \hline 
10.5044 & 0.040003065289686 & 0.900420995992488 & 30.876199420891700 & 477.632653223513000 & 11254.456630209200000\\  \hline 
20.9538 & 0.021304380884411 & 0.924772043678155 & 60.694395455665500 & 1784.286219086780000 & 79307.664262054600000\\  \hline 
35.7017 & 0.012897315290706 & 0.940258135564422 & 103.307086974352000 & 5068.138932269630000 & 374712.658866740000000\\  \hline 
51.9196 & 0.009013462373609 & 0.949470915666053 & 150.512127748210000 & 10638.361918298200000 & 1131556.682871470000000\\  \hline 
69.3583 & 0.006816023482166 & 0.955719638189442 & 201.500567239779000 & 18927.246550543200000 & 2673278.362712070000000\\  \hline 
87.8627 & 0.005418519284556 & 0.960299247611835 & 255.774202228416000 & 30335.614989561300000 & 5407202.467179070000000\\  \hline 
107.3506 & 0.004458053255881 & 0.963836463562330 & 313.064859152140000 & 45264.878166797100000 & 9832433.121974210000000\\  \hline 
127.8215 & 0.003759305734937 & 0.966679154371855 & 373.353836428638000 & 64172.139638711000000 & 16566692.535880100000000\\  \hline 
149.3660 & 0.003227635139863 & 0.969039351893033 & 436.897193397184000 & 87644.083923089700000 & 26402600.563888400000000\\  \hline 
172.1617 & 0.002807948458097 & 0.971054276427006 & 504.212697547382000 & 116472.927171035000000 & 40397270.423721200000000\\  \hline 
222.5187 & 0.002182090441513 & 0.974384926745674 & 653.142502031965000 & 194728.948425671000000 & 87151050.944645500000000\\  \hline 
280.9497 & 0.001734286200261 & 0.977095438492802 & 826.236162746333000 & 310705.515336457000000 & 175365719.699678000000000\\  \hline 
348.2858 & 0.001402947397976 & 0.979356411386679 & 1025.982804702670000 & 477935.192145890000000 & 334116453.208982000000000\\  \hline 
422.8567 & 0.001158190262783 & 0.981230916467000 & 1247.450242037150000 & 705138.107234676000000 & 598116579.333488000000000\\  \hline 
501.4654 & 0.000978443179650 & 0.982771788905577 & 1481.148639701370000 & 992496.232749923000000 & 997902240.046021000000000\\  \hline 
\end{tabular}
\end{flushright}
}
\end{landscape}


\newpage
  \setcounter{section}{8}
  \setcounter{equation}{0}

  \begin{center}
  {\textbf{\Large Appendix \Alph{section}:}}\\[2ex]
  \emph{\textbf{\Large Alternative Parametrization of Trial Amplitude $\tPhi(r)$}}
  \end{center}
  \myappendix{Alternative Parametrization of Trial Amplitude $\tPhi(r)$}
  \vspace{2ex}

  The normalization condition (\ref{eq:vi.2}) for the trial amplitude $\tPhi_b(r)$
  (\ref{eq:vi.1}) reduces the parameter space $\mathbb{R}^2$ of the two real-valued ansatz
  parameters $b_0, b_1$ to some one-dimensional compact subspace which is topologically
  equivalent to the circle $\mathcal{S}^1$. This entails ultimately that the energy
  function $\ETT(\nu, p_0,p_1)$ (\ref{eq:vi.26}) must be extremalized on the half-cylinder given
  by the constraint (\ref{eq:vi.2}) together with the condition $-\frac{1}{2} < \nu <
  \infty$. However, it may be instructive to use as the underlying parameter space also a
  half-plane $\left\{ -\frac{1}{2} < \nu < \infty, -\infty < p < +\infty \right\}$. In
  this sense, one parametrizes the generalized trial amplitude $\tPhib(r)$ (\ref{eq:vi.1})
  in the following alternative way
\begin{subequations}
\begin{align}
\label{eq:H.1a}
\tPhib(r) &= \Phi_*\,r^\nu \left( 1 + br \right) \e^{-\beta r} \\
\label{eq:H.1b}
\Phi_*^2(r) &= \frac{2}{\pi}\ \frac{\left( 2\beta \right)^{2\nu + 2}}{\Gamma(2\nu + 2)} \cdot \frac{1}{w^2(\nu, p)} \;,
\end{align}
\end{subequations}
where the denominator $w(\nu, p)$ in (\ref{eq:H.1b}) is given by
\begin{equation}
\label{eq:H.2}
w(\nu, p) = \sqrt{1 + 4(\nu + 1) \cdot p + 2(\nu+1)(2\nu + 3) \cdot p^2}
\end{equation}
and the variational parameter $p$ is related to the original ansatz parameter~$b$ through
\begin{equation}
\label{eq:H.3}
p \doteqdot \frac{b}{2\beta} \;.
\end{equation}
Substituting this back into the original ansatz (\ref{eq:H.1a}) for $\tPhib(r)$, one arrives at the following form of the trial amplitude
\begin{subequations}
\begin{align}
\label{eq:H.4a}
\tPhib(r) &= \sqrt{\frac{2}{\pi}}\ \frac{2\beta}{\sqrt{\Gamma(2\nu + 2)}} \cdot \frac{\tPhinuw(y)}{w(\nu, p)} \\
\label{eq:H.4b}
\tPhinuw(y) &\doteqdot y^\nu \left( 1 + p y \right) \e^{-\frac{y}{2}} \;.
\end{align}
\end{subequations}
Here, the original normalization condition (\ref{eq:iii.38}) on $\tPhib(r)$ transcribes to the present reduced amplitude $\tPhinuw(y)$ (\ref{eq:H.4b}) as
\begin{equation}
\label{eq:H.5}
\frac{1}{\Gamma(2\nu + 2) \cdot w(\nu, p)}\ \int\limits_0^\infty dy\,y\;\tPhinuw(y) = 1 \;.
\end{equation}
\vskip 2em

\begin{center}
  \large{\textit{Kinetic Energy}}
\end{center}

After the reparametrization of the trial amplitude is fixed now, one can next turn to the kinetic energy $\Ekin$ (\ref{eq:iii.42})--(\ref{eq:iii.44}) in terms of the new parameters $p, w$. Here, the radial kinetic energy $\rEkin$ (\ref{eq:iii.43}) emerges in terms of the new parametrization as
\begin{equation}
\label{eq:H.6}
\rEKIN = \frac{e^2}{2a_B} \left( 2\beta a_B \right)^2 \cdot \frac{\reKIN(\nu, p)}{w^2(\nu, p)}
\end{equation}
with the radial kinetic function $\reKIN(\nu, p)$ being defined by
\begin{equation}
\label{eq:H.7}
\reKIN(\nu, p) = \frac{1}{\Gamma(2\nu + 2)}\ \int\limits_0^\infty dy\,y\;\left[ \frac{d\,\tPhinuw(y)}{dy} \right]^2 \;.
\end{equation}
The explicit calculation of this integral by use of the dimensionless trial amplitude $\tPhinuw(y)$ (\ref{eq:H.4b}) lets then emerge this radial kinetic function in the following form
\begin{equation}
\label{eq:H.8}
\reKIN(\nu, p) = \frac{1}{4(2\nu + 1)} + \frac{\nu}{2\nu + 1} \cdot p + \frac{1}{2} (\nu + 1) \cdot p^2 \;,
\end{equation}
i.\,e. a quadratic function with respect to the additional parameter $p$ (\ref{eq:H.3}).

Of course, the kinetic energy is an observable quantity and therefore cannot depend upon
the special parametrization of the wave function $\tPhib(r)$. This means that both
parametrizations (\ref{eq:vi.1})-(\ref{eq:vi.3b}) and (\ref{eq:H.1a})--(\ref{eq:H.5}) must generate
the same kinetic energy, i.\,e. both results (\ref{eq:vi.5}) and (\ref{eq:H.6}) for the
radial type of energy must be identical
\begin{equation}
\label{eq:H.9}
\rEkin(\beta; \nu, p_0, p_1) \mustbe \rEKIN(\beta; \nu, p) \;.
\end{equation}
But this requirement says now that there must exist a diffeomorphism which relates the
half-cylinder, being defined by the constraint (\ref{eq:vi.2}) together with $-\frac{1}{2}
< \nu < \infty$, to the half-plane $\left\{ -\infty < p < \infty; -\frac{1}{2} < \nu <
  \infty \right\}$; and this then must guarantee the identification (\ref{eq:H.9}). The
desired map is specified by the following simple transformation
\begin{subequations}
\begin{align}
\label{eq:H.10a}
w &= \frac{1}{p_0} \\
\label{eq:H.10b}
p &= \frac{p_1}{p_0} \;,
\end{align}
\end{subequations}
since by this relationship both kinetic functions $\rekin(\nu, p_0, p_1)$
(\ref{eq:vi.6})--(\ref{eq:vi.7}) and $\reKIN(\nu, p)$ (\ref{eq:H.7})-(\ref{eq:H.8}) become linked through
\begin{equation}
\label{eq:H.11}
\rekin(\nu, p_0, p_1)\ \Rightarrow\ \frac{\reKIN(\nu, p)}{w^2(\nu, p)}
\end{equation}
so that the required numerical identity (\ref{eq:H.9}) can actually be true.

Similar arguments may also be used in order to set up the longitudinal kinetic energy $\thEKIN(\beta; \nu, p)$ (\ref{eq:iii.44})
\begin{align}
\label{eq:H.12}
\thEKIN(\beta; \nu, p) &\doteqdot \frac{\hbar^2}{2M}\, \lP^2 \cdot \frac{\pi}{2} \int\limits_0^\infty \frac{dr}{r}\ \tPhib^2(r) \\
&\doteqdot \frac{e^2}{2a_B}(2a_B \beta)^2\,\lP^2 \cdot \frac{\theKIN(\nu, p)}{w(\nu,p)^2}\ , \nonumber
\end{align}
with the longitudinal kinetic function being found as
\begin{equation}
\label{eq:H.13}
\theKIN(\nu, p) = \frac{1}{2\nu(2\nu + 1)} + \frac{2}{2\nu + 1} \cdot p + p^2 \;.
\end{equation}
Of course, this kinetic function arises again also through applying the parameter transformation (\ref{eq:H.10a})--(\ref{eq:H.10b}) to the former kinetic function $\thekin(\nu, p_0, p_1)$ (\ref{eq:vi.9})
\begin{equation}
\label{eq:H.14}
\thekin(\nu, p_0, p_1)\ \Rightarrow\ \frac{\theKIN(\nu, p)}{w^2(\nu, p)}  \;.
\end{equation}
Such a transformation must then also hold for the total kinetic function $\eKIN(\nu, p)$
\begin{align}
\label{eq:H.15}
\eKIN(\nu, p) &= \reKIN(\nu, p) + \theKIN(\nu, p) \cdot \lP^2 \\
&= \frac{\nu + 2\lP^2}{4\nu(2\nu + 1)} + \frac{\nu + 2\lP^2}{(2\nu + 1)} \cdot p + \left( \frac{\nu+1}{2} + \lP^2 \right) \cdot p^2 \;, \nonumber
\end{align}
i.\,e. the action of the parameter transformation (\ref{eq:H.10a})--(\ref{eq:H.10b}) on
the total kinetic function $\ekin(\nu, p_0, p_1)$ (\ref{eq:vi.11}) reads
\begin{equation}
\label{eq:H.16}
\ekin(\nu, p_0, p_1)\ \Rightarrow\ \frac{\eKIN(\nu, p)}{w^2(\nu, p)} \;.
\end{equation}
\vskip 2em

\begin{center}
  \large{\textit{Interaction Energy}}
\end{center}

Clearly, the intrinsic consistency of the reparametrization demands that an analogous
relationship must apply also to the potential energy $\ERee\ \left( = \tMMee \mathrm{c^2}
\right)$ (\ref{eq:vi.12}). The counterpart of this object ($\Ewee$, say) is built up by
the interaction potential $\peAw(r)$ as the counterpart of $\peAo(r)$ in the same way as
for the first parametrization, cf.~(\ref{eq:iii.45}), i.\,e.
\begin{equation}
\label{eq:H.17}
\Ewee = -\frac{\hbar\mathrm{c}}{\alpha_s} \int\limits_0^\infty dr\,r^2 \cdot \left( \frac{d\,\peAw(r)}{dr} \right)^2 \;.
\end{equation}
Here, the reparametrized potential $\peAw(r)$ is the counterpart of the former $\peAo(r)$ and is also the solution of the Poisson equation (\ref{eq:iii.36}), albeit with regard to the reparametrized source-term (\ref{eq:H.4a})--(\ref{eq:H.4b})
\begin{equation}
\label{eq:H.18}
\left( \frac{d^2}{dr^2} + \frac{2}{r}\,\frac{d}{dr} \right)\,\peAw(r) = -\frac{\pi}{2}\;\alpha_s\;\frac{\tPhib^2(r)}{r} \;,
\end{equation}
or rewritten in dimensionless form ($\peAw(r) \doteqdot 2 \beta \alpha_s \cdot \tanuw(y)$)
\begin{equation}
\label{eq:H.19}
\Delta_y\;\tanuw(y) = -\frac{1}{\Gamma(2\nu + 2) \cdot w^2(\nu, p)} \cdot \frac{\tPhinuw^2(y)}{y}
\end{equation}
where the dimensionless form $\tPhinuw(y)$ is already displayed by equation (\ref{eq:H.4b}). This
special form of the Poisson equation suggests again to compose its solution $\tanuw(y)$
from the more elementary potentials $a_\nu(y)$ (G.13a) as the solutions of the simplified
Poisson equation (G.5), i.\,e. we put again
\begin{equation}
\label{eq:H.20}
\tanuw(y) = - \frac{a_\nu(y) + 2p \cdot a_{\nu+\frac{1}{2}}(y) + p^2 \cdot a_{\nu+1}(y)}{\Gamma(2\nu + 2)\cdot w^2(\nu, p)}
\end{equation}
where the elementary potentials $a_\nu(y)$ obey the simplified Poisson equation (G.5) and therefore are given by equation (G.11) for integer $2\nu$, and by equation (G.13a) for arbitrary $\nu$ (but $\nu > -\frac{1}{2}$). Substituting this solution $\tanuw(y)$ (\ref{eq:H.20}) back into the electrostatic field energy $\Ewee$ (\ref{eq:H.17}) (or equivalently into its mass equivalent) lets then appear the latter objects in the following form
\begin{equation}
\label{eq:H.21}
\Ewee = \tMMee_w \mathrm{c^2} = -\frac{e^2}{a_B}(2\beta a_B) \cdot \frac{\ePOT(\nu, p)}{w^4(\nu, p)} \;,
\end{equation}
with the alternative potential function $\ePOT(\nu, p)$ being found as
\begin{equation}
\label{eq:H.22}
\ePOT(\nu, p) = \epot(\nu) + 4p \cdot \varepsilon_\mathrm{I}(\nu) + 4p^2 \cdot \varepsilon_\mathrm{II}(\nu) + 4p^3 \cdot \varepsilon_\mathrm{III}(\nu) + p^4 \cdot \varepsilon_\mathrm{IV}(\nu) \;.
\end{equation}

The crucial point is here that this alternative form of the potential function just
guarantees the numerical identity of both electrostatic field energies $\ERee$ (G.15) and
$\Ewee$ (\ref{eq:H.21}):
\begin{subequations}
\begin{align}
\label{eq:H.23a}
\ERee &\equiv \Ewee \\
\label{eq:H.23b}
\left( \tMMee \mathrm{c^2} \right. &\equiv \left. \tMMee_w \mathrm{c^2} \right) \;.
\end{align}
\end{subequations}
Indeed, applying the parameter transformation (\ref{eq:H.10a})--(\ref{eq:H.10b}) to the
original potential function $\epot(\nu, p_0, p_1)$ (\ref{eq:vi.13}) yields the transition
\begin{equation}
\label{eq:H.24}
\epot(\nu, p_0, p_1)\ \Rightarrow\ \frac{\ePOT(\nu, p)}{w^4(\nu, p)}
\end{equation}
and this is just what validates the claimed numerical identity (\ref{eq:H.23a}).

Finally, the value of the total energy functional $\EEeiv$ on the selected trial
configurations appears now again in the form of the following energy function
\begin{align}
\label{eq:H.25}
\ET(\beta, \nu, p) &= 2\EKIN(\beta, \nu, p) + \Ewee(\beta, \nu, p) \\
&= \frac{e^2}{a_B} \left\{ (2\beta a_B)^2 \cdot \frac{\eKIN(\nu, p)}{w^2(\nu, p)} - (2\beta a_B) \cdot \frac{\ePOT(\nu, p)}{w^4(\nu, p)} \right\} \nonumber
\end{align}
with the kinetic function $\eKIN(\nu, p)$ being specified by equation (\ref{eq:H.15}) and the potential function $\ePOT(\nu, p)$ by (\ref{eq:H.22}).
\vskip 2em

\begin{center}
  \large{\textit{Extremal Configurations}}
\end{center}

In order to determine the extremal values of the energy functional $\EEeiv$ on the
selected trial functions, one has again to determine the local minima of the total energy
$\ET(\beta, \nu, p)$ (\ref{eq:H.25}) for any~$\lP$. This is equivalent to looking for the
minima of the reduced energy function $\ETT(\nu, p)$
\begin{subequations}
\begin{align}
\label{eq:H.26a}
\ETT(\nu, p) &= - \frac{e^2}{4a_B} \cdot S_\mathcal{P}(\nu, p) \\
\label{eq:H.26b}
S_\mathcal{P}(\nu, p) &= \frac{\ePOT^2(\nu, p)}{\eKIN(\nu, p) \cdot w^6(\nu, p)} \;.
\end{align}
\end{subequations}
The local maxima of the spectral function $S_\mathcal{P}$ (\ref{eq:H.26b}) determine now
the binding energies as the local minima of the total energy $\ETT(\nu, p)$
(\ref{eq:H.26a}).

For a brief demonstration of the present approximation method, one may consider the
groundstate ($\lP = 0\ \Rightarrow\ n_\mathcal{P} = 1$). Obviously, the roughest
approximation is specified by putting both variational parameters $\nu$ and $p$ to
zero. For this situation, the kinetic function $\eKIN$ (\ref{eq:H.15}) becomes simplified
to
\begin{equation}
\label{eq:H.27}
\eKIN(0, 0) = \frac{1}{4} \;.
\end{equation}
Furthermore, the potential function $\ePOT$ (\ref{eq:H.22}) becomes reduced to
\begin{equation}
\label{eq:H.28}
\ePOT(0, 0) = \epot(0) = \frac{1}{2} \;,
\end{equation}
and the function $w(\nu, p)$ (\ref{eq:H.2}) collapses to unity
\begin{equation}
\label{eq:H.29}
w(0, 0) = 1 \;.
\end{equation}
Consequently, the spectral function $S_\mathcal{P}(\nu, p)$ (\ref{eq:H.26b}) adopts the unity
\begin{equation}
\label{eq:H.30}
S_\mathcal{P}(0, 0) = 1 \;,
\end{equation}
and therefore the positronium groundstate energy $\ETT(\nu,p)$ (\ref{eq:H.26a}) is found
in this roughest approximation as
\begin{equation}
\label{eq:H.31}
\ETT(0, 0) = -\frac{e^2}{4a_B} \simeq -6{,}8029\ldots\;\text{[eV]} \;.
\end{equation}

This is \emph{exactly} the value predicted by the conventional theory, cf. (\ref{eq:i.4});
but within the present RST framework it appears as an \emph{approximation}! Therefore one
expects that in the next higher approximation the corresponding RST prediction fore the
groundstate must yield some lower energy value; and indeed, putting $p$ to zero and thus
admitting only two variational parameters (i.\,e. $\beta, \nu$) yields a groundstate
prediction of $-7{,}23055\ldots\;\text{[eV]}$ in place of the conventional prediction
(\ref{eq:H.31}), see first line of \textbf{table~1} on p.~\pageref{table1}.

Alternatively, one could have taken as the next higher approximation also $\nu = 0$ and
leaving $p$ as the second variational parameter. In this case, the kinetic function
$\eKIN$ (\ref{eq:H.15}) is simplified for the groundstate ($\lP = 0$) to
\begin{equation}
\label{eq:H.32}
\eKIN(0, p) = \frac{1}{4} + \frac{1}{2} p^2
\end{equation}
and, furthermore, the potential function $\ePOT$ (\ref{eq:H.22}) becomes
\begin{equation}
\label{eq:H.33}
\ePOT(0, p) = \epot(0) + 4p \cdot \varepsilon_\mathrm{I}(0) + 4p^2 \cdot \varepsilon_\mathrm{II}(0) + 4p^3 \cdot \varepsilon_\mathrm{III}(0) + p^4 \cdot \varepsilon_\mathrm{IV}(0) \;,
\end{equation}
i.\,e. by use of the table in \textbf{App. G}
\begin{equation}
\label{eq:H.34}
\ePOT(0, p) = \frac{1}{2} + 3p + \frac{17}{2}p^2 + \frac{25}{2}p^3 + \frac{33}{4}p^4 \;.
\end{equation}
This together with $w(0, p)$ (\ref{eq:H.2}) builds up the spectral function $S_\mathcal{P}$ (\ref{eq:H.26b}) to the following form
\begin{equation}
\label{eq:H.35}
S_\mathcal{P}(0,p) \doteqdot S_\mathcal{P}^{[0]}(p) = \frac{\left\{ 1 + 6p + 17p^2 + 25p^3 + \frac{33}{2}p^4 \right\}^2}{\left( 1 + 2p^2 \right) \cdot \left[ 1 + 4p + 6p^2 \right]^3} \;.
\end{equation}

The present alternative parametrization serves as a technical consistency check for our
numerical program of approximately determining the positronium binding energies
$\EE_\mathcal{P}^{[n]} = \ET(\nu_*^{[n]},p_n)$; and it is very satisfying to see the
positronium energies $E_{1a}$ up to $E_{1d}$ (\ref{eq:vi.40a})-(\ref{eq:vi.40b}) of the
compact parametrization (\ref{eq:vi.36a})-(\ref{eq:vi.36b}) emerging also in the
alternative open parametrization (\ref{eq:H.10a})--(\ref{eq:H.10b}), see the \textbf{Fig.}
\textbf{H.I} below.

Concerning now the excited states, one has to relax the adhoc postulate~$\nu=0$ of the
simplified groundstate treatment and thus one has to admit any real value of~$\nu\,
(>-\frac{1}{2})$. This means that one has to look for the maximum of the spectral
function~$S_\wp(\nu,p)$ (\ref{eq:H.26b}) as a function of the two variables~$\nu$
and~$p$. This may be done by means of an appropriate computer program, for the first
excited state~$(\nP=2\Leftrightarrow\lP=1)$ one finds the
maximum~$S_\wp^{[2]}(*)=0,229683\ldots$ of the spectral function~$S_\wp(\nu,p)$
(\ref{eq:H.26b}) at the equilibrium point
\begin{subequations}
  \begin{align}
    \label{eq:H.40a}
    \nu_*^{[2]} &= 1,258290\ldots\\*
    \label{eq:H.40b}
    p_*^{[2]} &= 0,677982\ldots
  \end{align}
\end{subequations}
Thus, the energy~$E_\wp^{[2]}$ of the first excited state is obtained in the present
(spherically symmetric) approximation step as
\begin{equation}
  \label{eq:H.41}
  \begin{split}
    E_\wp^{[2]} =-\frac{e^2}{4\aB}\cdot S_\wp^{[2]}(x) &\simeq -6,8029\cdot 0,2296\ [eV]\\*
    &\simeq -1,56194\ [eV]
  \end{split}
\end{equation}
Comparing this to the corresponding first-order approximation (of the spherically
symmetric kind)~$\nrf{E}{T}(\nu_*^{[2]})$ (see second line of table~1 on p.(?))
\begin{equation}
  \label{eq:H.42}
  \nrf{E}{T}(\nu_*^{[2]}) = -1,55087\ [eV]
\end{equation}
one observes an improvement (of the deviation from the conventional
result~$\ru{E}{conv}^{(2)}$ (\ref{eq:i.4})) from 8,8\% (last column of table~1) to
presently 8,2\%:
\begin{equation}
  \label{eq:43}
  \frac {\ru{E}{conv}^{[2]}-\mathbb{E}_p^{[2]} }{\ru{E}{conv}^{[2]}} =
  \frac{1,70058-1,56194}{1,70058}\% = 8,2\%
\end{equation}
This is a relatively small improvement in view of those improvements due to the regard of
anisotropy which yielded a corresponding improvement from 8,8\% to 3,1\% (see first line
of table~2 on p.(x)). Similar magnitudes of improvement are also found for the higher
excited states (see table~3 on p.(x)). Thus the conclusion is that the first-order
anisotropy corrections (\textbf{Sect.V}) are much more important than the second-order
corrections of the spherically symmetric kind.

\begin{center}
\epsfig{file=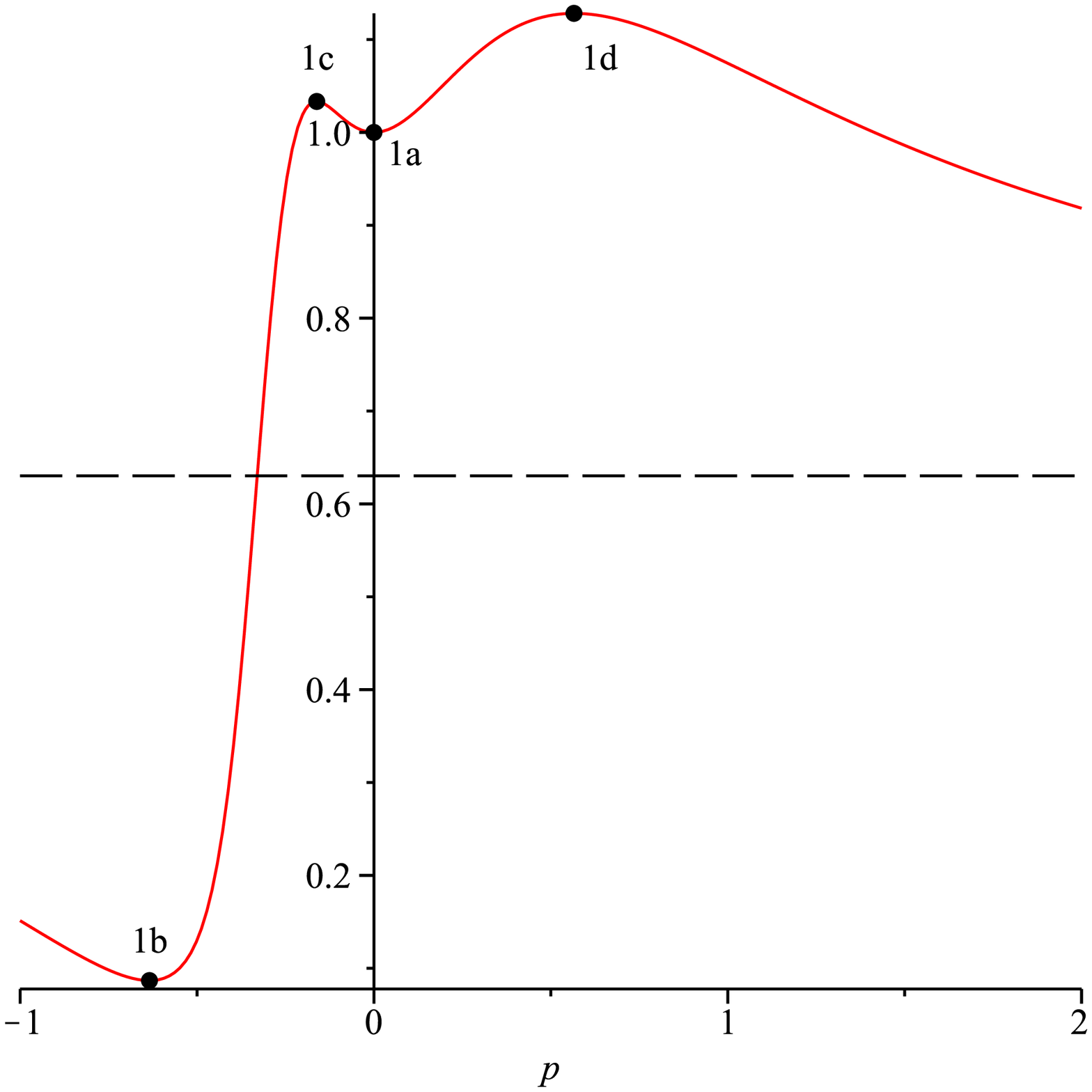,height=15cm}
\end{center}
{\textbf{Fig.~H.I}\hspace{5mm} \emph{\large\textbf{Spectral Function
      \boldmath$S^{[0]}_{\wp}(p)$ }}  }
\myfigure{{Fig.~H.I: Spectral Function \boldmath$S^{[0]}_{\wp}(p)$ }}
\indent

The spectral function~$S^{[0]}_{\wp}(p)$ (\ref{eq:H.35}) possesses two relative maxima
at~$p_{\rm 1c}=-0,161651\ldots$ and~$p_{\rm 1d}=0,565338\ldots$ with the corresponding
maximal values~$S^{[0]}_{\wp}|_{\rm 1c}=1,033319\ldots$ and $S^{[0]}_{\wp}|_{\rm
  1d}=1,128194\ldots$ which by means of (\ref{eq:H.26a}) yields the two energy
minima~$E_{\rm 1c}$ (\ref{eq:vi.40a}) and~$E_{\rm 1d}$ (\ref{eq:vi.40b}). The two relative
\emph{minima} 1a and 1b of~$S^{[0]}_{\wp}(p)$ are found at~$p_{\rm 1a}=0$ and~$p_{\rm
  1b}=-0,634469\ldots$ which yield the two energy maxima~$E_{\rm 1a}$ (\ref{eq:vi.39a})
and~$E_{\rm 1b}$ (\ref{eq:vi.39b}) of the angular parameterization in
\textbf{Sect.VI.3}. Both asymptotic values of~$S^{[0]}_{\wp}(p)$ for
infinity~$(p\to\pm\infty)$ do agree
\begin{equation}
  \label{eq:h36}
  \lim_{p\to\pm\infty} S^{[0]}_{\wp}(p) = \frac{121}{192}\simeq 0,6302\ldots
\end{equation}
so that the real line~$(-\infty<p<+\infty)$ of the present parameterization may be
compactified to the circle~$S^1\ (0\le\alpha\le2\pi)$ of the first parameterization in
\textbf{Sect.(VI.3)}.



\end{document}